\documentclass[aps,prc,amsfonts,preprintnumbers,superscriptaddress,showpacs,nofootinbib]{revtex4}
\pdfoutput=1
\usepackage{graphics}
\usepackage{amsmath}
\usepackage{amssymb}
\usepackage{psfrag}
\usepackage{epsfig}
\usepackage{epsf}
\usepackage{float}
\usepackage{slashed}
\allowdisplaybreaks

\newcommand{\fet}[1]{\mbox{\boldmath $#1$}}
\newcommand{\sfet}[1]{\mbox{\boldmath \scriptsize $#1$}}

\newcommand{\beq}{\begin{equation}}
\newcommand{\eeq}{\end{equation}}
\newcommand{\beqa}{\begin{eqnarray}}
\newcommand{\eeqa}{\end{eqnarray}}
\newcommand{\nn}{\nonumber \\ }

\newcommand{\pvec}[1]{\vec{#1}\mkern2mu\vphantom{#1}}

\numberwithin{equation}{section}
\renewcommand\theequation{\arabic{equation}}

\begin{document}

\title{Nuclear axial current operators to fourth order in chiral effective field theory}

\author{H.~Krebs}
\email[]{Email: hermann.krebs@rub.de}
\affiliation{Institut f\"ur Theoretische Physik II, Ruhr-Universit\"at Bochum,
  D-44780 Bochum, Germany}
\author{E.~Epelbaum}
\email[]{Email: evgeny.epelbaum@rub.de}
\affiliation{Institut f\"ur Theoretische Physik II, Ruhr-Universit\"at Bochum,
  D-44780 Bochum, Germany}
\affiliation{Kavli Institute for Theoretical Physics, University of
  California, Santa Barbara, CA 93016, USA}
\author{U.-G.~Mei{\ss}ner}
\email[]{Email: meissner@hiskp.uni-bonn.de}
\affiliation{Helmholtz-Institut~f\"{u}r~Strahlen-~und~Kernphysik~and~Bethe~Center~for~Theoretical~Physics,
~Universit\"{a}t~Bonn,~D-53115~Bonn,~Germany}
\affiliation{Institut~f\"{u}r~Kernphysik,~Institute~for~Advanced~Simulation,
and J\"{u}lich~Center~for~Hadron~Physics, Forschungszentrum~J\"{u}lich,~D-52425~J\"{u}lich,~Germany}
\affiliation{JARA~-~High~Performance~Computing,~Forschungszentrum~J\"{u}lich,~D-52425~J\"{u}lich,~Germany}
\date{\today}

\begin{abstract}
We present the \emph{complete} derivation of the nuclear axial 
charge and current operators as well as the pseudoscalar operators to fourth order in the chiral expansion relative to
the dominant one-body contribution using the method of unitary
transformation. We demonstrate that the unitary ambiguity in the
resulting operators can be eliminated by the requirement of renormalizability
and by matching of the pion-pole contributions to the
nuclear forces. We give expressions for the renormalized single-, two-
and three-nucleon contributions to the charge and current operators
and pseudoscalar operators
including the relevant relativistic corrections. We also verify explicitly
the validity of the continuity equation. 
\end{abstract}

\pacs{13.75.Cs,21.30.-x}

\maketitle

\vspace{-0.2cm}

\section{Introduction}
\def\theequation{\arabic{section}.\arabic{equation}}
\label{sec:intro}

The past quarter century has witnessed enormous progress towards
formulating low-energy nuclear physics within a systematically improvable and
well-founded framework of chiral effective field theory (EFT) as initiated
by Weinberg in the early 1990ties~\cite{Weinberg:1990rz}. It relies on the most general 
effective Lagrangian which incorporates the chiral symmetry of QCD and the various
patterns of its breaking. For the two-flavor case of the up- and 
down-quarks, the effective Lagrangian is written in terms of pions,
which are identified with (pseudo)-Goldstone bosons of the broken chiral
symmetry $\mbox{SU}(2)_L \times \mbox{SU} (2)_R \to \mbox{SU} (2)_V$, and the 
relevant matter fields such as the nucleons, complemented by various external sources that
parametrize e.g. the explicit chiral symmetry breaking.
The Goldstone-boson nature of the pions enables their perturbative 
treatment at low energy within the framework of chiral
perturbation theory. When processes involving two or more nucleons are
considered, chiral perturbation theory allows one to derive
interaction potentials and exchange current operators to be used in
the framework of the $A$-nucleon Schr\"odinger equation. 
The formulation thus reduces to the conventional quantum mechanical
many-body problem which can be efficiently dealt with numerically using a
variety of {\em ab initio} approaches developed over the past decades. In
addition to being firmly rooted in the symmetries of QCD, the important
advantages of nuclear chiral EFT in comparison with more
phenomenological approaches comprise its systematic improvability, 
the consistency between two- and many-body forces and exchange current
operators, a unified treatment of processes involving pions  and the intimate relation
between nucleon and nuclear structure~\cite{Meissner:2003dx}.

The chiral EFT approach outlined above and based on pions and nucleons as the
only active degrees of freedom has been extensively exploited to
derive nuclear forces and to analyze few- and many-nucleon systems,
see Refs.~\cite{Epelbaum:2008ga,Machleidt:2011zz} for review
articles. Recently, the description of the two-nucleon (2N) force has been
pushed to fifth order in the chiral expansion \cite{Epelbaum:2014sza,Entem:2014msa}, 
i.e.~the simultaneous expansion in powers of the nucleon three-momenta and pion masses, and
even most of the sixth-order terms have been derived
\cite{Entem:2015xwa}.  Three- (3N), four- (4N) and more-nucleon forces start
contributing at third, fourth and sixth orders,
respectively. They have been worked out completely up to fourth order in
the chiral expansion
\cite{vanKolck:1994yi,Epelbaum:2002vt,Epelbaum:2006eu,
Ishikawa:2007zz,Bernard:2007sp,Bernard:2011zr}, see also
Refs.~\cite{Krebs:2012yv,Krebs:2013kha,Epelbaum:2014sea,Girlanda:2011fh}
for the derivation of some of the fifth-order terms in the
three-nucleon force. A systematic investigation of the role of the
many-body forces is an important frontier in low-energy
nuclear physics \cite{Hammer:2012id}. With all these developments, coupled with on-going
efforts towards a reliable quantification of various sources of theoretical
uncertainties
\cite{Ekstrom:2014dxa,Furnstahl:2014xsa,Epelbaum:2014efa,Binder:2015mbz,Furnstahl:2015rha,Perez:2015bqa},
nuclear chiral EFT is now entering the precision era~\cite{Epelbaum:2015pfa}. 

Few- and many-nucleon reactions with external electroweak and pionic
probes have also been investigated in the framework of chiral EFT, see
e.g.~\cite{Epelbaum:2008ga,Marcucci:2013nsa,Bacca:2014tla} and references
therein. In particular, electromagnetic exchange current operators
were first discussed in this framework in the seminal
paper by Park et al.~\cite{Park:1995pn}, who, however only focused on
the near-threshold kinematics.  More recently, the electromagnetic currents were
re-derived for more general kinematical conditions by the JLab-Pisa
group~\cite{Pastore:2008ui,Pastore:2009is,Pastore:2011ip} using the
framework of time-ordered perturbation theory (TOPT) 
and by the Bochum-Bonn group \cite{Kolling:2009iq,Kolling:2011mt} within the method of unitary
transformation (MUT) \cite{TMO,Okubo:1954zz}, which will be described
below.  It is important to keep in
mind that nuclear forces and currents are not directly
observable. Contrary to the S-matrix, they are not uniquely defined
and can be changed by means of unitary transformations or,
equivalently, by changing the basis in the Fock space. It is,
therefore, important to maintain \emph{consistency} between nuclear
forces and current operators to ensure that all these objects
correspond to the same choice of the basis in the Fock space. Being
derived within the same theoretical approach,  
the exchange electromagnetic charge and current operators worked out
in Refs.~\cite{Kolling:2009iq,Kolling:2011mt} are, by construction,
consistent with the expressions for the nuclear forces given in
Refs.~\cite{Epelbaum:2002vt,Epelbaum:2006eu,Epelbaum:2007us, 
Bernard:2007sp,Bernard:2011zr,Krebs:2012yv,Krebs:2013kha,Epelbaum:2014sea}.   
Notice further that the unitary ambiguity associated with the freedom
in the choice of the basis in the Fock space has been shown in
\cite{Epelbaum:2006eu,Kolling:2011mt}  to be
strongly reduced by demanding renormalizability at the level of the
nuclear Hamiltonian and current operators. In particular,
exploiting this freedom in a systematic way was crucial to ensure 
cancellations between the ultraviolet divergences appearing in loop
contributions to the one-pion exchange electromagnetic current
operator and the corresponding counterterms in the effective
Lagrangian leading to finite matrix elements of the currents \cite{Kolling:2011mt}.   
Given the different choices for the unitary phases made by the
JLab-Pisa and Bochum-Bonn groups, it is not surprising that the
resulting expressions for the current operators exhibit strong
differences, see also \cite{Piarulli:2012bn} for a discussion. First
exploratory applications of (some of) the novel exchange current 
contributions derived in \cite{Kolling:2009iq,Kolling:2011mt} to the
$^2$H and $^3$He photodisintegration reactions yielded promising
results and have shown a significant sensitivity to the new terms in 
certain observables \cite{Rozpedzik:2011cx}, see also Ref.~\cite{Kolling:2012cs} for 
a study of the magnetic form factor of the deuteron. A more complete treatment using the last
generation of the chiral nucleon-nucleon potentials of Refs.~\cite{Epelbaum:2014efa,Epelbaum:2014sza} 
would require the implementation of the regularization scheme consistent
with the interactions. Work along these lines is in progress, see
Ref.~\cite{Skibinski:2016dve} for a first step in this direction. 

In this paper we focus on the iso-triplet axial-vector charge and current
operators which have been first considered in the framework of chiral
EFT in the pioneering paper by Park et al.~\cite{Park:1993jf}, see
also \cite{Park:2002yp}, who,
however, have ignored pion-pole terms and contributions from reducible-like 
diagrams involving purely nucleonic states (at least) in one of
  the time orderings. The resulting expressions have been
employed in a number of studies of few-nucleon processes and in 
nuclear structure calculations, see
\cite{Lazauskas:2009nw,Marcucci:2011jm,Marcucci:2013tda,Ekstrom:2014iya}
for recent examples. There are several reasons for a strong interest
in developing a precision theory for nuclear electroweak reactions in
the framework of chiral EFT. First, the lowest-order short-range
two-nucleon axial current operator depends on a low-energy
constant (LEC) which also contributes to the leading 3N 
force \cite{Epelbaum:2002vt,Gazit:2008ma}, the P-wave pion production operator in 2N collisions
\cite{Hanhart:2000gp,Baru:2009fm} and to pion photoproduction and radiative
capture reactions in the 2N system \cite{Lensky:2007zc,Gardestig:2005pp}. 
Thus, chiral EFT opens an intriguing possibility for ``bridging'' these very
different reactions within a unified theoretical approach \cite{Nakamura:2007vi}. 
Secondly, there is a strong interest in improving the accuracy and
reliability of theoretical predictions for nuclear reactions involving
neutrinos such as e.g.~the solar proton-proton fusion and the hep
processes which figure importantly in nuclear astrophysics. Clearly, this
requires a precise determination of the short-range part of the axial
current operator. This is the main motivation of the ongoing MuSun
experiment at PSI \cite{Andreev:2007wg} which aims at the determination of the rate
of muon capture on the deuteron with a precision of $1.5\%$. Notice
that the theoretical predictions for the doublet capture rate
$\Lambda_{1/2}$ show a significant spread, see
e.g.~\cite{Marcucci:2011tf,Adam:2011mr}. Last but not least, the
appearance of the same LEC in the leading 3N force and the short-range
part of the 2N weak current opens an exciting possibility to perform
nontrivial precision tests of chiral EFT in few-nucleon systems by
carrying out a simultaneous calculation of nucleon-deuteron scattering, 
the binding energy, radius and the precisely known half-life of $^3$H
and the muon capture rates on $^2$H and $^3$He, accompanied with a
careful analysis of the theoretical uncertainties along the lines of
Refs.~\cite{Epelbaum:2014efa,Binder:2015mbz}.  

Recently, the 2N contributions to the axial charge and current
operators at the leading one-loop order were re-derived by Baroni et
al.~\cite{Baroni:2015uza} using TOPT in the same
framework as used by the JLab-Pisa group to derive the electromagnetic
currents in Refs.~\cite{Pastore:2008ui,Pastore:2009is,Pastore:2011ip}. However, as
already pointed out, the resulting expressions for certain
contributions to the electromagnetic
currents differ from the ones of Refs.~\cite{Kolling:2009iq,Kolling:2011mt} 
worked out by our group using the MUT. Specifically, while the expressions
for the two-pion exchange terms derived by both groups agree with each
other, the one-pion exchange and contact operators are different.  
Notice further that JLab-Pisa group has not performed a complete renormalization of the one-pion
exchange current and charge operators. In \cite{Piarulli:2012bn}, the
origin of some 
of these differences was clarified. In particular, the one-loop
contributions to the short-range current operator re-derived in that
work were found to vanish in agreement with our results, while the
revised one-loop contributions to the charge operator still turned out 
to be different from those of our work \cite{Kolling:2011mt}. These
remaining differences can probably be explained by the unitary ambiguity
of the considered operators which manifests itself in a different
treatment of reducible-like diagrams in the two approaches. Notice
that while the unitary ambiguity of the nuclear Hamiltonian, charge
and current operators is systematically addressed in our approach 
by employing a broad class of unitary transformations (UTs) on the nucleonic
subspace of the Fock space compatible with the chiral order of the
calculation, the JLab-Pisa group has examined only the kind of
UTs associated with different off-the-energy-shell
extensions of the one-pion exchange potential \cite{Pastore:2011ip}, see also
Ref.~\cite{Friar:1999sj} for a related early work. In this context, it is important to emphasize 
that the inclusion of a broader class of unitary transformations was
found to be \emph{necessary} to renormalize the one-loop contributions to the
3N force \cite{Epelbaum:2006eu,Epelbaum:2007us} and to the one-pion
exchange current operator \cite{Kolling:2011mt}.   

\begin{figure}[tb]
\vskip 1 true cm
\includegraphics[width=15.5cm,keepaspectratio,angle=0,clip]{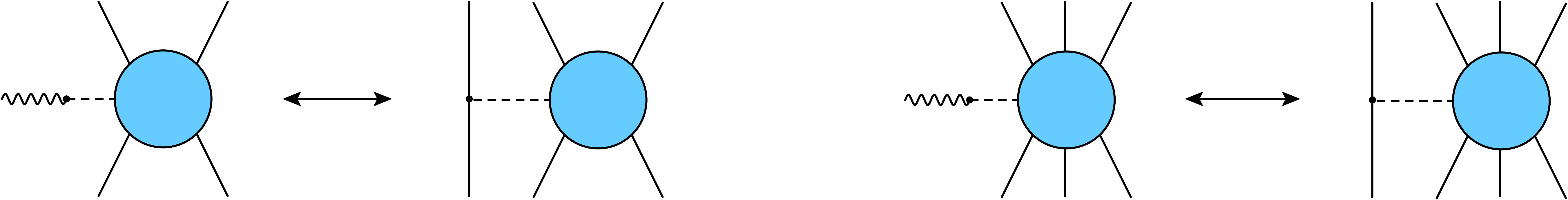}
    \caption{Matching of the pion-pole contributions to the axial
      current to the corresponding terms in the nuclear force as explained in the text. 
      Solid, dashed and wavy lines refer
      to nucleons, pions and the external axial sources,
      respectively. Solid dots denote the the lowest-order vertices
      from the effective Lagrangians $\mathcal{L}_\pi^{(2)}$ and
      $\mathcal{L}_{\pi N}^{(1)}$, 
       while shaded circles represent the irreducible parts of
      the corresponding amplitudes. 
\label{fig:factor} 
 }
\end{figure}

The above discussion provides a strong motivation to derive the
exchange axial charge and current operators using the MUT,
which was employed in the calculations 
of nuclear forces in Refs.~\cite{Epelbaum:2002vt,Epelbaum:2006eu,Epelbaum:2007us, 
Bernard:2007sp,Bernard:2011zr,Krebs:2012yv,Krebs:2013kha,Epelbaum:2014sea}. 
In this paper, we fill this gap and
consider the contributions to the nuclear axial charge and current
operators to fourth order in the chiral expansion relative to the
leading one-body terms, i.e.~to leading two-loop, one-loop and tree-level
order for the 1N, 2N and 3N terms,
respectively, which are, per construction, consistent
with the expressions for the nuclear forces in Refs.~\cite{Epelbaum:2002vt,Epelbaum:2006eu,Epelbaum:2007us, 
Bernard:2007sp,Bernard:2011zr,Krebs:2012yv,Krebs:2013kha,Epelbaum:2014sea}. We find a very high degree
of unitary ambiguity in the resulting operators which is parametrized 
by $33$ continuously varying parameters $\alpha_i^{ax}$ plus an
additional  phase $\beta_1^{ax}$. Such a richness of the possible UTs can be traced
back to the appearance of pion-pole contributions. In addition to the
renormalizability requirement, we demand 
 that the pion-pole contributions 
to the 1N, 2N and 3N 
axial current operators match the corresponding
expressions for the 2N, 3N and 4N forces, evaluated at the pion
pole, see Fig.~\ref{fig:factor}.
This choice is not only the most natural one
but is expected to be advantageous in calculations of observables
utilizing a finite cutoff. In particular, it will provide a simple way
to regularize the pion-pole contributions to the axial current
operator in exactly the same way as done in the chiral nuclear potentails. 
The resulting expressions for the exchange
axial charge and current operators turn out to be independent on the
phases of the considered UTs. Further, the same approach is applied to derive
the nuclear pseudoscalar currents to fourth order in the chiral
expansion relative to the dominant one-body contribution. This allows
us to perform a highly nontrivial check of our results by explicitly verifying
the continuity equation for all considered contributions to the charge
and current operators. 

Our paper is organized as follows. In section~\ref{sec:formalism} we
discuss in detail the formalism to derive nuclear forces and current
operators in chiral EFT using the MUT. In particular, we consider constraints imposed
by the chiral symmetry and work out the explicit form of the continuity
equations for the iso-triplet vector and axial current operators. We also
address the relation of the axial charge and current operators to the $S$-matrix and
work out constraints imposed by Poincar\'{e} invariance. Finally, we discuss in
this section the unitary ambiguity of the resulting operators and
specify our standard choice of the unitary phases fixed by the 
requirement of renormalizability and by matching of the
pion-pole contributions to the corresponding nuclear forces.
We then specify our notation in section~\ref{sec:Notation} and derive 
explicit expressions for one-, two- and three-nucleon axial charge and
current operators up to fourth  order in the chiral expansion relative
to the leading single-nucleon axial current contribution in 
sections~\ref{sec:singleN}, \ref{sec:2N} and \ref{sec:3N},
respectively. Our final results for the various
contributions to the axial charge and current operators are summarized
in section~\ref{sec:summaryCurrents}. Next, section \ref{sec:Pseudo}
is devoted to the derivation of the pseudoscalar current operator,
while our final results for the various contributions are summarized
in section \ref{sec:summaryPsCurrents}.  The
validity of the continuity equation for all derived operators is
verified in section~\ref{sec:results}. Next, in section \ref{sec:Baroni}, we
compare our expressions with those obtained by Baroni et al.~in
Refs.~\cite{Baroni:2015uza,Baroni:2016xll}. Finally, the  main results of our
paper are summarized in section~\ref{sec:summary}. Last but not least,
the appendices
\ref{BlockDiagonalizationProof}, \ref{sec:appen} and \ref{timederivOfaxialSourceAppendix}
contain a general proof of simultaneous block diagonalizability of the
generators of the Poincar\'{e} group in the Fock space and explicit
definitions of the employed additional unitary transformations.

\section{Chiral EFT for nuclear forces and currents: Foundations}
\def\theequation{\arabic{section}.\arabic{equation}}
\label{sec:formalism}

Throughout this work, we restrict ourselves to the two-flavor case of
the up- and down-quarks  and employ the heavy-baryon formulation of
chiral EFT based on pions and nucleons as the active degrees of
freedom. For the purpose of our work, the following terms in the
effective Lagrangian
describing the interactions between pions and nucleons in the presence of
external sources
are needed:
\beq
\label{lagr}
\mathcal{L}_{\rm eff} = \mathcal{L}_{\pi}^{(2)} + 
\mathcal{L}_{\pi}^{(4)} + \mathcal{L}_{\pi N}^{(1)} +
\mathcal{L}_{\pi N}^{(2)} + \mathcal{L}_{\pi N}^{(3)} +
\mathcal{L}_{NN}^{(0)} + \mathcal{L}_{\pi NN}^{(1)}\,, 
\eeq
where the superscripts refer to the number of derivatives and/or
insertions of the pion mass $M_\pi$.  Here and in what follows, we
employ the operator basis given in Refs.~\cite{Gasser:1987rb} and 
\cite{Fettes:2000gb} for the pionic and
pion-nucleon Lagrangians $\mathcal{L}_{\pi}$ and $\mathcal{L}_{\pi
  N}$, respectively. The relevant terms are also listed in
appendix A of Ref.~\cite{Baroni:2015uza} (except for vertices
involving pseudoscalar sources), where the same effective Lagrangian is
used.  For the 2N Lagrangian $\mathcal{L}_{NN}^{(0)}$, we use the standard
notation employed in the calculations of nuclear forces, see
e.g.~\cite{Epelbaum:2008ga}. It involves two terms with the corresponding LECs
denoted as $C_S$ and $C_T$. The Lagrangian $\mathcal{L}_{\pi
  NN}^{(1)}$ can also be found in \cite{Epelbaum:2008ga} and involves
just a single term with the corresponding LEC denoted by
$D$.\footnote{This LEC is usually expressed in terms of a dimensionless
constant $c_D$ \cite{Epelbaum:2002vt}.} 

In the pion and single-nucleon sectors, the effective chiral
Lagrangian can be used to calculate the scattering amplitude within
the chiral expansion in powers of $Q \in \{M_\pi / \Lambda_b, \;
p/\Lambda_b \}$, where $p$ refers to four- (three-)momenta of
external pions (nucleons) while $\Lambda_b$ denotes the breakdown
scale which is expected to be of the order of $\Lambda_b \equiv
\Lambda_\chi \sim M_\rho \sim 4 \pi F_\pi \sim 1$~GeV. As already
mentioned in the introduction, in the few-nucleon sector, the
perturbative chiral expansion is carried out for the kernel of the
corresponding dynamical equation such as e.g. the Lippmann-Schwinger
equation for the case of two nucleons or its generalizations for
systems involving three- and more nucleons. In the absence of external
sources, the kernel is identified with the (interacting part of the)
nuclear Hamiltonian while $A$-body terms describing the coupling to
the external vector and axial vector sources give rise to the
$A$-nucleon electroweak charge and current operators. 
The breakdown scale of the chiral expansion for nuclear forces was estimated
in Ref.~\cite{Epelbaum:2014efa} to be of the order of $\Lambda_b \sim
600$~MeV.  Throughout this work, we adopt the counting scheme for the
nucleon mass $m$ which is usually employed in few-nucleon calculations,
see \cite{Weinberg:1990rz,Epelbaum:2008ga,Epelbaum:2010nr} for more
details, namely $m \sim \Lambda_b^2/M_\pi $, instead of assigning $m
\sim \Lambda_b$ as done in the single-baryon sector. 

The derivation of the nuclear forces, charge and current operators
requires a subtraction of reducible contributions generated by
iterations of the dynamical equation. This can be achieved using a
variety of approaches including TOPT and
the MUT, see \cite{Epelbaum:2010nr} for a pedagogical
account of various techniques. We now briefly outline the main steps in the
derivation within the MUT, see Ref.~\cite{Kolling:2011mt} for a more
complete description.

\subsection{Interactions with time derivatives}
As the first step in the derivation of the nuclear forces and currents
in chiral EFT using the MUT, we employ the canonical formalism to
determine the Hamiltonian for pions, nucleons and external sources
from the heavy-baryon effective chiral Lagrangian. Due to the appearance of
derivative couplings in the effective Lagrangian, the derivation of the
Hamiltonian requires a careful treatment. Even at lowest orders
in the chiral expansion, one encounters 
interactions with time derivatives acting on the pion and nucleon
fields. At higher orders, there appear even second
or higher-order time derivatives. However, in order to derive the
Hamiltonian via a Legendre 
transformation in the usual way, we need a Lagrangian which includes at
most one time derivative of the fields. By using the equations of motion
(field redefinitions),  one can always
eliminate all second- and higher-order time derivatives of the pion 
fields and all time derivatives of the nucleon fields. The resulting 
modified Lagrangian represents an equally good starting point for our 
calculations as the original one since field redefinitions do not affect  on-shell
scattering amplitudes. With the modified Lagrangian, we can construct
the effective Hamiltonian in the usual way. Time derivatives of the nucleon fields
do not show up in the interaction terms in the static limit and are always
suppressed by inverse powers of the nucleon mass $m$. As an example,
consider the following
term in the effective Lagrangian\footnote{Notice that all quantities
  in the effective chiral Lagrangian are defined in the chiral
  limit. Our final results are, however, expressed in terms of
  physical masses and coupling constants. } 
\beq
\frac{1}{m^2}N^\dagger (D^2-(v\cdot D)^2) i\,v\cdot D N + {\rm h.c.}
\label{timederInt}\,,
\eeq
where $N$ denotes the nucleon field, $v$ is the nucleon four-velocity
and $D$ is the covariant derivative. We refer the reader to
Ref.~\cite{Fettes:2000gb} for more details on the notation and explicit expressions
for the pion-nucleon Lagrangian.  
The above term must be taken into account as it generates a
relativistic $1/m^2$-contribution to 
the single-nucleon axial vector current $\vec{\fet A}_{\rm 1N}$ at
order $Q$. When performing calculations within the MUT, it is
convenient to use the equation of motion for the nucleon field
\beqa
 i\,v\cdot D N = - g_A S\cdot u N + {\cal O}(Q^2),
\eeqa
or, equivalently, perform a field redefinition 
\beqa
N\to N -\frac{1}{m^2} (D^2-(v\cdot D)^2) N
\eeqa
to eliminate the term in Eq.~(\ref{timederInt}) in favor of the new
vertex 
\beq
-g_A\frac{1}{m^2}N^\dagger (D^2-(v\cdot D)^2) S\cdot u\, N + {\rm h.c.}\,,
\eeq
which does not involve the time derivative of the nucleon field but  a
series of higher-order terms which are irrelevant for our present application. 

For details concerning the treatment of the $\bar d_{22}$-vertex,
which involves a time derivative of the external axial vector source,
the reader is referred to appendix~\ref{timederivOfaxialSourceAppendix}.

\subsection{``Strong'' unitary transformations}
After the elimination of the various terms involving time derivatives as
described in the previous section,
the canonical formalism can be applied straightforwardly to derive the effective
Hamiltonian $H$ corresponding to the Lagrangian in Eq.~(\ref{lagr}),
which governs the pion-nucleon dynamics in the presence of 
external fields. In general, the chiral EFT Hamiltonian $H[a,v,s,p]$ depends on
external axial-vector, vector, scalar and pseudoscalar sources
$a_\mu$, $v_\mu$, $s$ and $p$, respectively, whose flavor
structure and transformation properties with respect to chiral
rotations will be specified in section \ref{sec:CS}.  
All these sources are  functions of space and time.

Next, one has to integrate out  the pion fields, i.e.~to decouple the
purely nucleonic subspace of the Fock space from the rest. This is
achieved via a suitably chosen unitary transformation on the Fock
space. Following Okubo~\cite{Okubo:1954zz}, the unitary operator
$U_{\rm Okubo}$
can be parametrized in terms of an operator $A = \lambda A \eta$. 
Here and in what follows, $\eta$  and $\lambda$
denote projection operators onto the purely nucleonic and the
remaining parts of the Fock space with the properties $\eta^2=\eta$, 
$\lambda^2 = \lambda$, $\eta \lambda = \lambda \eta = 0$ and $\eta +
\lambda =1$.  The requirement of decoupling of the $\eta$-subspace of
the Fock space, $\eta U_{\rm Okubo}^\dagger H_s U_{\rm Okubo} \lambda 
= \lambda U_{\rm Okubo}^\dagger H_s U_{\rm Okubo} \eta = 0$, 
leads to the nonlinear decoupling equation  for the operator $A$,
\beq
\lambda (H_s - [ A, \, H_s ] - A H_s A ) \eta = 0\,.\label{HBlockDiagonalizationRequirement}
\eeq
Here, $H_s$ is defined as 
\beq
H_s := H[0,0,s=m_q,0],
\eeq
where $m_q$ is the light quark mass, which we will express in terms of
the physical pion mass via the relation
\beq
2 B m_q=M_\pi^2 + {\cal O}(M_\pi^4).
\eeq
From here on, we work in the isospin limit $m_q = m_u = m_d$.
The LEC $B$ can be extracted from quark condensate in the
isospin  limit
\beq
\langle 0|\bar{u} u|0\rangle=\langle 0|\bar{d} d|0\rangle = - F^2 B
(1+{\cal O}(m_q))~,
\eeq
where $F$ denotes the pion decay constant in the chiral limit.
The solution of the decoupling equation and the
computation of the operators $U_{\rm Okubo}$ and $\eta U_{\rm
  Okubo}^\dagger H U_{\rm Okubo} \eta$ are
carried out perturbatively within the chiral expansion. This is
most easily achieved by counting the inverse powers of the \emph{hard}
scale as explained in Ref.~\cite{Epelbaum:2007us}. Specifically, the
various terms in the interaction part of the Hamiltonian $H_I$ can be classified according to
the inverse mass dimension $\kappa$ of the corresponding coupling
constants,  
\beq
\label{DefKappa}
H_I= \sum_{\kappa} H^{(\kappa)}  \quad \mbox{with} \quad
\kappa = d + \frac{3}{2} a + b + c - 4\,.
\eeq
Here, $d$ is the number of derivatives and/or insertions of $M_\pi$
while $a$, $b$ and $c$ refer to the number of nucleon fields, pion
fields and external sources, in order. 
Notice that we only
consider terms with, at most, a single coupling to the axial or
pseudoscalar source so
that $c=0$ or $1$. Further, one has $\kappa \ge 1$ for interaction
terms with $c=0$ and 
$\kappa \ge -1$ ($\kappa \ge -2$) for vertices involving a coupling to the axial
(pseudoscalar) source, i.e.~with $c=1$. It is easy to see \cite{Epelbaum:2007us} that the chiral
dimension $\nu$ of the resulting nuclear forces $V$, charge and current
operators $\fet A^0$ and $\vec{\fet A}$, which can be read off from
the terms in $\eta U_{\rm Okubo}^\dagger H U_{\rm Okubo} \eta$ involving a coupling to the
external axial source $\fet{a}^\mu$, is determined by the overall inverse mass dimension of the
coupling constants. Specifically, for the nuclear forces, one has 
\beq
\nu=-2 + \sum_i V_i \kappa_i,
\eeq
while the chiral dimension of the nuclear charge and current operators is
given by 
\beq
\label{PCkappa}
\nu=-3  + \sum_i V_i \kappa_i\,,
\eeq
where the shift relative to the dimension of the nuclear forces accounts for
the mass dimension of the external source. 
Here, $V_i$ denotes the number of vertices of a type $\kappa_i$. Using
these results, the decoupling equation can be solved recursively by
utilizing an expansion in powers of the inverse overall mass dimension
of the coupling constants, and the resulting contributions to $\eta
U_{\rm Okubo}^\dagger H U_{\rm Okubo}
\eta$ can be worked out to a desired order $\nu$, see
\cite{Epelbaum:2007us,Kolling:2009iq,Kolling:2011mt,Epelbaum:2010nr,Bernard:2007sp,Bernard:2011zr}
for more details and explicit expressions. Notice further that while 
formulating the power counting in terms of the inverse mass dimension $\kappa$ is particularly convenient for algebraic
calculations within the MUT, one can also express the chiral dimension
$\nu$ in terms of different variables such as the number of loops, see
Ref.~\cite{Weinberg:1990rz}. This notation is commonly employed in
chiral perturbation theory and is particularly convenient when 
using diagrammatic approaches. We refer the readers to Ref.~\cite{Epelbaum:2010nr}
for further details. 

The Okubo parametrization of the UT in terms of
the operator $A = \lambda A \eta$ does not describe the most general
possible UT.  In particular, one can subsequently perform additional
UTs on the $\eta$ space. Such additional UTs are, in fact,
required to achieve  renormalizability of the nuclear forces
\cite{Epelbaum:2006eu} (and
current operators). The additional UTs employed on the $\eta$-space
may or may not depend on the external sources. Here and in what
follows, we will denote the ``strong'' $\eta$-space
transformations, which do not depend on the external sources, by  
$U_\eta$. Since the corresponding generators, by definition, do not depend on the 
external sources, the operators $U_\eta$ are not explicitly time-dependent. 

\subsection{Unitary transformations involving external sources}
The second class of
unitary $\eta$ space transformations are the ones which explicitly
depend on external sources. We denote them by
$U[a,v,s,p]$. Since these transformations explicitly depend on
external sources, they also do explicitly depend on time $t$. In the
absence of  external sources, they, by definition, are required to
reduce to:
\beq
U[0,0,m_q,0]=1~.
\eeq

In general, a time-dependent unitary
transformation $U(t)$ of a given Hamiltonian $H$ (which might
have an explicit time dependence through external sources) is not just
given by  $U^\dagger(t)HU(t)$. This can be easily seen from 
the Schr\"odinger equation
\beq
i\frac{\partial}{\partial t}\Psi = H \Psi\,,
\eeq
that leads to
\beq
i \frac{\partial}{\partial t}U(t)U^\dagger(t)\Psi = U(t) i
\frac{\partial}{\partial t}U^\dagger(t)\Psi +
\left(i\frac{\partial}{\partial t}U(t)\right) U^\dagger(t)\Psi= H
U(t)U^\dagger(t)\Psi \,.
\eeq
Multiplying both sides by $U^\dagger(t)$ and bringing the term with the
time-derivative of the unitary transformation on the right-hand side,
we obtain the Schr\"odinger equation for the transformed state
$\Psi^\prime=U^\dagger(t)\Psi$
in the form
\beq
i\frac{\partial}{\partial t}\Psi^\prime =\left[U^\dagger(t) H
U(t)-U^\dagger(t)\left(i\frac{\partial}{\partial
    t}U(t)\right)\right]\Psi^\prime ~. 
\eeq
Thus, the unitary transformation of the Hamiltonian $H$ is given by
\beq
H\to U^\dagger(t) H U(t) +\left(i\frac{\partial}{\partial
    t}U^\dagger(t)\right)U(t).
\eeq
We see that in the case of a time-dependent UT,
there is an additional term which depends on the time-derivative of
the operator $U (t)$. For this reason, the transformed
Hamiltonian in our case depends on external sources and their time derivatives:
\beqa
&&H_{\rm eff}[a,{\dot a}, v, {\dot v}, s, {\dot s}, p, {\dot
  p}]\,=\,\nn
&&\eta U^\dagger[a,v,s,p] U_\eta^\dagger U_{\rm Okubo}^\dagger H[a,v,s,p] U_{\rm
  Okubo} U_\eta U[a,v,s,p]\eta + \eta \left(i\frac{\partial}{\partial
    t}U^\dagger[a,v,s,p]\right)U[a,v,s,p]\eta.
\eeqa
For $a=v=p=0$ and $s=m_q$ we obtain the nuclear potential
\beq
V:=H_{\rm eff}
[a=0,\dot a=0, v=0, \dot v = 0, s=m_q, \dot s =0, p=0, \dot p = 0] - H_0,
\eeq
where $H_0$ is the free nucleon Hamiltonian.
The individual contributions to $V$ have a form similar to
those obtained in TOPT and are given by a sequence of vertices and the
corresponding energy denominators. For example, the leading and
subleading contributions to the nuclear force $V^{(Q^0)}$ and $V^{(Q^2)}$
constructed solely from the
lowest order $\pi N$ coupling proportional to the nucleon axial-vector
constant $g_A$ from $\mathcal{L}_{\pi N}^{(1)}$ have the form 
\beqa
V^{(Q^0)} &=& - \eta H_{2,1}^{(1)} \frac{\lambda^1}{E_\pi}  H_{2,1}^{(1)}
\eta\,,\nn
V^{(Q^2)} &=& - \frac{1}{2} \eta H_{2,1}^{(1)} \frac{\lambda^1}{E_\pi}
H_{2,1}^{(1)} \frac{\lambda^2}{E_\pi}
H_{2,1}^{(1)} \frac{\lambda^1}{E_\pi}
H_{2,1}^{(1)} \eta
+ 
\frac{1}{2} \eta H_{2,1}^{(1)} \frac{\lambda^1}{E_\pi^2}
H_{2,1}^{(1)} \eta
H_{2,1}^{(1)} \frac{\lambda^1}{E_\pi}
H_{2,1}^{(1)} \eta + \mbox{h.c.}\,,
\eeqa
where we have adopted the notation introduced in \cite{Epelbaum:2007us} with
$H^{(\kappa)}_{a, b}$ denoting an interaction from the Hamiltonian
with $a$ nucleon and $b$ pion fields. Further, $\lambda^i$ denotes a
projection operator onto states with $i$ pions while $E_\pi = \sum_i
\omega_i = \sum_i \sqrt{\vec p_i\, ^2 + M_\pi^2}$ is the pion kinetic
energy. The operator $V^{(Q^0)}$ contributes to the nucleon self-energy
and gives rise to the one-pion exchange 2N potential while the terms in $V^{(Q^2)}$
contribute to the nucleon self-energy, renormalization of the one-pion
exchange 2N potential, the leading two-pion exchange
2N potential and the tree-level two-pion exchange 3N forces (which,
however, turn out to vanish). The explicit form of these contributions
is easily obtained by substituting the explicit expressions for the vertices $H_{2,1}^{(1)}$,
written in second quantization, and performing the algebra.  Here and
in what follows, all loop integrals are calculated in the standard way
using dimensional regularization. Finally, we emphasize that the
contributions which do not involve reducible topologies can be
more efficiently calculated using the Feynman graph technique. This
is, in fact, the way some of the presented results are obtained.

\subsection{Chiral symmetry constraints and the continuity equations}
\label{sec:CS}

Under chiral SU(2)$_L \times$SU(2)$_R$ rotations, the external 
sources transform as
\beqa
r_\mu&\to& r_\mu^\prime=R\, r_\mu R^\dagger + i R\, \partial_\mu
R^\dagger~,\nn
l_\mu&\to& l_\mu^\prime=L \,l_\mu L^\dagger + i L\, \partial_\mu
L^\dagger~,\nn
s + i\,p &\to&s^\prime + i\,p^\prime= R (s + i\,p) L^\dagger~,\nn
 s - i\,p &\to&s^\prime - i\,p^\prime= L (s - i\,p) R^\dagger~.
\eeqa
The vector and axial-vector sources can be expressed as a linear
combination of the left- and right-handed sources:
\beq
v_\mu=\frac{1}{2}\left(r_\mu + l_\mu\right)\quad {\rm and }\quad a_\mu=\frac{1}{2}\left(r_\mu - l_\mu\right).
\eeq
In the above expressions, $R$ and $L$ denote independent chiral SU(2)
transformations which can be parametrized in the exponential form 
\beq
R=\exp\left(\frac{i}{2}{\fet \tau}\cdot{\fet \epsilon}_R (\vec{x},t)\right)\quad {\rm
  and}\quad L=\exp\left(\frac{i}{2}{\fet \tau}\cdot{\fet
    \epsilon}_L(\vec{x},t)\right). 
\eeq
Using the standard parametrization of the external sources in terms of
the isoscalar and isovector components \cite{Gasser:1983yg}, 
\beqa
v_\mu=v_\mu^{(s)}+\frac{1}{2}{\fet \tau}\cdot{\fet v}, \quad a_\mu=\frac{1}{2}{\fet
  \tau}\cdot{\fet a}, \quad s=s_0+{\fet \tau}\cdot{\fet s}, \quad
p=p_0 + {\fet \tau}\cdot{\fet p}\,,
\eeqa
the transformation properties of the sources with respect to
infinitesimal $\mbox{SU(2)}_L \times \mbox{SU(2)}_R$  rotations have the form 
\beqa
{\fet v}_\mu&\to& {\fet v}_\mu^\prime = {\fet v}_\mu + {\fet
  v}_\mu\times{\fet \epsilon}_V + {\fet a}_\mu\times{\fet \epsilon}_A
+ \partial_\mu{\fet \epsilon}_V,\nn
{\fet a}_\mu&\to&{\fet a}_\mu^\prime ={\fet a}_\mu + {\fet
  a}_\mu\times{\fet \epsilon}_V + {\fet v}_\mu\times{\fet \epsilon}_A
+\partial_\mu{\fet \epsilon}_A,\nn
s_0&\to&s_0^\prime=s_0 - {\fet p}\cdot{\fet \epsilon}_A, \nn
{\fet s}&\to&{\fet s}^\prime={\fet s} + {\fet s}\times{\fet \epsilon}_V - p_0
{\fet \epsilon}_A,\nn
i\,p_0&\to&i\,p_0^\prime = i (p_0 +{\fet s}\cdot{\fet \epsilon}_A),\nn
i\,{\fet p}&\to&i\,{\fet p}^\prime = i({\fet p}+{\fet p}\times{\fet
  \epsilon}_V+ s_0\,{\fet \epsilon}_A),
\eeqa
where
\beq
{\fet \epsilon}_V = \frac{1}{2}\left({\fet \epsilon}_R + {\fet
  \epsilon}_L\right) \quad {\rm and}\quad {\fet \epsilon}_A = \frac{1}{2}\left({\fet \epsilon}_R - {\fet
  \epsilon}_L\right).
\eeq
Notice that as it is well known, the singlet axial-vector current is
not conserved as the U(1)$_A$ is anomalously broken. 

We now proceed similar to Ref.~\cite{Kotlyar:1995wg}. Starting with the
original Schr\"odinger equation\footnote{In general, also second and
  higher order time-derivatives of external sources can appear in the
  Hamiltonian. These  terms, however, are only relevant at higher chiral
orders beyond the accuracy of the current work.} 
\beq
i \frac{\partial}{\partial t}\Psi = H_{\rm eff}[a, \dot a, v, \dot v,
s, \dot s, p, \dot p]\Psi ,
\eeq
we expect that there is an (in general, time-dependent) unitary transformation $U$  on the Fock space
such that
\beq
i \frac{\partial}{\partial t} U^\dagger \Psi = H_{\rm eff}[a^\prime, \dot a^\prime, v^\prime, \dot v^\prime,
s^\prime, \dot s^\prime, p^\prime, \dot p^\prime\,]U^\dagger\Psi,
\eeq
which means that observables are not affected by chiral
rotations. In other words, we expect that the Hamiltonians $H_{\rm eff}[a^\prime, 
\dot a^\prime, v^\prime, \dot v^\prime,
s^\prime, \dot s^\prime, p^\prime, \dot p^\prime\,]$ and $H_{\rm eff}[a, \dot a, v, \dot v,
s, \dot s, p, \dot p]$ are unitary equivalent:
\beq
H_{\rm eff}[a^\prime, \dot a^\prime, v^\prime, \dot v^\prime,
s^\prime, \dot s^\prime, p^\prime, \dot p^\prime\,] = U^\dagger  H_{\rm eff}[a, \dot a, v, \dot v,
s, \dot s, p, \dot p] U + \left(i\frac{\partial}{\partial
    t}U^\dagger\right) U. \label{symmconstr}
\eeq
We make an ansatz for the unitary transformation $U$ by writing it in
the form 
\beq
U=\exp\left(i\int d^3 x\big[{\fet R}_0^v(\vec{x})\cdot{\fet \epsilon}_V(\vec{x},t) 
+ {\fet R}_1^v(\vec{x})\cdot {\dot{\fet \epsilon}}_V(\vec{x},t)  
+ {\fet R}_0^a(\vec{x})\cdot{\fet \epsilon}_A(\vec{x},t)  + {\fet
    R}_1^a(\vec{x})\cdot{\dot{ \fet \epsilon}}_A(\vec{x},t) \big]
\right),
\eeq
with $R_{0,1}^{v,a}(\vec{x})$ being some Hermitian
field operators in the Schr\"odinger picture.
Eq.~(\ref{symmconstr}) can be used to derive the continuity
equation for the currents.  Setting $v=\dot v=a=\dot a=p=\dot p=\dot
s={\fet s}=0$ and $s_0=m_q=(m_u+m_d)/2$ on the right-hand side of
Eq.~(\ref{symmconstr}) and keeping only terms linear in ${\fet
  \epsilon}_V,{\fet \epsilon}_A$ and their time derivatives, we obtain
\beqa
&&H_{\rm eff}[a^\prime, \dot a^\prime, v^\prime, \dot v^\prime,
s^\prime, \dot s^\prime, p^\prime, \dot p^\prime\,]|_{v=\dot v=a=\dot a=p=\dot p=\dot
s=0, 
s=m_q} = \nn
&&W + \int d^3 x\Big(i\big[W,{\fet R}_0^v(\vec{x})\big]\cdot{\fet \epsilon}_V(\vec{x},t) +
i\big[W,{\fet R}_1^v(\vec{x})\big]\cdot{\dot{\fet \epsilon}}_V(\vec{x},t)
+ i\big[W,{\fet R}_0^a(\vec{x})\big]\cdot{\fet
  \epsilon}_A(\vec{x},t) \label{lhsHeffExpanded}\\
&&+
i\big[W,{\fet R}_1^a(\vec{x})\big]\cdot{\dot{\fet
    \epsilon}}_A(\vec{x},t) 
+{\fet R}_0^v(\vec{x})\cdot{\dot{\fet \epsilon}}_V(\vec{x},t) 
+{\fet R}_1^v(\vec{x})\cdot {\ddot{\fet \epsilon}}_V(\vec{x},t)  
+{\fet R}_0^a(\vec{x})\cdot{\dot{\fet \epsilon}}_A(\vec{x},t)  + {\fet
    R}_1^a(\vec{x})\cdot{\ddot{ \fet \epsilon}}_A(\vec{x},t)\Big), \nonumber
\eeqa 
with $W \equiv H_0+V$. On the other hand, we can directly expand the
left-hand side of
Eq.~(\ref{symmconstr}) in ${\fet\epsilon}_V,{\fet \epsilon}_A$ and their time derivatives to get
\beqa
&&H_{\rm eff}[a^\prime, \dot a^\prime, v^\prime, \dot v^\prime,
s^\prime, \dot s^\prime, p^\prime, \dot p^\prime\,]|_{v=\dot v=a=\dot a=p=\dot p=\dot
s={ \sfet s }=0, s_0=m_q} = \nn
&&W + \int d^3 x\Big({\fet
  V}_\mu^{(0)}(\vec{x})\cdot\partial^\mu{\fet\epsilon}_V(\vec{x},t) +
{\fet
  V}_\mu^{(1)}(\vec{x})\cdot\partial^\mu\dot{\fet\epsilon}_V(\vec{x},t)
+{\fet
  A}_\mu^{(0)}(\vec{x})\cdot\partial^\mu{\fet\epsilon}_A(\vec{x},t) +
{\fet
  A}_\mu^{(1)}(\vec{x})\cdot\partial^\mu\dot{\fet\epsilon}_A(\vec{x},t)\nn
&&+m_q\,{\fet P}^{(0)}(\vec{x})\cdot{\fet\epsilon}_A(\vec{x},t)
+m_q\,{\fet
  P}^{(1)}(\vec{x})\cdot{\dot{\fet\epsilon}}_A(\vec{x},t)\Big),
\label{rhsHeffExpanded}
\eeqa
where the vector, axial-vector and pseudoscalar currents are defined
by
\beqa
&&V_\mu^{(0) j}(\vec{x}):=\frac{\delta H_{{\rm eff}}}{\delta
  v_j^\mu(\vec{x},t)},\quad V_\mu^{(1) j}(\vec{x}):=\frac{\delta H_{{\rm eff}}}{\delta
  \dot{v}_j^\mu(\vec{x},t)},\quad
A_\mu^{(0) j}(\vec{x}):=\frac{\delta H_{{\rm eff}}}{\delta
  a_j^\mu(\vec{x},t)}, \nn
&&A_\mu^{(1) j}(\vec{x}):=\frac{\delta H_{{\rm eff}}}{\delta
  \dot{a}_j^\mu(\vec{x},t)}, \quad P_j^{(0)}(\vec{x}):=\frac{\delta H_{{\rm eff}}}{\delta
  p_j(\vec{x},t)}, \quad P_j^{(1)}(\vec{x}):=\frac{\delta H_{{\rm eff}}}{\delta
  \dot{p}_j(\vec{x},t)},
\eeqa
with $j=1,2,3$ an isospin index. In all these expressions, the
functional derivatives are taken at $v=\dot v=a=\dot a=p=\dot p=\dot
s={\fet s}=0$ and $s_0=m_q$. Matching 
Eqs.~(\ref{lhsHeffExpanded}) and (\ref{rhsHeffExpanded}) with respect to
$\ddot{\fet\epsilon}_V(\vec{x},t)$ and 
$\ddot{\fet\epsilon}_A(\vec{x},t)$, we read off
\beq
{\fet R}_1^v(\vec{x})={\fet V}_0^{(1)}(\vec{x}), \quad {\fet R}_1^a(\vec{x})={\fet A}_0^{(1)}(\vec{x})~.
\eeq
Next, matching the coefficients in front of the first derivatives $\dot{\fet\epsilon}_V(\vec{x},t)$ and 
$\dot{\fet\epsilon}_A(\vec{x},t)$ yields
\beqa
{\fet R}_0^v(\vec{x}) +i\,\big[W,{\fet R}_1^v(\vec{x})\big]&=&{\fet
  V}_0^{(0)}(\vec{x})-\vec{\nabla}\cdot\vec{\fet V}^{(1)}(\vec{x}),\nn
{\fet R}_0^a(\vec{x}) +i\,\big[W,{\fet R}_1^a(\vec{x})\big]&=&{\fet
  A}_0^{(0)}(\vec{x})-\vec{\nabla}\cdot\vec{\fet A}^{(1)}(\vec{x}) + 
m_q {\fet P}^{(1)}(\vec{x}).
\eeqa
Finally, matching  the coefficients in front of ${\fet\epsilon}_V(\vec{x},t)$ and 
${\fet\epsilon}_A(\vec{x},t)$ gives
\beqa
i\big[W,{\fet R}_0^v(\vec{x})\big]&=&-\vec{\nabla}\cdot\vec{\fet
  V}^{(0)}(\vec{x}),\nn
i\big[W,{\fet R}_0^a(\vec{x})\big]&=&-\vec{\nabla}\cdot\vec{\fet
  A}^{(0)}(\vec{x})+ m_q {\fet P}^{(0)}(\vec{x}).
\eeqa
Combining these relations,  we obtain the continuity equations
\beqa
i\,\big[W,{\fet V}_0^{(0)}(\vec{x}) - \vec{\nabla}\cdot\vec{\fet
  V}^{(1)}(\vec{x}) - i\,\big[W,{\fet V}_0^{(1)}(\vec{x})\big]\big] &=&-\vec{\nabla}\cdot\vec{\fet
  V}^{(0)}(\vec{x}),\nn
i\,\big[W,{\fet A}_0^{(0)}(\vec{x}) - \vec{\nabla}\cdot\vec{\fet
  A}^{(1)}(\vec{x}) - i\,\big[W,{\fet A}_0^{(1)}(\vec{x})\big]+m_q {\fet P}^{(1)}(\vec{x})\big] &=&-\vec{\nabla}\cdot\vec{\fet
  A}^{(0)}(\vec{x})+m_q {\fet P}^{(0)}(\vec{x}).\label{continuityeqcoordspace}
\eeqa
Notice that the form of the continuity equation we obtain differs from
the usual one with ${\fet V}_\mu^{(1)}(\vec{x})={\fet
  A}_\mu^{(1)}(\vec{x})={\fet P}^{(1)}(\vec{x})=0$.  
In our case,
these terms originate from the additional unitary transformations
involving external sources, and they cannot be discarded.\footnote{It has
  already been mentioned in the literature that the continuity
  equation gets modified if one uses time-dependent unitary
  transformations~\cite{Ohta:1989bsh}.} 
Clearly, ${\fet V}_\mu^{(1)}(\vec{x}), {\fet
  A}_\mu^{(1)}(\vec{x})$ and ${\fet P}^{(1)}(\vec{x})$ are
proportional to the unitary phases so that the resulting contributions
affect only the off-shell behavior of the current operators. In order to
avoid the introduction of a set of axial-vector $\big\{ {\fet A}_\mu^{(0)},\,  {\fet
  A}_\mu^{(1)}\big\}$ and vector $\big\{ {\fet V}_\mu^{(0)}, \, {\fet
  V}_\mu^{(1)}\big\} $ currents which are not four-vectors individually, we
will combine the two current operators into a single one required to
be a four-vector, see sections~\ref{Poincare} 
and \ref{boost} for more details. In momentum
space, the current can be defined by
\beqa
&&\tilde{V}_\mu^j(\vec{k},k_0):=\frac{\delta H_{\rm eff}}{\delta \tilde{v}_j^\mu(\vec{k},k_0)}, \quad 
\tilde{A}_\mu^j(\vec{k},k_0):=\frac{\delta H_{\rm eff}}{\delta
  \tilde{a}_j^\mu(\vec{k},k_0)}, \quad
\tilde{P}^j(\vec{k},k_0):=\frac{\delta H_{\rm eff}}{\delta \tilde{p}^j(\vec{k},k_0)},
\label{CurrentsDefMomentumSpace}
\eeqa
where $H_{\rm eff}$ is taken at $t=0$ and the functional derivatives are
taken at $v=\dot v=a=\dot a=p=\dot p=\dot
s={\fet s}=0, s_0=m_q$. The Fourier transform for the sources is
defined by
\beq
v_\mu^j(x)=:\int d^4 q \exp( -i q\cdot x)\tilde{v}_\mu^j(q),\quad
a_\mu^j(x)=:\int d^4 q \exp( -i q\cdot x)\tilde{a}_\mu^j(q),
\quad p^j(x)=:\int d^4 q \exp( -i q\cdot x)\tilde{p}^j(q).
\eeq 
The currents in Eq.~(\ref{CurrentsDefMomentumSpace}) can be expressed
as a linear combination of the previously defined currents:
\beqa
\tilde{V}_\mu^j(\vec{k},k_0)&=&\tilde{V}_\mu^{(0) j}(\vec{k}) -i\, k_0
\tilde{V}_\mu^{(1) j}(\vec{k}),\nn
\tilde{A}_\mu^j(\vec{k},k_0)&=&\tilde{A}_\mu^{(0) j}(\vec{k}) -i\, k_0
\tilde{A}_\mu^{(1) j}(\vec{k}),\nn
\tilde{P}^j(\vec{k},k_0)&=&\tilde{P}^{(0) j}(\vec{k}) -i\, k_0
\tilde{P}^{(1) j}(\vec{k}),
\eeqa
where 
\beqa
\tilde{V}_\mu^{(l)j}(\vec{k}) &=& \int d^3 x \exp(i \vec{k}\cdot\vec{x}) V_\mu^{(l)j}(\vec{x}),\quad
\tilde{A}_\mu^{(l)j}(\vec{k}) = \int d^3 x \exp(i
\vec{k}\cdot\vec{x}) A_\mu^{(l)j}(\vec{x}),\nn
\tilde{P}^{(l)j}(\vec{k}) &=& \int d^3 x \exp(i \vec{k}\cdot\vec{x}) P^{(l)j}(\vec{x}),\quad l=0,1~.
\eeqa
A linear appearance of the energy transfer $k_0$ is an off-shell
effect and is found to be unavoidable if the currents are to be
renormalized.   It emerges as a consequence of additional unitary
transformations involving external sources, which are needed to
maintain renormalizability. The continuity equations given in 
Eq.~(\ref{continuityeqcoordspace}) in coordinate space can be rewritten in momentum space as
\beqa
\label{continuityeqmomspacevector}
\big[W,\tilde{\fet V}_0(\vec{k},0) - \frac{\partial}{\partial k_0}\vec{k}\cdot\vec{\tilde{\fet
  V}}(\vec{k},k_0) + \frac{\partial}{\partial k_0}\,\big[W,\tilde{\fet V}_0(\vec{k},k_0)\big]\big] &=&\vec{k}\cdot\vec{\tilde{\fet
  V}}(\vec{k},0),\nn
\big[W,\tilde{\fet A}_0(\vec{k},0) - \frac{\partial}{\partial k_0}\vec{k}\cdot\vec{\tilde{\fet
  A}}(\vec{k},k_0) + \frac{\partial}{\partial k_0}\,\big[W,\tilde{\fet A}_0(\vec{k},k_0)\big]+m_q \,i\frac{\partial}{\partial k_0}\tilde{\fet P}(\vec{k},k_0)\big] &=&\vec{k}\cdot\vec{\tilde{\fet
  A}}(\vec{k},0)-m_q\,i\, \tilde{\fet P}(\vec{k},0). \quad\label{continuityeqmomspace}
\eeqa  
The continuity equations are direct consequences of the chiral
symmetry and provide non-trivial tests of the derived current operators. Obviously, similar
expressions can be found for the singlet vector current related
to the $U(1)_V$ symmetry (baryon number conservation).

\subsection{Relation to the $S$-matrix}
It is instructive to analyze in detail  the relation between the
current operators defined in the previous section to $S$-matrix
elements. Here and in what follows, we restrict our discussion to 
the axial currents which are the main focus of our paper. 

We begin with the definition of $S$-matrix in the Heisenberg representation,
\beq
S=T \exp\left(-i\,S_A[a]\right),\quad {\rm with}\quad S_A[a]=\int d^4 x \left[{\fet A}_\mu^{(0)H}(x)\cdot{\fet
      a}^\mu(x)+{\fet A}_\mu^{(1)H}(x)\cdot \dot{\fet
      a}^\mu(x)\right]~. \label{SADefinition}
\eeq
Here, $T$ denotes the time ordering operator and 
\beq
{\fet A}_\mu^{(j)H}(x):=\exp(i\,W x_0) {\fet A}_\mu^{(j)}(\vec{x})
\exp(-i\,W x_0), \quad j=0,1.
\eeq
In linear approximation we can drop the $T$ operator. Sandwiching  the
$S$-matrix operator  between the final and initial states $|\alpha\rangle$ and
$|\beta\rangle$, respectively, we get
\beq
\langle\alpha|S|\beta\rangle=\langle\alpha|\beta\rangle-i
\langle\alpha|S_A[a]|\beta\rangle.
\eeq
Using the eigenvalue relations 
\beq
W |\alpha\rangle = E_\alpha |\alpha\rangle\quad{\rm and}\quad
W |\beta\rangle = E_\beta |\beta\rangle \,,
\eeq
we obtain 
\beq
\langle\alpha|S|\beta\rangle=\langle\alpha|\beta\rangle-i\int d^4x
\exp\left(i\,(E_\alpha-E_\beta)x_0\right)
\left(\langle\alpha|{\fet A}_\mu^{(0)}(\vec{x}) \cdot 
  {\fet
    a}^\mu(x_0,\vec{x})|\beta\rangle +
\langle\alpha|{\fet A}_\mu^{(1)}(\vec{x}) \cdot \dot{\fet a}^\mu(x_0,\vec{x})|\beta\rangle 
\right).
\eeq
Taking the functional derivative with respect to the sources in momentum space we get
\beqa
\frac{\delta}{\delta \tilde{a}^{j \mu}(k_0,\vec{k})}
\langle\alpha|S|\beta\rangle&=&
-i\int d^4x
\exp\left(i\,(E_\alpha-E_\beta - k_0)x_0\right) 
\exp\big(i\,\vec{k}\cdot\vec{x}\big)
\left(\langle\alpha|A_\mu^{(0)j}(\vec{x}) |\beta\rangle -i\, k_0
\langle\alpha|A_\mu^{(1)j}(\vec{x})|\beta\rangle 
\right)\nn
&=&-i\,2\pi \delta(E_\alpha-E_\beta-k_0)
\left(\langle\alpha|\tilde{A}_\mu^{(0)j}(\vec{k}) |\beta\rangle -i\, k_0
\langle\alpha|\tilde{A}_\mu^{(1)j}(\vec{k})|\beta\rangle
\right)
\nn
&=&-i \,2\pi \delta(E_\alpha-E_\beta-k_0) \langle\alpha|\tilde A_\mu^{j}(k_0,\vec{k})|\beta\rangle.
\eeqa
Thus, we see that the current operator, defined in the Schr\"odinger
picture as described above, is indeed identical to the $S$-matrix contribution of the
axial-vector current on the energy shell.

Notice that Eq.~(\ref{SADefinition}) can be rewritten upon performing a partial
integration in time
\beq
S_A[a]=\int d^4 x \left({\fet
    A}_\mu^{(0)H}(x)-\frac{\partial}{\partial t}{\fet
    A}_\mu^{(1)H}(x)\right)\cdot{\fet a}^\mu(x)=\int d^4 x \left({\fet
    A}_\mu^{(0)H}(x)-i\,\left[W,{\fet
    A}_\mu^{(1)H}(x)\right]\right)\cdot{\fet a}^\mu(x).
\eeq
This would lead to the following modification of the current
\beq
{\fet A}_\mu^H(x)\to {\fet A}_\mu^{\prime H}(x)={\fet
    A}_\mu^{(0)H}(x)-i\,\left[W,{\fet
    A}_\mu^{(1)H}(x)\right].\label{APrimeCurrentDefinition}
\eeq
Notice that the resulting current operator ${\fet A}_\mu^{\prime}(\vec{x})= {\fet A}_\mu^{\prime H}(\vec{x},x_0=0)$ satisfies
the usual, non-modified continuity equation
\beqa
i\,\big[W,{\fet A}_0^{\prime}(\vec{x})\big]&=&-\vec{\nabla}\cdot\vec{\fet
  A}^{\prime}(\vec{x})+2 m_q {\fet P}^{\prime}(\vec{x}),\label{Usualcontinuityeqcoordspace}
\eeqa
where, similar to the axial-vector current, we have introduced the
corresponding modified version of the pseudoscalar
current
\beq
{\fet P}^{\prime}(\vec{x})={\fet P}^{(0)}(\vec{x})-i\,\big[W,{\fet P}^{(1)}(\vec{x})]. \label{PPrimeCurrentDefinition}
\eeq
Eq.~(\ref{Usualcontinuityeqcoordspace}) is obtained straightforwardly by inserting
the definitions of Eqs.~(\ref{APrimeCurrentDefinition}) and
(\ref{PPrimeCurrentDefinition}) in Eq.~(\ref{continuityeqcoordspace}).
The current  ${\fet A}_\mu^{\prime H}$ is, however, identical to the
current without additional unitary transformations involving external
fields. To see this we consider the axial-vector current contribution
to the effective Hamiltonian before applying the unitary transformation
involving external fields
\beq
U_\eta^\dagger U_{\rm Okubo}^\dagger H[a] U_{\rm Okubo}U_\eta = W +
\int d^3 x {\fet B}_\mu(\vec{x})\cdot{\fet a}^\mu(x_0, \vec{x}).
\eeq
Parametrizing an additional unitary transformation with external
sources by
\beq
U[a]=1-\int d^3 x \,{\fet C}_\mu(\vec{x})\cdot{\fet
  a}^\mu(x_0,\vec{x}) + {\cal O}(a^2),
\eeq
where ${\fet C}_\mu$ is an antihermitian field operator, we get 
\beqa
\eta U^\dagger[a]U_\eta^\dagger U_{\rm Okubo}^\dagger H[a] U_{\rm
  Okubo}U_\eta U[a] \eta + i\,\eta\left(\frac{\partial}{\partial x_0}
  U^\dagger[a]\right)U[a]\eta
&=& W + \int d^3 x
\Big[\left( {\fet
  B}_\mu(\vec{x}) + \left[{\fet
  C}_\mu(\vec{x}),W\right]\right) \cdot{\fet a}^\mu(x_0, \vec{x}) \nn
&+& i\, {\fet C}_\mu(\vec{x})\cdot\dot{\fet
  a}^\mu(x_0,\vec{x})\Big].
\eeqa
We then read off
\beq
{\fet A}_\mu^{(0)}(\vec{x})={\fet
  B}_\mu(\vec{x}) + \left[{\fet
  C}_\mu(\vec{x}),W\right]\quad {\rm and}\quad {\fet A}_\mu^{(1)}(\vec{x})=i\, {\fet C}_\mu(\vec{x})\,,
\eeq
and conclude
\beq
{\fet A}_\mu^{\prime H}(x)={\fet
  B}_\mu^H(x) + \left[{\fet
  C}_\mu^H(x),W\right] + \left[W, {\fet C}_\mu^H(x)\right]={\fet
  B}_\mu^H(x).
\eeq
As already pointed out above, additional unitary transformations
involving external sources are needed
for the renormalization of the current. Since their effects are switched
off in the current ${\fet A}_\mu^\prime$, we prefer to work with the
current ${\fet A}_\mu$ in order to have properly renormalized current operators.

\subsection{Poincar\'{e} invariance constraints}
\label{Poincare}
It is instructive to analyze the four-vector constraint on ${\fet A}_\mu$. 
In the Heisenberg picture, this means 
\beq
\exp\left(-i\,\vec{e}\cdot\vec{K}\theta\right){\fet A}_\mu^{H}(x)
\exp\left(i\,\vec{e}\cdot\vec{K}\theta\right)
= \Lambda_\mu^{\,\,\,\,\nu}(\theta){\fet A}_\nu^H\left(\Lambda^{-1}(\theta)x\right),
\eeq
where $\vec K$ is a boost operator, $\vec{e}$ is a boost direction
and $\Lambda$ is a $4\times 4$ boost matrix which depends on the boost
direction $\vec{e}$ and a boost angle $\theta$:
\beq
\Lambda(\theta)\left(
\begin{matrix}
x_0 \\
\vec{x}
\end{matrix}\right)=\left(
\begin{matrix}
x_0\, \cosh(\theta) + \vec{e}\cdot\vec{x} \sinh(\theta)\\
\vec{x} + \vec{e}\, x_0 \sinh(\theta) + \vec{e}\,(\cosh(\theta) -1)\vec{e}\cdot\vec{x}
\end{matrix}
\right) = 
\left(
\begin{matrix}
x_0\\
\vec{x} 
\end{matrix}
\right) +
\theta \left(
\begin{matrix}
\vec{e}\cdot\vec{x} \\
\vec{e}\, x_0 
\end{matrix}
\right) + {\cal O}\big(\theta^2\big).
\eeq
For a given four-vector $x = (x^0,\vec{x})$, we can introduce  an orthogonal four-vector via
\beq
x^\perp = (\vec{e}\cdot\vec{x},\vec{e}\, x_0)\,,
\eeq
so that a Lorentz transformation has the form
\beq
\Lambda(\theta) x = x + \theta \,x^\perp + {\cal O}\big(\theta^2).
\eeq
Since in coordinate space the current ${\fet A}_\mu^{H}(x)$ is given
in terms of ${\fet A}_\mu^{(0)}$ and ${\fet A}_\mu^{(1)}$, we rewrite
this relation as
\beq
\exp\left(-i\,\vec{e}\cdot\vec{K}\theta\right)S_A[a]\exp\left(i\,\vec{e}\cdot\vec{K}\theta\right)
= S_A[a^\prime],\quad
{\fet a}^{\prime \mu}(x)=(\Lambda^{-1})^\mu_{\,\,\,\,\nu}(\theta){\fet a}^\nu(\Lambda(\theta) x),
\eeq
where $S_A[a]$ is defined in Eq.~(\ref{SADefinition}). The time derivative
of ${\fet a}^\prime$ is given by
\beq
\dot{\fet a}^{\prime \mu}(x)=(\Lambda^{-1})^\mu_{\,\,\,\,\nu}(\theta)
\frac{\partial}{\partial y^\alpha}{\fet
  a}^\nu(y)\bigg|_{y=\Lambda(\theta)x} \; 
\Lambda^\alpha_{\,\,\,\,0}(\theta).
\eeq
We now make the
substitution $x\to \Lambda^{-1}(\theta)x$ in the integral appearing in
the definition of $S_A[a^\prime]$
to obtain
\beqa
&&\exp\left(-i\,\vec{e}\cdot\vec{K}\theta\right)S_A[a]\exp\left(i\,\vec{e}\cdot\vec{K}\theta\right)\nn
&&= \Lambda_\mu^{\,\,\,\,\nu}(\theta)\int d^4x \bigg({\fet A}_\nu^{(0)H}(\Lambda^{-1}(\theta)x)\cdot {\fet
  a}^{\mu}(x) + {\fet A}_\nu^{(1)H}(\Lambda^{-1}(\theta)x)\cdot \frac{\partial}{\partial x^\alpha}{\fet a}^\mu(x)
\Lambda^\alpha_{\,\,\,\,0}(\theta)\bigg). \label{LorentzTrSA}
\eeqa
For an infinitesimally small $\theta$, this leads to
\beqa
\frac{\partial}{\partial x^\alpha}{\fet a}^\mu(x)
\Lambda^\alpha_{\,\,\,\,0}(\theta)&=&(\Lambda^{-1})_0^{\,\,\,\,\alpha}(\theta)
\frac{\partial}{\partial x^\alpha}{\fet a}^\mu(x)=\dot{\fet
  a}^\mu(x)+\theta\,\vec{e}\cdot\vec{\nabla}{\fet a}^\mu(x),\nn
 {\fet A}_\nu^{(l)H}(\Lambda^{-1}(\theta)x)&=&{\fet
   A}_\nu^{(l)H}(x)-\theta\, x_\alpha^\perp\frac{\partial}{\partial
  x_\alpha}{\fet
   A}_\nu^{(l)H}(x).
\eeqa
Applying this to Eq.~(\ref{LorentzTrSA}) we get
\beqa
-i\,\big[\vec{e}\cdot\vec{K},S_A[a]\big]&=&
  S_{A^\perp}[a] -\int d^4x\Big[ x_\nu^\perp\left(\frac{\partial}{\partial
  x_\nu}{\fet A}_\mu^{(0)H}(x)\right)\cdot{\fet a}^\mu(x)+x_\nu^\perp\left(\frac{\partial}{\partial
  x_\nu}{\fet A}_\mu^{(1)H}(x)\right)\cdot\dot{\fet a}^\mu(x)\nn
&-&{\fet A}_\mu^{(1)H}(x)\cdot\left(\vec{e}\cdot\vec{\nabla}{\fet a}^\mu(x)\right)\Big], \label{fourvectortest}
\eeqa
with 
\beq
S_{A^\perp}[a]=\int d^4 x \Big({\fet A}_\mu^{(0)H\perp}(x)\cdot{\fet
  a}^\mu(x) +{\fet A}_\mu^{(1)H\perp}(x)\cdot\dot{\fet
  a}^\mu(x)\Big).
\eeq
After a partial integration in the spatial components, one obtains
\beqa
-i\,\big[\vec{e}\cdot\vec{K},S_A[a]\big]&=&\int d^4x\Bigg[\left({\fet A}_\mu^{(0)H\perp}(x)-x_\nu^\perp\frac{\partial}{\partial
  x_\nu}{\fet A}_\mu^{(0)H}(x) - \vec{e}\cdot\vec{\nabla}{\fet
  A}_\mu^{(1)H}(x)\right)\cdot{\fet a}^\mu(x)\nn
&+&\left({\fet A}_\mu^{(1)H\perp}(x)
-x_\nu^\perp\frac{\partial}{\partial
  x_\nu}{\fet A}_\mu^{(1)H}(x)\right)\cdot\dot{\fet a}^\mu(x)\Bigg]\nn
&=&\int d^4x\Bigg[\left({\fet A}_\mu^{(0)H\perp}(x)-
  i\,\vec{e}\cdot\vec{x}\big[W,{\fet A}_\mu^{(0)H}(x)\big]
-x_0\vec{e}\cdot\vec{\nabla}{\fet A}_\mu^{(0)H}(x) - \vec{e}\cdot\vec{\nabla}{\fet
  A}_\mu^{(1)H}(x)\right)\cdot{\fet a}^\mu(x)\nn
&+&\left({\fet A}_\mu^{(1)H\perp}(x)
-i\,\vec{e}\cdot\vec{x}\big[W,{\fet A}_\mu^{(1)H}(x)\big]
-x_0\vec{e}\cdot\vec{\nabla}{\fet A}_\mu^{(1)H}(x)\right)\cdot\dot{\fet a}^\mu(x)\Bigg].\label{CommutatorWithBoost}
\eeqa
To proceed further, we need the following commutation relations
\beq
\big[W,K_i\big]=-i\,P_i\quad {\rm and}\quad \big[W,P_i\big]=0
\eeq
from the Poincar\'{e} algebra.
Using Hadamard's lemma 
\beq
\exp(A)B\exp(-A)=B +
\big[A,B\big]+\frac{1}{2}\big[A,\big[A,B\big]\big]+\frac{1}{3!}\big[A,\big[A,\big[A,B\big]\big]\big]
+ \ldots,
\eeq
which is valid for any square matrices $A$ and $B$, we get
\beq
\exp(-i\,W x_0)\vec{e}\cdot\vec{K}\exp(i\,W x_0)=\vec{e}\cdot\vec{K} -
i\,x_0\big[W,\vec{e}\cdot\vec{K}\big] = \vec{e}\cdot\vec{K} - x_0 \,\vec{e}\cdot\vec{P}. 
\eeq
Notice that all higher commutators vanish due to $\big[W,P_i\big]=0$. We conclude that
\beqa
\label{tempo1}
-i\,\big[\vec{e}\cdot\vec{K},{\fet A}_\mu^{(l)H}(x)]&=&-i\,\exp(i\,W
x_0)\big[\exp(-i\,W x_0)\vec{e}\cdot\vec{K}\exp(i\,W x_0), {\fet
  A}_\mu^{(l)}(\vec{x})\big]\exp(-i\,W x_0)\nn
&=&-i\,\exp(i\,W x_0)\big[\vec{e}\cdot\vec{K}-x_0 \vec{e}\cdot\vec{P} , {\fet
  A}_\mu^{(l)}(\vec{x})\big]\exp(-i\,W x_0).
\eeqa
Using the relation \beq
\big[i\,\vec{e}\cdot\vec{P},{\fet
  A}_\mu^{(l)}(\vec{x})\big]=-\vec{e}\cdot\vec{\nabla}{\fet
  A}_\mu^{(l)}(\vec{x}),
\eeq
which follows from a general translation
\beq
{\fet
  A}_\mu^{(l)}(\vec{x})=\exp(-i\,\vec{P}\cdot\vec{x}) {\fet
  A}_\mu^{(l)}(0) \exp(i\,\vec{P}\cdot\vec{x}),
\eeq
we can rewrite Eq.~(\ref{tempo1}) as
\beqa
-i\,\big[\vec{e}\cdot\vec{K},{\fet A}_\mu^{(l)H}(x)]&=&
-i\,\exp(i\,W x_0)\big[\vec{e}\cdot\vec{K}, {\fet
  A}_\mu^{(l)}(\vec{x})\big]\exp(-i\,W x_0) - x_0\,\vec{e}\cdot\vec{\nabla} {\fet
  A}_\mu^{(l)H}(x).
\eeqa
With this relation, we can finally bring Eq.~(\ref{CommutatorWithBoost}) into the
form
\beqa
&&-i\,\int d^4x \exp(i\,W x_0)\Big(\big[\vec{e}\cdot\vec{K}, {\fet
  A}_\mu^{(0)}(\vec{x})\big]\cdot {\fet a}^\mu(x)  +\big[\vec{e}\cdot\vec{K}, {\fet
  A}_\mu^{(1)}(\vec{x}) \cdot\dot{\fet a}^\mu(x)\big]\Big)\exp(-i\,W
x_0)\nn
&&=\int d^4x\Bigg[\left({\fet A}_\mu^{(0)H\perp}(x)-
  i\,\vec{e}\cdot\vec{x}\big[W,{\fet A}_\mu^{(0)H}(x)\big]
- \vec{e}\cdot\vec{\nabla}{\fet
  A}_\mu^{(1)H}(x)\right)\cdot{\fet a}^\mu(x)\nn
&&\quad+\left({\fet A}_\mu^{(1)H\perp}(x)
-i\,\vec{e}\cdot\vec{x}\big[W,{\fet A}_\mu^{(1)H}(x)\big]
\right)\cdot\dot{\fet a}^\mu(x)\Bigg].
\eeqa
Sandwiching this relation between the final and initial states
$|\alpha\rangle$ and $|\beta\rangle$, respectively, and taking
the functional derivative with respect to the axial-vector source in momentum space on 
both sides, we end up with our final result
\beqa
\label{FourVectorCondMomentumSpaceOnShell}
&&2\pi\,\delta(E_\alpha-E_\beta-k_0)\langle\alpha|\big[-i\,\vec{e}\cdot\vec{K},\tilde{\fet
  A}_\mu^{(0)}(\vec{k})-i\,k_0 \,\tilde{\fet
  A}_\mu^{(1)}(\vec{k})\big]|\beta\rangle\\
&=&2\pi\,\delta(E_\alpha-E_\beta-k_0) \langle\alpha|\Big(\tilde{\fet
  A}_\mu^{(0)\perp}(\vec{k})-i\,k_0\,\tilde{\fet
  A}_\mu^{(1)\perp}(\vec{k})-\vec{e}\cdot\vec{\nabla}_k\big[W,\tilde{\fet
A}_\mu^{(0)}(\vec{k})-i\,k_0\,\tilde{\fet
A}_\mu^{(1)}(\vec{k})\big]) + i\,\vec{e}\cdot\vec{k}\,\tilde{\fet
A}_\mu^{(1)}(\vec{k})\Big)|\beta\rangle.\nonumber
\eeqa
We can also write this relation in the operator form as
\beq
-i\,\big[\vec{e}\cdot\vec{K},\tilde{\fet A}_\mu(k)\big]=\tilde{\fet
  A}_\mu^\perp(k)-\vec{e}\cdot\vec{\nabla}_k\big[W,\tilde{\fet
  A}_\mu(k)\big] -\,\vec{e}\cdot\vec{k}\frac{\partial}{\partial k_0}\tilde{\fet
  A}_\mu(k) + i\,\big[W,{\fet X}_\mu\big]-i\,k_0 {\fet X}_\mu,\label{FourVectorCondMomentumSpace}
\eeq
where $X_\mu$ is an arbitrary operator satisfying
\beq
\lim_{k_0\to E_\beta - E_\alpha} (k_0 + E_\alpha - E_\beta)\langle
\beta|{\fet X}_\mu|\alpha\rangle = 0.
\eeq

\subsection{Effective boost operator}
\label{boost}
In order to explicitly verify  the four-vector condition of
Eq.~(\ref{FourVectorCondMomentumSpaceOnShell}) we need to construct the boost
operator $\vec{K}$. As usual, we start from the classical conserved
Noether current which is a reflection of the fact that proper
orthochronous Lorentz transformations represent the symmetry of any
relativistic field theory. An infinitesimal proper
orthochronous Lorentz transformation, which is a combination of a
rotation and a boost, is given by
\beq
x_\mu\to x_\mu + \epsilon_{\mu\nu}x^\nu,
\eeq
with $\epsilon_{\mu\nu}=-\epsilon_{\nu\mu}$ 
an antisymmetric infinitesimal angle. Consider a field transformation 
\beqa
\pi^a(x)&\to&\pi^{\prime a}(x)=\pi^a(\Lambda^{-1}
x)=\pi^a(x-\epsilon\cdot x+{\cal O}(\epsilon^2)),\nn
\partial_\mu \pi^a(x)&\to&\partial_\mu\pi^{\prime a}(x)=
\left(\frac{\partial}{\partial y^\mu}\pi^a(y) - \epsilon_{\alpha\mu}\frac{\partial}{\partial
  y_\alpha}\pi^a(y)  +{\cal O}(\epsilon^2)\right)_{y=x-\epsilon\cdot x+ {\cal O}(\epsilon^2)},\nn
N_{i}(x)&\to&N_{i}^\prime(x)=\left(1 +
  \frac{1}{2}\epsilon_{\mu\nu}\Sigma^{\mu\nu}+{\cal
    O}(\epsilon^2)\right)N_i(x-\epsilon\cdot x+{\cal
  O}(\epsilon^2)),\nn
\partial_\mu N_{i}(x)&\to&\partial_\mu
N_{i}^\prime(x)=\left(\frac{\partial}{\partial
    y^\mu}N_i(y)
+\frac{1}{2}\epsilon_{\alpha\beta}
\Sigma^{\alpha\beta}\frac{\partial}{\partial
    y^\mu}N_i(y)
-\epsilon_{\alpha\mu}\frac{\partial}{\partial
    y_\alpha}N_i(y)+{\cal O}(\epsilon^2))\right)_{y=x-\epsilon\cdot x+ {\cal O}(\epsilon^2))},
\eeqa
where $i$ is  the  isospin index of the nucleon, 
\beq
\Sigma^{\mu\nu}=\frac{i}{2}\big[\gamma^\mu,\gamma^\nu\big],
\eeq
and the $\gamma^\mu$ are the Dirac matrices. If we set all external sources
to zero (or to  $m_q$ in the case of the scalar source), the effective chiral Lagrangian density 
does not depend explicitly on $x$, and we get
\beq
\frac{\partial}{\partial \epsilon^{\nu\sigma}} {\cal L}_{\rm
  ChPT}({\fet\pi}^\prime,N^\prime) \Big |_{\epsilon=0}=-(x_\sigma\partial_\nu-x_\nu\partial_\sigma)
{\cal L}_{\rm ChPT}({\fet\pi},N)=\partial_\mu\left(x_\nu\,
  g^{\mu}_{\,\,\,\,\sigma}-x_\sigma\, g^{\mu}_{\,\,\,\,\nu}\right)
{\cal L}_{\rm ChPT}({\fet\pi},N)=:\partial_\mu F^\mu_{\nu\sigma},
\eeq
where we assume that the chiral Lagrangian density is a Lorentz scalar:
\beqa
\partial_\mu \pi^a(x)&\to&\partial_\mu \pi^a(x) -
\epsilon_{\alpha\mu}\partial^\alpha\pi^a(x) + {\cal O}(\epsilon^2),\nn
N_i(x)&\to&N_i(x) +
\frac{1}{2}\epsilon_{\mu\nu}\Sigma^{\mu\nu}N_i(x) + {\cal O}(\epsilon^2),\nn
\partial_\mu N_i(x)&\to&\partial_\mu N_i(x) +
\frac{1}{2}\epsilon_{\alpha\beta}\Sigma^{\alpha\beta}\partial_\mu N_i(x)-\epsilon_{\alpha\mu}\partial^\alpha
N_i(x)+{\cal O}(\epsilon^2).\label{LorentzInvarianceOfLagrangian}
\eeqa
The conserved classical Noether current is given by
\beq
{\cal M}_c^{\mu\alpha\beta}=\frac{\partial {\cal L}_{\rm
    ChPT}}{\partial \partial_\mu\pi^a(x)}D^{\alpha\beta}\pi^a(x)+\frac{\partial
  {\cal L}_{\rm ChPT}}{\partial \partial_\mu N_i(x)}D^{\alpha\beta} N_i(x)
+D^{\beta\alpha}\bar{N}_i(x)\frac{\partial
  {\cal L}_{\rm ChPT}}{\partial \partial_\mu \bar{N}_i(x)} - F^{\mu\alpha\beta}(x),
\eeq
where the subscript ``$c$'' stays for canonical and
\beqa
D^{\alpha\beta}\pi^a(x) &=& \frac{\partial}{\partial
  \epsilon_{\alpha\beta}}\pi^{\prime
  a}(x) \Big|_{\epsilon=0}=x^\alpha\partial^{\beta}\pi^a(x) - x^\beta\partial^{\alpha}\pi^a(x),
\nn
D^{\alpha\beta}N_i(x)&=&\frac{\partial}{\partial
  \epsilon_{\alpha\beta}}N_i^{\prime a}(x)\Big|_{\epsilon=0}=\left(\Sigma^{\alpha\beta}
+x^\alpha\partial^{\beta}-x^\beta\partial^{\alpha}\right)N_i(x).
\eeqa
In terms of the energy-momentum tensor
\beqa
T^{\mu\nu} = \frac{\partial {\cal L}_{\rm
    ChPT}}{\partial \partial_\mu\pi^a(x)}\partial^\nu \pi^a(x) +  \frac{\partial {\cal L}_{\rm
    ChPT}}{\partial \partial_\mu N_{i}(x)}\partial^\nu N_{i}(x)
+\partial^\nu \bar{N}_{i}(x)\frac{\partial {\cal L}_{\rm
    ChPT}}{\partial \partial_\mu \bar{N}_{i}(x)}- g^{\mu\nu} {\cal L}_{\rm
    ChPT},
\eeqa
${\cal M}_c^{\mu\alpha\beta}$ is given by
\beq
{\cal M}_c^{\mu\alpha\beta}=x^\alpha T^{\mu\beta}-x^\beta T^{\mu\alpha}
+ \frac{\partial
  {\cal L}_{\rm ChPT}}{\partial \partial_\mu N_i(x)}\Sigma^{\alpha\beta} N_i(x)
+\bar{N}_i(x)\Sigma^{\beta\alpha}\frac{\partial
  {\cal L}_{\rm ChPT}}{\partial \partial_\mu \bar{N}_i(x)}.
\eeq
Due to current conservation 
\beq
\partial_\mu{\cal M}_c^{\mu\alpha\beta}=0~,
\eeq
there are six conserved charges: the boost
\beq
K_c^j=\int d^3 x\, {\cal M}_c^{0j0}=\int d^3 x\left(x^j T^{0
    0}-x^0 T^{0j} +
\frac{\partial
  {\cal L}_{\rm ChPT}}{\partial \partial_0 N_i(x)}\Sigma^{j 0} N_i(x)+\bar{N}_i(x)\Sigma^{0 j}\frac{\partial
  {\cal L}_{\rm ChPT}}{\partial \partial_0 \bar{N}_i(x)}\right)
\eeq 
and the angular momentum 
\beq
J_c^k=\frac{1}{2}\epsilon^{i j k}\int d^3 x\, {\cal M}_c^{0 i
  j}=\frac{1}{2}\epsilon^{i j k}\int d^3 x 
\left(x^i T_c^{0 j} -x^j T_c^{0 i} + \frac{\partial
  {\cal L}_{\rm ChPT}}{\partial \partial_0 N_n(x)}\Sigma^{i j} N_n(x)
+\bar{N}_n(x)\Sigma^{j i}\frac{\partial
  {\cal L}_{\rm ChPT}}{\partial \partial_0 \bar{N}_n(x)}\right).
\eeq
In this notation, the boost appears to explicitly depend on the spin
$\Sigma^{0 j}$. One can, however, give a simpler form of the boost if
one uses the Belinfante energy-momentum tensor
\beq
\Theta^{\mu\nu}(x)=T^{\mu\nu}(x)+\frac{1}{2}\partial_\alpha X^{\alpha\mu\nu}(x),
\eeq
which is assumed to be symmetric under the interchange $\mu\leftrightarrow\nu$. The tensor
$X^{\alpha\mu\nu}$ has to be antisymmetric under the
$\alpha\leftrightarrow\mu$ interchange,
\beq
X^{\alpha\mu\nu}(x)=-X^{\mu\alpha\nu}(x),
\eeq
in order to maintain the current conservation relation
\beq
\partial_\mu \Theta^{\mu\nu}=\partial_\mu T^{\mu\nu} = 0~.
\eeq
The symmetry requirement under the interchange $\mu\leftrightarrow\nu$ gives
a constraint on the tensor $X^{\lambda\mu\nu}$
\beq
0=\Theta^{\mu\nu}(x)-\Theta^{\nu\mu}(x)=T^{\mu\nu}(x)-T^{\nu\mu}(x)+\frac{1}{2}\partial_\alpha\left(X^{\alpha\mu\nu}(x)-X^{\alpha\nu\mu}(x)\right).\label{EnergyMomentumSymmetricContstr}
\eeq
Due to the invariance of the Lagrangian under the transformation in 
Eq.~(\ref{LorentzInvarianceOfLagrangian}), we get
\beqa
T^{\mu\nu}-T^{\nu\mu}&=&-\frac{\partial
  {\cal L}_{\rm ChPT}}{\partial \partial_\alpha N_i(x)}\Sigma^{\mu
  \nu}\partial_\alpha N_i(x)
-\bar{N}_i(x)\overleftarrow{\partial}_\alpha\Sigma^{\nu \mu}\frac{\partial
  {\cal L}_{\rm ChPT}}{\partial \partial_\alpha \bar{N}_i(x)} - 
\frac{\partial
  {\cal L}_{\rm ChPT}}{\partial N_i(x)}\Sigma^{\mu
  \nu} N_i(x)
-\bar{N}_i(x)\Sigma^{\nu \mu}\frac{\partial
  {\cal L}_{\rm ChPT}}{\partial \bar{N}_i(x)}\nn
&=&
-\partial_\alpha\left(\frac{\partial
  {\cal L}_{\rm ChPT}}{\partial \partial_\alpha N_i(x)}\Sigma^{\mu
  \nu} N_i(x) +
\bar{N}_i(x)\Sigma^{\nu \mu}\frac{\partial
  {\cal L}_{\rm ChPT}}{\partial \partial_\alpha \bar{N}_i(x)} \right),\label{LorentzInvarianceOfLagrConstr}
\eeqa
where we have used the equation of motion in the second line. Matching
Eqs.~(\ref{EnergyMomentumSymmetricContstr}) and
(\ref{LorentzInvarianceOfLagrConstr}), we obtain the relation
\beq
X^{\alpha\mu\nu}-X^{\alpha\nu\mu}=2 \left(\frac{\partial
  {\cal L}_{\rm ChPT}}{\partial \partial_\alpha N_i(x)}\Sigma^{\mu
  \nu} N_i(x) +
\bar{N}_i(x)\Sigma^{\nu \mu}\frac{\partial
  {\cal L}_{\rm ChPT}}{\partial \partial_\alpha \bar{N}_i(x)} \right).\label{ConstraintOnXamunu}
\eeq
The ansatz
\beqa
X^{\alpha\mu\nu} &=& \frac{\partial
  {\cal L}_{\rm ChPT}}{\partial \partial_\alpha N_i(x)}\Sigma^{\mu
  \nu} N_i(x) - \frac{\partial
  {\cal L}_{\rm ChPT}}{\partial \partial_\mu N_i(x)}\Sigma^{\alpha
  \nu} N_i(x) + \frac{\partial
  {\cal L}_{\rm ChPT}}{\partial \partial_\nu N_i(x)}\Sigma^{\mu
  \alpha} N_i(x)\nn
&+&
\bar{N}_i(x)\Sigma^{\nu \mu}\frac{\partial
  {\cal L}_{\rm ChPT}}{\partial \partial_\alpha \bar{N}_i(x)}-
\bar{N}_i(x)\Sigma^{\nu \alpha}\frac{\partial
  {\cal L}_{\rm ChPT}}{\partial \partial_\mu \bar{N}_i(x)}
+\bar{N}_i(x)\Sigma^{\alpha \mu}\frac{\partial
  {\cal L}_{\rm ChPT}}{\partial \partial_\nu \bar{N}_i(x)}
\eeqa
solves Eq.~(\ref{ConstraintOnXamunu}). Given that
$X^{\alpha\mu\nu}=X^{\alpha\nu\mu}$ by construction,  only the
antisymmetric part survives on the right-hand side of Eq.~(\ref{ConstraintOnXamunu}). In terms
of $X^{\alpha\mu\nu}$, we can rewrite the angular momentum tensor as
\beq
{\cal M}_c^{\mu\alpha\beta}=x^\alpha T^{\mu\beta}-x^\beta
T^{\mu\alpha} + \frac{1}{2}\left(X^{\mu\alpha\beta}-X^{\mu\beta\alpha}\right).
\eeq
One can now redefine the angular momentum tensor as
\beqa
{\cal M}^{\mu\alpha\beta}&=&{\cal
  M}_c^{\mu\alpha\beta}+\frac{1}{2}\partial_\lambda\left(x^\alpha
  X^{\lambda\mu\beta}-x^\beta X^{\lambda \mu\alpha}\right)\nn
&=&
{\cal
  M}_c^{\mu\alpha\beta} +\frac{1}{2}\left(x^\alpha\partial_\lambda
  X^{\lambda\mu\beta}-x^\beta \partial_\lambda X^{\lambda\mu\alpha}\right)+
\frac{1}{2}\left(X^{\alpha\mu\beta}-X^{\beta\mu\alpha}\right),
\eeqa
which leads to 
\beq
{\cal M}^{\mu\alpha\beta}=x^\alpha \Theta^{\mu\beta}-x^\beta \Theta^{\mu\alpha}.
\eeq
The current ${\cal M}^{\mu\alpha\beta}$ is conserved since
\beq
\partial_\mu {\cal M}^{\mu\alpha\beta} = \frac{1}{2}\partial_\mu\partial_\lambda \left(x^\alpha
  X^{\lambda\mu\beta}-x^\beta X^{\lambda\mu\alpha}\right)=0.
\eeq
The last equation is valid since the tensor in the bracket is antisymmetric under 
the interchange $\lambda\leftrightarrow\mu$.
We obtain the time-independent boost function
\beq
K^j=\int d^3 x \left(x^j \Theta^{0
    0} - x^0 \Theta^{0j}\right)
\eeq
from the conserved angular momentum tensor ${\cal M}^{\mu\alpha\beta}$,
while the angular momentum function has the form
\beq
J^k=
\frac{1}{2}\epsilon^{i j k}\int d^3 x\, {\cal M}^{0 i
  j}=\frac{1}{2}\epsilon^{i j k}\int d^3 x 
\left(x^i \Theta^{0 j} -x^j \Theta^{0 i}\right).
\eeq
Notice that due to antisymmetry of $X^{\lambda\mu\beta}$ under the
interchange $\lambda\leftrightarrow\mu$, one gets
\beq
\partial_0\left(x^\alpha
  X^{0 0\beta}-x^\beta X^{0 0\alpha}\right)=0.
\eeq
For this reason, 
\beq
K^j=K_c^j\quad{\rm and}\quad J^k=J_c^k.
\eeq
We see that with the symmetric energy-momentum tensor, the definition
of the boost
function becomes more compact. This, however, is just a matter of
notation. In either case, at $x_0=0$ the boost operator has a contribution from
the Hamiltonian density $T^{0 0}$ and and from spin part proportional
to $\Sigma^{0 j}$.

Once we defined the boost function, we take the same definition in
quantum field theory by promoting fields to operators. In the final 
step, we need to block diagonalize the boost operator $K^j$ in Fock space. As
has been shown in Ref.~\cite{Kruger:1997aw,Gloeckle:1981js} the Hamiltonian and boost
operators get block diagonalized simultaneously by the Okubo unitary
transformation. The proof in Ref.~\cite{Kruger:1997aw} was, however, restricted to
Yukawa-like interactions that are linear in meson fields.
This is certainly not enough for chiral EFT. Fortunately, the algebraic
considerations of that work, which are based on commutation relations of the
Poincar\'{e} algebra, can be generalized to any interaction. We give
the somewhat lengthy proof of this statement in
Appendix~\ref{BlockDiagonalizationProof}. In order to calculate an effective
boost operator acting on the nucleonic part of the Fock space, we
follow our usual strategy by first applying  the Okubo transformation on the  
chiral boost operator and subsequently performing additional unitary transformations on the
$\eta$ space, which do not depend on the external sources. The resulting effective
boost operator is given by
\beq
K_{\rm eff}^j = \eta U_\eta^\dagger U_{\rm Okubo}^\dagger K^j U_{\rm
  Okubo}U_\eta^{} \eta. 
\eeq
Due to translational invariance of the unitary transformations $U_\eta$ and
$U_{\rm Okubo}$, we can write the boost as
\beq
\langle \alpha|K^j|\beta\rangle =
(2\pi)^3\left(i\,\frac{\partial}{\partial
    P_\alpha^j}\delta(\vec{P}_\alpha-\vec{P}_\beta)\right)
\langle\alpha|\eta U_\eta^\dagger U_{\rm Okubo}^\dagger \Theta^{0 0}(0) U_{\rm
  Okubo}^{}U_\eta \eta|\beta\rangle.
\eeq
Here $|\alpha\rangle$ and $|\beta\rangle$ are chosen to be eigenstates of the 
free Hamiltonian $H_0$, and the total momentum is an eigenvalue of the momentum operator
\beq
P^j|\alpha\rangle = P_\alpha^j |\alpha\rangle, \quad
P^j|\beta\rangle=P_\beta^j|\beta\rangle, \quad H_0 |\alpha\rangle =
E_\alpha |\alpha\rangle, \quad H_0 |\beta\rangle = E_\beta |\beta\rangle.
\eeq
In this notation, the boost and Hamiltonian matrix elements look very
similar. Indeed, the effective Hamiltonian is given by
\beq
\langle \alpha|H_0 + V|\beta\rangle =
(2\pi)^3\delta(\vec{P}_\alpha-\vec{P}_\beta)
\langle\alpha|\eta U_\eta^\dagger U_{\rm Okubo}^\dagger \Theta^{0 0}(0) U_{\rm
  Okubo}U_\eta \eta|\beta\rangle.
\eeq
So we see that both for the boost and Hamiltonian, we
need to calculate the matrix elements 
\beq
\langle\alpha|\eta U_\eta^\dagger U_{\rm Okubo}^\dagger \Theta^{0 0}(0) U_{\rm
  Okubo}^{}U_\eta^{} \eta|\beta\rangle.\label{MatrixElHamBoost}
\eeq
The only difference between the Hamiltonian and the boost is
  that the Hamiltonian matrix element is given by
  (\ref{MatrixElHamBoost}) multiplied with the momentum-conserving delta
  function, while the boost matrix element is given by
  (\ref{MatrixElHamBoost}) multiplied with  a derivative of the delta
  function.
A particularly convenient  way to calculate this matrix element is by 
introducing a coupling with an external source $\rho(x)$ in the
original chiral pion-nucleon Hamiltonian via 
\beq
H_\rho=\int d^3 x\, \rho(\vec{x}\,)\, \Theta^{0 0}(x)\Big|_{x_0=0}~.
\eeq
Taking the Fourier transform
\beq
\rho(\vec{x}\,)=\int d^3 q \exp(i\, \vec{q}\cdot \vec{x})\tilde\rho(\vec{q}\,),
\eeq
we can calculate
\beq
\frac{\delta H_\rho}{\delta \tilde \rho(\vec{q}\,)} = \int d^3 x \exp(i\, \vec{q}\cdot \vec{x})\, \Theta^{0 0}(x)\Big|_{x_0=0}.
\eeq
After application of a unitary transformation to this operator, we
obtain the desired matrix element by 
integrating over $\vec{q}$: 
\beq
\int \frac{d^3 q}{(2\pi)^3} \langle\alpha|\eta U_\eta^\dagger U_{\rm
  Okubo}^\dagger \frac{\delta H_\rho}{\delta \tilde \rho(\vec{q}\,)}U_{\rm
  Okubo}^{}U_\eta^{} \eta|\beta\rangle = 
\langle\alpha|\eta U_\eta^\dagger U_{\rm
  Okubo}^\dagger \Theta^{0 0}(0)U_{\rm
  Okubo}^{}U_\eta^{} \eta|\beta\rangle.\label{PracticalCalcOfBoostGeneral}
\eeq
For the calculation of the left-hand side, we proceed in a similar way
as for the calculation of the current operator with an incoming momentum 
$\vec{q}$. Replacing finally the momentum transfer by
\beq
\vec{q}=\vec{P}_\alpha - \vec{P}_\beta,
\eeq
and dropping the momentum-conserving delta function (which is
equivalent to performing an integration over $\vec{q}\,$) yields  the
right-hand side of Eq.~(\ref{PracticalCalcOfBoostGeneral}).

\subsection{Constraints on unitary phases}

The derived matrix elements for the current operators involve loop
integrals which are ultraviolet divergent and need to be
renormalized. This is carried out in the usual way by rewriting the bare LECs in terms of
the renormalized ones and the corresponding counterterms, which have
to cancel the ultraviolet (UV) divergences appearing in the loop
integrals. Contrary to the scattering amplitude which is evaluated
on shell, there is no guarantee that all UV divergences in the
expressions for the nuclear potentials and exchange charge and current
operators are cancelled by the counterterms.\footnote{When evaluating
  the on-shell scattering amplitude, the remaining
  divergences have to cancel against the ones emerging from iterations of
  the dynamical equation.} In fact, already the leading one-loop contributions
to the 3N force calculated by using the Okubo transformation cannot be
renormalized \cite{Epelbaum:2006eu}. Notice, however, that the
parametrization of the unitary transformation in terms of the
operator $\lambda A \eta$ corresponds to one particular choice of
basis in the Fock space. The unitary ambiguity of the nuclear forces
and currents can be systematically accounted for by invoking
additional unitary transformations on the $\eta$-subspace of the Fock
space. It was shown in \cite{Epelbaum:2006eu} by introducing six 
additional unitary transformations, expressed in terms of 
continuously varying parameters $\alpha_i$, $i=1, \ldots 6$, 
that the renormalization of the 3N force can be achieved for certain
choices of the phases $\alpha_i$.   Interestingly, the static part of the
resulting nuclear potentials at order $Q^4$ was found to be
independent on the unitary phases $\alpha_i$. On the other hand, the non-static
contributions to the 2N and 3N forces at the same order do exhibit
some degree of unitary ambiguity which manifests itself in the dependence on two
(arbitrary) phases $\bar \beta_{8,9}$, see \cite{Kolling:2011mt,Bernard:2011zr} for more details
and \cite{Friar:1994zz,Friar:1999sj} for an earlier discussion. 
Notice further that we had to invoke yet additional unitary
transformations depending on the external electromagnetic source and
leaving the results in the strong sector unaffected  in order to renormalize the one-loop
contributions to the one-pion exchange 2N current operator
\cite{Kolling:2011mt}. Similarly, for the axial charge and current
operators considered in this paper, we introduce all possible
unitary transformations on the $\eta$-subspace of the Fock space,
whose generators involve a single insertion of the external axial
source. Clearly, such transformations do not affect the results we
have obtained in the strong and electromagnetic sectors. In appendix~\ref{sec:appen}, 
we provide an explicit list of various unitary transformations 
which  may contribute to the axial charge and
current operators up to order $Q$  parametrized in terms
of $33$ phases $\alpha_{1, \ldots , 15}^{ax}$,  $\alpha_{16}^{ax, \rm
  LO}$, $\alpha_{16}^{ax, \rm  Static}$, $\alpha_{16}^{ax, 1/m}$,
$\alpha_{16}^{ax, \rm Tadpole}$,  $\alpha_{17, \ldots ,
  30}^{ax}$ and an additional phase $\beta_1^{ax}$ related to the
$d_{22}$-vertex with time derivative acting on the axial source as
detailed in appendix~\ref{timederivOfaxialSourceAppendix}. The very high degree of 
unitary ambiguity as compared to
the nuclear forces and electromagnetic currents at the same chiral
order can be traced back to the appearance of the pion-axial-source interaction with
$\kappa=-1$, which enters $30$ out of the $33$ generators listed in  
appendix~\ref{sec:appen}. Here and in what follows, we impose the
following three conditions on the phases of
the unitary transformations: 
\begin{enumerate}
\item We require the one-loop contributions to the axial-vector current to be
  expressible in the form of 4-dimensional integrals with heavy-baryon
  propagators. This requirement is necessary for factorizability of the
  exchanged pions which itself is necessary (but not sufficient) for the
  renormalization of the axial-vector current. It leads
  to the following $14$ constraints
\beqa
\alpha_{6}^{ax}&=&-\alpha_{4}^{ax},\nn
\alpha_{7}^{ax}&=&-\alpha_{4}^{ax},\nn
\alpha_{8}^{ax}&=&\alpha_{5}^{ax},\nn
\alpha_{9}^{ax}&=&-\alpha_{4}^{ax}+\alpha_{5}^{ax},\nn
\alpha_{11}^{ax}&=&\frac{1}{2}\big(\alpha_{4}^{ax}+\alpha_{5}^{ax}\big),\nn
\alpha_{13}^{ax}&=&\alpha_{4}^{ax}-\alpha_{5}^{ax}-\alpha_{7}^{ax}+\alpha_{10}^{ax},\nn
\alpha_{14}^{ax}&=&\frac{1}{2}\big(\alpha_{4}^{ax}-\alpha_{5}^{ax}\big),\nn
\alpha_{15}^{ax}&=&-\alpha_{4}^{ax} - \alpha_{7}^{ax} +
\alpha_{12}^{ax},\nn
\alpha_{20}^{ax}&=&-\alpha_{17}^{ax}-\alpha_{18}^{ax}-\alpha_{19}^{ax},\nn
\alpha_{24}^{ax}&=&-\alpha_{21}^{ax}-\alpha_{22}^{ax}-\alpha_{23}^{ax},\nn
\alpha_{27}^{ax}&=&\alpha_{25}^{ax},\nn
\alpha_{28}^{ax}&=&\alpha_{25}^{ax}-\alpha_{26}^{ax},\nn
\alpha_{29}^{ax}&=&\alpha_{25}^{ax}-\alpha_{26}^{ax},\nn
\alpha_{30}^{ax}&=&-\alpha_{26}^{ax}.\label{standard1}
\eeqa

\item The resulting expressions for the
axial charge and current operators are required to be properly
renormalized. For the one-loop contributions to
the one-pion exchange charge and current operators, this requires that
all UV divergences are cancelled by expressing the relevant bare LECs
$l_i$ from $\mathcal{L}_\pi^{(4)}$ and $d_i$ from $\mathcal{L}_{\pi N}^{(3)}$ in terms of the renormalized
ones via 
\beqa
l_i &=& l_i^r (\mu )  + \gamma_i \lambda  =: \frac{1}{16 \pi^2} \bar l_i
+ \gamma_i \lambda + \frac{\gamma_i}{16 \pi^2} \ln
\bigg(\frac{M_\pi}{\mu} \bigg)\,, \nn
d_i &=& d_i^r (\mu )  + \frac{\beta_i}{F^2}\lambda
=: \bar d_i + \frac{\beta_i}{F^2} \lambda + \frac{\beta_i}{16 \pi^2 F^2}\ln
\bigg(\frac{M_\pi}{\mu} \bigg)\,,\label{physicalcouplingsrenorm}
\eeqa 
where $\mu$ is the scale introduced by dimensional regularization. 
Further, the quantity $\lambda$ is defined as 
\beq
\lambda = \frac{\mu^{d-4}}{16 \pi ^2}
\bigg( \frac{1}{d-4} + \frac{1}{2} ( \gamma_{\rm E} - \ln 4 \pi - 1) \bigg)\,,
\eeq
with $\gamma_{\rm E} = - \Gamma ' (1) \simeq 0.577$  the Euler constant and $d$ 
the number of space-time dimensions. The
$\beta$- and $\gamma$-functions appearing in the above expressions are
well known \cite{Gasser:1987rb,Gasser:2002am}. For the sake of completeness, we list below the
expressions relevant to our calculations: 
\beqa
\gamma_3 &=& - \frac{1}{2}\,, \nn 
\gamma_4 &=& 2 \,, \nn
\beta_2 =  -2 \beta_5 = \frac{1}{2} \beta_6 &=& - \frac{1}{12}  (1 + 5 g_A^2)\,, \nn
\beta_{15} =  \beta_{18} = \beta_{22} = \beta_{23} &=& 0 \,, \nn
\beta_{16} &=&  \frac{1}{2} g_A + g_A^3 \,.
\eeqa
A cancellation of the UV divergences in the loop
contributions to the single nucleon and to the one-pion exchange axial charge operators by the
counterterms fixed by the $\beta$-functions listed above  yields 
additional constraints on the unitary phases: 
\beqa
\alpha_{10}^{ax}+\alpha_{11}^{ax}&=&-\frac{1}{2},\nn
\alpha_{16}^{ax,{\rm Tadpole}}\,=\,\alpha_{16}^{ax,{\rm Static}}
&=&\alpha_{16}^{ax,{\rm LO}}\,=\,-1. \label{standard2Prelim}
\eeqa
A more general requirement of the cancellation of power-low divergences in
addition to the logarithmic ones, which are taken care of in
dimensional regularization,  implies that the current should 
be renormalizable in $d=3$ dimensions. This yields   an additional constraint 
\beq
\alpha_1^{ax}=0.\label{ax1const}
\eeq
Thus, in total, we obtain $19$ constraints from the
requirement that the axial-vector current is renormalizable.

\item 
In addition to the renormalizability constraints specified above, we
require the irreducible two- and three-nucleon pion-production
amplitudes appearing in the expressions for the current operators to
match the corresponding ones which appear in the expressions for the three-
and four-nucleon forces, respectively, as  visualized
schematically earlier in Fig.~\ref{fig:factor}.  More precisely, we require
the following matching condition between the one-pion exchange contributions to the nuclear
forces and the corresponding axial current operator at the pion pole: 
\beq
\lim_{k^2 \to -M_\pi^2} (k^2 + M_\pi^2) \, 
W \big|_{k^2=-M_\pi^2}= \lim_{k^2 \to -M_\pi^2} (k^2 + M_\pi^2) \, \tilde A_\mu^{b}(k)\tilde a^{\mu b}(k).
\eeq 
Here, $k \equiv | \vec k \,|$ refers to the momentum of the exchanged pion,
while $k \cdot\tilde a^{b}(k)$ denotes the leading pion-nucleon vertex rather
than the leading axial-source-pion interaction. This
matching condition does not only represent a very natural choice of 
unitary phases as it makes the consistency between nuclear forces and
current operators explicit, but is also expected to be advantageous from the practical point of 
view. In particular, it allows one to regularize the pion-pole
contributions to the current operators in the way consistent with the
regularization of the nuclear potentials. 
Notice further that
in the actual
calculation, the matching requirement, in combination with the
renormalizability condition, leads to the vanishing
of some pion-pole terms in the modified charge, i.e.~in the
quantity which enters the commutator on the left-hand side of 
Eq.~(\ref{continuityeqmomspace}),
\beq
\tilde{\fet A}_0(\vec{k},0) - \frac{\partial}{\partial k_0}\vec{k}\cdot\vec{\tilde{\fet
  A}}(\vec{k},k_0) + \frac{\partial}{\partial k_0}\,\big[W,\tilde{\fet
A}_0(\vec{k},k_0)\big]+ m_q \,i\frac{\partial}{\partial
k_0}\tilde{\fet P}(\vec{k},k_0)\Big. \,.\label{modifiedchargedef}
\eeq
More precisely, we see that after the requirement of the
renormalizability and matching to the 3N force, the modified
charge 
has no pion poles in the static
limit. With the same constraints on the unitary phases, relativistic
corrections to the modified charge have no  second-order pion poles,
i.e.~no contributions proportional to $1/(k^2+M_\pi^2)^2$ (but do have
simple poles). 
For this reasons, the continuity Eq.~(\ref{continuityeqmomspace}) 
simplifies for this choice of the unitary phases.

The required matching conditions lead to a number of constraints on
the unitary phases 
which are summarized below. 
First, matching the two-pion-exchange two-nucleon current operator to the
  corresponding two-pion-one-pion contributions to the
  three-nucleon force (3NF) at N$^3$LO yields the three constraints
\beqa
\alpha_{10}^{ax}&=&1-2\alpha_{4}^{ax}+\alpha_{5}^{ax},\nn
\alpha_{26}^{ax}&=&1-\alpha_{25}^{ax}.\nn
\alpha_{12}^{ax}&=&0. \label{standard4}
\eeqa
Next, matching the $1/m$-corrections to the two-nucleon
current involving the contact interactions  with the
corresponding N$^3$LO three-nucleon forces gives an additional constraint
\beqa
\alpha_{23}^{ax}&=&1-\alpha_{21}^{ax}-\alpha_{22}^{ax}. \label{standard5}
\eeqa
Finally, matching the $1/m$-corrections of the one-pion exchange axial-vector current with
the corresponding $1/m$-corrections to the N$^3$LO
3NFs gives two more constraints:
\beqa
\alpha_{17}^{ax}+\alpha_{18}^{ax}+\alpha_{19}^{ax}&=&\frac{1}{2}\Big(1+2\bar{\beta}_8\Big),\nn
\alpha_{16}^{1/m}&=&-\frac{1}{2}\Big(1+2\bar{\beta}_9\Big).\label{standardoneoverm}
\eeqa

As detailed in the previous sections, the expressions for the charge
and current operators do, in general, involve contributions which
depend on the energy transfer $k_0$ and arise 
 from the explicit time
dependence of the employed unitary transformations. 
In the Breit-frame with $k_0=0$, any choice of the 
non-locality for the nuclear force, which at  N$^3$LO is parametrized
in terms of the phases $\bar{\beta}_8$ and
$\bar{\beta}_9$, is consistent with the matching condition. In a
general frame, one encounters pion-pole contributions to the
axial-vector 2N
current operator proportional to the energy transfer $k_0$, which 
match  the corresponding $1/m$-corrections of the 3NFs only if one chooses
\beq
\bar{\beta}_8=\frac{1}{2}.\label{beta8constr}
\eeq
Similarly, the $k_0/m$-contributions
to the single-nucleon current can be required to match the
corresponding $1/m$-corrections of
one-pion exchange NN potential. This finally leads to
\beq
\bar{\beta}_9=\frac{1}{2}.\label{beta9constr}
\eeq
Note that with the last constraint, all $\alpha_{16}^{ax}$ phases are
equal 
\beqa
\alpha_{16}^{ax,{\rm Tadpole}}&=&\alpha_{16}^{ax,{\rm Static}}\,=\,\alpha_{16}^{ax,{\rm LO}}\,=
\,\alpha_{16}^{ax,1/m}\,=\,-1. \label{standard2}
\eeqa
In the following sections, we will show results for an arbitrary choice of
$\bar{\beta}_8$ and $\bar{\beta}_9$. One should, however, keep in mind
that the relativistic corrections to the axial current operator
proportional to the energy transfer $k_0$ match 
the corresponding nuclear forces only for the particular choice of the
phases $\bar \beta_8$ and  $\bar \beta_9$ specified in  Eqs.~(\ref{beta8constr})
and (\ref{beta9constr}).  
Matching all remaining pion-pole contributions to the nuclear axial
current operator, not explicitly mentioned above, with the
corresponding terms in the nuclear potentials does not introduce
any additional constraints on the phases. 
Here and in what follows, Eqs.~(\ref{standard1}), (\ref{standard2Prelim}),
(\ref{ax1const}), (\ref{standard4}), (\ref{standard5}) and
(\ref{standardoneoverm}) will be referred to as our standard choice of the
unitary phases. 
\end{enumerate}
With these constraints the currents have no further ambiguities (with
respect to the considered class of unitary transformations), and the expressions
become unique modulo the phases $\bar \beta_8$ and $\bar \beta_9$. 
Notice further that the continuity equation (\ref{continuityeqmomspace}),
being a manifestation of the symmetry, is valid independently on the choice of the
unitary phases and thus does not lead to any constraints. Similarly,
Eq.~(\ref{FourVectorCondMomentumSpaceOnShell}) represents a constraint on the axial current operator due
to Poincar\'{e} invariance and is valid for all choices of the unitary
phases, too. On the other hand, one may ask whether the unitary phases can
be chosen in such a way, that Eq.~(\ref{FourVectorCondMomentumSpaceOnShell}) remains valid off the
energy shell, i.e.~the operator equation (\ref{FourVectorCondMomentumSpace}) holds true for $\fet
X_\mu = 0$. While we found this indeed to be possible, the resulting
constraints on the unitary phases appear to be incompatible with the
renormalizability requirement.
Both Eqs.~(\ref{continuityeqmomspace}) and (\ref{FourVectorCondMomentumSpaceOnShell}) represent extremely useful tools to
check the expressions for the derived operators. We have explicitly
verified that these equations are satisfied for the axial charge and
current operators we have derived, see section \ref{sec:results} for a detailed
discussion of the continuity equation.

\section{Notation for the current operators}
\def\theequation{\arabic{section}.\arabic{equation}}
\label{sec:Notation}

Throughout this work, we employ the standard nuclear physics
convention based on the nonrelativistic normalization for the
nucleon states
\beq
\langle \pvec p ' | \vec
p  \, \rangle = \delta(\pvec{p} ' - \vec{p}\,).\label{nonrelativisticnorm}
\eeq
In relativistic quantum field theory calculations, one
usually uses 
\beq
\langle  p' | 
p  \rangle = (2\pi)^3 2 E_p\delta(\pvec{p}^\prime - \vec{p}\,),
\eeq
with the nucleon energy given by 
\beq
E_p=\sqrt{\pvec{p}^2 + m^2},
\eeq
and factorizes the
momentum conservation term $(2\pi)^3\delta({\rm sum\, of\, incoming\,
  momenta})$ out of the  connected contributions to the scattering
amplitude. Using the normalization as given in Eq.~(\ref{nonrelativisticnorm}), it is
convenient to factorize the factors of $(2 \pi )^{-3A+3}$ out of the
expressions for $A$-body operators. Specifically, we define 
\beqa
\langle \pvec p_1 ' | \hat A^{\mu ,  a}_{\rm 2N} | \vec
p_1  \rangle &=:& \delta ( \pvec p_1 '  - \vec p_1
- \vec k )    A^{\mu ,  a}_{\rm 1N} \,, \nn
\langle \pvec p_1 ' \pvec p_2 ' | \hat A^{\mu ,  a}_{\rm 2N} | \vec
p_1 \vec p_2 \rangle &=:& (2\pi)^{-3}\delta ( \pvec p_1  ' + \pvec p_2 ' - \vec p_1
- \vec p_2 - \vec k )    A^{\mu ,  a}_{\rm 2N} \,, \nn
  \langle \pvec p_1 ' \pvec p_2 ' \pvec p_3 ' | \hat A^{\mu ,  a}_{\rm 3N} | \vec
p_1 \vec p_2 \vec p_3\rangle &=:& (2\pi)^{-6}\delta ( \pvec p_1  ' + \pvec p_2 ' + \pvec p_3 '- \vec p_1
- \vec p_2 - \vec p_3 - \vec k )  A^{\mu ,  a}_{\rm 3N} \,,
\eeqa
where $\vec p_i$ ($\pvec p_i '$) denotes the incoming (outgoing)
momentum of nucleon $i$, $\vec k$ is the momentum of the external
axial source while $a$ is an isospin index. Further, $\hat { \fet  A}^\mu_{\rm nN}$ means
that $\fet A^\mu_{\rm nN}$ is to be understood as an $n$-nucleon \emph{operator}.
Notice that the phase factors of $1/\sqrt{2
  E_p}$ for every nucleon field with momentum $p$ are kept in the
expressions for the currents $\fet A^\mu$.  This is particularly
important for the calculation of $1/m^2$-corrections to the single-nucleon
current. Here and in what follows, we suppress the ``$\sim$'' over
$\fet A^\mu$ to simplify the notation. It should be understood that
all expressions for the axial charge and current
operators given in the next sections are defined in momentum space according to
Eq.~(\ref{CurrentsDefMomentumSpace}). 
We will also use the same notation for the 
pseudoscalar current operators $\fet P_{\rm nN}$. 
In the following, we derive the contributions to the single-, two- and
three-nucleon operators  $A^{\mu ,
  a}_{\rm 1N} \equiv (A^{0 ,  a}_{\rm 1N}, \; \vec A^{a}_{\rm 1N})$, 
$A^{\mu ,
  a}_{\rm 2N} \equiv (A^{0 ,  a}_{\rm 2N}, \; \vec A^{a}_{\rm 2N})$
and $A^{\mu ,  a}_{\rm 3N} \equiv (A^{0 ,  a}_{\rm 3N}, \; \vec A^{a}_{\rm
  3N})$ up to order $Q$ in the chiral expansion.   

Last but not least, it is important to specify the
dynamical equation since it affects the form of the relativistic
corrections to both the current operators and nuclear forces \cite{Friar:1999sj}. Here and
in what follows, we employ the $A$-nucleon Schr\"odinger equation with the
relativistic expression for the kinetic-energy term,
i.e.:\footnote{Using the relativistic expression for the kinetic
  energy instead of its $1/m$-expanded form is a matter of practical
  convenience. 
In the power counting scheme we employ, the standard nonrelativistic
  approximation of the kinetic energy is valid up-to-and-including
  N$^2$LO, i.e.~order-$Q^3$ relative to the dominant terms in the
  nuclear force and current operator.
} 
\beq
\label{SE}
\bigg[\sum_i \Big(\sqrt{p_i^2 + m_i^2} - m_i \Big) + V  \bigg]\Psi = E \Psi\,,
\eeq
where $p_i \equiv |\vec p_i \,|$ and $m_i$ are the three-momentum  and
mass of the nucleon $i$. Notice that in the two-nucleon sector, it is
customary to rewrite this expression in a form which resembles that
of the usual nonrelativistic equation. This, in fact, is the
convention adopted in the new chiral potentials of
Refs.~\cite{Epelbaum:2014efa,Epelbaum:2014sza}. Here and in what
follows, we refrain from using this convention since we will also consider
three-nucleon operators.

\section{Single-nucleon axial charge and current operators}
\def\theequation{\arabic{section}.\arabic{equation}}
\label{sec:singleN}

We now turn to the derivation of the contributions to the nuclear
axial charge and current operators using the MUT and discuss the various
classes of the contributions grouped according to the number of
nucleons involved and the range of the interaction. 
To facilitate a comparison with calculations using
different methods, we will explicitly indicate the dependence of the various
terms upon the unitary phases, even though our final expressions are
only given for the standard choice corresponding to Eqs.~(\ref{standard1}), (\ref{standard2Prelim}),
(\ref{ax1const}), (\ref{standard4}), (\ref{standard5}) and
(\ref{standardoneoverm}).

\begin{figure}[tb]
\vskip 1 true cm
\includegraphics[width=0.2\textwidth,keepaspectratio,angle=0,clip]{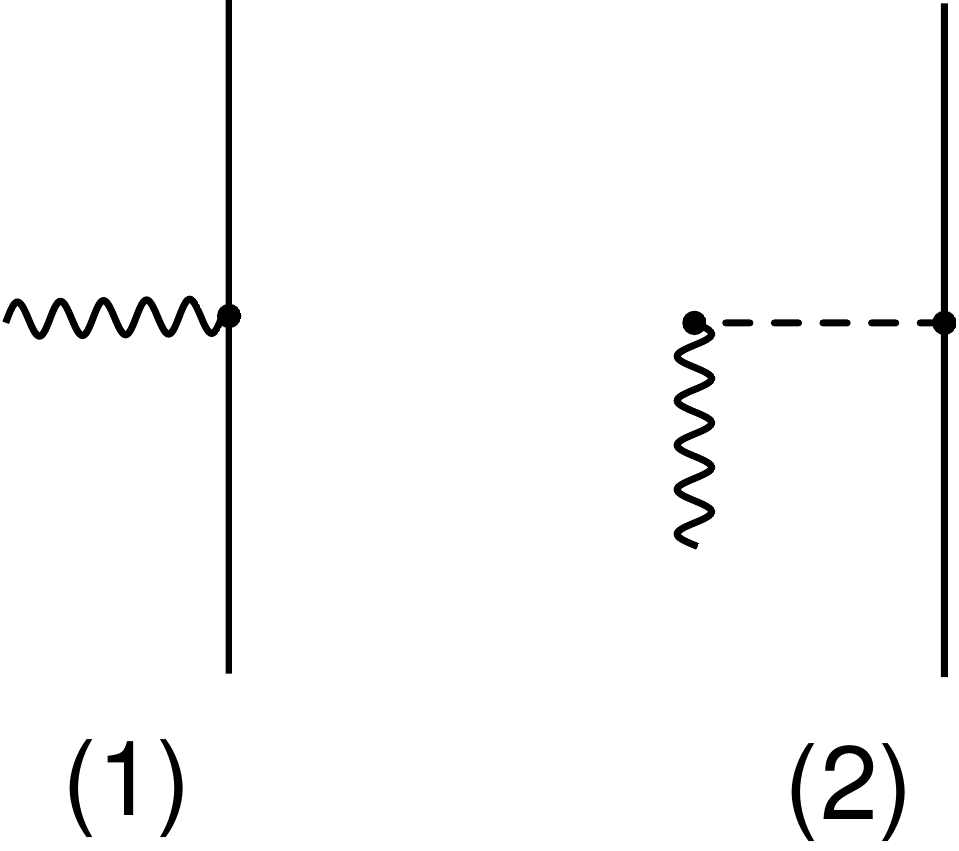}
    \caption{Diagrams generating the lowest-order contribution $\vec
      {\fet A}_{\rm 1N}^{(Q^{-3})}$. Diagram (2) also
      contributes to ${\fet A}_{\rm 1N}^{0 \; (Q^{-1})}$ as explained
      in the text. Solid dots denote vertices from
        ${\cal L}_{\pi N}^{(1)}$. Diagrams resulting from the application of the time reversal operation
      are not shown. 
For remaining notation see Fig.~\ref{fig:factor}.
\label{fig:singleNuclLO} 
 }
\end{figure}

The first contribution to the single-nucleon axial vector current
appears at order
$Q^{-3}$ from the well-known diagrams shown in Fig.~\ref{fig:singleNuclLO}.
The chiral order of the corresponding contributions can be read off
from Eq.~(\ref{PCkappa}) by noticing that the leading-order $a_\mu NN$,
$a_\mu \pi$ and $\pi NN$ vertices have, according to Eq.~(\ref{DefKappa}), the dimensions of $\kappa = 0$,
$\kappa = -1$ and $\kappa = 1$, respectively.
While there is no dependence on unitary phases at order $Q^{-3}$,
diagram $(2)$ of Fig.~\ref{fig:singleNuclLO} also generates a
contribution  to the charge operator 
at order $Q^{-1}$:
\beq
(2)\sim k_0 \frac{\alpha_{16}^{ax, {\rm LO}}}{k^2+M_\pi^2}.
\eeq
Here and in what follows, $\sim X$ means that the corresponding
expressions involve terms proportional to $X$. 
If $k_0$ would be counted as a quantity  of order $Q$, the
contribution of this diagram would appear 
already at leading order. However, on shell, we associate $k_0$ with
the difference of kinetic energies such that it counts as order $Q^3$ 
in our scheme. Thus, diagram $(2)$ starts to contribute  at
order $Q^{-1}$. The well-known leading-order result for the axial charge and
current operators has the form
\beqa
A^{0,  a \, (Q^{-3})}_{{\rm 1N: \, static}}  &=& 0,\nn
\vec{A}^{ a \, (Q^{-3})}_{{\rm 1N: \, static}}  &=&
-\frac{g_A}{2}\tau_i^a\bigg(\vec{\sigma}_i - \frac{\vec{k}\,
  \vec{k}\cdot\vec{\sigma}_i}{k^2+M_\pi^2}\bigg).\label{LOSingleNCurrentExpr}
\eeqa
Notice that in the notation we are using, the chiral dimension of an
$n$-nucleon contribution to the current operator can be easily read
off from the corresponding expression by simply counting the powers of the
soft scale (i.e.~the three-momenta of the nucleons and external
sources and $M_\pi$) and adding to the resulting dimension the factor
of $3 (n-2)$ to account for
the different normalization of $n$-nucleon states. 

There are only vanishing contributions at order $Q^{-2}$:
\beq
A^{0,  a \, (Q^{-2})}_{{\rm 1N: \, static}}  = 0,\quad \quad
\vec{A}^{ a \, (Q^{-2})}_{{\rm 1N: \, static}}  = 0.
\eeq

\begin{figure}[tb]
\vskip 1 true cm
\includegraphics[width=0.53\textwidth,keepaspectratio,angle=0,clip]{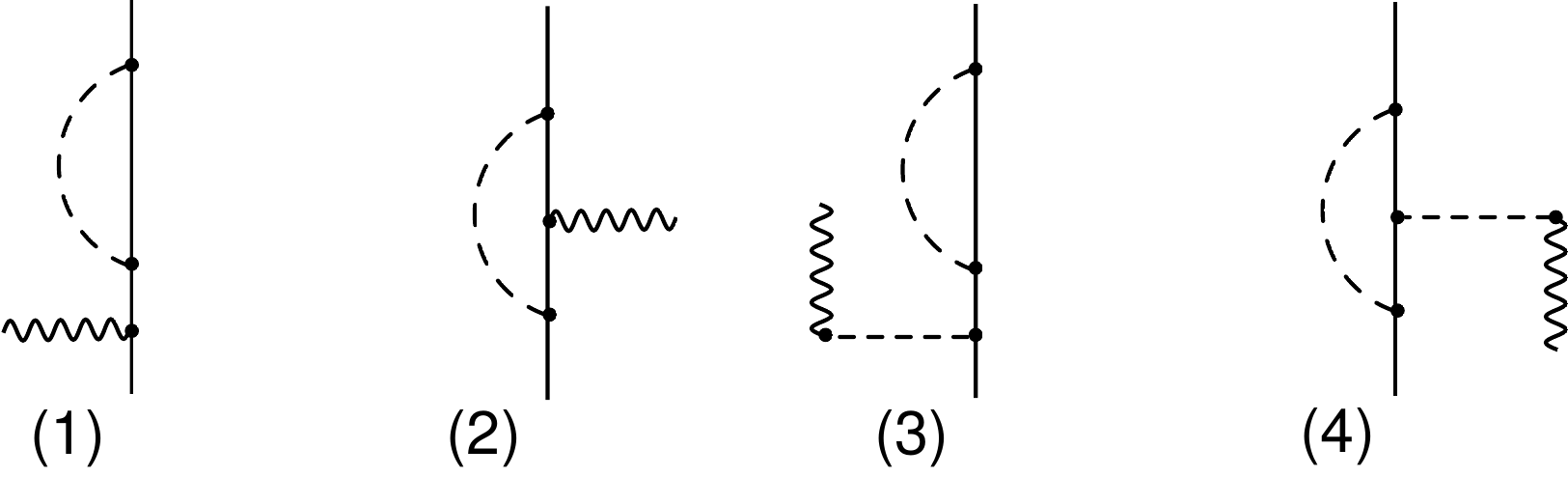}
    \caption{One-loop diagrams which yield non-vanishing contributions
      to $\vec {\fet A}_{\rm 1N}^{(Q^{-1})}$ and ${\fet A}_{\rm 1N}^{0 \; (Q)}$. For notation see Fig.~\ref{fig:singleNuclLO}.
\label{fig:singleNuclOneLoop} 
 }
\end{figure}

\begin{figure}[tb]
\vskip 1 true cm
\includegraphics[width=0.3\textwidth,keepaspectratio,angle=0,clip]{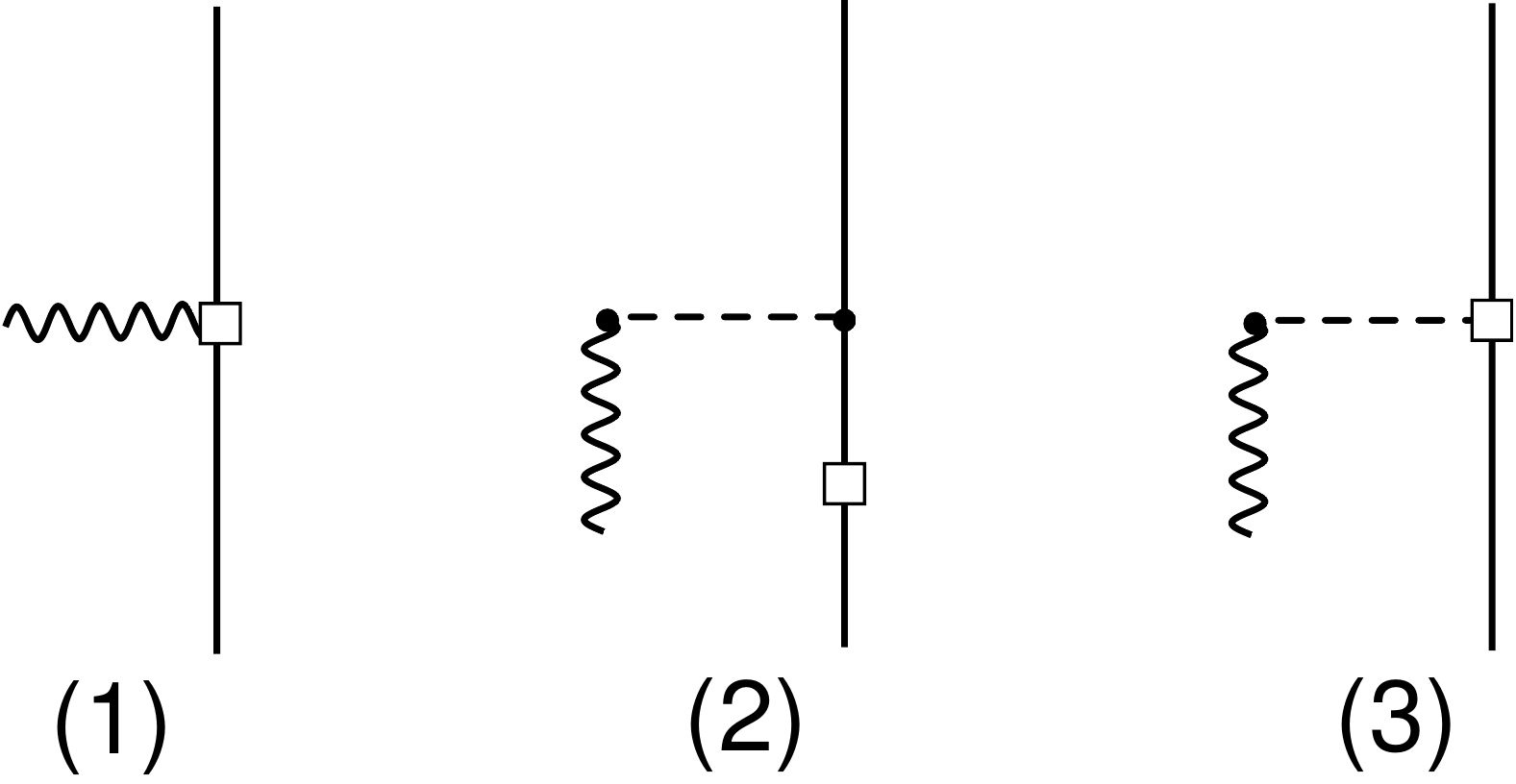}
    \caption{Relativistic $1/m$-contributions to  ${\fet A}_{\rm
        1N}^{0 \; (Q^{-1})}$ and $\vec
      {\fet A}_{\rm 1N}^{(Q)}$. Solid dots (open rectangles) denote vertices from
        ${\cal L}_{\pi N}^{(1)}$ ($1/m$-corrections from ${\cal
          L}_{\pi N}^{(2)}$). For remaining notation see Fig.~\ref{fig:singleNuclLO}.
\label{fig:singleNucleonOneOvermNucl} 
 }
\end{figure}

At order $Q^{-1}$, we encounter three kinds of
corrections. First, as already mentioned, there are terms emerging
from the time-dependence of unitary transformations in
diagram $(2)$ of Fig.~\ref{fig:singleNuclLO}, which have the form 
\beqa
\label{SingleNChargeUTPrimeQToMinus1}
A^{0,  a \, (Q^{-1})}_{{\rm 1N: \, UT^\prime}}  &=& \frac{g_A}{2} 
\frac{k_0}{k^2+M_\pi^2}\vec{k}\cdot\vec{\sigma}_i \tau_i^a,\\
\vec{A}^{ a \, (Q^{-1})}_{{\rm 1N: \, UT^\prime}}  &=&0.
\eeqa
Secondly, we encounter the leading one-loop contributions together
with tree-level diagrams with a single insertion of 
the ${\cal L}_{\pi N}^{(3)}$ vertices. There are no loop contributions
to the charge operator at this order. 
More precisely, the leading one-loop contributions to the
charge operator turn out to be proportional to the energy transfer and are thus
shifted to order $Q$. In Fig.~\ref{fig:singleNuclOneLoop}, we show the
non-vanishing loop diagrams that contribute
to the current (charge) operator at order $Q^{-1}$ ($Q$).
The explicit dependence on the unitary
phases for the charge operator is given by 
\beq
(3),(4)\sim
k_0\frac{\alpha_{10}^{ax}+\alpha_{11}^{ax}-\alpha_{13}^{ax}}{k^2+M_\pi^2}.
\label{temp1}
\eeq
\begin{figure}[tb]
\vskip 1 true cm
\includegraphics[width=1.0\textwidth,keepaspectratio,angle=0,clip]{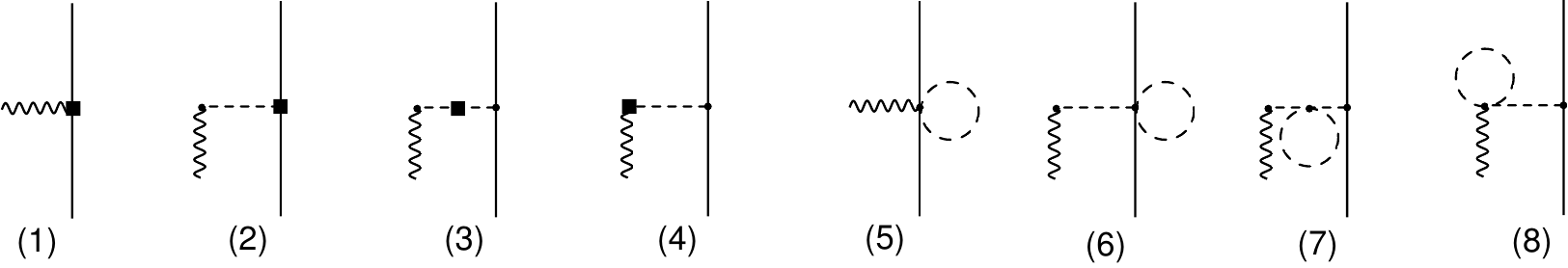}
    \caption{Tree-level and tadpole diagrams yielding non-vanishing
      contributions to ${\fet A}_{\rm 1N}^{\mu \; (Q^{-1})}$.  Solid dots (filled squares) denote vertices from
        ${\cal L}_{\pi N}^{(1)}$ or ${\cal L}_{\pi \pi}^{(2)}$ (${\cal
          L}_{\pi N}^{(3)}$ or ${\cal L}_{\pi \pi}^{(4)}$). 
For remaining notation see Fig.~\ref{fig:singleNuclLO}.
\label{fig:singleNuclCounterTerms} 
 }
\end{figure}
One-loop contributions to the current operator are, on the other hand,
independent of the unitary phases. Further, in
Fig.~\ref{fig:singleNuclCounterTerms}, we show tree-level and tadpole
diagrams which generate nonvanishing contributions to the current
operator, which turn out to be independent on the unitary phases. The final result 
for the charge and
current operators at order $Q^{-1}$ in the static limit is given by
\beqa
A^{0,  a \, (Q^{-1})}_{{\rm 1N: \, static}}  &=& 0,\nn
\vec{A}^{a \, (Q^{-1})}_{{\rm 1N: \, static}}  &=&
\frac{1}{2}\bar{d}_{22}\Big(\vec{\sigma}_i k^2-\vec{k}\,
\vec{k}\cdot\vec{\sigma}_i\Big)\tau_i^a - \bar{d}_{18} M_\pi^2\tau_i^a
\frac{\vec{k}\,\vec{k}\cdot\vec{\sigma}_i}{k^2+M_\pi^2} \,.\label{SingleNQtoMinus1StaticExpr}
\eeqa
Finally, we have to take into account the leading relativistic
$1/m$-corrections which in our counting scheme start contributing at 
order $Q^{-1}$. The corresponding diagrams are shown
in Fig.~\ref{fig:singleNucleonOneOvermNucl}. Diagram (1) contributes
only to the charge operator at order $Q^{-1}$. On the other hand, diagram (2) 
contributes to the charge at order $Q^{-1}$ but also yields a
correction to  the current operator which is proportional to $k_0$
and, therefore, contributes at 
order $Q$. We find the following dependence of the
resulting contributions on the unitary phases for the charge
\beq
(2)\sim\frac{1+\alpha_{16}^{ax, {\rm LO}}}{k^2+M_\pi^2},
\eeq  
and current operator
\beqa
(2)&\sim&
\frac{k_0}{m}\frac{\alpha_{17}^{ax}+\alpha_{18}^{ax}+\alpha_{19}^{ax}}{(k^2+M_\pi^2)^2},\nn
(3)&\sim&
\frac{k_0}{m}\frac{\alpha_{16}^{ax, 1/m}}{k^2+M_\pi^2}. \label{k0OvermNuclUnitaryPhases}
\eeqa  
The final result for the $1/m$ corrections to the 1N axial-vector current
for our standard choice of unitary phases reads
\beqa
A^{0,  a \, (Q^{-1})}_{{\rm 1N:} \, 1/m}  &=& - \frac{g_A}{2 m }
\tau_i^a \, \vec \sigma_i \cdot \vec k_i \,,\label{SingleNChargeOneOvermQToMinus1}\\
\vec{A}^{ a \, (Q^{-1})}_{{\rm 1N: \, 1/m}}  &=& 0\,,
\eeqa
where 
\beq
\vec{k}=\vec{p_i}^{\prime} - \vec{p}_i, \quad \vec{k}_i=\frac{\vec{p_i}^{\prime} + \vec{p}_i}{2}.
\eeq
This completes the derivation of the 1N terms at order  $Q^{-1}$. 

There are no corrections to the 1N charge and current operators at
order $Q^0$. In particular, the absence of $c_i/m$-corrections and 
$k_0$-dependent contributions can be understood from the fact that
there are no order-$Q^{-2}$ terms while the contributions of one-loop diagrams 
with a single insertion of subleading interactions vanish after
renormalization. 
\begin{figure}[tb]
\vskip 1 true cm
\includegraphics[width=0.6\textwidth,keepaspectratio,angle=0,clip]{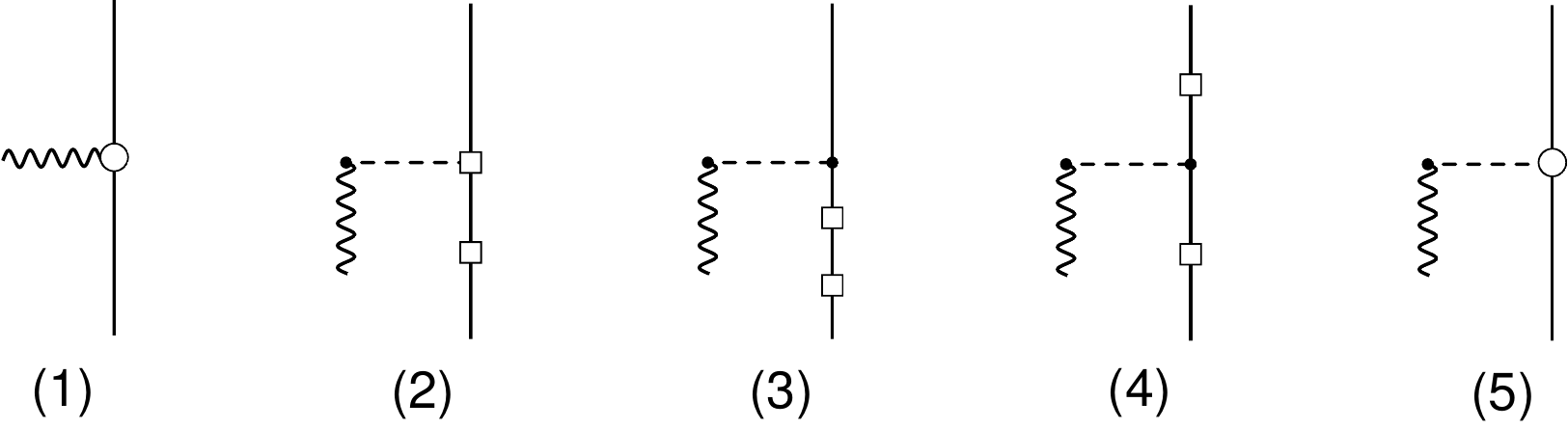}
    \caption{Relativistic $1/m^2$-contributions to $\vec
      {\fet A}_{\rm 1N}^{(Q)}$.  Solid dots (open rectangles) denote vertices from
        ${\cal L}_{\pi N}^{(1)}$ ($1/m$-corrections from ${\cal
          L}_{\pi N}^{(2)}$). Open circle denotes $1/m^2$-corrections from ${\cal
          L}_{\pi N}^{(3)}$. For remaining notation see Fig.~\ref{fig:singleNuclLO}.
\label{fig:singleNuclOneOvermNuclTo2} 
 }
\end{figure}

Finally, there are various contributions at order $Q$. We begin
with terms proportional to $k_0/m$ which emerge from diagrams $(2)$ and
$(3)$ of  Fig.~\ref{fig:singleNucleonOneOvermNucl}. Their dependence
on unitary phases is given by Eq.~(\ref{k0OvermNuclUnitaryPhases}). 
For the standard choice of unitary phases, the explicit expressions
have the form   
\beqa
A^{0,  a \, (Q)}_{{\rm 1N:\,}1/m, {\rm UT^\prime}}  &=&0,\nn
\vec{A}^{ a \, (Q)}_{{\rm 1N:\,}1/m, {\rm UT^\prime}}  &=&
-\frac{g_A k_0}{8
  m}\frac{\vec{k}}{k^2+M_\pi^2}\tau_i^a
\bigg(
2(1 + 2 \,\bar{\beta}_9)\vec \sigma_i \cdot \vec k_i - (1 + 2 \,\bar{\beta}_8)\vec{k}\cdot\vec{\sigma}_i\frac{p_i^{\prime\,2}-p_i^2}{k^2+M_\pi^2}\bigg).\label{k0OvermNuclcorr}
\eeqa
For the static part which is proportional to $k_0$, we get
nonvanishing contributions from diagrams $(3)$ and $(4)$ of
Fig.~\ref{fig:singleNuclOneLoop} as well as from the diagrams shown in 
Fig.~\ref{fig:singleNuclCounterTerms}. For the standard choice of unitary
phases we obtain after renormalization 
\beqa
\label{SingleNChargeStaticUTPrimeQTo1}
A^{0,  a \, (Q)}_{{\rm 1N:\,static,\, UT^\prime}}  &=&-k_0  \frac{\tau_i}{2}
\vec{k}\cdot \vec{\sigma}_i \bigg[\bar{d}_{22}+ \frac{2
\bar{d}_{18} M_\pi^2}{k^2+M_\pi^2}\bigg]\,,\\
\vec{A}^{ a \, (Q)}_{{\rm 1N:\,static,\, UT^\prime}}  &=& 0.
\eeqa
So far, the $k_0$ dependence in our expressions emerged entirely from
the time dependence of the unitary
transformations. The contribution proportional to $\bar{d}_{22}$ in 
Eq.~(\ref{SingleNChargeStaticUTPrimeQTo1}), however, originates 
directly from the interaction term
\beq
-d_{22} N^\dagger S_\mu
\Big(\partial^2\fet{a}^{\mu}-\partial^\mu\partial_\nu\fet{a}^\nu\Big)\cdot
\fet{\tau} N=d_{22} N^\dagger S_\mu
(\partial^\mu\dot{\fet{a}}^0)\cdot{\fet \tau} N+ \ldots\,,
\eeq
of the Lagrangian ${\cal L}_{\pi N}^{(3)}$ when evaluating the
corresponding Feynman diagram. In the formulation
presented so far, we have not included time-derivatives of the axial
vector source in the interaction. Rather, their time-derivatives were eliminated
from the action by performing partial integration in time. In
Appendix~\ref{timederivOfaxialSourceAppendix}, we 
show that the term $\propto \bar d_{22}$ indeed emerges from a
corresponding unitary transformations
if the time-derivative of the axial source is eliminated via partial time-integration.

The second class of order-$Q$ contributions involves relativistic
$1/m^2$-corrections. The corresponding
non-vanishing diagrams are visualized in
Fig.~\ref{fig:singleNuclOneOvermNuclTo2}. They contribute only to the
current operator. 
The dependence on the 
unitary phases of the contributions from diagrams 
in Fig.~\ref{fig:singleNuclOneOvermNuclTo2} is given by
\beqa
(2)&\sim&1+\alpha_{16}^{ax, 1/m}, \nn
(3),(4)&\sim&1-\alpha_{17}^{ax}-\alpha_{18}^{ax}-\alpha_{19}^{ax},
\eeqa
while diagrams $(1)$ and $(5)$ do not depend on unitary
phases. Our final result for $1/m^2$-corrections at order $Q$ reads
\beqa
A^{0,  a \, (Q)}_{{\rm 1N:} \, 1/m^2}  &=&0,\nn
\vec{A}^{ a \, (Q)}_{{\rm 1N: \, 1/m^2}}  &=&\frac{g_A}{16
  m^2}\tau_i^a\bigg(\vec{k}\,\vec{k}\cdot\vec{\sigma}_i(1-2\bar{\beta}_8)\frac{(p_i^{\prime\,2}-p_i^2)^2}{(k^2+M_\pi^2)^2}
-2 \vec{k} \frac{(p_i^{\prime\, 2} +
  p_i^2)\vec{k}\cdot\vec{\sigma}_i-2 \bar{\beta}_9 (p_i^{\prime\, 2} -
  p_i^2)\vec{k}_i\cdot\vec{\sigma}_i }{k^2+M_\pi^2}\nn
&+&2\,i\,[\vec{k}\times\vec{k}_i]+\vec{k}\,\vec{k}\cdot\vec{\sigma}_i
- 4\,\vec{k}_i\,\vec{k}_i\cdot\vec{\sigma}_i + \vec{\sigma}_i\Big(2(p_i^{\prime\,2}+p_i^2)-k^2\Big)\bigg)\,.\label{oneovermTo2corr}
\eeqa
In order to obtain this result,  we have calculated the contributions
of all diagrams in Fig.~\ref{fig:singleNuclOneOvermNuclTo2} using the
MUT.  Notice that we have also included in
Eq.~(\ref{oneovermTo2corr}) the corrections accounting for the
nonrelativistic normalization of the nucleon fields which amounts to
multiplying the expressions for the current operator with the factors
of $\sqrt{m/E_p}$ for every external nucleon with a momentum $p$, 
see \cite{Epelbaum:2014efa,Friar:1999sj} for details. 
Expanding the result in $1/m$ we obtain, in addition to the expressions
for diagrams of Fig.~\ref{fig:singleNuclOneOvermNuclTo2}, the contribution
\beq
\delta A^{\mu,
  a,\,(Q)}_{{\rm 1N}: 1/m^2}=\sqrt{\frac{m}{E_{p_i^\prime}}}A^{\mu,
  a,\,(Q^{-3})}_{{\rm 1N}: \,{\rm static}} \sqrt{\frac{m}{E_{p_i}}}-A^{\mu,
  a,\,(Q^{-3})}_{{\rm 1N}: \,{\rm
    static}}=-\frac{p_i^{\prime\,2}+p_i^2}{4 m^2} A^{\mu,
  a,\,(Q^{-3})}_{{\rm 1N}: \,{\rm static}} + {\cal O}(1/m^4),
\eeq
which is already accounted for in Eq.~(\ref{oneovermTo2corr}).

\begin{figure}[tb]
\vskip 1 true cm
\includegraphics[width=0.6\textwidth,keepaspectratio,angle=0,clip]{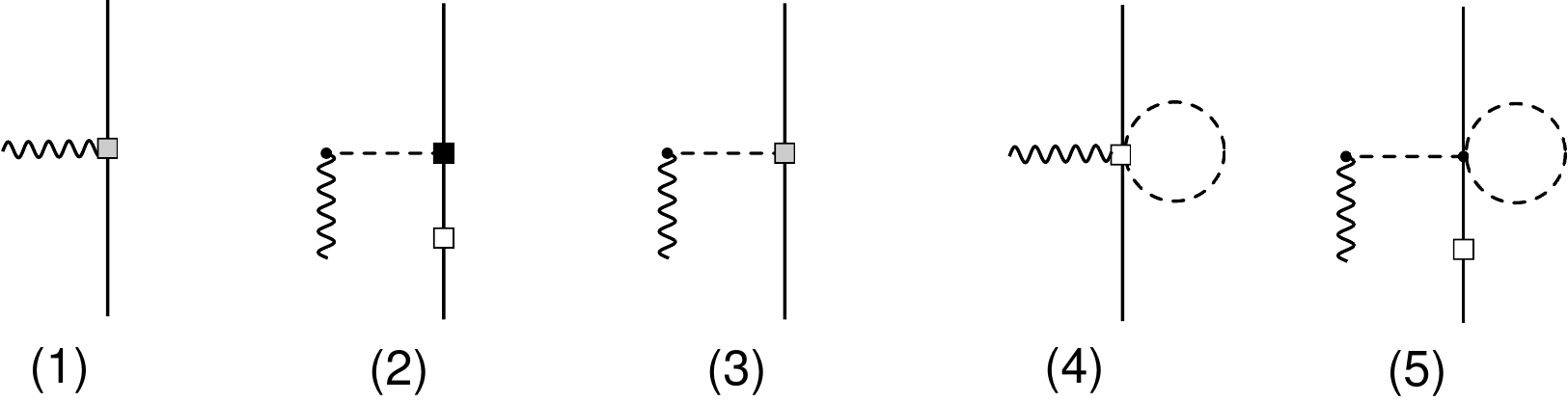}
    \caption{Relativistic $1/m$-contributions to ${\fet A}_{\rm 1N}^{\mu
        \; (Q)}$ from tree-level and tadpole diagrams.  Solid dots (open rectangles) denote vertices from
        ${\cal L}_{\pi N}^{(1)}$ ($1/m$-corrections from ${\cal
          L}_{\pi N}^{(2)}$). Shaded rectangles denote $1/m$-corrections from ${\cal
          L}_{\pi N}^{(3)}$.  Black rectangles denote vertices from ${\cal
          L}_{\pi N}^{(3)}$ in the static limit. For remaining notation see Fig.~\ref{fig:singleNuclLO}.
\label{fig:singleNuclOneOvermNuclCounterTerms} 
 }
\end{figure}

\begin{figure}[tb]
\vskip 1 true cm
\includegraphics[width=0.9\textwidth,keepaspectratio,angle=0,clip]{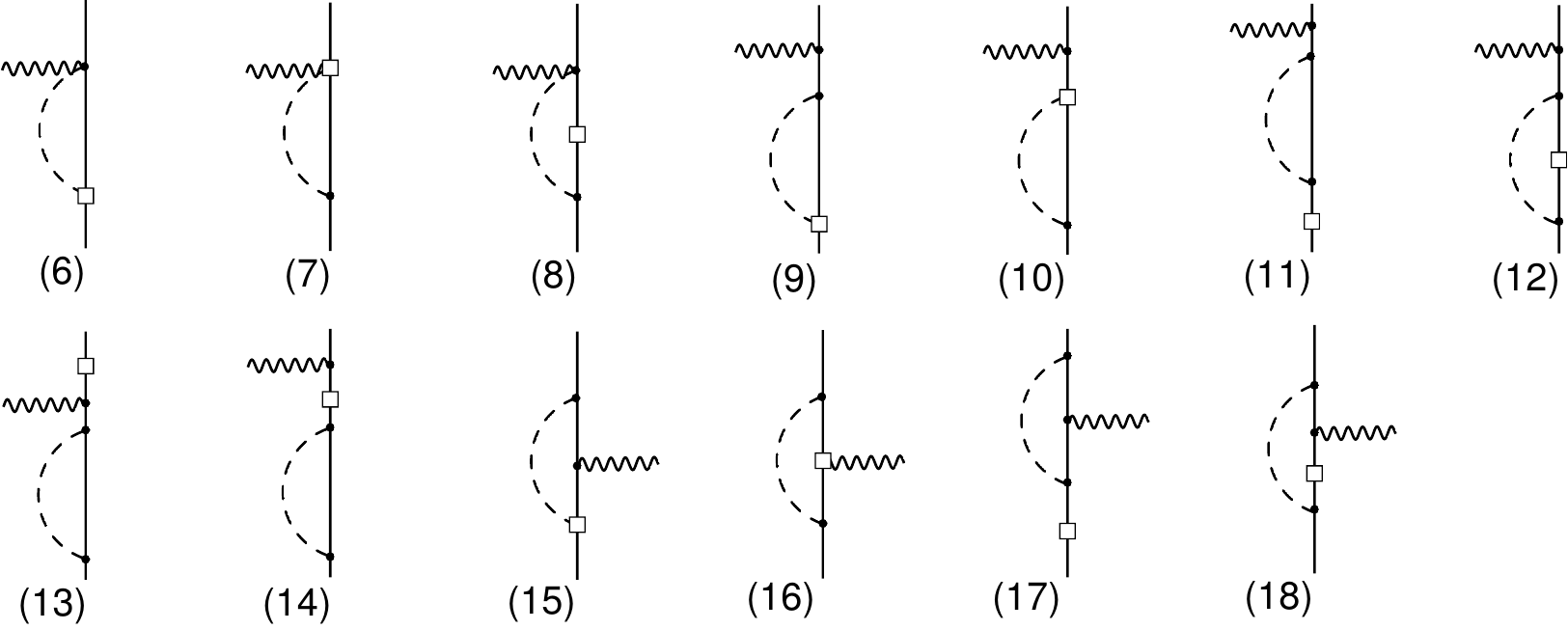}
    \caption{Non-pion-pole relativistic $1/m$-contributions to ${\fet A}_{\rm 1N}^{\mu
        \; (Q)}$ from one-loop diagrams.  Solid dots (open rectangles) denote vertices from
        ${\cal L}_{\pi N}^{(1)}$ ($1/m$-corrections from ${\cal
          L}_{\pi N}^{(2)}$). For remaining notation see Fig.~\ref{fig:singleNuclLO}.
\label{fig:singleNuclOneOvermNuclOneLoopNonPoleTerms} 
 }
\end{figure}

\begin{figure}[tb]
\vskip 1 true cm
\includegraphics[width=0.8\textwidth,keepaspectratio,angle=0,clip]{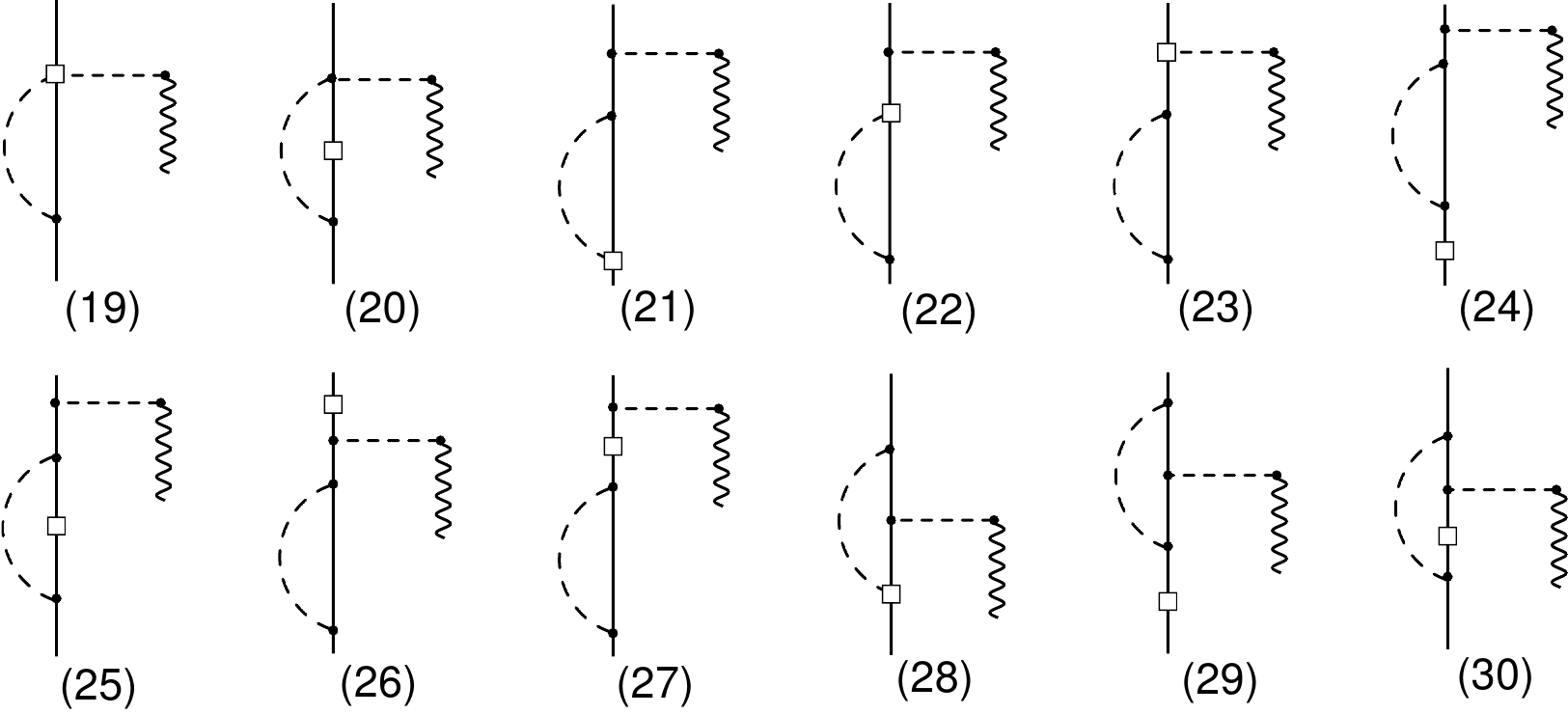}
    \caption{Pion-pole relativistic $1/m$-contributions to ${\fet A}_{\rm 1N}^{\mu
        \; (Q)}$ from one-loop diagrams.   Solid dots (open rectangles) denote vertices from
        ${\cal L}_{\pi N}^{(1)}$ ($1/m$-corrections from ${\cal
          L}_{\pi N}^{(2)}$). For remaining notation see Fig.~\ref{fig:singleNuclLO}.
\label{fig:singleNuclOneOvermNuclOneLoopPionPoleTerms} 
 }
\end{figure}

The third kind of order-$Q$ contributions emerges from relativistic
$1/m$-corrections to the leading one-loop terms. The corresponding
non-vanishing diagrams are shown in
Figs.~\ref{fig:singleNuclOneOvermNuclCounterTerms}, \ref{fig:singleNuclOneOvermNuclOneLoopNonPoleTerms}
and~\ref{fig:singleNuclOneOvermNuclOneLoopPionPoleTerms}.
The dependence on the unitary phases has the form 
given by
\beqa
(2) &\sim& 1+\alpha_{16}^{ax, {\rm Static}},\nn 
(5) &\sim& 1+\alpha_{16}^{ax, {\rm Tadpole}},\nn 
(21),(22),(25),(27) &\sim&
1+\alpha_{16}^{ax, {\rm LO}}, \nn 
(24)&\sim& 6 +
2(\alpha_{10}^{ax}+\alpha_{11}^{ax}-\alpha_{13}^{ax})+3\alpha_{16}^{ax,{\rm
    LO}},\nn
(26),(29) &\sim& 3 + 2(\alpha_{10}^{ax}+\alpha_{11}^{ax}-\alpha_{13}^{ax}),
\eeqa
for the charge operator and 
\beqa
(9),(21)&\sim& 1+2\bar{\beta}_9, \nn 
(10),(22)&\sim& 1 - 2\bar{\beta}_9,\nn 
(11)&\sim& 1+2\bar{\beta}_8 + 2\alpha_1^{ax}, \nn 
(13)&\sim&
\alpha_{1}^{ax}, \nn  
(14)&\sim& 1-2 \bar{\beta}_8, \nn
(23)&\sim& 1+\alpha_{16}^{ax,1/m},  \nn 
(24) &\sim& (1 +
2\bar{\beta}_8- 2\alpha_{12}^{ax})\frac{1}{k^2+M_\pi^2}\big[ \ldots \big]  +
(1-\alpha_{13}^{ax})\frac{1}{(k^2+M_\pi^2)^2}\big[ \ldots \big] , \nn
(26)&\sim&\alpha_{12}^{ax}\frac{1}{k^2+M_\pi^2}\big[ \ldots \big]  +
(2-\alpha_{13}^{ax}-\alpha_{17}^{ax}-\alpha_{18}^{ax}-\alpha_{19}^{ax})
\frac{1}{(k^2+M_\pi^2)^2}\big[ \ldots \big] ,
\nn 
(27)&\sim& (1-2\bar{\beta}_8)\frac{1}{k^2+M_\pi^2}\big[ \ldots \big]  +
(1-\alpha_{17}^{ax}-\alpha_{18}^{ax}-\alpha_{19}^{ax})
\frac{1}{(k^2+M_\pi^2)^2}\big[ \ldots \big] 
\eeqa
for the current operators, where  $\big[ \ldots \big]$ denote spin-isospin-momentum structures which do
  not depend on the unitary phases.
All other diagrams do not induce any dependence on the unitary phases. For the
standard choice of the phases we obtain the following simple result after renormalization:
\beqa
\label{SingleNChargeOneOvermNuclQTo1}
A^{0,  a \, (Q)}_{{\rm 1N:} \, 1/m}  &=&\bar{d}_{22}\vec{k}_i\cdot\vec{\sigma}_i\tau_i^a\frac{k^2}{2 m},\\
\vec{A}^{ a \, (Q)}_{{\rm 1N: \, 1/m}}  &=& 0\,.
\eeqa

Before discussing the final class of corrections at order $Q$
corresponding to static contributions at two-loop level, it is
instructive to compare our results with the
on-shell expressions for the three-point function with an external
axial-vector source, which can be parametrized in terms of the 
form factors $G_A(t)$ and $G_P(t)$. The relativistic parametrization is given by
\beq
\langle N(p_i^\prime)|A^{\mu, a}(0)|N(p_i)\rangle =
\frac{1}{2m}\bar{u}(p_i^\prime\,)\bigg[\gamma^\mu\gamma_5 G_A(t)+\frac{k^\mu}{2
  m}\gamma_5 G_P(t)\bigg]\frac{\tau_i^a}{2}u(p_i),
\eeq 
where
\beq
t=k_0^2-k^2, \quad (\mbox{with } k \equiv | \vec k \,|)
\eeq
and the spinors are normalized as
\beq
\bar{u}(p_i)u(p_i)=2m.
\eeq
The chiral expansion of the form factors can be found
 e.g.~in~\cite{Bernard:1996cc,Fearing:1997dp}, 
see also~\cite{Bernard:1993bq,Schindler:2006it} for results obtained within Lorentz-invariant
formulations. Rewritten in our notation, the chiral expansion of the axial form factor is given
by
\beqa
\label{tempGA}
G_A(t)&=&g_A+(\bar{d}_{22}+f_0^A M_\pi^2) t + f_1^A t^2 + G_A^{(Q^4)}(t) + {\cal O}(Q^5),
\eeqa
where $f_i^A$ are LECs of dimension ${\rm GeV}^{-4}$ and 
\beq
G_A^{(Q^4)}(t) = \frac{t^3}{\pi}\int_{9 M_\pi^2}^\infty \frac{{\rm
    Im}\,G_A^{(Q^4)}(t^\prime\,)}{t^{\prime\,3}(t^\prime - t - i \epsilon)}d t^\prime\,,\label{TwoLoopGA}
\eeq
with the imaginary part calculated utilizing the Cutkosky rules~\cite{Bernard:1996cc}
\beqa
{\rm
    Im}\,G_A^{(Q^4)}(t)&=&\frac{g_A}{192 \pi^3 F_\pi^4}\int_{z^2 < 1} d
  \omega_1 d \omega_2\bigg[6 g_A^2 (\sqrt{t}\, \omega_1 -
  M_\pi^2)\Big(\frac{l_2}{l_1} + z\Big)\frac{\arccos(-z)}{\sqrt{1-z^2}}\nn
&+&2 g_A^2\Big(M_\pi^2 - \sqrt{t}\,\omega_1 -\omega_1^2\Big) + M_\pi^2 -
\sqrt{t}\,\omega_1 + 2\omega_1^2\bigg],
\eeqa
where 
\beq
\omega_i=\sqrt{l_i^2 + M_\pi^2}\quad{\rm with}\quad i=1,2, \quad{\rm and}\quad z=
  \hat{l}_1\cdot\hat{l}_2=\frac{\omega_1
    \omega_2-\sqrt{t}(\omega_1+\omega_2)+\frac{1}{2}(t+M_\pi^2)}{l_1 l_2}\,.
\eeq
Here and in what follows, $l_i \equiv | \vec l_i |$, while $\hat l_i \equiv \vec
l_i/l_i$.  
In the chiral limit, the expression for ${\rm Im}\,G_A^{(Q^4)}(t)$
simplifies to
\beqa
\label{CL1}
{\rm Im}\,G_A^{(Q^4)}(t)\Big|_{M_\pi=0}&=&\frac{2 \mathring{g}_A
  t^2}{9 (16\pi)^3
  F^4}\bigg[\mathring{g}_A^2\bigg(\frac{64}{35}\pi^2+1\bigg) - 1\bigg],
\eeqa
where $\mathring{g}_A$ and $F$ are  the axial coupling and pion decay
constants in the chiral limit, respectively.
Note that ${\rm
    Im}\,G_A^{(Q^4)}(t)$ grows at least quadratically in $t$~\cite{Bernard:1996cc}, and for this
  reason one has to introduce three subtractions in the dispersion
  integral of Eq.~(\ref{TwoLoopGA}). The linear combination
  $\bar{d}_{22}+f_0^A M_\pi^2$ is related to the axial radius of the
  nucleon via
\beq
\bar{d}_{22}+f_0^A M_\pi^2 = \frac{g_A}{6} \langle r_A^2\rangle.
\eeq
The LECs $f_i^A$ refer to the corresponding linear combinations of LEC's from ${\cal L}_{\pi
  N}^{(5)}$. Notice that there are no
contributions to $f_i^A$ from $1/m^2$
corrections since the latter emerge only from 
loop diagrams and for this reason start contributing at higher
orders. Indeed, the tree-level relativistic corrections associated
with diagram $(1)$ of Fig.~\ref{fig:singleNuclLO} are entirely given
by the $1/m$-expansion of
$\bar{u}(p^\prime\,)\gamma^\mu\gamma_5 u(p)/2m$:
\beqa
\bar{u}(p^\prime\,)\gamma^0\gamma_5
u(p)/2m&=&\frac{\vec{k}_i\cdot\vec{\sigma}_i}{m}+{\cal O}(m^{-3}),\\
\bar{u}(p^\prime\,)\gamma^j\gamma_5
u(p)/2m&=&\sigma_i^j-\frac{1}{16 m^2}\bigg[4 i\,
[\vec{k}\times\vec{k}_i]^j + 2 k^j\,
\vec{k}\cdot\vec{\sigma}_i-\sigma^j\,\Big(k^2+2(p_{i}^{\prime\,2}+p_i^2-2k_i^2)\Big)-8
k_i^j\, \vec{k}_i\cdot\vec{\sigma}_i\bigg]+{\cal O}(m^{-4})\,.\nonumber
\eeqa 
The pseudoscalar form-factor is given up to order $Q^4$ by~\cite{Kaiser:2003dr}
\beq
G_P(t)=\frac{4m g_{\pi N} F_\pi}{M_\pi^2-t}-\frac{2}{3}g_A m^2
\langle r_A^2\rangle + m^2
f_1^P t
+ m^2 G_P^{(Q^2)}(t) +
{\cal O}(Q^3),\label{PseudoScalarFormFactorUpToQTo4}
\eeq
where $f_i^P$ denotes the corresponding linear combinations of the  LECs of dimension ${\rm GeV}^{-4}$ from ${\cal L}_{\pi
  N}^{(5)}$ and 
\beqa
G_P^{(Q^2)}(t)&=&\frac{t^2}{\pi}\int_{9 M_\pi^2}^\infty \frac{{\rm
    Im}\,G_P^{(Q^2)}(t^\prime\,)}{t^{\prime\,2}(t^\prime - t - i \epsilon)}d t^\prime\,,\label{TwoLoopGP}
\eeqa
with the imaginary part calculated by the Cutkosky
rules~\cite{Kaiser:2003dr}
\beqa
{\rm Im}\,G_P^{(Q^2)}(t)&=&{\rm Im}\,G_P^{(1)}(t) + {\rm Im}\,G_P^{(2)}(t)
\eeqa
and
\beqa
{\rm Im}\,G_P^{(1)}(t)&=&\frac{g_A}{8\pi^3 F_\pi^4}\int_{z^2<1}
d\omega_1
d\omega_2\bigg[\frac{1}{18}-\frac{M_\pi^4}{12(t-M_\pi^2)^2}+\frac{4\omega_1^2-M_\pi^2}{6
  t}+\frac{\omega_1^2(3 M_\pi^2-t)}{(t-M_\pi^2)^2}+\frac{2 M_\pi^2
  \omega_1 \omega_2 z}{t(t-M_\pi^2)} \frac{l_2}{l_1}\bigg],\nn
{\rm Im}\,G_P^{(2)}(t)&=&\frac{g_A^3}{8\pi^3 F_\pi^4
  t}\int_{z^2<1} d\omega_1
d\omega_2\bigg[(M_\pi^2-\sqrt{t}\omega_1)\bigg(z+\frac{l_2}{l_1}\bigg)\frac{\arccos(-z)}{\sqrt{1-z^2}}
+ \frac{l_1^2}{3}+\frac{t}{9}\nn
&+&\frac{M_\pi^2}{t-M_\pi^2}\bigg(\frac{7}{8}\sqrt{t}-\omega_1-\omega_2\bigg)\bigg(2\omega_1
z
\frac{l_2}{l_1}+\sqrt{t}+\Big((t+M_\pi^2)(4\omega_1-\sqrt{t})-4\sqrt{t}\omega_1\omega_2\Big)\frac{\arccos(-z)}{2
  l_1 l_2 \sqrt{1-z^2}}\bigg)\bigg]\,.\quad\quad
\eeqa
In the chiral limit, one finds
\beqa
\label{CL2}
{\rm Im} \, G_P^{(Q^2)}(t)\Big|_{M_\pi=0}&=&\frac{\mathring{g}_A t}{9 (8\pi)^3 F^4}\bigg[1-\mathring{g}_A^2\bigg(1+\frac{64}{35}\pi^2\bigg)\bigg],
\eeqa
where we have already used the Goldberger-Treiman relation.
Due to linear growth of ${\rm Im}\,G_P^{(Q^2)}(t)$ in $t$~\cite{Kaiser:2003dr}, one needs
two subtractions in the dispersion integral of Eq.~(\ref{TwoLoopGP}).

Similarly to the case of the axial form-factor, there are no
$1/m^2$-corrections in $f_i^P$. In a relativistic calculation, diagram $(2)$
of Fig.~\ref{fig:singleNuclLO} generates the contribution  
\beq
\frac{4m g_{\pi N} F_\pi}{M_\pi^2-t}
\eeq
to the pseudoscalar form factor. The nonrelativistic expansion then emerges 
entirely from the expansion of
$\bar{u}(p^\prime\,)\gamma_5 u(p)/2m$ in inverse powers of the nucleon
mass,
\beqa
\bar{u}(p^\prime\,)\gamma_5
u(p)/2m=-\frac{\vec{k}\cdot\vec{\sigma}_i}{2m}+{\cal O}(1/m^3).\label{uBarGamma4uOneOvermNuclExpansion}
\eeqa

We are now in the position to compare our results for the
single-nucleon axial-vector current operator with the on-shell 
results up-to-and-including the one-loop corrections. 
Notice that given the absence of the iterative contributions to the 
single-nucleon scattering amplitude,\footnote{It is important to keep
  in mind that diagrams like the ones shown  in Fig.~\ref{fig:singleNuclOneOvermNuclTo2}
  can not be interpreted as iterations with the effective potential
  since the non-relativistic kinetic energy and its corrections at
  higher orders in the $1/m$-expansion are  not part of the potential,
  see Eq.~(\ref{SE}).}
we do expect the derived expressions for the charge and current
operator to match the three-point function with an external axial
vector source on the energy shell. Using the Goldberger-Treiman discrepancy
\beq
g_{\pi N}=\frac{g_A m}{F_\pi}\bigg(1-\frac{2 M_\pi^2 \bar{d}_{18}}{g_A}\bigg),
\eeq
one observes that our results for the charge operator agree with the on-shell expressions
even when the energy-conserving delta-function is dropped.
For the current operator, the results for the off-shell kinematics agree only
up-to-and-including $1/m$-corrections, while the disagreement starts first
at the level of the $1/m^2$- and $k_0/m$-terms. The difference is given by
\beqa
&&\bigg[\sqrt{\frac{m}{E_{p_i^{\prime}}}}\frac{1}{2m}\bar{u}(p_i^\prime\,)\bigg[\vec
\gamma
\gamma_5 G_A(t)+\frac{\vec k}{2
  m}\gamma_5 G_P(t)\bigg]\frac{\tau_i^a}{2}u(p_i)
\sqrt{\frac{m}{E_{p_i}}}\bigg]_{1/m^2- \;{\rm and}\,k_0/m-{\rm
parts}}+\vec{A}^{ a \, (Q)}_{{\rm 1N:} \, 1/m^2}+\vec{A}^{  a \,
  (Q)}_{{\rm 1N: \, 1/m, UT^\prime}}\nn
&=&-\bigg(k_0-\frac{p_i^{\prime \,2}}{2 m}+\frac{p_i^2}{2
  m}\bigg)\frac{g_A}{8
  m}\frac{\vec{k}}{k^2+M_\pi^2}\tau_i^a\bigg[-(1+2\bar{\beta}_8)(p_i^{\prime\,2}-p_i^2)\frac{\vec{k}\cdot\vec{\sigma}_i}{k^2+M_\pi^2}
+2(1+2\bar{\beta}_9)\vec{k}_i\cdot\vec{\sigma}_i\bigg],
\eeqa
which, given the on-shell relation $k_0={p_i^{\prime \,2}}/{2 m}-{p_i^2}/{2m}$,  is indeed an off-the-energy-shell effect. 
Notice that the off-shell difference disappears for
$\bar\beta_8=\bar\beta_9=-1/2$, which is, however, incompatible
with the matching condition to the nuclear forces except for the Breit frame. 
It is remarkable that even the static terms proportional to $k_0$, which are
parametrized by unitary phases and for this reason describe off-shell
effects, agree with the on-shell result for our standard choice of the
phases. In particular, if $\alpha_{16}^{ax,
  {\rm LO}}$ were not fixed, the leading contribution to
the charge operator at order $Q^{-1}$ would read 
\beqa
A^{0,  a \, (Q^{-1})}_{{\rm 1N: \, UT^\prime}}
&=&-\alpha_{16}^{ax,{\rm LO}} \frac{g_A}{2} 
\frac{k_0}{k^2+M_\pi^2}\vec{k}\cdot\vec{\sigma}_i \tau_i^a.
\eeqa
On the other hand, the leading contribution to the on-shell charge
comes from the pseudoscalar form factor given by
\beqa
-\frac{1}{2m}\bar{u}(p_i^\prime\,)\frac{k^0}{2
  m}\gamma_5 G_P(t)\frac{\tau_i^a}{2}u(p_i)&=&\frac{g_A}{2} 
\frac{k_0}{k^2+M_\pi^2}\vec{k}\cdot\vec{\sigma}_i \tau_i^a,
\eeqa 
where we have used
Eqs.~(\ref{uBarGamma4uOneOvermNuclExpansion}),
(\ref{PseudoScalarFormFactorUpToQTo4}) with
$t=-k^2$ along with the Goldberger-Treiman relation
\beq
g_{\pi N}=\frac{g_A m}{F_\pi}.
\eeq
Obviously for $\alpha_{16}^{ax, {\rm LO}}\neq -1$ the two results
would disagree. The renormalizability condition, however, dictates the
choice $\alpha_{16}^{ax,{\rm LO}}=-1$, which leads to the agreement with the
on-shell result. 

Based on the above results, we conjecture that the \emph{static}
two-loop contributions to the axial current operator can be obtained
by taking the on-shell result for the corresponding form factors and
dropping the energy-conserving delta-function. This allows us to give
the last missing piece in the current operator at order $Q$ without
explicitly evaluating it using the method of UT:
\beqa
A^{0,  a \, (Q)}_{{\rm 1N: \, static}}  &=& 0,\nn
\vec{A}^{a \, (Q)}_{{\rm 1N: \, static}}
&=&-\frac{1}{2}\tau_i^a\vec{\sigma}_i \Big(
- f_0^A M_\pi^2 k^2 + 
f_1^A k^4 +
G_A^{(Q^4)}(-k^2)\Big)+\frac{1}{8}\vec{k}\,\vec{k}\cdot\vec{\sigma}_i
\tau_i^a\Big(
-4 f_0^A M_\pi^2 
- f_1^P k^2 + G_P^{(Q^2)}(-k^2)\Big)\,.\quad\quad\label{SingleNtwoLoopExpr}
\eeqa

To summarize, our final result for the single-nucleon axial charge and
current operators up to order $Q$ can be expressed in terms of the
nucleon form factors the following compact form
\beqa
A_{\rm 1N}^{0, a} &=& -
\frac{G_A (-k^2)}{2 m} \tau_i^a \vec k_i 
\cdot \vec \sigma_i \, + \, \frac{G_P (- k^2)}{8 m^2} \tau_i^a k_0 \, \vec k
\cdot \vec \sigma_i \,, \nn
\vec A_{\rm 1N}^{a} &=& - \frac{G_A (-k^2)}{2} \tau_i^a \vec \sigma_i
\, + \, \frac{G_P (-k^2)}{8 m^2}  \tau_i^a  \vec k \, \vec k \cdot \vec
\sigma_i \, + \,  \vec{A}^{ a \, (Q)}_{{\rm 1N:\,}1/m, {\rm
    UT^\prime}} \, + \, 
\vec{A}^{ a \, (Q)}_{{\rm 1N: \, 1/m^2}} \,,
\label{OneNucleonFF}
\eeqa
where the last two terms are specified in Eqs.~(\ref{k0OvermNuclcorr}) and (\ref{oneovermTo2corr}).

\section{Two-nucleon axial charge and current operators}
\def\theequation{\arabic{section}.\arabic{equation}}
\label{sec:2N}

\subsection{Contributions at orders $Q^{-1}$ and $Q^0$}

\begin{figure}[tb]
\vskip 1 true cm
\includegraphics[width=0.9\textwidth,keepaspectratio,angle=0,clip]{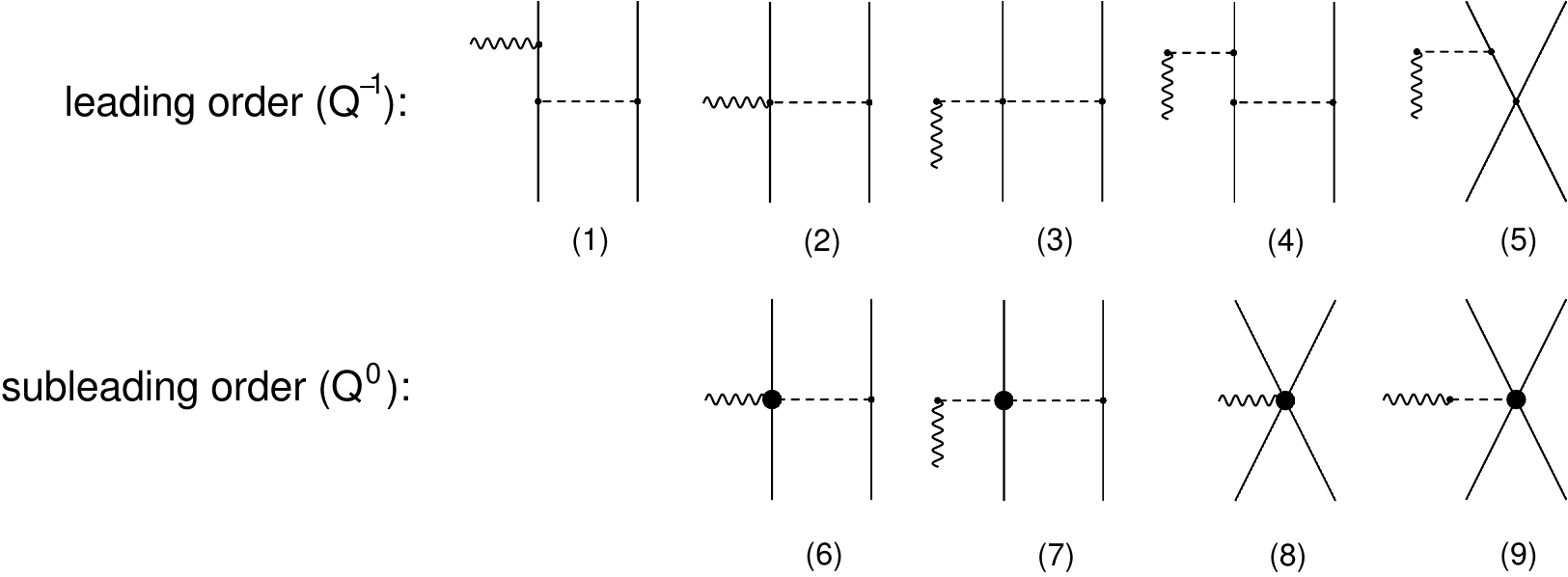}
    \caption{Diagrams leading to the lowest-order contributions to the
      2N axial charge ${\fet A}_{\rm 2N}^{0 \; (Q^{-1})}$
      (upper line) and current operator $\vec {\fet A}_{\rm 2N}^{(Q^{0})}$ (lower line).  Filled circles
      denote the subleading vertices from the effective Lagrangians
      $\mathcal{L}_{\pi N}^{(2)}$ and $\mathcal{L}_{\pi
        NN}^{(1)}$. Diagrams resulting from the interchange of the
      nucleon lines and/or application of the time reversal operation
      are not shown. 
For remaining notation see Fig.~\ref{fig:factor}.
\label{fig:tree} 
 }
\end{figure}

As already mentioned above, the chiral expansion for
the 2N axial four-current operator starts at order $Q^{-1}$. The
relevant diagrams generating the dominant contributions are shown in the first line of Fig.~\ref{fig:tree}. 
As should be clear from the previous sections, all diagrams shown here
and in the following are to be understood as representing the
irreducible (i.e.~non-iterative) pieces of the corresponding
amplitudes.  
For the charge operator, the last two diagrams in the first line of
Fig.~\ref{fig:tree} depend on the lowest-order unitary
transformation
\beq  
(4), (5) \sim 1 + \alpha_{16}^{ax, \rm LO}\,.
\eeq
For the current operator, the contributions of the leading-order
tree-level graphs depend on unitary phases as follows:
\beqa
(1)&\sim& k_0 \alpha_1^{ax},\nn
(3)&\sim& k_0(\alpha_{25}^{ax}+\alpha_{26}^{ax}), \nn 
(4)&\sim& \frac{k_0
  \alpha_{12}^{ax}}{k^2+M_\pi^2}\big[ \ldots \big] +\frac{k_0
  \alpha_{13}^{ax}}{(k^2+M_\pi^2)^2}\big[ \ldots \big] , \nn 
(5)&\sim& k_0(\alpha_{21}^{ax}+\alpha_{22}^{ax}+\alpha_{23}^{ax}).
\label{tempo2}
\eeqa
As already pointed out, we adopt the choice of unitary phases
$\alpha_1^{ax}=\alpha_{12}^{ax}=0$,
$\alpha_{13}^{ax}=\alpha_{25}^{ax}+\alpha_{26}^{ax}=\alpha_{21}^{ax}+\alpha_{22}^{ax}+\alpha_{23}^{ax}=1$
and $\alpha_{16}^{ax, {\rm LO}}=-1$ which is dictated by
renormalizability and by matching to the nuclear force.  
With this choice, the expressions for the
one-pion-exchange contributions take the form 
\beqa
 A_{{\rm 2N:\,}1\pi }^{0, a \; ({Q^{-1}})} &=&
-\frac{i g_A 
\vec{q}_1\cdot \vec{\sigma}_1 [{\fet \tau}_1\times{\fet \tau}_2]^a}{4 
F_\pi^2 \left(q_1^2 +M_\pi^2\right)}\; + \; 1 \leftrightarrow
2\,,\label{ChargeTreeOPE}\\
\vec { A}_{{\rm 2N:\,}1\pi }^{a \; ({Q^{-1}})} &=& 0,\label{VectorTreeOPE}
\eeqa
where $\vec q_i = \pvec p_i  ' - \vec p_i$ ($\fet \tau_i$) denotes the momentum
transfer (Pauli isospin matrix) of nucleon $i$ and $q_i \equiv | \vec
q_i \, |$. For the standard choice of the unitary phases, the short-range
contribution of the last diagram  in the first line of
Fig.~\ref{fig:tree} vanishes: 
\beqa
A_{{\rm 2N:\,cont} }^{0, a \; ({Q^{-1}})}
&=&0,\label{ChargeTreeContact}\\
\vec {A}_{{\rm 2N:\,cont}}^{a \; ({Q^{-1}})} &=&0. \label{VectorTreeContact}
\eeqa
Notice that the choice $\alpha_{16}^{ax,{\rm
    LO}}=\alpha_{16}^{ax,{\rm Static}}=\alpha_{16}^{ax,{\rm
    Tadpole}}=-1$ switches off all pion-pole contributions to the
charge. This choice is dictated by (off-shell) renormalizability
of the single-nucleon charge operator, where terms proportional to $k_0$
are required to be finite. 

Next, at order $Q^0$, one encounters the contributions to the
2N axial current operator only, which originate from diagrams shown in the second
line of Fig.~\ref{fig:tree}. There are no charge contributions at this order.
Again, the corresponding expressions are
well-known
and have the form
\beqa
\label{CurrentTree1}
\vec {A}_{{\rm 2N:}\,  1\pi }^{a \; ({Q^{0}})} &=& \frac{g_A}{2 F_\pi^2}
\frac{\vec \sigma_1 \cdot \vec q_1}{q_1^2 + M_\pi^2} \bigg\{
\tau_1^a \bigg[ - 4 c_1 M_\pi^2 \frac{\vec k}{k^2 + M_\pi^2}
+ 2 c_3 \bigg( \vec q_1 - \frac{\vec k \, \vec k \cdot \vec q_1}{k^2 +
  M_\pi^2} \bigg) \bigg]+ c_4 [ \fet \tau_1 \times \fet \tau_2 ]^a \bigg( \vec q_1 \times
\vec \sigma_2 - \frac{\vec k \, \vec k \cdot \vec q_1 \times \vec
  \sigma_2}{k^2 + M_\pi^2} \bigg) \nn
&-&\frac{\kappa_v}{4 m}  [ \fet \tau_1 \times \fet \tau_2 ]^a
\vec{k}\times\vec{\sigma}_2\bigg\} + \; 1 \leftrightarrow 2\,, \\
\label{CurrentTree2}
\vec {A}_{{\rm 2N:}\,  \rm cont}^{a \; ({Q^{0}})} &=& -\frac{1}{4} D\,
\tau_1^a \bigg(\vec{\sigma}_1-\frac{\vec{k}\, \vec{\sigma}_1\cdot\vec{k}}{k^2+M_\pi^2}\bigg) + \; 1 \leftrightarrow 2\,,
\eeqa
where $c_i$ and $D$ denote the LECs from $\mathcal{L}_{\pi N}^{(2)}$
and $\mathcal{L}_{\pi NN}^{(1)}$, respectively, while 
$\kappa_v$ is the isovector anomalous magnetic moment of the nucleon. Further, we
use the notation with $k \equiv | \vec k \, |$. It is easy to verify that the pion-pole
contributions to the axial current fulfill the matching relations
\beq
- \frac{g_A}{2 F_\pi^2} \sum_a \tau_3^a \vec \sigma_3 \cdot \vec A_{{\rm
    2N:}\,  1\pi }^{a \; ({Q^{0}})} \Big|_{\vec k = - \vec q_3} =
\left[V_{\rm TPE}^{\rm 3NF} \right]_{13}\,, \quad \quad 
- \frac{g_A}{2 F_\pi^2} \sum_a \tau_3^a \vec \sigma_3 \cdot \vec A_{{\rm
    2N:}\, \rm cont}^{a \; ({Q^{0}})} \Big|_{\vec k = - \vec q_3} =
\left[V_{\rm OPE}^{\rm 3NF} \right]_{12},
\eeq
where $\left[V_{\rm TPE}^{\rm 3NF} \right]_{13}$ ($\left[V_{\rm OPE}^{\rm 3NF} \right]_{12}$) denotes the part of
the order-$Q^3$ two-pion exchange 3N force in Eq.~(2) (Eq.~(10)) of \cite{Epelbaum:2002vt} 
symmetric with respect to the interchange of nucleons $1$ and $3$ ($1$
and $2$).

\subsection{One-pion-exchange contributions at order $Q$}
\label{sec:OnePionQ}

\begin{figure}[tb]
\vskip 1 true cm
\includegraphics[width=15.5cm,keepaspectratio,angle=0,clip]{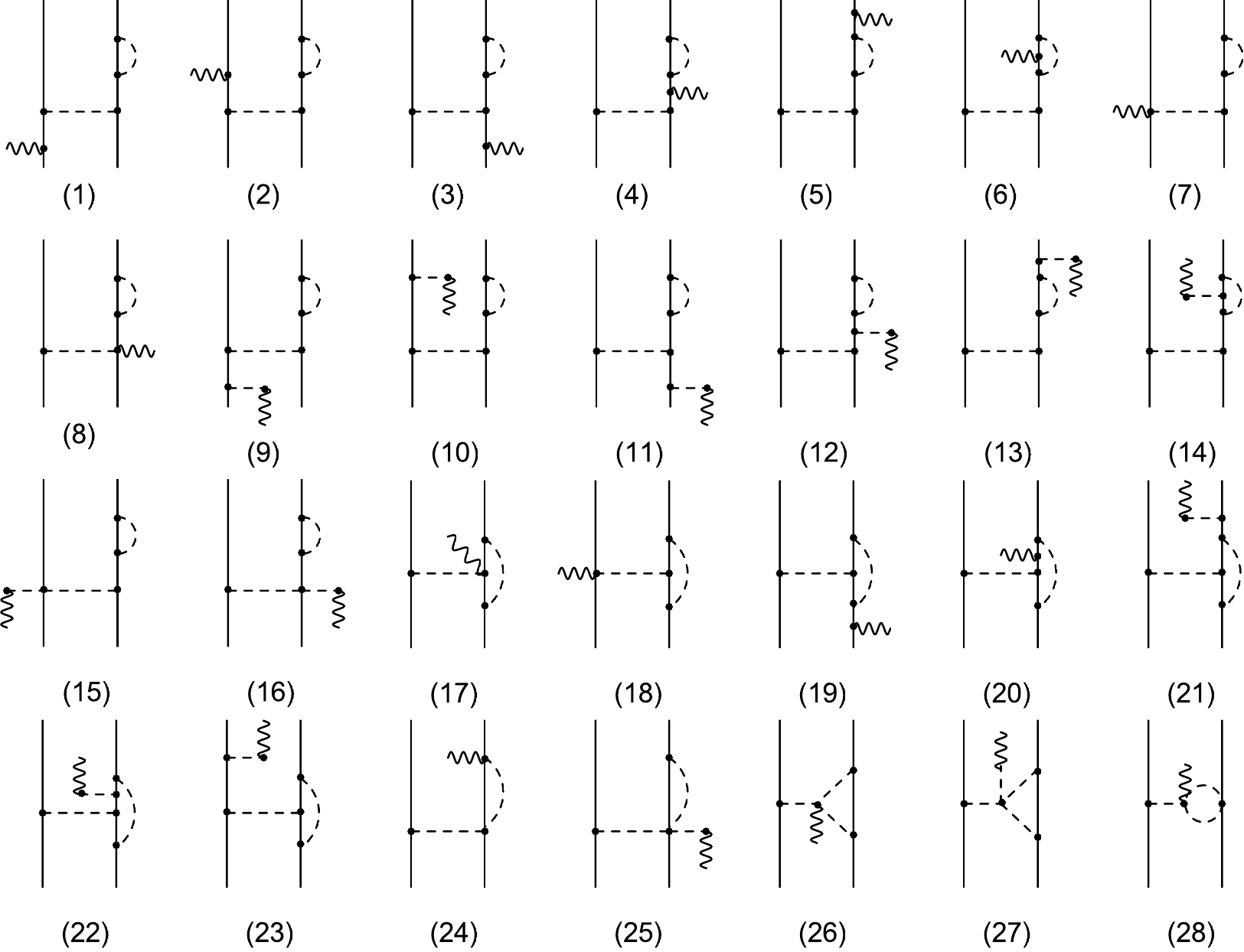}
    \caption{
         Non-tadpole one-loop one-pion-exchange diagrams contributing
         to ${\fet A}_{\rm 2N}^{\mu \; (Q)}$. 
For notation see Fig.~\ref{fig:tree}.
\label{fig:ope} 
 }
\end{figure}

In Fig.~\ref{fig:ope} we show all one-loop diagrams of non-tadpole
type with a single
pion being exchanged between the nucleons, which produce non-vanishing
contributions to the axial charge and/or current operators. 
Specifically, we found that diagrams
(1), (2), (7), (8), (11-18), (21), (23), (24), (26) and (28) generate
non-vanishing contributions to the axial charge, from which 
those of the diagrams (1), (2), (11-14), (21), (23) do explicitly
depend on the unitary phases in the following way:
\beqa
(1), (2), (11),(21),(23) &\sim& 1+\alpha_{16}^{ax, {\rm LO}}, \nn 
(12), (14) &\sim& 1-\alpha_4^{ax}+\alpha_5^{ax}, \nn 
(13)&\sim& 2+\alpha_{16}^{ax, {\rm
    LO}} - \alpha_4^{ax}+\alpha_5^{ax}.
\eeqa

For the axial current, the diagrams (1-6), (9-15), (19-22) and (25-27) give non-vanishing
contributions, from which those of  graphs (1-5), (9-11), (13), (15)
turn out to depend on the unitary phases: 
\beqa
(1),(2),(5)&\sim& 1+\alpha_1^{ax}, \nn 
(3),(4)&\sim& \alpha_1^{ax}, \nn
(15)&\sim& 2-\alpha_{25}^{ax}-\alpha_{26}^{ax},\nn
(9),(10),(13)&\sim&(1-2\alpha_4^{ax}+\alpha_5^{ax}-\alpha_{10}^{ax})\frac{1}{(k^2+M_\pi^2)^2}
\big[ \ldots \big]
+(1-\alpha_{12}^{ax})\frac{1}{k^2+M_\pi^2}\big[ \ldots \big] \,,\nn
(11),(12)&\sim&(1-2\alpha_4^{ax}+\alpha_5^{ax}-\alpha_{10}^{ax})\frac{1}{(k^2+M_\pi^2)^2}\big[ \ldots \big]
+\alpha_{12}^{ax}\frac{1}{k^2+M_\pi^2}\big[ \ldots \big]\,.
\eeqa
One observes that 
the contributions to the axial current operator from diagrams (3) and (4)
vanish for the standard choice of the unitary phases. For the charge
operator, diagrams $(1), (2), (11),(21),(23)$ also turn out to yield vanishing
contributions for the standard choice of the unitary phases.

\begin{figure}[tb]
\vskip 1 true cm
\includegraphics[width=14.0cm,keepaspectratio,angle=0,clip]{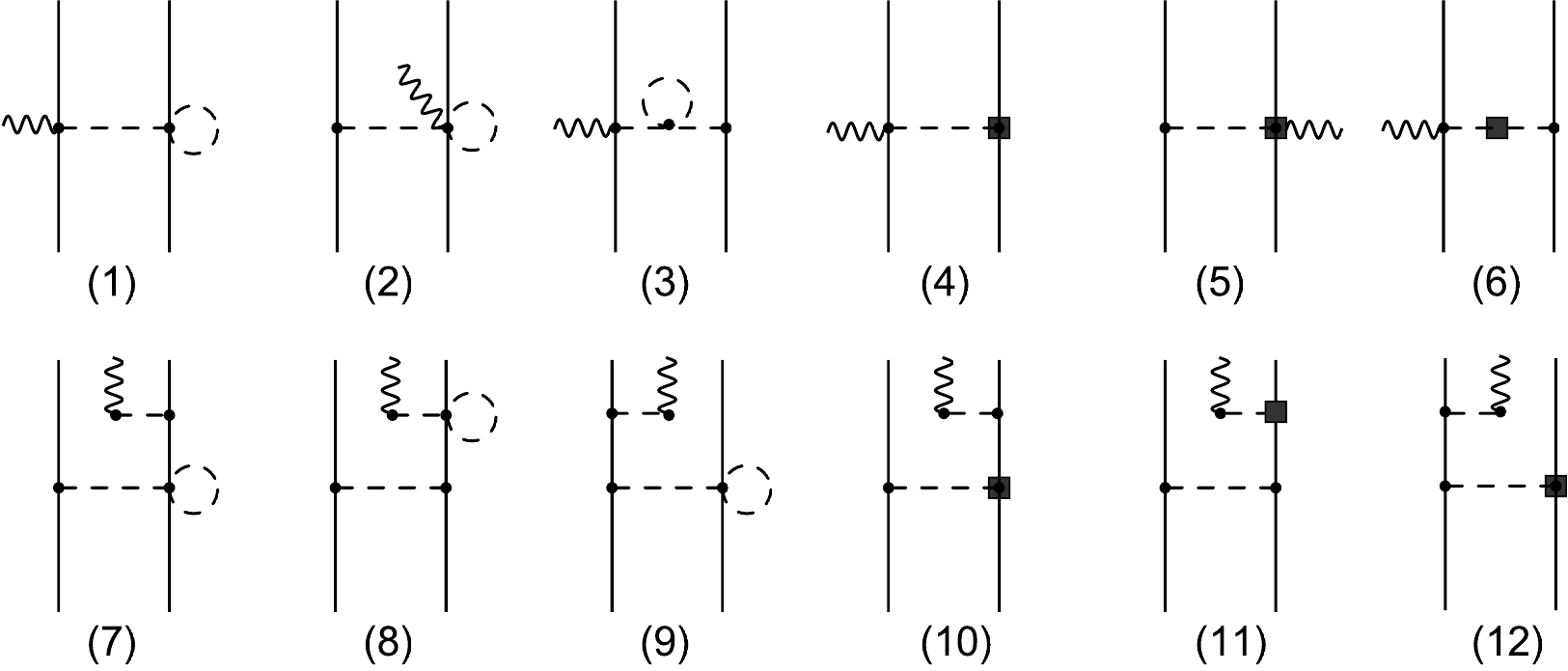}
    \caption{One-pion-exchange
tadpole and tree-level dragrams involving $d_i$-vertices from ${\cal
  L}_{\pi N}^{(3)}$ (denoted by filled squares) which contribute to ${\fet A}_{\rm 2N}^{0 \; (Q)}$.
For remaining notation see Fig.~\ref{fig:tree}.
\label{fig:tadpoles} 
 }
\end{figure}

Next, in Fig.~\ref{fig:tadpoles} we show all non-vanishing one-pion-exchange
tadpole diagrams and tree graphs involving $d_i$-vertices from ${\cal
  L}_{\pi N}^{(3)}$ and $l_i$-vertices from ${\cal  L}_{\pi}^{(4)}$. 
We found that these diagrams contribute only to the axial charge
operator. Further, graphs (7-12) yield contributions which  depend
on the unitary phases as
\beqa
(7),(9),(10),(12) &\sim& 1+\alpha_{16}^{ax,{\rm LO}},\nn 
(8) &\sim&
1+\alpha_{16}^{ax,{\rm Tadpole}}, \nn 
(11) &\sim& 1+\alpha_{16}^{ax,{\rm Static}} \,,
\eeqa
and vanish for the standard choice of the phases.
Evaluating the contributions from the diagrams depicted in
Figs.~\ref{fig:ope} and \ref{fig:tadpoles} for our standard choice of
the unitary phases, replacing all bare LECs $l_i$ and $d_i$ in terms of
their renormalized values as defined in Eq.~(\ref{physicalcouplingsrenorm}), and expressing
the results in terms of physical parameters $F_\pi$, $M_\pi$ and
$g_A$, see e.g.~\cite{Kolling:2011mt}, 
leads to our final result for the static order-$Q$ contributions to
the 2N one-pion-exchange axial current and charge operators:   
\beqa
\label{Current1pi}
\vec {A}_{{\rm 2N:} \, 1\pi}^{a \, (Q)}&=& \frac{4 F_\pi^2}{g_A} \frac{\vec q_1
 \cdot \vec \sigma_1}{q_1^2 + M_\pi^2} \Big\{ [ \fet \tau_1 \times
\fet \tau_2 ]^a \Big( [\vec q_1 \times \vec \sigma_2 ] \, h_1(q_2) + [\vec
q_2 \times \vec \sigma_2 ] \, h_2(q_2)\Big) + \fet \tau_1^a \big(\vec q_1 -
\vec q_2\big) h_3(q_2) \Big\}\nn
&+& \frac{4 F_\pi^2}{g_A}\frac{\vec q_1
 \cdot \vec \sigma_1 \,\vec k }{(k^2 + M_\pi^2)(q_1^2 + M_\pi^2)}\Big\{
\fet \tau_1^a h_4(q_2) +   [ \fet \tau_1 \times
\fet \tau_2 ]^a \vec q_1 \cdot [\vec q_2
  \times \vec \sigma_2] h_5(q_2)\Big\} \; + \; 1
\leftrightarrow 2, \\  [4pt]
\label{Charge1piQTo1}
A_{{\rm 2N:} \, 1\pi}^{0, a \, (Q)}&=& i \frac{4 F_\pi^2}{g_A} \frac{\vec q_1
  \cdot \vec \sigma_1}{q_1^2 + M_\pi^2} \Big\{
[ \fet \tau_1 \times
\fet \tau_2 ]^a \, \big( h_6(q_2)+ k^2 h_7(q_2) 
\big) 
+ \tau_1^a \, \vec q_1 \cdot [ \vec q_2 \times
\vec \sigma_2 ] \,h_8(q_2) 
\Big\} \; + \; 1
\leftrightarrow 2\,, 
\eeqa
where the scalar functions $h_i (q_2)$ are given by 
\bigskip
\beqa
h_1(q_2)&=&-\frac{g_A^6M_\pi}{128\pi F_\pi^6}, \nn
h_2(q_2)&=&\frac{g_A^4M_\pi}{256\pi F_\pi^6} + \frac{g_A^4 A(q_2)\left(4M_\pi^2+q_2^2
\right)}{256\pi F_\pi^6},\nn
h_3(q_2)&=&\frac{g_A^4\left(g_A^2+1
\right)M_\pi}{128\pi F_\pi^6} +\frac{g_A^4A(q_2)\left(2M_\pi^2+q_2^2
\right)}{128\pi F_\pi^6}, \nn
h_4(q_2)&=&\frac{g_A^4}{256\pi F_\pi^6}\left(A(q_2)\left(2M_\pi^4+5M_\pi^2q_2^2+
2 q_2^4\right)+\left(4g_A^2+1\right)M_\pi^3+2\left(g_A^2+1
\right)M_\pi q_2^2\right),\nn
h_5(q_2)&=&-\frac{g_A^4}{256\pi F_\pi^6}\left(A(q_2)\left(4M_\pi^2+q_2^2\right)+
\left(2g_A^2+1\right)M_\pi\right),\nn
h_6(q_2)&=&\frac{g_A^2\left(3\left(64+128 g_A^2 
\right)M_\pi^2+8\left(19 g_A^2+5\right)q_2^2\right)}{36864\pi^2F_\pi^6}
-\frac{g_A^2}{768\pi^2F_\pi^6}  L(q_2)\left(\left(8g_A^2+4\right)M_\pi^2+\left(5g_A^2+1
\right)q_2^2\right)\nn
&+&\frac{\bar{d}_{18}g_A
  M_\pi^2}{8 F_\pi^4}-\frac{g_A^2(2\bar{d}_{2}+
\bar{d}_{6})\left(M_\pi^2+q_2^2
\right)}{16F_\pi^4}-\frac{\bar{d}_{5}g_A^2M_\pi^2}{2F_\pi^4},\nn
h_7(q_2)&=&\frac{g_A^2(2\bar{d}_{2}-\bar{d}_{6})}{16F_\pi^4},\nn
h_8(q_2)&=&-\frac{g_A^2(\bar{d}_{15}-2\bar{d}_{23})}{8F_\pi^4}. \label{OPEhiDefenition}
\eeqa
Here, the loop functions $L(q)$ and $A(q)$ are defined as
\beq
L(q) = \frac{\sqrt{q^2 + 4 M_\pi^2}}{q} \ln \bigg(\frac{\sqrt{q^2 + 4
    M_\pi^2}+q}{2 M_\pi} \bigg) \quad \quad 
\mbox{and} \quad \quad 
A(q) = \frac{1}{2 q} \arctan \bigg( \frac{q}{2 M_\pi} \bigg) \,.
\eeq
Notice that as desired,  the pion-pole contributions to the current
operator are directly related to the two-pion exchange contributions
to the order-$Q^4$ (N$^3$LO) 3N
force. In particular, the following relation holds true 
\beq
h_4(q_2)\,=\,{\cal A}^{(4)}(q_2),\quad h_5(q_2)\,=\,{\cal B}^{(4)}(q_2),
\eeq
where the scalar functions ${\cal A}^{(4)}(q_2)$ and ${\cal
  B}^{(4)}(q_2)$ entering the 3N force are defined in 
Eq.~(3.4) of \cite{Krebs:2012yv}.

\begin{figure}[tb]
\vskip 1 true cm
\includegraphics[width=15cm,keepaspectratio,angle=0,clip]{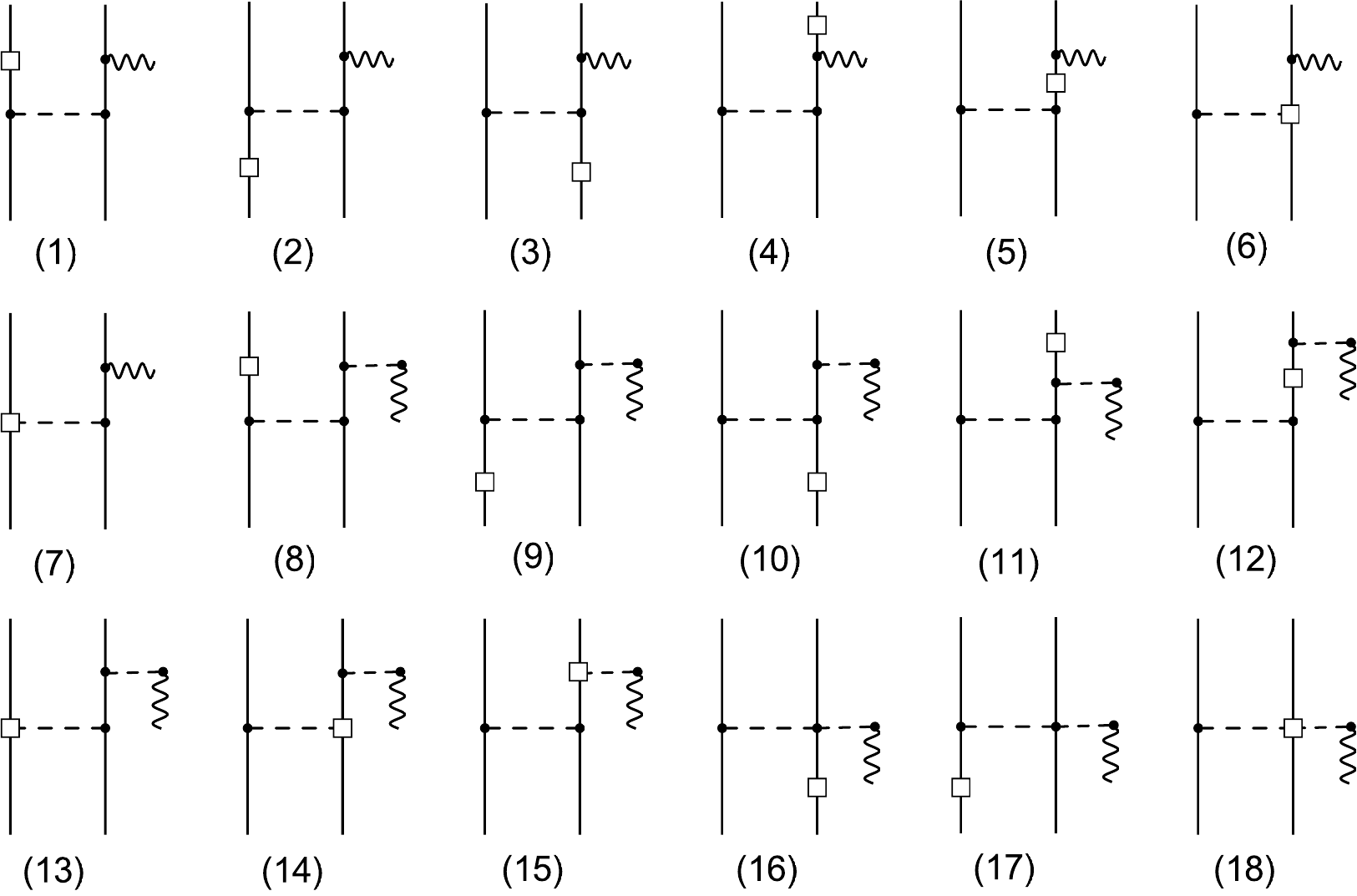}
    \caption{One-pion exchange diagrams leading to non-vanishing
      $1/m$-contributions to $\vec {\fet A}_{\rm 2N}^{(Q)}$. Open rectangles refer to $1/m$-vertices from ${\cal
          L}_{\pi N}^{(2)}$. For remaining notation see Fig.~\ref{fig:tree}.
\label{fig:relativistic_corrections} 
 }
\end{figure}

\begin{figure}[tb]
\vskip 1 true cm
\includegraphics[width=0.75\textwidth,keepaspectratio,angle=0,clip]{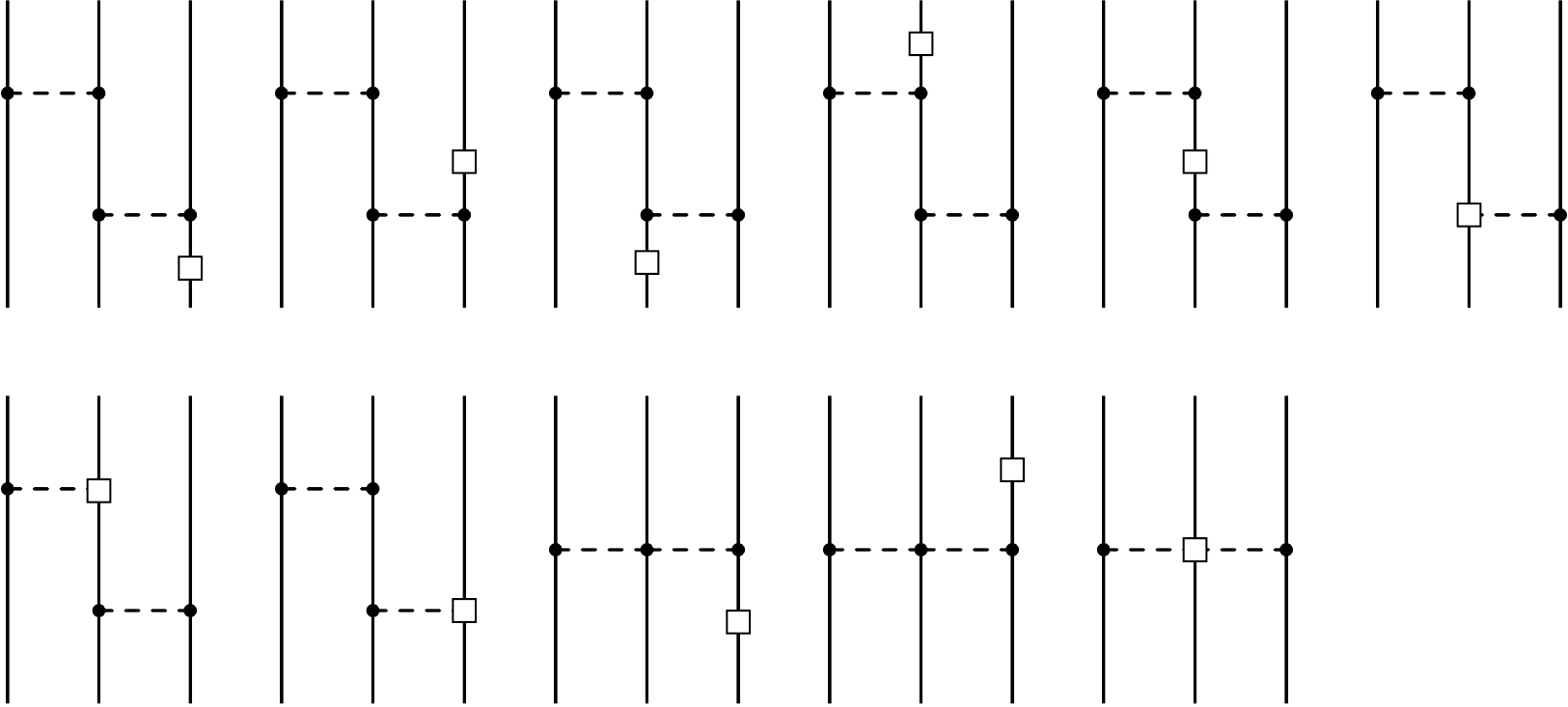}
    \caption{Diagrams leading to non-vanishing relativistic
      corrections to the two-pion exchange 3N force at N$^3$LO which
      do not include the $1/m$-vertices from ${\cal
          L}_{\pi N}^{(2)}$ (shown by open squares) at the leftmost
      nucleon line. Time reversed diagrams are not shown. For
      remaining notation see Fig.~\ref{fig:factor}.
\label{fig:one_over_m_3nf_2pe} 
 }
\end{figure}

Finally, apart from the static contributions, we need to take
into account for the leading relativistic corrections emerging from tree-level
diagrams with a single insertion of $1/m$-vertices from the
Lagrangian $\mathcal{L}_{\pi N}^{(2)}$. We stress again that due to
the employed counting for the nucleon mass with $m \sim \Lambda_b^2/M_\pi$, 
these contributions are shifted one order higher
relative to the ones emerging from tree-level diagrams with a single
insertion of the $c_i$-vertices from  $\mathcal{L}_{\pi N}^{(2)}$
shown in the second line of Fig.~\ref{fig:tree}. 
In Fig.~\ref{fig:relativistic_corrections}, we show all diagrams leading
to  non-vanishing contributions to the axial current operator. 
Notice that no
$1/m$-corrections to the 2N axial charge operator appear at this order.   
Diagrams (1-17) turn out to induce a dependence 
on the unitary phases in the following way:
\beqa
(1),(2)&\sim& 1+2\bar{\beta}_8+2\alpha_1^{ax},\nn 
(3)&\sim& -1+2\bar{\beta}_8+2\alpha_1^{ax},\nn 
(4)&\sim& \alpha_1^{ax}, \nn
(5)&\sim& -1+2\bar{\beta}_8, \nn 
(6),(14) &\sim& -1 + 2\bar{\beta}_9, \nn
(7),(13)&\sim& 1+2\bar{\beta}_9, \nn 
(8),(9),(10)&\sim&
(-1+2\alpha_4^{ax}-\alpha_5^{ax}+\alpha_{10}^{ax})\frac{1}{(k^2+M_\pi^2)^2}
\big[\ldots \big]
+ (-1-2\bar{\beta}_8+2\alpha_{12}^{ax})\frac{1}{k^2+M_\pi^2}\big[\ldots \big],\nn
(11)&\sim&
(-2+2\alpha_4^{ax}-\alpha_5^{ax}+\alpha_{10}^{ax}+\alpha_{17}^{ax}+\alpha_{18}^{ax}+\alpha_{19}^{ax})\frac{1}{(k^2+M_\pi^2)^2}\big[\ldots \big]
+ \alpha_{12}^{ax}\frac{1}{k^2+M_\pi^2}\big[\ldots \big],\nn
(12)&\sim&
(-1+\alpha_{17}^{ax}+\alpha_{18}^{ax}+\alpha_{19}^{ax})\frac{1}{(k^2+M_\pi^2)^2}\big[\ldots \big]
+ (-1+2\bar{\beta}_8)\frac{1}{k^2+M_\pi^2}\big[\ldots \big],\nn 
(15)&\sim&
1+\alpha_{16}^{ax,1/m},\nn
(16)&\sim& -1 + \alpha_{25}^{ax}+\alpha_{26}^{ax}, \nn 
(17)&\sim& -2 +  \alpha_{25}^{ax}+\alpha_{26}^{ax} \, .
\eeqa
Again, our standard choice of the unitary phases leads to some simplifications. In particular,  
it eliminates the contributions from
diagrams (4) and (16).
The explicit results for the $1/m$-corrections to the
one-pion-exchange current operator have the form 
\beqa
\label{Current1piRel}
\vec {A}_{{\rm 2N:} \, 1\pi , \, 1/m}^{a \, (Q)}&=&\frac{g_A}{16 F_\pi^2 m}\bigg\{i [ \fet \tau_1 \times
\fet \tau_2 ]^a
\bigg[\frac{1}{(q_1^2+M_\pi^2)^2}\bigg(\vec{B}_1 - \frac{\vec k \, \vec k
  \cdot \vec{B}_1}{k^2+M_\pi^2} 
\bigg) + \frac{1}{q_1^2+M_\pi^2}\bigg(\frac{\vec{B}_2}{(k^2+M_\pi^2)^2} 
+ \frac{\vec{B}_3}{k^2+M_\pi^2} + \vec{B}_4\bigg)  \bigg]\nn
&+& \tau_1^a\bigg[ \frac{1}{(q_1^2+M_\pi^2)^2}\bigg(\vec{B}_5 -
\frac{\vec k \, \vec k \cdot \vec{B}_5}{k^2+M_\pi^2}
\bigg) + \frac{1}{q_1^2+M_\pi^2}\bigg(\frac{\vec{B}_6}{(k^2+M_\pi^2)^2} 
+ \frac{\vec{B}_7}{k^2+M_\pi^2} + \vec{B}_{8}\bigg)\bigg] \bigg\} +
1 \; \leftrightarrow \; 2\,,
\eeqa
where the vector-valued quantities $\vec B_i$ depend on various momenta and
the Pauli spin matrices and are given by 
\beqa
\vec{B}_1&=&g_A^2\vec{q}_1\cdot\vec{\sigma}_1[
-2 (1+2\bar\beta_8)\vec{q}_1\,\vec{k}_1\cdot\vec{q}_1
-(1-2\bar\beta_8)(2\vec{q}_1\,\vec{k}_2\cdot\vec{q}_1-i\,
\vec{q}_1\times\vec{\sigma}_2\,\vec{k}\cdot\vec{q}_1],\nn
\vec{B}_2&=&(1-2\bar{\beta}_8) g_A^2
\vec{k}\,\vec{k}\cdot\vec{q}_1\vec{q}_1\cdot\vec{\sigma}_1[2\vec{k}\cdot\vec{k}_2
-i \vec{k}\cdot\vec{q}_1\times\vec{\sigma}_2],\nn
\vec{B}_3&=&2\vec{k}\Big[-g_A^2((1+2\bar\beta_9) \vec{k}\cdot\vec{q}_1\vec{k}_1
\cdot\vec{\sigma}_1+(1-2\bar\beta_9)\vec{q}_1\cdot\vec{\sigma}_1(
\vec{k}\cdot\vec{k}_2+\vec{k}_2\cdot\vec{q}_1))\nn
&+&\vec{q}_1\cdot\vec{\sigma}_1(
\vec{k}\cdot\vec{k}_2+i\,
\vec{k}\cdot\vec{q}_1\times\vec{\sigma}_2-\vec{k}_1\cdot\vec{q}_1+\vec{k}_2\cdot\vec{q}_1)
\Big],\nn
\vec{B}_4&=&g_A^2[2(1+2\bar\beta_9)\vec{q}_1\vec{k}_1\cdot\vec{\sigma}_1+(1-2\bar\beta_9)
\vec{q}_1\cdot\vec{\sigma}_1(2\vec{k}_2
-i\vec{k}\times\vec{\sigma}_2)]-2\vec{q}_1\cdot\vec{\sigma}_1(i
\vec{q}_1\times\vec{\sigma}_2 - i \vec{k}\times\vec{\sigma}_2 + 2 \vec{k}_2),\nn
\vec{B}_5&=&g_A^2\vec{q}_1\cdot\vec{\sigma}_1\Big[(1-2\bar\beta_8)(
\vec{q}_1\, \vec{k}\cdot\vec{q}_1-2i \,\vec{q}_1\times\vec{\sigma}_2\vec{k}_2\cdot\vec{q}_1
)-2i(1+2\bar\beta_8) \vec{q}_1\times\vec{\sigma}_2\,\vec{k}_1\cdot\vec{q}_1
\Big],\nn
\vec{B}_6&=&-(1-2\bar{\beta}_8) g_A^2
\vec{k}\,\vec{q}_1\cdot\vec{\sigma}_1[(\vec{k}\cdot\vec{q}_1)^2
-2i\, \vec{k}\cdot\vec{k}_2
\vec{k}\cdot\vec{q}_1\times\vec{\sigma}_2],\nn
\vec{B}_7&=&g_A^2\vec{k}\Big[(1-2\bar\beta_9)\vec{q}_1\cdot\vec{\sigma}_1(
-2i(\vec{k}\cdot\vec{k}_2\times\vec{\sigma}_2
+\vec{k}_2\cdot\vec{q}_1\times\vec{\sigma}_2)+k^2+q_1^2)
-2i (1+2\bar\beta_9)
\vec{k}_1\cdot\vec{\sigma}_1\vec{k}\cdot\vec{q}_1\times\vec{\sigma}_2\Big],\nn
\vec{B}_{8}&=&-g_A^2[(1-2\bar\beta_9)\vec{q}_1\cdot\vec{\sigma}_1(\vec{k}
-2i\,\vec{k}_2\times\vec{\sigma}_2)
-2i(1+2\bar\beta_9) \vec{q}_1\times\vec{\sigma}_2\,\vec{k}_1\cdot\vec{\sigma}_1].\label{OPEOneOvermNuclFFs}
\eeqa
It is not quite straightforward to make a connection between the derived relativistic
corrections to the axial current operator and the corresponding
$1/m$-terms appearing in the 3N force at N$^3$LO. This is because
the later ones also receive contributions from diagrams involving an insertion of the
$1/m$-vertex at the nucleon line, which we regard as being attributed
to the axial-vector source (i.e.~the leftmost nucleon line in the 3N
force shown in Fig.~\ref{fig:factor}), and which is connected with the two-nucleon
system via one-pion exchange. Thus, to establish the
connection, we have to consider only those topologies in the 3N force, which do not include such
contributions. In Fig.~\ref{fig:one_over_m_3nf_2pe} we show all
relevant diagrams which generate non-vanishing terms in the 3N
force. We have calculated the resulting contributions 
$[V^{\rm 3NF}_{{\rm TPE,} \, 1/m}]_{\rm modified}$
and  verified the validity of the 
relation 
\beqa
[V^{\rm 3NF}_{{\rm TPE,} \, 1/m}]_{\rm modified}&=&-\frac{g_A}{2
  F_\pi^2}\sum_{a}\tau_3^a\vec{\sigma}_3\cdot\vec{A}_{{\rm 2N:} \, 1\pi
  , \, 1/m}^{a\, (Q)}
\bigg|_{\vec{k}=-\vec{q}_3,k^2=-M_\pi^2}
\; +\;  {\cal O}\Big((k^2+M_\pi^2)^0 \Big).
\eeqa

\begin{figure}[tb]
\vskip 1 true cm
\includegraphics[width=0.35\textwidth,keepaspectratio,angle=0,clip]{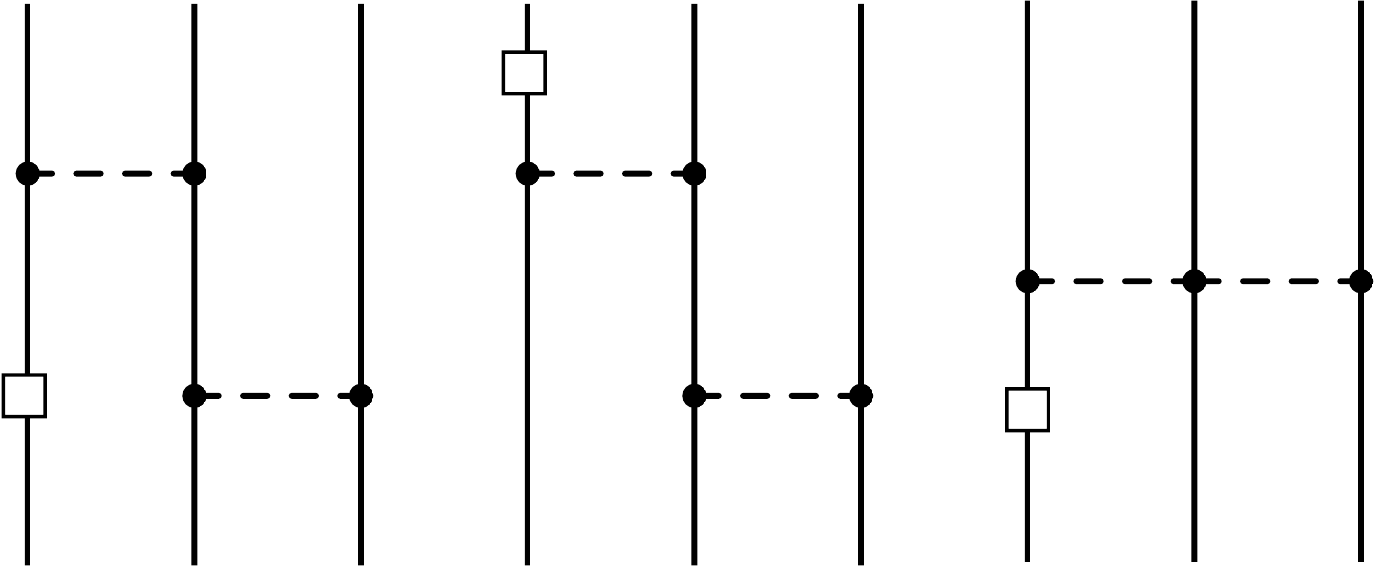}
    \caption{Diagrams leading to non-vanishing contributions to the two-pion exchange 3N force at N$^3$LO which include the $1/m$-vertices from ${\cal
          L}_{\pi N}^{(2)}$ (shown by open squares) at the leftmost
      nucleon line. The indicated kinetic energy insertions at the leftmost nucleon
      line can be identically expressed in terms of the energy transfer $k_0$ of Eq.~(\ref{k0From3N}). Time reversed diagrams are not shown. 
\label{fig:k0_3nf_2pe} 
 }
\end{figure}

Last but not least, there are also contributions proportional to the
energy transfer $k_0$ stemming from time-derivatives of the unitary
transformations in diagrams shown in
Fig.~\ref{fig:tree}. As already
mentioned earlier, $k_0$ counts as a quantity of order $Q^3$ 
so that
the contributions from diagrams $(1),(3)$ and $(4)$ of
Fig.~\ref{fig:tree} are shifted from order $Q^{-1}$ to order $Q$.  For the standard choice
of unitary phases we obtain 
\beqa
A_{{\rm 2N:\,}1\pi, {\rm UT^\prime} }^{0, a \; ({Q})} &=& 0\,,\\
\vec {A}_{{\rm 2N:\,}1\pi, {\rm UT^\prime} }^{a \; ({Q})} &=&
- i \frac{g_A}{8 F_\pi^2}\frac{k_0\, \vec k \, \vec{q}_1\cdot\vec{\sigma}_1}{(k^2+M_\pi^2)(q_1^2+M_\pi^2)}\bigg(
[{\fet \tau}_1\times{\fet 
\tau}_2]^a\bigg(1-\frac{2
g_A^2\vec{k}\cdot\vec{q}_1}{k^2+M_\pi^2}\bigg)-\frac{2 g_A^2\tau_1^a \vec{k}\cdot[\vec{q}_1\times\vec{\sigma}_2]}{k^2+M_\pi^2}\bigg)
+ \; 1 \leftrightarrow 2\,. \quad\quad
\label{NNCurrent1piUTPrime}
\eeqa
The current contribution in Eq.~(\ref{NNCurrent1piUTPrime}) involves 
the pion production operator, which, again, can be matched to the
corresponding expressions in the 3N force. The energy transfer $k_0$ can
then be written as the difference of the initial and final kinetic
energies of the third nucleon,
\beqa
k_0&=&\frac{p_3^2}{2 m}-\frac{p_3^{\prime\,2}}{2 m}.\label{k0From3N}
\eeqa
Notice that $k_0$ refers to the outgoing (incoming) energy transfer of the third nucleon
(subsystem of the nucleons $1$ and $2$). Thus, we need to consider a
subset of diagrams generating relativistic corrections to
the two-pion exchange 3N force with the kinetic-energy insertions at
the third nucleon, i.e. at the leftmost nucleon lines in Fig.~\ref{fig:k0_3nf_2pe}. 
The explicit expressions for the corresponding 3N force contributions 
are given by
\beqa
[V^{\rm 3NF}_{{\rm TPE,} \, k_0}]_{\rm modified}&=&- i \frac{g_A^2}{16 F_\pi^4}\frac{k_0\, \vec{q}_3\cdot\vec{\sigma}_3 \, \vec{q}_1\cdot\vec{\sigma}_1}{(q_3^2+M_\pi^2)(q_1^2+M_\pi^2)}\bigg[
{\fet \tau}_3\cdot[{\fet \tau}_1\times{\fet 
\tau}_2]\bigg(1+(1+2\bar{\beta}_8)\frac{
g_A^2\vec{q}_3\cdot\vec{q}_1}{q_3^2+M_\pi^2}\bigg)\nn
&+&(1+2\bar{\beta}_8)\frac{g_A^2{\fet \tau}_1\cdot{\fet \tau}_3 \vec{q}_3\cdot[\vec{q}_1\times\vec{\sigma}_2]}{q_3^2+M_\pi^2}\bigg]
+ \; 1 \leftrightarrow 2\,, \quad\quad
\label{V3Nk0TPE}
\eeqa
where $k_0$ is specified in Eq.~(\ref{k0From3N}). We then find that
the difference between the 3N force in the above equation and the
contribution reconstructed from the axial current in Eq.~(\ref{NNCurrent1piUTPrime}) is given by 
\beqa
&&[V^{\rm 3NF}_{{\rm TPE,} \, k_0}]_{\rm modified}+\frac{g_A}{2
  F_\pi^2}\sum_{a}\tau_3^a\vec{\sigma}_3\cdot
\vec {A}_{{\rm 2N:\,}1\pi, {\rm UT^\prime} }^{a \; ({Q})}
\bigg|_{\vec{k}=-\vec{q}_3,k^2=-M_\pi^2}\,=\,\nn
&& (1-2\bar{\beta}_8) i \frac{g_A^4}{16 F_\pi^4}\frac{k_0\, \vec{q}_3\cdot\vec{\sigma}_3 \, \vec{q}_1\cdot\vec{\sigma}_1}{(q_3^2+M_\pi^2)^2(q_1^2+M_\pi^2)}\bigg(
{\fet \tau}_3\cdot[{\fet \tau}_1\times{\fet 
\tau}_2]
\vec{q}_3\cdot\vec{q}_1
+{\fet \tau}_1\cdot{\fet \tau}_3 \vec{q}_3\cdot[\vec{q}_1\times\vec{\sigma}_2]\bigg)
+ \; 1 \leftrightarrow 2\,. \quad\quad\label{k0MatchingExplicit}\,.
\eeqa
This shows that the matching of $\vec {A}_{{\rm 2N:\,}1\pi, {\rm UT^\prime} }^{a \; ({Q})}$
to the 3N force is only possible for $\bar{\beta}_8=1/2$.
 
\subsection{Two-pion-exchange contributions}
\label{sec:2N2pi}
\begin{figure}[tb]
\vskip 1 true cm
\includegraphics[width=14.5cm,keepaspectratio,angle=0,clip]{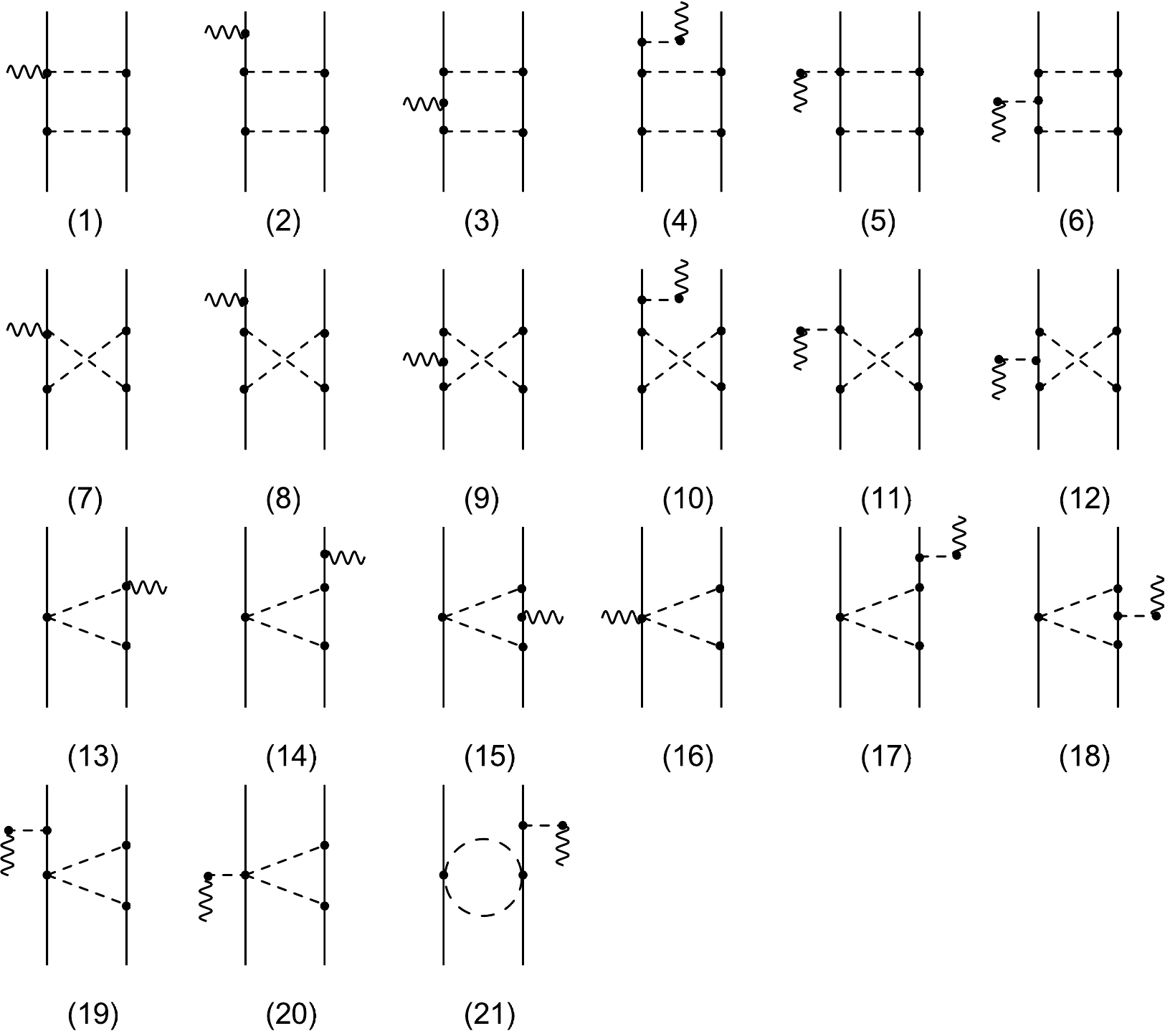}
    \caption{
         Two-pion-exchange diagrams contributing to ${\fet A}_{\rm
           2N}^{\mu \; (Q)}$. For notation see Fig.~\ref{fig:tree}.  
\label{fig:tpe} 
 }
\end{figure}

We now turn to the two-pion exchange contributions. 
In Fig.~\ref{fig:tpe}, we show all diagrams yielding non-vanishing
results for the axial charge and/or current operator with two
exchanged pions. 
For the axial charge, diagrams
(1), (4), (5), (7), (10), (13), (17), (19), (21) give non-vanishing
contributions, from which those of graphs 
(4), (10), (17), (19), (21) appear to depend on the unitary phases via
\beq 
(4),(10),(17),(19),(21)\sim 1+\alpha_{16}^{ax,{\rm LO}}.
\eeq
Clearly, all these contributions vanish for our standard choice.
For the axial current operator, we find non-vanishing results from
diagrams (2-6), (8-12), (14-18), (20), which in the case of graphs
(2-6) depend on the unitary phases according to 
\beqa
(2),(3) &\sim& 1+2\alpha_{1}^{ax}, \nn 
(5)&\sim& -3 + 2\alpha_{25}^{ax}+2\alpha_{26}^{ax},\nn
(4),(6)&\sim&
(-1+2\alpha_{4}^{ax}-\alpha_5^{ax}+\alpha_{10}^{ax})\frac{1}{(k^2+M_\pi^2)^2}
\big[ \ldots \big] +(-1+2\alpha_{12}^{ax})\frac{1}{k^2+M_\pi^2}.
\eeqa
Notice that the contributions involving second-order pion-pole terms
resulting from  diagrams (4), (6) vanish for our standard choice
of the unitary phases.  The final results for the two-pion exchange
operators read
\beqa
\label{Current2pi}
\vec {A}_{{\rm 2N:} \, 2\pi}^{a\, (Q)}&=& \frac{2 F_\pi^2}{g_A}\frac{\vec{k}}{k^2+M_\pi^2}
\bigg\{
\tau_1^a \Big(-\vec{q}_1\cdot \vec{\sigma}_2\, \vec{q}_1\cdot\vec{k}\,
g_1(q_1) + \vec{q}_1\cdot \vec{\sigma}_2\, g_2(q_1) - 
\vec{k}\cdot \vec{\sigma}_2\, g_3(q_1)\Big) \; + \; 
\tau_2^a \Big(-\vec{q}_1\cdot \vec{\sigma}_1\, \vec{q}_1\cdot\vec{k}\,
g_4(q_1)
\nn
&-&\vec{k}\cdot \vec{\sigma}_1 \,g_5(q_1) -\vec{q}_1\cdot
\vec{\sigma}_2\, 
\vec{q}_1\cdot\vec{k}\,g_6(q_1)
+ \vec{q}_1\cdot \vec{\sigma}_2\,
g_7(q_1) + \vec{k}\cdot
\vec{\sigma}_2\, 
\vec{q}_1\cdot\vec{k}\,g_8(q_1)
-\vec{k}\cdot \vec{\sigma}_2\, g_9(q_1) \Big)\nn
&+&
 [ \fet \tau_1 \times
\fet \tau_2 ]^a \Big( - \vec q_1 \cdot [ \vec \sigma_1 \times
\vec \sigma_2 ] \, \vec{q}_1\cdot\vec{k} \,g_{10}(q_1)+\vec q_1 \cdot [ \vec \sigma_1 \times
\vec \sigma_2 ] \, g_{11}(q_1) -  \vec q_1 \cdot
\vec \sigma_2 \,
\vec{q}_1\cdot[\vec q_2 \times \vec \sigma_1 ]\,
g_{12}(q_1)\Big)\bigg\}\nn
&+&\frac{2 F_\pi^2}{g_A}\bigg\{
\vec{q}_1\Big(
\tau_2^a \,\vec{q}_1\cdot \vec{\sigma}_1\,g_{13}(q_1)
+\tau_1^a \,\vec{q}_1\cdot \vec{\sigma}_2\, g_{14}(q_1)\Big)
-\tau_1^a \,\vec{\sigma}_2\, g_{15}(q_1) - \tau_2^a \,\vec{\sigma}_2\,
g_{16}(q_1) -\tau_2^a \,\vec{\sigma}_1\, g_{17}(q_1)
\bigg\}\nn [4pt]
& + & 1
\leftrightarrow 2\,, \\ [7pt]
\label{Charge2pi}
A_{{\rm 2N:} \, 2\pi}^{0, a \, (Q)}&=& i \frac{2 F_\pi^2}{g_A} \bigg\{
[ \fet \tau_1 \times
\fet \tau_2 ]^a
\vec q_1 \cdot \vec \sigma_2\,  g_{18}(q_1)
+ \tau_2^a \vec q_1 \cdot [ \vec \sigma_1 \times
\vec \sigma_2 ] 
g_{19}(q_1)
\bigg\} \; + \; 1
\leftrightarrow 2\,, 
\eeqa
where the scalar functions $g_i(q_1)$ are defined as 
\beqa
g_1(q_1)&=&\frac{g_A^4A(q_1)\left(\left(8g_A^2-4
\right)M_\pi^2+\left(g_A^2+1\right)q_1^2\right)}{256\pi 
F_\pi^6q_1^2}-\frac{g_A^4M_\pi\left(\left(8g_A^2-4
\right)M_\pi^2+\left(3g_A^2-1\right)q_1^2\right)}{256\pi 
F_\pi^6q_1^2\left(4M_\pi^2+q_1^2\right)}\nn
g_2(q_1)&=&\frac{g_A^4A(q_1)\left(2M_\pi^2+q_1^2\right)}{128\pi 
F_\pi^6}+\frac{g_A^4M_\pi}{128\pi F_\pi^6} 
,\nn
g_3(q_1)&=&-\frac{g_A^4A(q_1)\left(
\left(8g_A^2-4\right)M_\pi^2+\left(3g_A^2-1\right)q_1^2
\right)}{256\pi F_\pi^6}-\frac{\left(3g_A^2-1\right)g_A^4M_\pi}{256\pi 
F_\pi^6},\nn
g_4(q_1)&=&-\frac{g_A^6A(q_1)}{128\pi F_\pi^6},\nn 
g_5(q_1)&=&-q_1^2\,g_4(q_1),\nn [4pt]
g_6(q_1)&=& g_8(q_1)\,=\, g_{10}(q_1)\,=\, g_{12}(q_1)\,=\, 
0,\nn [4pt]
g_7(q_1)&=&\frac{g_A^4A(q_1)\left(2M_\pi^2+q_1^2\right)}{128\pi 
F_\pi^6}+\frac{\left(2g_A^2+1\right)g_A^4M_\pi}{128\pi F_\pi^6},\nn 
g_9(q_1)&=&\frac{g_A^6M_\pi}{64\pi F_\pi^6},\nn
g_{11}(q_1)&=&-\frac{g_A^4A(q_1)\left(4M_\pi^2+q_1^2\right)}{512\pi 
F_\pi^6}-\frac{g_A^4M_\pi}{512\pi F_\pi^6},\nn
g_{13}(q_1)&=&
-\frac{g_A^6A(q_1)}{128\pi F_\pi^6},\nn
g_{14}(q_1)&=&\frac{g_A^4A(q_1)\left(\left(8g_A^2-4
\right)M_\pi^2+\left(g_A^2+1\right)q_1^2\right)}{256\pi 
F_\pi^6q_1^2}+\frac{g_A^4M_\pi\left(\left(4-8g_A^2
\right)M_\pi^2+\left(1-3g_A^2\right)q_1^2\right)}{256\pi 
F_\pi^6q_1^2\left(4M_\pi^2+q_1^2\right)}\nn
g_{15}(q_1)&=&\frac{g_A^4A(q_1)\left(\left(8g_A^2-4
\right)M_\pi^2+\left(3g_A^2-1\right)q_1^2\right)}{256\pi 
F_\pi^6}+\frac{\left(3g_A^2-1\right)g_A^4M_\pi}{256\pi F_\pi^6},\nn
g_{16}(q_1)&=&\frac{g_A^4A(q_1)\left(
2M_\pi^2+q_1^2\right)}{64\pi F_\pi^6}+\frac{g_A^4M_\pi}{64\pi F_\pi^6},\nn
g_{17}(q_1)&=&-\frac{g_A^6q_1^2A(q_1)}{128\pi F_\pi^6},\nn
g_{18}(q_1)&=&\frac{g_A^2L(q_1)\left(\left(4-8g_A^2
\right)M_\pi^2+\left(1-3g_A^2\right)q_1^2
\right)}{128\pi^2F_\pi^6\left(4M_\pi^2+q_1^2\right)}, \nn
g_{19}(q_1)&=&\frac{g_A^4L(q_1)}{32\pi^2F_\pi^6}. \label{giOfq1TPEStrF}
\eeqa
Our standard choice of the unitary phases
ensures that the 2N irreducible pion production amplitude entering the
pion-pole contributions to the axial current operator equals 
the one appearing in the one-pion-two-pion-exchange 3N force at 
N$^3$LO. This manifests itself in the relations 
\beq
g_i(q_1) \,=\,F_i(q_1),\quad
i=1,\ldots ,12,
\label{tpeope::tpe::3nf::current}
\eeq
where $F_i(q_1)$ are the scalar functions entering the corresponding
3N force and defined in Eq. ~(3.2) of
\cite{Krebs:2013kha}.\footnote{Notice that in
  \cite{Krebs:2013kha}, we have only shown explicitly non-polynomial
  contributions to the scalar functions
  $F_i$ since the polynomial terms, which for dimensional reasons have
  to be momentum-independent and proportional to $M_\pi$, only lead to finite shifts of the
  LEC $c_D$. Eq.~(\ref{tpeope::tpe::3nf::current}) is valid both for polynomial
  and non-polynomial parts.
  }
Notice further that the loop contributions to the current operator are
finite in dimensional regularization, whereas the divergences in the
loop integrals appearing in the axial charge are absorbed into
redefinition of LECs accompanying the contact operators to be
specified below.

\subsection{Short-range contributions at order $Q$}

We first consider static contributions and begin with tree-level
diagrams, which emerge from the terms in the effective Lagrangian ${\cal L}_{NN}^{(2)}$
involving one derivative and one insertion of the axial vector source.  
While there are no
contributions to the current at this order, four independent
structures appear in the charge operator. Using the notation
of Ref.~\cite{Baroni:2015uza}, the tree-level contributions read: 
\beqa
\label{CurrentCont}
\vec {A}_{\rm 2N: \, cont}^{a \, (Q)}&=& 0,\\
A_{\rm 2N: \, cont}^{0, a \, (Q)}&=& i z_1 [\fet \tau_1 \times \fet
\tau_2]^a \, \vec \sigma_1 \cdot \vec q_2 \; + \; 
 i z_2 [\fet \tau_1 \times \fet
\tau_2]^a \, \vec \sigma_1 \cdot \vec q_1\;  +\;  
 i z_3 \tau_1^a \, \vec q_2 \cdot \vec \sigma_1 \times \vec \sigma_2 \nn
& + & z_4 (\tau_1^a - \tau_2^a) (\vec \sigma_1 - \vec \sigma_2) \cdot \vec
k_1 \; + \; 1 \leftrightarrow 2\,,\label{ChargeContQTo1}
\eeqa
with $z_i$ denoting the corresponding LECs. 

\begin{figure}[tb]
\vskip 1 true cm
\includegraphics[width=16.5cm,keepaspectratio,angle=0,clip]{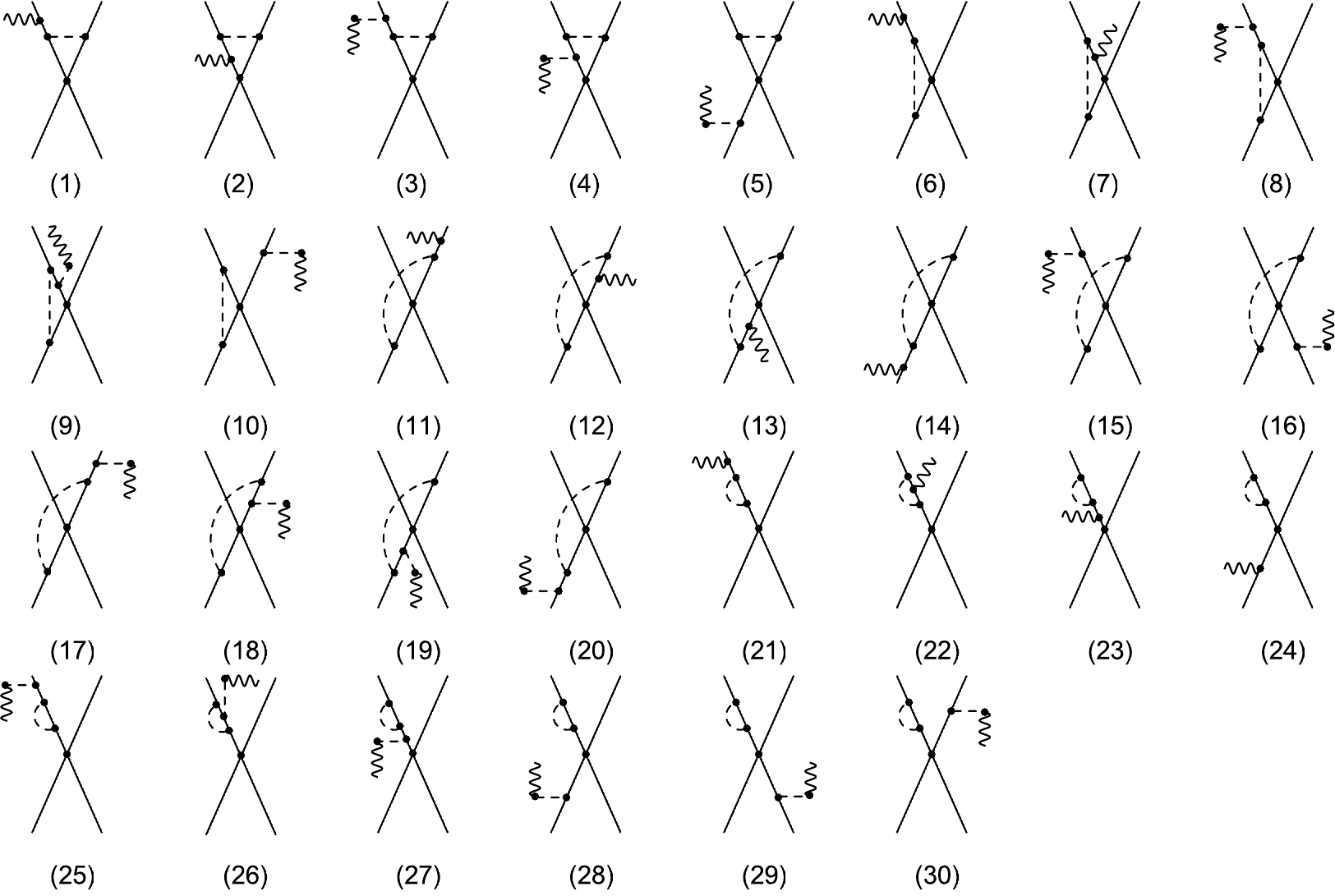}
    \caption{
         Loop diagrams with contact interactions contributing to ${\fet A}_{\rm
           2N}^{\mu \; (Q)}$. Solid dots denote vertices from
        ${\cal L}_{\pi N}^{(1)}$, ${\cal L}_{\pi}^{(2)}$ or  ${\cal
          L}_{NN}^{(0)}$. For remaining notation see Fig.~\ref{fig:tree}. 
\label{fig:contact} 
 }
\end{figure}

Next, we show in Fig.~\ref{fig:contact} all non-vanishing one-loop diagrams
involving a single insertion of the leading contact interactions
from $\mathcal{L}_{NN}^{(0)}$.  
For the axial charge, diagrams
(3), (5), (8), (10), (15-17), (20), (25-30) are found to give non-vanishing
contributions, from which those of graphs
(3), (5), (8), (10), (15-17), (20), (25-30) depend on the choice of
the unitary phases:
\beqa
 (3),(5),(8),(10),(15-17),(20),(28-30)&\sim& 1+\alpha_{16}^{ax,{\rm
    LO}},\nn 
(25)&\sim& 2 -\alpha_4^{ax}+\alpha_5^{ax} + \alpha_{16}^{ax,{\rm
    LO}}, \nn
(26),(27)&\sim& -1 + \alpha_4^{ax}-\alpha_5^{ax}. 
\eeqa
For our standard choice of the unitary phases, we obtain a vanishing
result for the pion-pole terms in the axial charge.
All ultraviolet divergences in the loop contributions to the axial
charge operator are cancelled by
the corresponding counterterms upon  expressing the bare LECs $z_i$ in
terms of their renormalized values $\bar z_i$ via 
\beqa
z_i &=&\overline{z}_i +\beta_{z_i}\frac{1}{F_\pi^4}\bigg(\lambda +\frac{1}{16 \pi^2 }
\ln\left(\frac{M_\pi}{\mu}\right)\bigg)\,.
\eeqa
The corresponding $\beta$-functions read: 
\beqa
\beta_{z_1}&=&\frac{(3 g_A^2-1) g_A}{4},\nn
\beta_{z_2}&=&-\frac{(5 g_A^2+1) g_A}{12},\nn
\beta_{z_3}&=&-g_A^3,\nn
\beta_{z_4}&=&0.
\eeqa
The remaining finite contributions to the charge operator at order $Q$
can be absorbed into a redefinition of the renormalized LECs $\bar z_i$ via the following shifts: 
\beqa
\bar{z}_1&\to&\bar{z}_1 + \frac{(1-g_A^2)g_A}{128 \pi^2 F_\pi^4},\nn
\bar{z}_2&\to&\bar{z}_2 + \frac{g_A}{4 F_\pi^2}(2 \bar{d}_2
+\bar{d}_6),\nn
\bar{z}_3&\to&\frac{g_A^3}{32 \pi^2 F_\pi^4}.
\eeqa

For the current operator, diagrams
(1-9), (11-14), (17-30) are found to yield  non-vanishing
contributions, from which those of graphs 
(1-5), (21), (23-25), (27-30) depend on the unitary phases:
\beqa
(1),(2)&\sim& 1+ 2 \alpha_1^{ax}, \nn 
(3)&\sim& (-1 + 2\alpha_4^{ax} -
\alpha_5^{ax}+\alpha_{10}^{ax})\frac{1}{(k^2+M_\pi^2)^2} \big[\ldots \big] +
(-1+2\alpha_{12}^{ax})\frac{1}{k^2+M_\pi^2} \big[\ldots \big] ,\nn
(4)&\sim& (-2 + 2 \alpha_4^{ax} - \alpha_5^{ax} + \alpha_{10}^{ax}
+\alpha_{21}^{ax}+\alpha_{22}^{ax} + \alpha_{23}^{ax})
\frac{1}{(k^2+M_\pi^2)^2} \big[\ldots \big]  + (-1 +
2\alpha_{12}^{ax})\frac{1}{k^2+M_\pi^2} \big[\ldots \big] ,\nn
(5),(24),(28-30)&\sim& -1 + \alpha_{21}^{ax}+\alpha_{22}^{ax} + \alpha_{23}^{ax},
\nn 
(21)&\sim& 1+\alpha_1^{ax},\nn 
(23)&\sim& \alpha_1^{ax}, \nn
(25)&\sim& (-1 + 2 \alpha_4^{ax} - \alpha_5^{ax} +
\alpha_{10}^{ax})\frac{1}{(k^2+M_\pi^2)^2} \big[\ldots \big]  + (-1 +
\alpha_{12}^{ax})\frac{1}{k^2 + M_\pi^2} \big[\ldots \big] , \nn
(27)&\sim& (-2 + 2 \alpha_4^{ax} - \alpha_5^{ax} + \alpha_{10}^{ax}
+\alpha_{21}^{ax}+\alpha_{22}^{ax} + \alpha_{23}^{ax})
\frac{1}{(k^2+M_\pi^2)^2} \big[\ldots \big]  +
\alpha_{12}^{ax}\frac{1}{k^2+M_\pi^2} \big[\ldots \big] .
\eeqa
For our standard choice, we find a vanishing result for the short-range
current operator after renormalization. This is consistent with the
matching condition to the nuclear forces, since the static one-pion-contact 3NF's
vanish after antisymmetrization. Thus, after
renormalization and antisymmetrization, there are no static contributions
to the short-range axial current operator at order $Q$, while those to
the charge operator are given by Eq.~(\ref{CurrentCont}).

\begin{figure}[tb]
\vskip 1 true cm
\includegraphics[width=16.5cm,keepaspectratio,angle=0,clip]{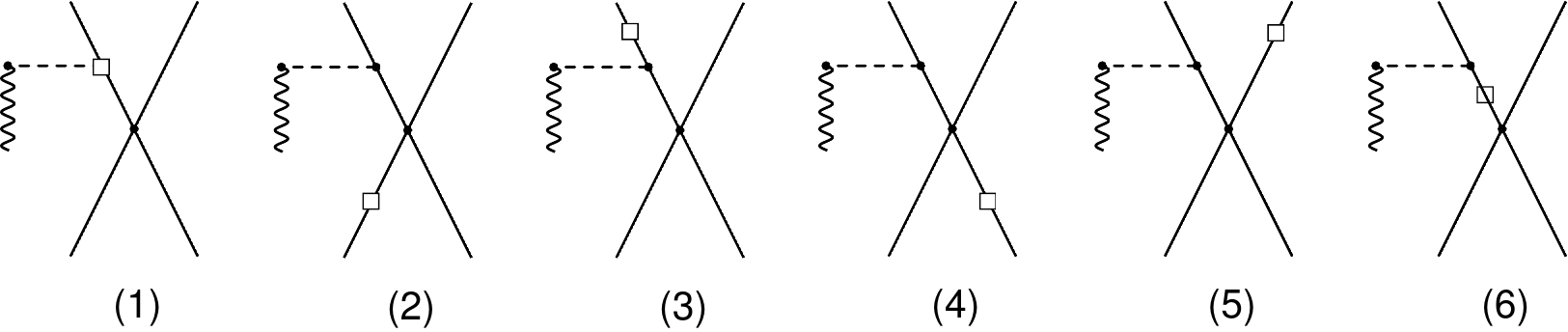}
    \caption{Tree-level diagrams with contact interactions leading to
      $1/m$-contributions to $\vec {\fet A}_{\rm 2N}^{(Q)}$. Solid
         dots (open squares) denote vertices from
        ${\cal L}_{\pi N}^{(1)}$ $(1/m-$corrections from ${\cal
          L}_{\pi N}^{(2)}$). For remaining notation see Fig.~\ref{fig:tree}.
\label{fig:relativistic_contact} 
 }
\end{figure} 

\begin{figure}[tb]
\vskip 1 true cm
\includegraphics[width=7.5cm,keepaspectratio,angle=0,clip]{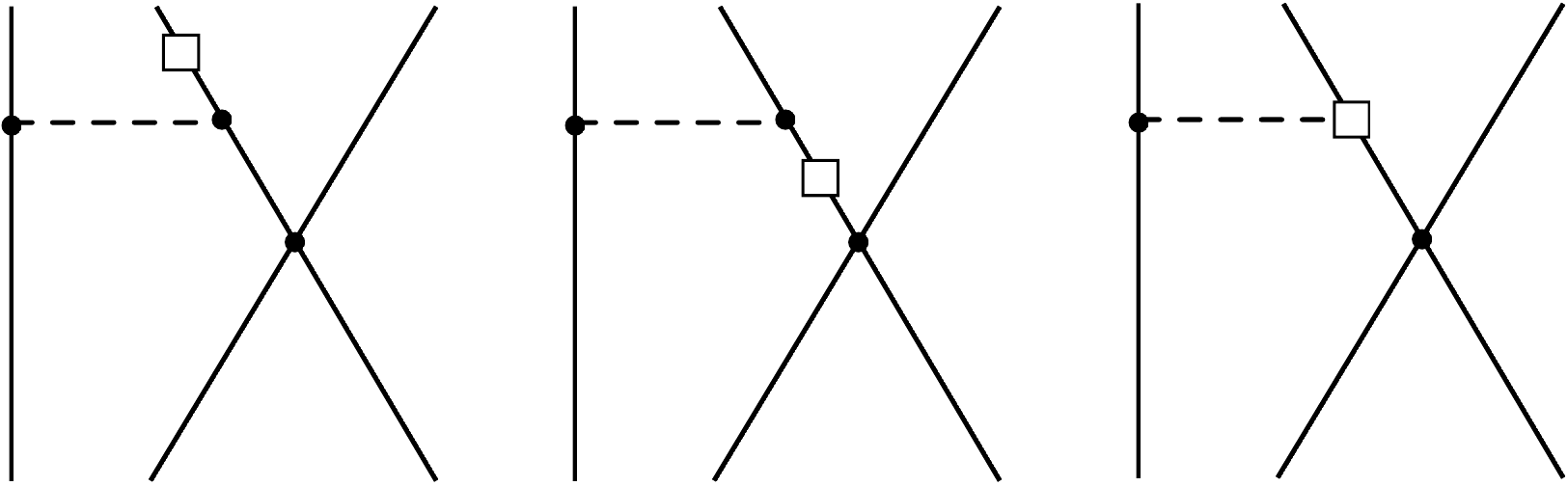}
    \caption{Diagrams leading to non-vanishing relativistic
      corrections to the one-pion-exchange-contact 3N force at N$^3$LO which
      do not include the $1/m$-vertices from ${\cal
          L}_{\pi N}^{(2)}$ (shown by open squares) at the leftmost
      nucleon line. Time reversed diagrams are not shown. For
      remaining notation see Fig.~\ref{fig:factor}.
\label{fig:one_over_m_3nf_ope_contact} 
 }
\end{figure}

In addition to the static terms considered above, one also encounters relativistic $1/m$ corrections involving a
single insertion of the LO contact interactions from
$\mathcal{L}_{NN}^{(0)}$.  We find that the
contributions to the axial charge disappear regardless of the
choice of the unitary phases. For the vector current, non-vanishing 
$1/m$-corrections emerge from the diagrams shown in Fig.~\ref{fig:relativistic_contact}. The unitary phase
dependence of these diagrams is given by
\beqa
(1)&\sim& 1+\alpha_{16}^{ax, 1/m},\nn 
(2),(4),(5)&\sim&
\alpha_{21}^{ax}+\alpha_{22}^{ax}+\alpha_{23}^{ax} - 1, \nn 
(3)&\sim&
\alpha_{17}^{ax}+\alpha_{18}^{ax}+\alpha_{19}^{ax}
+\alpha_{21}^{ax}+\alpha_{22}^{ax}+\alpha_{23}^{ax} - 2,\nn
(6)&\sim& \alpha_{17}^{ax}+\alpha_{18}^{ax}+\alpha_{19}^{ax} - 1.
\eeqa 
For our standard choice of the unitary phases, we find 
the following  result
for the axial current operator:
\beqa
\label{CurrentContRel}
\vec {A}_{{\rm 2N: \, cont,}\, 1/m}^{a \, (Q)}&=&-\frac{g_A }{4m}\frac{\vec{k}}{k^2+M_\pi^2}
\tau_1^a\bigg\{(1-2\bar{\beta}_9)\Big(C_S\vec{q}_2\cdot\vec{\sigma}_1
+C_T(\vec{q}_2\cdot\vec{\sigma}_2
+2i\,\vec{k}_1\cdot\vec{\sigma}_1\times\vec{\sigma}_2)\Big)\nn
&-&\frac{1-2\bar{\beta}_8}{k^2+M_\pi^2} \Big(C_S\vec{k}\cdot\vec{q}_2\vec{k}\cdot\vec{\sigma}_1
+C_T(\vec{k}\cdot\vec{q}_2\vec{k}\cdot\vec{\sigma}_2+2i \,\vec{k}\cdot\vec{k}_1
\vec{k}\cdot\vec{\sigma}_1\times\vec{\sigma}_2)\Big)\bigg\} \; +\;
1\leftrightarrow 2\,.
\eeqa
To verify the matching condition with the 3N force, we have calculated the
contribution $[V^{\rm 3NF}_{1\pi-{\rm cont,} \, 1/m}]_{\rm modified}$ of the diagrams shown in
Fig.~\ref{fig:one_over_m_3nf_ope_contact}, which do not involve
$1/m$-corrections at the leftmost nucleon line.
We have then verified that the following relation is indeed valid:
\beqa
[V^{\rm 3NF}_{1\pi-{\rm cont,} \, 1/m}]_{\rm modified} &=&-\frac{g_A}{2
  F_\pi^2}\sum_{a}\tau_3^a\vec{\sigma}_3\cdot\vec A_{{\rm 2N: \,
    cont,}\, 1/m}^{a \, (Q)} \bigg|_{\vec{k}=-\vec{q}_3,k^2=-M_\pi^2}
+ {\cal O} \Big((k^2+M_\pi^2)^0 \Big).
\eeqa

Finally, there are also contributions from diagram $(5)$ of Fig.~\ref{fig:tree},
which are proportional to the energy transfer $k^0$. For the standard
choice of unitary phases we obtain the following result:
\beqa
A_{{\rm 2N:\,cont,\,UT^\prime} }^{0, a \; ({Q})}
&=&0,\nn
\vec {A}_{{\rm 2N:\,cont,\,UT^\prime}}^{a \; ({Q})} &=&-i\,k_0 \vec{k}\frac{g_A C_T
\vec{k}\cdot \vec{\sigma}_1 [{\fet 
\tau}_1\times{\fet \tau}_2]^a}{
\left(k^2+M_\pi^2\right)^2}
\; + \; 1 \leftrightarrow 2\,.\label{NNCurrentContUTPrime}
\eeqa
Similar to the one-pion exchange current, this expression matches the
corresponding terms in the 3N force for the choice of $\bar \beta_8 =
1/2$.  

\section{Three-nucleon axial currents}
\def\theequation{\arabic{section}.\arabic{equation}}
\label{sec:3N}

\begin{figure}[tb]
\vskip 1 true cm
\includegraphics[width=16.5cm,keepaspectratio,angle=0,clip]{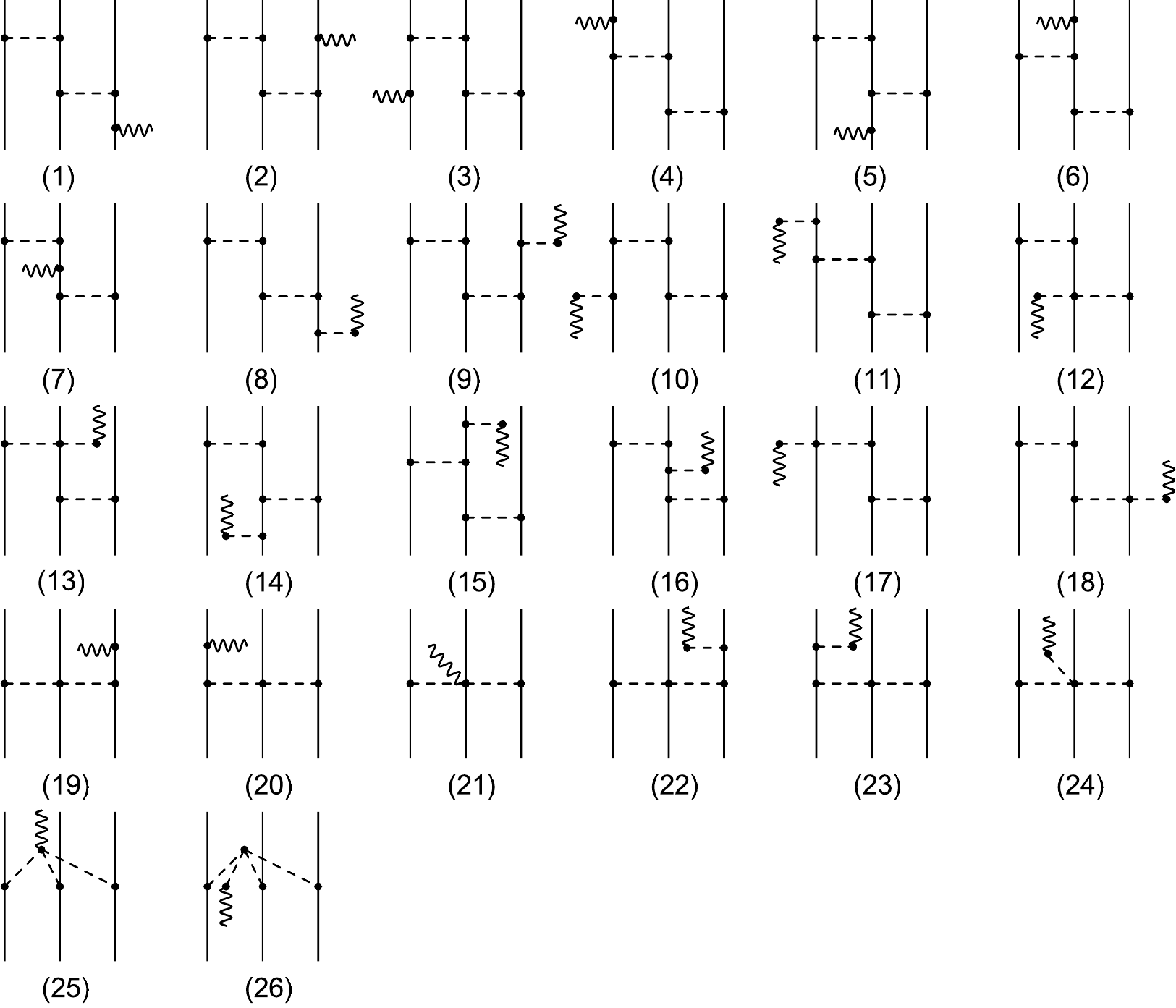}
    \caption{Tree-level two-pion exchange diagrams leading to non-vanishing
         contributions to $\vec {\fet A}_{\rm 3N}^{(Q)}$. For notation see Fig.~\ref{fig:tree}.
\label{fig:tpe_3N} 
 }
\end{figure}

We now turn to the 3N  axial current
operators, whose dominant contributions appear at order $Q$ 
from tree-level diagrams constructed solely from the lowest-order
vertices.
In Fig.~\ref{fig:tpe_3N}, we show all graphs which
yield non-vanishing contributions to the axial current,
and that do not involve contact interactions. 
Interestingly, we find only vanishing contributions to the 3N charge operator
at this order.  Out of  the 26 diagrams shown in Fig.~\ref{fig:tpe_3N},
graphs (1-18) yield expressions which depend on the unitary phases as follows:
\beqa
(1-4)&\sim& 1+\alpha_1^{ax}, \nn 
(5-7)&\sim& \alpha_1^{ax}, \nn
(12),(13)&\sim& -1+\alpha_{25}^{ax}+\alpha_{26}^{ax}, \nn 
(17),(18)&\sim& -2+\alpha_{25}^{ax}+\alpha_{26}^{ax}, \nn
(8-11)&\sim&
(-1+2\alpha_4^{ax}-\alpha_5^{ax}+\alpha_{10}^{ax})\frac{1}{(k^2+M_\pi^2)^2}
\big[ \ldots \big] + (-1+\alpha_{12}^{ax})\frac{1}{k^2+M_\pi^2}\big[ \ldots \big] ,\nn
(14-16)&\sim&
(-1+2\alpha_4^{ax}-\alpha_5^{ax}+\alpha_{10}^{ax})\frac{1}{(k^2+M_\pi^2)^2}
\big[ \ldots \big]  + \alpha_{12}^{ax}\frac{1}{k^2+M_\pi^2}\big[ \ldots \big]  \,.
\eeqa
For our standard choice of the unitary phases, the contributions from diagrams
 (5-7) and (12-16) as well as all expressions involving second-order
 pion-pole terms are found to vanish.
In order to facilitate a comparison with the four-nucleon force at
N$^3$LO, we write the resulting expression for the 3N axial current $\vec A_{{\rm 3N:}\,
  \pi}^{a \, (Q)}$ in the form 
\beq
\label{Current3NPi}
\vec {A}_{{\rm 3N:}\,
  \pi}^{a \, (Q)} =- \frac{2 F_\pi^2}{g_A} \sum_{i=1}^{8}   \vec C_i^a
+
5\,{\rm permutations}, 
\eeq
where 
\beqa
\vec C_1^a&=&\frac{g_A^6}{16 F_\pi^6} \vec{q}_2\cdot \vec{\sigma}_2 \bigg[\tau_2^a \Big[
\vec{q}_3 ( (\vec{q}_2\cdot \vec{q}_3+q_2^2) 
(\fet\tau_1\cdot \fet
\tau_3-\vec{\sigma}_1\cdot \vec{\sigma}_3)+\vec{q}_2\cdot \vec{\sigma}_1 (\vec{q}_2\cdot 
\vec{\sigma}_3+\vec{q}_3\cdot \vec{\sigma}_3))\nn
&-&
\vec{q}_2 (\vec{q}_3\cdot \vec{\sigma}_1 (\vec{q}_2\cdot 
\vec{\sigma}_3+\vec{q}_3\cdot \vec{\sigma}_3)-(\vec{q}_2\cdot 
\vec{q}_3+q_3^2) \vec{\sigma}_1\cdot 
\vec{\sigma}_3-(\vec{q}_2\cdot \vec{q}_3+q_2^2) 
\fet\tau_1\cdot \fet\tau_3)\nn
&-&\vec{\sigma}_3 
((\vec{q}_2\cdot \vec{q}_3+q_3^2) 
\vec{q}_2\cdot \vec{\sigma}_1-(\vec{q}_2\cdot 
\vec{q}_3+q_2^2) \vec{q}_3\cdot 
\vec{\sigma}_1)\Big]- \left[\fet\tau_2\times \fet\tau_3\right]^a (\vec{q}_2
\times \vec{\sigma}_1+\vec{q}_3\times \vec{\sigma}_1) (\vec{q}_2\cdot \vec{q}_3+q_2^2)\nn
&-&(
\vec{q}_2+\vec{q}_3) ([\fet\tau_1\times 
\fet\tau_2]^a \vec{q}_2\cdot \vec{q}_3\times \vec{\sigma}_3+\tau_3^a \fet
\tau_1\cdot \fet\tau_2
(\vec{q}_2\cdot \vec{q}_3+q_2^2))\bigg]\frac{1}{ [q_2^2 +M_\pi^2]
[(\vec{q}_1-\vec{k})^2 +M_\pi^2]^2},\nn
\vec C_2^a&=&\frac{g_A^4}{32 F_\pi^6} \frac{\vec{q}_2\cdot \vec{\sigma}_2 \big[[\fet\tau_2\times \fet\tau_3]^a 
(\vec{ k}\times \vec{\sigma}_1-\vec{q}_1
\times \vec{\sigma}_1)+(\vec{k}- \vec{q}_1) (
\tau_3^a \fet\tau_1\cdot \fet\tau_2-\tau_2^a \fet\tau_1\cdot \fet
\tau_3)\big]}{ [q_2^2 +M_\pi^2]
[(\vec{q}_1-\vec{k})^2+M_\pi^2]},\nn
\vec C_3^a&=&\frac{g_A^4}{32 F_\pi^6}\frac{ (\vec{ k}-2 
\vec{q}_3)\tau_3^a \fet\tau_1\cdot \fet\tau_2\vec{q}_1\cdot \vec{\sigma}_1 \vec{q}_2\cdot 
\vec{\sigma}_2 (\vec{k}\cdot \vec{\sigma}_3-2 \vec{q}_1\cdot 
\vec{\sigma}_3)}{ [q_1^2 +M_\pi^2] [
q_2^2 +M_\pi^2] [q_3^2 +M_\pi^2]},\nn
\vec C_4^a&=&\frac{g_A^4 }{32 F_\pi^6}\frac{\vec{\sigma}_1 \vec{q}_2\cdot 
\vec{\sigma}_2 \vec{q}_3\cdot \vec{\sigma}_3 (\tau_1^a \fet
\tau_2\cdot \fet\tau_3-\tau_3^a \fet\tau_1\cdot \fet\tau_2)}{ 
[q_2^2 + M_\pi^2] [q_3^2 +M_\pi^2]},\nn
\vec C_5^a&=&\frac{g_A^6}{16 F_\pi^6} \vec{k} \,\vec{q}_1\cdot \vec{\sigma}_1 
\bigg[-\tau_1^a \vec{q}_1\cdot \vec{q}_2\times 
\vec{\sigma}_2 (\vec{k}\cdot \vec{q}_1\times \vec{\sigma}_3+\vec{k}\cdot 
\vec{q}_2\times \vec{\sigma}_3)-\left[\fet
\tau_1\times \fet\tau_3\right]^a\vec{q}_1\cdot \vec{q}_2\times \vec{\sigma}_2 (\vec{k}\cdot \vec{q}_1+\vec{k}\cdot 
\vec{q}_2) \nn
&+&\left[\fet
\tau_1\times \fet\tau_2\right]^a (\vec{q}_1\cdot 
\vec{q}_2+q_1^2) (\vec{k}\cdot \vec{q}_1\times 
\vec{\sigma}_3+\vec{k}\cdot \vec{q}_2\times \vec{\sigma}_3)
-\left(\fet\tau_2\cdot \fet\tau_3 \tau_1^a-\fet\tau_1\cdot \fet\tau_3
  \tau_2^a\right) 
\nn
&\times& (\vec{q}_1\cdot 
\vec{q}_2+q_1^2) (\vec{k}\cdot \vec{q}_1+\vec{k}\cdot 
\vec{q}_2)
\bigg]\frac{1}{ [k^2+M_\pi^2] [q_1^2 +M_\pi^2
] [(\vec{q}_1+\vec{q}_2)^2 +M_\pi^2]^2},\nn
\vec C_6^a &=&-\frac{g_A^4}{32 F_\pi^6}\vec{k}\, \vec{q}_1\cdot \vec{\sigma}_1 \bigg[(\vec{k}
\cdot \vec{q}_1\times \vec{\sigma}_3+\vec{k}\cdot \vec{q}_2\times 
\vec{\sigma}_3) \left[\fet\tau_1\times \fet\tau_2\right]^a-(\vec{k}\cdot 
\vec{q}_1+\vec{k}\cdot \vec{q}_2+\vec{q}_1\cdot 
\vec{q}_2+q_1^2) \nn
&\times&\left(\fet\tau_2\cdot \fet
\tau_3 \tau_1^a-\fet\tau_1\cdot \fet\tau_3
\tau_2^a\right)- \vec{q}_1\cdot \vec{q}_2\times \vec{\sigma}_2 
\left[\fet\tau_1\times \fet\tau_3\right]^a\bigg]\frac{ 1}{ 
[k^2+M_\pi^2] [q_1^2 +M_\pi^2]
[(\vec{q}_1+\vec{q}_2)^2 +M_\pi^2]},\nn
\vec C_7^a&=&-\frac{g_A^4}{32 F_\pi^6}\frac{\vec{ k}\, \vec{q}_1\cdot \vec{\sigma}_1 \vec{q}_2
\cdot \vec{\sigma}_2 \vec{q}_3\cdot \vec{\sigma}_3 \fet\tau_1\cdot 
\fet\tau_2 \tau_3^a\left(M_\pi^2+2 \vec{q}_1\cdot \vec{q}_2+q_1^2+q_2^2\right) }{ [k^2+M_\pi^2]
[q_1^2 +M_\pi^2] [q_2^2 +M_\pi^2]
[q_3^2 +M_\pi^2]},\nn
\vec C_8^a&=&-\frac{g_A^4}{64 F_\pi^6}\frac{ \vec{ k}\, \vec{q}_2\cdot \vec{\sigma}_2 
\vec{q}_3\cdot \vec{\sigma}_3 \left(\fet\tau_2\cdot \fet\tau_3 \tau_1^a
(\vec{q}_2\cdot \vec{\sigma}_1+\vec{q}_3\cdot \vec{\sigma}_1) 
-2 \fet\tau_1\cdot \fet\tau_2 \tau_3^a
(\vec{q}_1\cdot \vec{\sigma}_1+\vec{q}_2\cdot \vec{\sigma}_1) 
\right)}{ [k^2+M_\pi^2] 
[q_2^2 +M_\pi^2] [q_3^2 +M_\pi^2]}\,. \label{Ci3NDefinition}
\eeqa
The pion-pole terms can be related to the corresponding contributions
to the leading four-nucleon force
at N$^3$LO 
if we multiply the axial current operator by $- g_A/(2 F_\pi^2) \vec{\sigma_4}\tau_4^a$ and replace
$\vec{k}\to - \vec{q}_4$:
\beqa
V_{{\rm
    class-I}}&=&\frac{1}{2}\sum_{a}\tau_4^a\vec{\sigma}_4\cdot\vec{C}_{5}^a
\Big|_{\vec{k}=-\vec{q}_4}+23\,{\rm
  permutations},\nn
V_{{\rm
    class-II}}^1&=&\frac{1}{2}\sum_{a}\tau_4^a\vec{\sigma}_4\cdot\vec{C}_{6}^a
\Big|_{\vec{k}=-\vec{q}_4}+23\,{\rm
  permutations}, \nn
V_{{\rm
    class-II}}^2&=&\frac{1}{4}\sum_{a}\tau_4^a\vec{\sigma}_4\cdot\vec{C}_{7}^a
\Big|_{\vec{k}=-\vec{q}_4} +
\frac{1}{3}\sum_{a}\tau_4^a\vec{\sigma}_4\cdot\vec{C}_{8}^a \Big|_{\vec{k}=-\vec{q}_4}+23\,{\rm
  permutations},
\eeqa
where the expressions for the four-nucleon force contributions 
$V_{{\rm    class-I}}$, $V_{{\rm   class-II}}^1$ and $V_{{\rm  class-II}}^2$
 are taken from~\cite{Epelbaum:2007us}.

\begin{figure}[tb]
\vskip 1 true cm
\includegraphics[width=17.5cm,keepaspectratio,angle=0,clip]{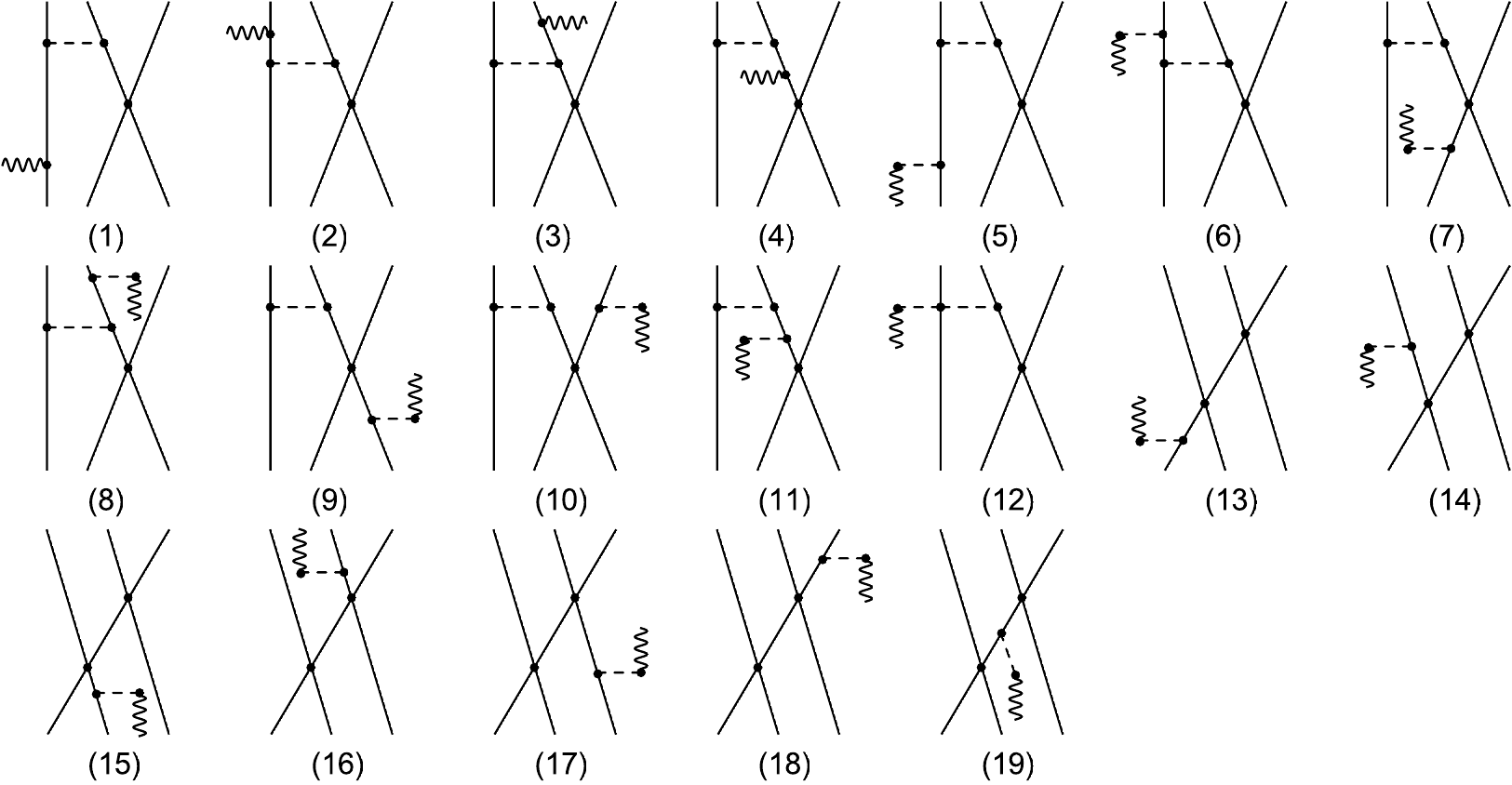}
    \caption{
Tree-level diagrams involving one or two insertions of the leading contact interactions  from
         ${\cal L}_{NN}^{(0)}$ which lead to non-vanishing
         contributions to $\vec {\fet A}_{\rm 3N}^{(Q)}$. For notation see Fig.~\ref{fig:tree}.
\label{fig:contact_3N} 
 }
\end{figure}

Finally, in Fig.~\ref{fig:contact_3N}, we show all non-vanishing diagrams which
contribute to the 3N axial vector current and include one or
more insertions of the leading 2N contact interactions. These
diagrams also do not contribute to the axial charge. 
All these diagrams yield contributions which depend on the unitary phases:
\beqa
(1),(2)&\sim& 1+\alpha_1^{ax}, \nn 
(3),(4)&\sim& \alpha_1^{ax}, \nn
(7),(9),(10),(13-19)&\sim& -1 + \alpha_{21}^{ax} + \alpha_{22}^{ax} + \alpha_{23}^{ax},
\nn
(5),(6)&\sim&
(-1+2\alpha_4^{ax}-\alpha_5^{ax}+\alpha_{10}^{ax})\frac{1}{(k^2+M_\pi^2)^2}
\big[ \ldots \big] + (-1+\alpha_{12}^{ax})\frac{1}{k^2+M_\pi^2}\big[ \ldots \big] , \nn
(8)&\sim&
(-1+2\alpha_4^{ax}-\alpha_5^{ax}+\alpha_{10}^{ax})\frac{1}{(k^2+M_\pi^2)^2}
\big[ \ldots \big]  + \alpha_{12}^{ax}\frac{1}{k^2+M_\pi^2}\big[ \ldots \big] , \nn 
(12)&\sim& -2 + \alpha_{25}^{ax} + \alpha_{26}^{ax}, \\
(11)&\sim& (-2+2\alpha_4^{ax}-\alpha_5^{ax}+\alpha_{10}^{ax} + \alpha_{21}^{ax} + \alpha_{22}^{ax} + \alpha_{23}^{ax})\frac{1}{(k^2+M_\pi^2)^2}
\big[ \ldots \big]  + \alpha_{12}^{ax}\frac{1}{k^2+M_\pi^2}\big[
\ldots \big] . \nonumber
\eeqa
For our standard choice of the unitary phases, the contributions of graphs
(3), (4), (7-11), (13-19)  and all expressions involving second-order pion-pole terms vanish 
yielding:
\beq
\label{Current3NCont}
\vec {A}_{\rm 3N:\, cont}^{a \, (Q)} =- \frac{2 F_\pi^2}{g_A} \sum_{i=1}^{3}   \vec D_i^a
+
5\,{\rm permutations}, 
\eeq
with 
\beqa
\vec D_1^a&=&-\frac{g_A^4 C_T}{4 
F_\pi^4} \big[(\vec{k}-\vec{q}_1) 
[\fet\tau_1\times \fet\tau_3]^a (\vec{k}\cdot \vec{\sigma}_2\times 
\vec{\sigma}_3-\vec{q}_1\cdot \vec{\sigma}_2\times \vec{\sigma}_3)-(
\tau_2^a-\tau_3^a) ((
\vec{k}-\vec{q}_1)\vec{\sigma}_1\cdot \vec{\sigma}_2 (\vec{k}\cdot \vec{\sigma}_3-\vec{q}_1\cdot 
\vec{\sigma}_3)\nn
&+&\vec{\sigma}_3 ((\vec{k}\cdot 
\vec{\sigma}_1-\vec{q}_1\cdot \vec{\sigma}_1) (\vec{k}\cdot \vec{\sigma}_2-
\vec{q}_1\cdot \vec{\sigma}_2)+(2 \vec{k}\cdot
\vec{q}_1-k^2-q_1^2) \vec{\sigma}_1\cdot
\vec{\sigma}_2))\big]\frac{1}{[(\vec{q}_1-\vec{k})^2+M_\pi^2]^2},\nn
\vec{D}_{2}^a&=&\frac{g_A^4 C_T}{4 F_\pi^4}\frac{ \vec{ k}\, (\vec{q}_1\cdot
  \vec{\sigma}_1
\times \vec{\sigma}_3+\vec{q}_3\cdot \vec{\sigma}_1\times 
\vec{\sigma}_3) \left(\tau_3^a \vec{k}\cdot \vec{q}_2
\times \vec{\sigma}_2+\left[\fet\tau_2
\times \fet\tau_3\right]^a (k^2-\vec{k}\cdot \vec{q}_2) \right)}{ [k^2+M_\pi^2]
[(\vec{q}_1+\vec{q}_3)^2 +M_\pi^2]^2},\nn
\vec{D}_{3}^a&=&-\frac{g_A^2 C_T}{8 F_\pi^4 }\frac{ \vec{ k}\, (\vec{q}_1\cdot
  \vec{\sigma}_1
\times \vec{\sigma}_3+\vec{q}_3\cdot \vec{\sigma}_1\times 
\vec{\sigma}_3) \left[\fet\tau_2\times \fet\tau_3\right]^a}{
[k^2+M_\pi^2] [(\vec{q}_1+\vec{q}_3)^2 +M_\pi^2]}. \label{DiFF3NDefinition}
\eeqa
The connection to the four-nucleon force manifests itself through the
relations 
\beqa
V_{{\rm class-IV}}&=&\sum_{a}\tau_4^a\vec{\sigma}_4\cdot\vec{D}_{2}^a \Big|_{\vec{k}=-\vec{q}_4}+23\,{\rm
  permutations},\nn
V_{{\rm class-V}}&=&\sum_{a}\tau_4^a\vec{\sigma}_4\cdot\vec{D}_{3}^a \Big|_{\vec{k}=-\vec{q}_4}+23\,{\rm
  permutations},
\eeqa
where the expressions for the four-nucleon force contributions 
$V_{{\rm    class-IV}}$  and $V_{{\rm  class-V}}^2$
 are given in~\cite{Epelbaum:2007us}.

\section{Summary of the derived contributions}
\def\theequation{\arabic{section}.\arabic{equation}}
\label{sec:summaryCurrents}

In this section we provide a summary of the derived contributions to
the nuclear axial charge and current operators
\beq
{\fet A}^{\mu} = {\fet A}^{\mu}_{\rm 1N}   + {\fet
    A}^{\mu}_{\rm 2N}   +  {\fet A}^{\mu}_{\rm 3N}   + \ldots \,.
\eeq  
The chiral power counting implies that
$n$-nucleon operators are, in general, suppressed by two powers of the expansion
parameter relative to $n-1$-nucleon operators. Thus, to LO we have:
\beq
 {\fet A}^{\mu}_{\rm 1N}   \sim \mathcal{O}(Q^{-3}), \quad \quad 
 {\fet A}^{\mu}_{\rm 2N}   \sim \mathcal{O} (Q^{-1}), \quad  \quad
{\fet A}^{\mu}_{\rm 3N}   \sim \mathcal{O} (Q)\,, \quad \quad \ldots \,.
\eeq
Therefore, up to N$^3$LO ($Q$) in the chiral expansion, i.e.~up
to order $Q^4$ relative to the dominant contribution at order
$Q^{-3}$, it is necessary to include  single-,
two- and three-nucleon operators. 

\begin{table}[t]
\caption{Chiral expansion of the nuclear axial current operator up to N$^3$LO. 
\label{tab_sum_current}}
\smallskip
\begin{ruledtabular}
\begin{tabular}{@{\extracolsep{\fill}}lrrr}
\noalign{\smallskip}
 order &  single-nucleon  &  two-nucleon  &
                                                            three-nucleon  
\smallskip
 \\
\hline \hline
&&& \\[-7pt]
LO ($Q^{-3}$) & $\vec{A}^{a}_{{\rm 1N:  \, static}}, \;$
                Eq.~(\ref{LOSingleNCurrentExpr})  & --- & --- \\ [5pt] \hline
&&& \\[-9pt]
NLO ($Q^{-1}$) & $\vec{A}^{a}_{{\rm 1N:  \, static}}, \;$
                Eq.~(\ref{SingleNQtoMinus1StaticExpr})  & --- & --- \\
[5pt] \hline
&&& \\[-9pt]
N$^2$LO ($Q^{0}$) & --- & $\vec{A}^{a}_{{\rm 2N:  \, 1\pi}}, \;$ Eq.~(\ref{CurrentTree1})  & --- \\ [2pt]
& & \hskip -0.35 true cm $+ \; \vec{A}^{a}_{{\rm 2N:  \, cont}}, \;$
    Eq.~(\ref{CurrentTree2})  &\\ [5pt] \hline
&&& \\[-9pt]
N$^3$LO ($Q$) & $\vec{A}^{a}_{{\rm 1N:  \, static}}, \;$
                Eq.~(\ref{SingleNtwoLoopExpr}) & $\vec{A}^{a}_{{\rm
                                                 2N:  \, 1\pi}}, \;$
                                                 Eq.~(\ref{Current1pi})
                                            &  $\vec{A}^{a}_{{\rm 3N:
                                              \, \pi}}, \;$
                                              Eq.~(\ref{Current3NPi})
  \\ [2pt]
& \hskip -0.35 true cm $+ \; \vec{A}^{ a}_{{\rm 1N:\,}1/m, {\rm
    UT^\prime}}, \;$  Eq.~(\ref{k0OvermNuclcorr}) &
 \hskip -0.35 true cm $+ \;  \vec {A}_{{\rm 2N:\,}1\pi, {\rm
                                                    UT^\prime} }^{a} ,
                                                    \;$
                                                    Eq.~(\ref{NNCurrent1piUTPrime})
                                            &
  \hskip -0.35 true cm $+ \;  \vec A^{a}_{{\rm
    3N: \, cont}}, \;$ Eq.~(\ref{Current3NCont})   \\ [2pt]
& \hskip -0.35 true cm $+ \; \vec{A}^{ a }_{{\rm 1N: \, 1/m^2}} , \;$
  Eq.~(\ref{oneovermTo2corr}) &
  \hskip -0.35 true cm $+ \;  \vec A^{a}_{{\rm 2N:} \, 1\pi , \, 1/m },
                                                    \;$
                                Eq.~(\ref{Current1piRel}) & \\ [2pt]
&&  \hskip -0.35 true cm $+ \;  \vec A^{a}_{{\rm 2N:} \, 2\pi },
                                                    \;$
                                Eq.~(\ref{Current2pi}) & \\ [2pt]
&&  \hskip -0.35 true cm $+ \;    \vec {A}_{{\rm 2N:\,cont,\,UT^\prime}}^{a} ,
                                                    \;$
                                Eq.~(\ref{NNCurrentContUTPrime}) & \\
  [2pt]
&&  \hskip -0.35 true cm $+ \;    \vec {A}_{{\rm 2N:\,cont,\, 1/m}}^{a} ,
                                                    \;$
                                Eq.~(\ref{CurrentContRel}) &  \\ &&& \\[-11pt]
\end{tabular}
\end{ruledtabular}
\end{table}

\begin{table}[t]
\caption{Chiral expansion of the nuclear axial charge operator up to N$^3$LO. 
\label{tab_sum_charge}}
\smallskip
\begin{ruledtabular}
\begin{tabular}{@{\extracolsep{\fill}}lrrr}
\noalign{\smallskip}
 order &  single-nucleon  &  two-nucleon  &
                                                            three-nucleon  
\smallskip
 \\
\hline \hline
&&& \\[-7pt]
LO ($Q^{-3}$) & ---  & --- & --- \\ [5pt] \hline
&&& \\[-9pt]
NLO ($Q^{-1}$) & $A_{{\rm 1N: \,UT^\prime}}^{0, a} , \; $
                Eq.~(\ref{SingleNChargeUTPrimeQToMinus1})  & $A^{0, a}_{{\rm 2N:}\,
  1\pi}, \;$ Eq.~(\ref{ChargeTreeOPE})  & --- \\ [2pt]
& \hskip -0.35 true cm $+ \; A_{{\rm 1N: \,}1/m}^{0, a} , \;$
  Eq.~(\ref{SingleNChargeOneOvermQToMinus1}) &  & \\ [5pt]
\hline
&&& \\[-9pt]
N$^2$LO ($Q^{0}$) &---  & ---  &---   \\ [5pt] \hline
&&& \\[-9pt]
N$^3$LO ($Q$) & $A_{{\rm 1N:\,static, \,UT^\prime}}^{0, a} , \;$
                Eq.~(\ref{SingleNChargeStaticUTPrimeQTo1}) & $A^{0,
                                                             a}_{{\rm
                                                             2N:}\,
                                                             1\pi},
                                                             \;$
                                                             Eq.~(\ref{Charge1piQTo1})
                                          & ---\\ [2pt]
& \hskip -0.35 true cm $+ \;  A_{{\rm 1N:\,} 1/m}^{0, a }    , \;$ Eq.~(\ref{SingleNChargeOneOvermNuclQTo1}) &
\hskip -0.35 true cm $+ \;  A^{0, a}_{{\rm 2N:}\, 2\pi}   , \;$
                                                                                                               Eq.~(\ref{Charge2pi}) &\\ [2pt]
&&
\hskip -0.35 true cm $+ \;  A^{0, a }_{\rm 2N:\, cont}  , \;$
                                                                                                               Eq.~(\ref{ChargeContQTo1})
                          & \\ &&& \\[-11pt]
\end{tabular}
\end{ruledtabular}
\end{table}

We have worked out \emph{all} contributions to the
nuclear axial current up to N$^3$LO. The chiral expansion of the
current and charge operators is summarized in Tables
\ref{tab_sum_current} and \ref{tab_sum_charge}, respectively.  We
distinguish explicitly between the static contributions, terms which are
proportional to the energy transfer (which are labelled with
``UT$^\prime$'') and relativistic corrections ($1/m$ and
$1/m^2$). Further, the contributions to the exchange operator are
classified according to the range of the interaction between the
nucleons (``$1 \pi$'', ``$2 \pi$'' and ``cont''). It is important to
keep in mind that the expansion pattern of the
derived operators and their explicit form are, to a large extent, specific to the employed choice
of unitary transformations and to the chosen counting scheme for the
nucleon mass. To illustrate the importance of consistency between
nuclear forces and current operators, we notice that the contributions
to the two- and three-nucleon forces and the single- and two-nucleon
axial current operators at N$^3$LO depend on the phases $\bar
\beta_{8,9}$, which parametrize the unitary ambiguity of the leading
relativistic corrections to the corresponding operators. The
approximate independence of the calculated observables on these
unitary phases can only be achieved using \emph{consistent} nuclear
forces and current operators.

Last but not least, we would like to comment on isospin-breaking (IB)
contributions to the nuclear axial charge and current operators. While
we have not explicitly considered IB corrections in our analysis, the
only IB contributions to the exchange operators at the considered
order within the power counting scheme of Ref.~\cite{Epelbaum:2005fd}
can be accounted for by simply using the proper pion masses in the propagator of the
leading 2N charge operator $A^{0, a \; (Q^{-1})}_{\rm 2N:\, 1\pi}$ in
Eq.~(\ref{ChargeTreeOPE}). IB corrections to the 2N current operator
start contributing at order $Q^2$ which is not considered in our
work.

\section{Pseudoscalar currents}
\label{sec:Pseudo}

Chiral symmetry connects different physical processes via chiral Ward
identities. In particular, the axial-vector
current is connected to pseudoscalar current via the
continuity equation~(\ref{continuityeqmomspace}). Thus, the
pseudoscalar current can be determined (up to unitary ambiguity) from the
axial current using Eq.~(\ref{continuityeqmomspace}). Instead of
following this approach, we use chiral EFT in combination with the MUT
as explained in the previous sections to independently derive the  pseudoscalar current up-to-and-including
order $Q^4$ relative to the dominant one-body term. This will allow us
to perform a highly nontrivial check of our results by explicitly
verifying the validity of the  continuity equation~(\ref{continuityeqmomspace}). 
Notice that the
derived expressions might also be useful in connection with ongoing and future direct-detection
experiments of weakly-interacting massive particles (WIMP) through
scattering off nuclei, see Ref.~\cite{Hoferichter:2015ipa} for a related discussion. 

Throughout this section, we follow the notation of
Ref.~\cite{Gasser:1983yg} and introduce  
a LEC $G_\pi$ defined as
\beqa
\langle 0|\bar{q}\, i \gamma_5 \tau^a q|\pi^b\rangle&=&\delta_{a b}G_\pi,
\eeqa
and related to the physical pion mass via
\beqa
\label{tempo21}
F_\pi M_\pi^2&=& m_q G_\pi.
\eeqa
Notice that the current operators describing nuclear response
on external pseudoscalar source which couples to the QCD density
$\bar{q}\, i \gamma_5 \tau^a q$ are renormalization-scale
dependent. This manifests itself in all resulting expressions for the
current operator being proportional to the LEC $G_\pi$, whose value
depends on the choice of renormalization conditions in
QCD.
Notice that in the continuity equation, pseudoscalar current
operators always occur in a renormalization-scale invariant
combination with the quark mass $m_q$, see
Eqs.~(\ref{continuityeqmomspace}) and (\ref{tempo21}). As an alternative convention,
one could directly absorb the quark mass into the definition of the
pseudoscalar quark density. This would result in renormalization-scale
invariant expressions for the pseudoscalar operators, which can be
obtained from our expressions by making the replacement $G_\pi \to F_\pi
M_\pi^2$ and shifting the contributions two orders higher in the
chiral expansion.  
In the following we will express all pseudoscalar currents in terms of the
quark mass $m_q$.

The diagrammatic structure of the various contributions to 
the pseudoscalar current is the same as in the calculation of the
axial-vector current. 
For this reason we can use the set of diagrams
introduced in sections~\ref{sec:singleN}-\ref{sec:3N}, with 
axial-vector sources being replaced by pseudoscalar ones, also for the calculation of
the pseudoscalar current. Notice, however, that certain vertices contribute to the axial-vector current 
but vanish when the axial-vector
source is replaced by the pseudoscalar one. This leads to a simplification such that
only a subset of diagrams discussed in the previous sections will have
to be considered. The same statement is also
valid for contributions involving additional unitary
transformations. In order to write down the most
general set of additional unitary transformations with a pseudoscalar
source we simply take the unitary transformations listed in Appendix~\ref{sec:appen} and
replace all vertices involving axial-vector sources by the
corresponding ones with pseudoscalar sources
\beq
A_{a,b}^{(\kappa_i)}\to P_{a,b}^{(\kappa_j)},
\eeq
where $P_{a,b}^{(\kappa_j )}$ denotes an interaction from the Hamiltonian with one pseudoscalar
current, $a$ nucleons, $b$ pion fields and $\kappa_j$ defined by
Eq.~(\ref{DefKappa}). 
Since the unitary phases of the additional unitary
transformations with pseudoscalar sources may, in general, be different
from the unitary phases of the similar transformations with axial
vector sources, we also make a replacement 
\beq
\alpha_{i}^{ax}\to\alpha_{i}^{p}
\eeq
in Eq.~(\ref{add:unitary:transf}) and consider $\alpha_{i}^{p}$ as
independent parameters. By this
procedure we obtain, for example, 
\beq
S_{1}^{ax}\,=\,\alpha_{1}^{ax} \eta  A_{2,0}^{(0)}  \eta  H_{2,1}^{(1)}  
\lambda^1  \frac{1}{E_\pi^{3}}  H_{2,1}^{(1)}  \eta \to 
S_{1}^{p}\,=\,\alpha_{1}^{p} \eta  P_{2,0}^{(0)}  \eta  H_{2,1}^{(1)}  
\lambda^1  \frac{1}{E_\pi^{3}}  H_{2,1}^{(1)}  \eta,
\eeq
for the first generator of the unitary transformation from
Eq.~(\ref{add:unitary:transf}). Similar replacements are performed for
all transformations of Appendix~\ref{sec:appen}. In this way, we end
up with $33$
additional phases which we fix by applying the same requirements as in
the case of the axial-vector case. 
There is, however, a simplification due to the absence of
certain interactions with pseudoscalar sources, namely
\beq
P_{2,0}^{(0)}=P_{2,1}^{(1)}=P_{2,2}^{(2)}=0,
\eeq
which implies 
\beq
S_1^{p}=S_2^{p}=S_3^{p}=0.
\eeq
For this reason, we only need to consider $30$ unitary phases
$\alpha_i^p$ with $i = 4, \ldots , 15, 17, \ldots , 30$, $\alpha_{16}^{p,\, {\rm LO}}$, $\alpha_{16}^{p,\, {\rm
    Static}}$, $\alpha_{16}^{p, 1/m}$ and $\alpha_{16}^{p, \,{\rm Tadpole}}$.
    
\subsection{Unitary phases of pseudoscalar transformations}
\def\theequation{\arabic{section}.\arabic{equation}}

To fix the values of the additional
unitary transformations involving a pseudoscalar source, we proceed as in the
case of the axial vector current: 
\begin{enumerate}
\item We require the one-loop contributions to the pseudoscalar current to be
  expressible in the form of 4-dimensional integrals with heavy-baryon
  propagators. This requirement is necessary for factorizability of the
  exchanged pions which is necessary (but not sufficient) for the
  renormalization of the pseudoscalar current. It leads
  to the following $7$ constraints
\beqa
\alpha_{8}^{p}&=&\alpha_{5}^{p},\nn
\alpha_{9}^{p}&=&-\alpha_{4}^{p}+\alpha_{5}^{p}+\alpha_{6}^{p}-\alpha_{7}^{p},\nn
\alpha_{14}^{p}&=&\alpha_{4}^{p}-\alpha_{11}^{p},\nn
\alpha_{27}^{p}&=&\alpha_{25}^{p},\nn
\alpha_{28}^{p}&=&\alpha_{25}^{p}-\alpha_{26}^{p},\nn
\alpha_{29}^{p}&=&\alpha_{25}^{p}-\alpha_{26}^{p},\nn
\alpha_{30}^{p}&=&-\alpha_{26}^{p}.\label{standardp1}
\eeqa
\item The resulting expressions for the
pseudoscalar current operators are required to be properly
renormalized. 
However, we found that this requirement does not yield any further constraints on
the pseudoscalar unitary phases.
\item 
We require the pion-pole part of the pseudoscalar current to match the corresponding
  expressions for the nuclear forces upon replacing the leading
  pseudoscalar source-pion interaction by the leading pion-nucleon
  vertex. Matching the two-pion-exchange current to the
  corresponding two-pion-one-pion contributions of N$^3$LO
  3NFs gives three
  constraints
\beqa
\alpha_{13}^{p}&=&2+\alpha_{5}^{p}+\alpha_{6}^{p}-\alpha_{10}^{p}-\alpha_{11}^{p}-\alpha_{14}^{p},\nn
\alpha_{15}^{p}&=&-\alpha_{12}^{p},\nn
\alpha_{26}^{p}&=&1-\alpha_{25}^{p}. \label{standardp4}
\eeqa
Finally, matching the $1/m$-corrections of the pseudoscalar current with
the corresponding $1/m$-corrections to the N$^3$LO
  3NFs yields the last three constraints:
\beqa
\alpha_{20}^{p}&=&-1-2\bar{\beta}_8+\alpha_{17}^{p}+\alpha_{18}^{p}+\alpha_{19}^{p},\nn
\alpha_{24}^{p}&=&-2 + \alpha_{21}^{p}+ \alpha_{22}^{p}+ \alpha_{23}^{p},\nn
\alpha_{16}^{p, 1/m}&=&-\frac{1}{2}\Big(1+2\bar{\beta}_9\Big).\label{standardponeoverm}
\eeqa
Matching the  pion-pole contributions to the three-nucleon
pseudoscalar current operator with the corresponding terms in the
four-nucleon force does not produce any additional constraints.
The same is true for the single-nucleon current operator.  
\end{enumerate}
With these constraints, the pseudoscalar current has no further unitary ambiguity
and becomes unique (modulo the phases $\bar \beta_8$ and $\bar
\beta_9$) at the considered chiral order.  Furthermore, we have
verified that the pion-pole contributions to the single-nucleon (two-nucleon) pseudoscalar current
operator proportional to the energy transfer $k_0$ match the
corresponding terms in the two-nucleon (three-nucleon) force for
the choice of $\bar \beta_9 = 1/2$ ($\bar \beta_8 = 1/2$). 
Here and in what follows, Eqs.~(\ref{standardp1}), (\ref{standardp4}) and
(\ref{standardponeoverm}) are referred to as the standard choice of the
unitary phases for the pseudoscalar current operator. 

We are now in the position to present explicit results for the
pseudoscalar nuclear currents up to fourth order in the chiral
expansion. Due to the direct relations between pion pole contributions to
the pseudoscalar and axial-vector current as described below, 
the corresponding matching relations with the nuclear forces can be taken
from sections \ref{sec:2N}, \ref{sec:3N} and will not be given explicitly in the following.  

\subsection{Single-nucleon pseudoscalar current operator}
\def\theequation{\arabic{section}.\arabic{equation}}
\label{sec:singleNP}


The single-nucleon pseudoscalar current starts to contribute at order
$Q^{-4}$. The dominant contribution emerges from diagram $(2)$ of Fig.~\ref{fig:singleNuclLO}.
At this order, there is no dependence on the unitary phases. The
leading-order expression for the pseudoscalar current is given by 
\beqa
P^{ a \, (Q^{-4})}_{{\rm 1N: \, static}}  &=& i\,\frac{M_\pi^2 g_A}{2 m_q} \tau_i^a
\frac{\vec{k}\cdot\vec{\sigma}_i}{k^2+M_\pi^2} = - i
\frac{1}{m_q} \, \vec k \cdot \vec {A}_{\rm 1N:\, static }^{a \; (Q^{-3})} 
,\label{LOSingleNCurrentPExpr}
\eeqa
with $\vec {A}_{\rm 1N:\, static }^{a \; (Q^{-3})} $ given in
Eq.~(\ref{LOSingleNCurrentExpr}). 

At order $Q^{-3}$, there are only vanishing contributions, i.e.
\beqa
P^{ a \, (Q^{-3})}_{{\rm 1N: \, static}}  &=& 0.
\eeqa
Next, at order $Q^{-2}$, we have to account for two kinds of corrections. 
The first one comes from the leading one-loop contributions, along with
the corresponding counter-terms from the ${\cal L}_{\pi N}^{(3)}$
Lagrangian. One-loop diagrams that contribute
to the pseudoscalar current at order $Q^{-2}$ are shown in
Fig.~\ref{fig:singleNuclOneLoop}. Only diagrams $(3)$ and $(4)$ yield  
non-vanishing contributions.
There is no explicit dependence on unitary
phases for pseudoscalar current at this order. 
Non-vanishing counterterms
and tadpoles are given by diagrams $(1-4)$ and $(6-8)$ 
of Fig.~\ref{fig:singleNuclCounterTerms}. None
of the resulting contributions depend on unitary phases. 
The final result for the pseudoscalar
current operator at order $Q^{-2}$ in the static limit is given by 
\beqa
P^{  a \, (Q^{-2})}_{{\rm 1N: \, static}}  &=&  i\frac{M_\pi^2 \bar{d}_{18}}{m_q}
k^2 \tau_i^a\frac{\vec{k}\cdot\vec{\sigma}_i}{k^2+M_\pi^2} = - i
\frac{1}{m_q} \, \vec k \cdot \vec {A}_{\rm 1N:\, static }^{a \; (Q^{-1})} 
\,,\label{SingleNPQtoMinus2StaticExpr}
\eeqa
with $\vec {A}_{\rm 1N:\, static }^{a \; (Q^{-1})} $ given in Eq.~(\ref{SingleNQtoMinus1StaticExpr}).

The second kind of contributions at order $Q^{-2}$ could emerge from
relativistic $1/m$-corrections. However, the corresponding
diagrams shown
in Fig.~\ref{fig:singleNucleonOneOvermNucl} only generate operators  
which are proportional to $k_0$ and thus do not contribute at order
$Q^{-2}$. 
More precisely, the non-vanishing contributions 
are generated by
diagrams $(2)$ and $(3)$ of
Fig.~\ref{fig:singleNucleonOneOvermNucl}. Their unitary phase
dependence is given by
\beqa
(2)&\sim&
\frac{k_0}{m}\frac{\alpha_{17}^{p}+\alpha_{18}^{p}+\alpha_{19}^{p}-\alpha_{20}^{p}}{(k^2+M_\pi^2)^2},\nn
(3)&\sim&
\frac{k_0}{m}\frac{\alpha_{16}^{p, 1/m}}{k^2+M_\pi^2}. \label{k0OvermNuclPUnitaryPhases}
\eeqa  
Given that these terms contribute at order $Q^0$, we find 
\beqa
P^{  a \, (Q^{-2})}_{{\rm 1N:} \, 1/m}  &=&0\,.\label{SingleNPChargeOneOvermQToMinus1}
\eeqa

At order $Q^{-1}$, there are no $c_i/m$-corrections and no
$k_0$-dependent contributions to the 1N pseudoscalar
current, which can be traced back to vanishing $Q^{-3}$
terms. We also found only vanishing contributions  from 
one-loop diagrams with a single insertion of subleading
vertices (after renormalization). Thus, there are no contributions to the single-nucleon
pseudoscalar current at order $Q^{-1}$.  

Finally there are various corrections at order $Q^0$. We begin
with terms proportional to $k_0/m$ which emerge from diagrams $(2)$ and
$(3)$ shown in  Fig.~\ref{fig:singleNucleonOneOvermNucl}. Their unitary phase dependence
is given by Eq.~(\ref{k0OvermNuclPUnitaryPhases}), and the explicit
expressions for the standard choice of  the unitary phases are given
by
\beqa
P^{ a \, (Q^0)}_{{\rm 1N:\,}1/m, {\rm UT^\prime}}  &=&
i \frac{M_\pi^2}{m_q k^2} 
\, \vec k \cdot \vec {A}_{{\rm 1N:\,}1/m, {\rm UT^\prime}}^{a \; (Q)} \,,
\label{k0OvermNuclcorrP}
\eeqa
with $\vec {A}_{{\rm 1N:\,}1/m, {\rm UT^\prime}}^{a \; (Q)}$ given in
Eq.~(\ref{k0OvermNuclcorr}). 

The second kind of corrections could emerge from static terms
proportional to the energy transfer $k_0$. We, however, find a
vanishing result for such contributions. 


The third kind of order-$Q^0$ contributions is given by relativistic
$1/m^2$-corrections. The corresponding
diagrams yielding non-vanishing results are $(2)$, $(3)$, $(4)$ and $(5)$ of
Fig.~\ref{fig:singleNuclOneOvermNuclTo2}. The
unitary phase dependence of these graphs is given by
\beqa
(2)&\sim&1+\alpha_{16}^{p, 1/m}, \nn
(3),(4)&\sim&2-\alpha_{17}^{p}-\alpha_{18}^{p}-\alpha_{19}^{p}+\alpha_{20}^{p}.
\eeqa
The contribution of diagram $(5)$ does not depend on unitary
phases. Our 
final result for $1/m^2$-corrections at order $Q^0$ for the
standard choice of the phases is given by
\beqa
P^{ a \, (Q^0)}_{{\rm 1N: \, 1/m^2}}  &=& i \frac{M_\pi^2 g_A}{16 m_q
  m^2}\tau_i^a\bigg(\vec{k}\cdot\vec{\sigma}_i(1-2\bar{\beta}_8)\frac{(p_i^{\prime\,2}-p_i^2)^2}{(k^2+M_\pi^2)^2}
-2 \frac{(p_i^{\prime\, 2} +
  p_i^2)\vec{k}\cdot\vec{\sigma}_i-2 \bar{\beta}_9 (p_i^{\prime\, 2} -
  p_i^2)\vec{k}_i\cdot\vec{\sigma}_i }{k^2+M_\pi^2}\bigg)\,.\label{oneovermTo2corrP}
\eeqa
Notice that this expression is related to the pion-pole
terms in the corresponding axial current operator $\vec {A}_{{\rm 1N: \, 1/m^2}}^{a \; (Q)} $, whose expression is given in
Eq.~(\ref{oneovermTo2corr}),  via 
\beq
P^{ a \, (Q^0)}_{{\rm 1N: \, 1/m^2}} = i \frac{M_\pi^2}{m_q k^2} 
\, \vec k \cdot \vec {A}_{{\rm 1N: \, 1/m^2}}^{a \; (Q)} \bigg|_{\rm
  pion-pole \; terms} \,.
\eeq

The fourth kind of order-$Q^0$ contributions are coming from relativistic
$1/m$-corrections to the leading one-loop current. The corresponding
non-vanishing diagrams are $(2)$ and $(3)$ of
Fig.~\ref{fig:singleNuclOneOvermNuclCounterTerms},
and all diagrams of Fig.~\ref{fig:singleNuclOneOvermNuclOneLoopPionPoleTerms}.
The unitary phase dependence of these diagrams is given by 
\beqa
(2) &\sim& 1+\alpha_{16}^{p, {\rm Static}},\nn 
(21)&\sim& 1+2\bar{\beta}_9, \nn 
(22)&\sim& 1 - 2\bar{\beta}_9,\nn 
(23)&\sim& 1+\alpha_{16}^{p,1/m},  \nn 
(24) &\sim& (1 +
2\bar{\beta}_8- \alpha_{12}^{p}-\alpha_{15}^{p})\frac{1}{k^2+M_\pi^2}\big[ \ldots \big] +
(2-\alpha_{4}^{p}+\alpha_{5}^{p}+\alpha_{6}^{p}-\alpha_{10}^{p}-\alpha_{13}^{p})\frac{1}{(k^2+M_\pi^2)^2}
\big[ \ldots \big], \nn
(26)&\sim&(\alpha_{12}^{p}+\alpha_{15}^{p})\frac{1}{k^2+M_\pi^2}\big[ \ldots \big] +
(4-\alpha_{4}^{p}+\alpha_{5}^{p}+\alpha_{6}^{p}-\alpha_{10}^{p}-\alpha_{13}^{p}
-\alpha_{17}^{p}-\alpha_{18}^{p}-\alpha_{19}^{p}+\alpha_{20}^{p})
\frac{1}{(k^2+M_\pi^2)^2}\big[ \ldots \big],
\nn 
(27)&\sim& (1-2\bar{\beta}_8)\frac{1}{k^2+M_\pi^2}\big[ \ldots \big] +
(2-\alpha_{17}^{p}-\alpha_{18}^{p}-\alpha_{19}^{p}+\alpha_{20}^{p})
\frac{1}{(k^2+M_\pi^2)^2}\big[ \ldots \big],
\eeqa
while the remaining graphs do not depend on the unitary phases. For our
standard choice of  the phases we get a vanishing result after
renormalization, 
\beqa
\label{SingleNChargeOneOvermNuclQTo0P}
P^{ a \, (Q^0)}_{{\rm 1N: \, 1/m}}  &=& 0\,.
\eeqa

Before specifying the last missing kind of corrections emerging from
static contributions at the two-loop level, it is instructive 
to compare our results with the
on-shell expressions for pseudoscalar amplitude, which can be written
in terms of the form factor $G_{\pi N}(t)$ as follows 
\beqa
\langle N(p_i^\prime)|P^{a}(0)|N(p_i)\rangle&\equiv&\langle
N(p_i^\prime)|\bar{q}\,i\,\gamma_5\,\tau^a q|N(p_i)\rangle\,=:\,
\frac{G_\pi}{M_\pi^2-t}\, G_{\pi N}(t)\frac{1}{2m}\bar{u}(p_i^\prime\,)i\,\gamma_5\,\tau_i^a 
u(p_i)\nn
&=&i\,\tau_i^a \frac{G_\pi}{2m}\frac{G_{\pi
    N}(t)}{M_\pi^2-t}\bigg(-\vec{k}\cdot\vec{\sigma}_i+{\cal
  O}(1/m^2)\bigg).\label{GPiNParametrization}
\eeqa
The second equation in the first line of (\ref{GPiNParametrization})
serves as a definition of the form factor $G_{\pi N}(t)$, which for a
particular choice of pion interpolating fields can be identified with
the pion-nucleon form factor. 
As it is well known, the form factor $G_{\pi N}(t)$ is
related to the axial and pseudoscalar form factors $G_A(t)$ and
$G_P(t)$ via the Partially Conserved Axial Current (PCAC) relation:
\beqa
2 m G_A(t)+\frac{t}{2 m} G_P(t)=2\frac{M_\pi^2 F_\pi}{M_\pi^2-t}G_{\pi N}(t).
\eeqa 
Using this relation along with Eqs.~(\ref{tempGA}) and
(\ref{PseudoScalarFormFactorUpToQTo4}), we can read off the chiral expansion of the form-factor
$G_{\pi N}(t)$: 
\beqa
G_{\pi N}(t)&=&\frac{m}{F_\pi}\bigg(g_A - 2 \bar{d}_{18}
t
+
\bigg[1-\frac{t}{M_\pi^2}\bigg]\bigg[G_A^{(Q^4)}(t)+\frac{t}{4}G_P^{(Q^2)}(t)\bigg]\bigg)
+ {\cal O}(Q^5).\label{GPiNExpressionUpToQTo4}
\eeqa
Notice that since 
\beqa
G_A^{(Q^4)}(t)+\frac{t}{4}G_P^{(Q^2)}(t)&=&M_\pi^2 f(t),
\eeqa
where the function $f(t)$ is finite in the chiral limit, the form
factor $G_{\pi N}(t)$ is clearly well-defined in the chiral limit. 
Plugging Eq.~(\ref{GPiNExpressionUpToQTo4}), truncated at order $Q^4$,  into the parametrization
(\ref{GPiNParametrization}), we read off the on-shell expression for
the static contribution to the single-nucleon pseudoscalar current
operator  $P^{a\, (Q^0)}_{{\rm 1N:\,
    static}}$ from the relation  
\beqa
\langle N(p_i^\prime)|P^{a}(0)|N(p_i)\rangle + P^{ a \, (Q^{-4})}_{{\rm 1N: \, static}} + P^{  a \,
  (Q^{-2})}_{{\rm 1N: \, static}} + P^{a\, (Q^0)}_{{\rm 1N:\,
    static}}&=&\mathcal{O} (1/m^2)
\eeqa
to be 
\beqa
P^{a\, (Q^0)}_{{\rm 1N:\, static}}&=& i\frac{1}{8
  m_q}\vec{k}\cdot\vec{\sigma}_i \tau_i^a
\Big(4 \,G_A^{(Q^4)}(-k^2)-k^2 \,G_P^{(Q^2)}(-k^2)\Big)
.\label{PseudoscalarTwoLoopConjecture}
\eeqa
As in the case of the axial-vector
current, we conjecture that the static contribution to the
single-nucleon pseudoscalar current calculated using the MUT at the
two-loop level coincides with the above on-shell expression. 

It is also instructive to verify the consistency between our results
for the current operator and the parametrization
(\ref{GPiNParametrization}) for $1/m^2$-corrections. Taking into
account the nonrelativistic normalization of the nucleon states in the
expressions for the current, we obtain the relation 
\beqa
&&\bigg[
\frac{G_\pi}{M_\pi^2-t}\, G_{\pi N}(t)\frac{1}{2m}\sqrt{\frac{m}{E_{p_i^{\prime}}}}\bar{u}(p_i^\prime\,)i\,\gamma_5\,\tau_i^a
u(p_i)\sqrt{\frac{m}{E_{p_i}}}\bigg]_{1/m^2 - {\rm
part}}+P^{ a \, (Q)}_{{\rm 1N:} \, 1/m^2}+P^{  a \,
  (Q)}_{{\rm 1N: \, UT^\prime}}\nn
&=&-i\bigg(k_0-\frac{p_i^{\prime \,2}}{2 m}+\frac{p_i^2}{2
  m}\bigg)\frac{g_A M_\pi^2}{8
  m_q  m}\frac{1}{k^2+M_\pi^2}\tau_i^a\bigg[-(1+2\bar{\beta}_8)(p_i^{\prime\,2}-p_i^2)\frac{\vec{k}\cdot\vec{\sigma}_i}{k^2+M_\pi^2}
+2(1+2\bar{\beta}_9)\vec{k}_i\cdot\vec{\sigma}_i\bigg].
\eeqa
As expected, we observe that our results for the current operator
agree with the  parametrization
(\ref{GPiNParametrization}) on the energy shell, i.e.~with $k_0={p_i^{\prime \,2}}/{2 m}-{p_i^2}/{2m}$.
As in the case of the axial-vector current, the two expressions agree
even off the energy shell for a particular choice of the phases $\bar
\beta_{8,9}$, namely  
$\bar\beta_8=\bar\beta_9=-1/2$.

To conclude, to order $Q^0$, the single nucleon pseudoscalar current
operator can be written in terms of the axial and and induced
pseudoscalar form factors of the nucleon as 
\beq
P^{ a}_{{\rm 1N}}  = i \frac{1}{2 m_q} \tau_i^a \vec k
\cdot \vec \sigma_i \bigg( G_A (-k^2) -
\frac{k^2}{4 m^2} G_P (-k^2) \bigg) +P^{ a \, (Q^0)}_{{\rm 1N:\,}1/m,
  {\rm UT^\prime}} + P^{ a \, (Q^0)}_{{\rm 1N: \, 1/m^2}}\,,
\eeq
where the expressions for the last two terms are given in Eqs.~(\ref{k0OvermNuclcorrP}),
and (\ref{oneovermTo2corrP}). As already pointed out
above, $P^{ a}_{{\rm 1N}}$ is well-defined in the chiral limit (except
at the kinematical point $k=0$).

\subsection{Two-nucleon pseudoscalar operators}
\def\theequation{\arabic{section}.\arabic{equation}}
\label{sec:2NP}

\subsubsection{Contributions at orders $Q^{-2}$ and $Q^{-1}$}

We now turn to the two-nucleon contributions to the pseudoscalar
current operator, whose chiral expansion is expected to start at
order $Q^{-2}$.  The relevant diagrams are $(3), (4)$ and $(5)$ of
Fig.~\ref{fig:tree}. All of them yield contributions which
depend on the unitary phases: 
\beqa
\label{temp22}
(3)&\sim& k_0(\alpha_{25}^{p}+\alpha_{26}^{p}), \nn 
(4)&\sim& \frac{k_0
  (\alpha_{12}^{p}+\alpha_{15}^{p})}{k^2+M_\pi^2}\big[ \ldots \big] +\frac{k_0
  (\alpha_{4}^{p}-\alpha_{5}^{p}-\alpha_{6}^{p}+\alpha_{10}^{p}+\alpha_{13}^{p})}{(k^2+M_\pi^2)^2}\big[ \ldots \big], \nn 
(5)&\sim& k_0(\alpha_{21}^{p}+\alpha_{22}^{p}+\alpha_{23}^{p}-\alpha_{24}^{p}).
\eeqa
The resulting
expressions are
proportional to the energy transfer $k_0$, so that their actual
contributions are shifted to order $Q^0$. For this reason, we find a vanishing result for
both the one-pion-exchange and contact interaction $2N$ currents at the order $Q^{-2}$:
\beq
 P_{{\rm 2N:\,}1\pi }^{a \; ({Q^{-2}})} = P_{{\rm 2N:\,cont}}^{a \;
   ({Q^{-2}})}  = 0,\label{VectorTreePeffOPE}
\eeq

At order $Q^{-1}$, one encounters various contributions to the
2N pseudoscalar 
current operator stemming from diagrams shown in the second
line of Fig.~\ref{fig:tree}, whose explicit expressions have the form
\beqa
P_{{\rm 2N:}\,  1\pi }^{a \; ({Q^{-1}})} &=& - i \frac{1}{m_q
  } \, \vec k \cdot \vec {A}_{{\rm 2N:}\,  1\pi }^{a \;
  (Q^{0})} \,, \label{PsCurrentTreeOPE} \\
P_{{\rm 2N:}\,  \rm cont}^{a \; ({Q^{-1}})} &=&
- i \frac{1}{m_q} \, \vec k \cdot \vec {A}_{{\rm 2N:}\,  \rm cont}^{a \;
  (Q^{0})} \,, \label{PsCurrentTreeCont}
\eeqa
with  $\vec {A}_{{\rm 2N:}\,  1\pi }^{a \;
  (Q^{0})}$ and $\vec {A}_{{\rm 2N:}\,  \rm cont}^{a \;
  (Q^{0})} $ given in Eqs.~(\ref{CurrentTree1}) and
(\ref{CurrentTree2}), respectively. 

\subsubsection{One-pion-exchange contributions at order $Q^0$}

We consider first one-loop graphs of non-tadpole type which are
visualized in Fig.~\ref{fig:ope}. 
For the pseudoscalar current, only diagrams  $(9-14)$, $(21-22)$ and $(25-27)$ 
generate nonvanishing contributions, from which those of diagrams $(9-13)$
turn out to depend on the unitary phases:
\beqa
(9),(10),(13)&\sim&(2-\alpha_4^{p}+\alpha_5^{p}+\alpha_6^{p}-\alpha_{10}^{p}-\alpha_{13}^{p})\frac{1}{(k^2+M_\pi^2)^2}\big[ \ldots \big]
+(2-\alpha_{12}^{p}-\alpha_{15}^{p})\frac{1}{k^2+M_\pi^2}\big[ \ldots \big] \,,\nn
(11),(12)&\sim&(2-\alpha_4^{p}+\alpha_5^{p}+\alpha_6^{p}-\alpha_{10}^{p}-\alpha_{13}^{p})\frac{1}{(k^2+M_\pi^2)^2}\big[ \ldots \big]
+(\alpha_{12}^{p}+\alpha_{15}^{p})\frac{1}{k^2+M_\pi^2}\big[ \ldots \big] \,.
\eeqa
Furthermore, all one-pion-exchange
tadpole diagrams and tree graphs involving $d_i$-vertices from ${\cal
  L}_{\pi N}^{(3)}$ and $l_i$-vertices from ${\cal
  L}_{\pi}^{(4)}$ from Fig.~\ref{fig:tadpoles} lead to vanishing contributions. 
After renormalization of the LECs, we find our final result for the static order-$Q^0$ contributions to
the 2N one-pion-exchange pseudoscalar current operator for our
standard choice of unitary phases: 
\beqa
\label{CurrentPs1pi}
P_{{\rm 2N:} \, 1\pi}^{a \, (Q^0)}&=& 
i\frac{4\,M_\pi^2 F_\pi^2}{g_A m_q}\frac{\vec q_1
 \cdot \vec \sigma_1  }{q_1^2 + M_\pi^2}\bigg\{\tau_1^a \bigg[h_1^P(q_2)+
\frac{h_4(q_2) }{k^2 + M_\pi^2}\bigg] +   [ \fet \tau_1 \times
\fet \tau_2 ]^a \vec q_1 \cdot [\vec q_2
  \times \vec \sigma_2] \frac{h_5(q_2) }{k^2 + M_\pi^2}\bigg\} \; + \; 1
\leftrightarrow 2,   
\eeqa
where the scalar functions $h_4 (q_2)$ and $h_5 (q_2)$ are defined in
Eq.~(\ref{OPEhiDefenition}), while 
$h_1^P(q_2)$ is given by
\bigskip
\beqa
h_1^P(q_2)&=&\frac{g_A^4}{256\pi F_\pi^6}\Big((1-2 g_A^2) M_\pi + (2
M_\pi^2 + q_2^2)A(q_2)\Big).
\eeqa
It is easy to verify that the pion-pole contributions match the
corresponding expressions in the N$^3$LO 3NF. Notice further the
relation between the pion-pole terms in the pseudoscalar and the
corresponding axial current operator:
\beq
P_{{\rm 2N:} \, 1\pi}^{a \, (Q^0)} \bigg|_{\rm pion-pole \; terms}= i \frac{M_\pi^2}{m_q k^2} \, \vec
k \cdot \vec A_{{\rm 2N:} \, 1\pi}^{a \, (Q)} \bigg|_{\rm pion-pole \; terms}\,,
\eeq
where $A_{{\rm 2N:} \, 1\pi}^{a \, (Q)} $ is specified in
Eq.~(\ref{Current1pi}). 

Finally, apart from the static contributions, we need to 
account for the leading relativistic corrections emerging from tree-level
diagrams with a single insertion of $1/m$-vertices from the
Lagrangian $\mathcal{L}_{\pi N}^{(2)}$. From the corresponding
diagrams shown in Fig.~\ref{fig:relativistic_corrections} , only
graphs $(8-18)$ yield nonvanishing contributions. Furthermore,  
diagrams $(8-17)$ turn out to induce the dependence 
on the unitary phases in the following way:
\beqa
(8),(9),(10) &\sim&
(-2+\alpha_4^{p}-\alpha_5^{p}-\alpha_6^{p}+\alpha_{10}^{p}+\alpha_{13}^{p})\frac{1}{(k^2+M_\pi^2)^2}
\big[ \ldots \big]
+ (-1-2\bar{\beta}_8+\alpha_{12}^{p}+\alpha_{15}^{p})\frac{1}{k^2+M_\pi^2}\big[ \ldots \big],\nn
(11)&\sim&
(-4+\alpha_4^{p}-\alpha_5^{p}-\alpha_6^{p}+\alpha_{10}^{p}+\alpha_{13}^{p}+\alpha_{17}^{p}+\alpha_{18}^{p}+\alpha_{19}^{p}-\alpha_{20}^{p})\frac{1}{(k^2+M_\pi^2)^2}\big[
\ldots \big] \nn
&+& (\alpha_{12}^{p}+\alpha_{15}^{p})\frac{1}{k^2+M_\pi^2}\big[ \ldots \big],\nn
(12)&\sim&
(-2+\alpha_{17}^{p}+\alpha_{18}^{p}+\alpha_{19}^{p}-\alpha_{20}^{p})\frac{1}{(k^2+M_\pi^2)^2}\big[ \ldots \big]
+ (-1+2\bar{\beta}_8)\frac{1}{k^2+M_\pi^2}\big[ \ldots \big],\nn 
(13)&\sim& 1+2\bar{\beta}_9,\nn
(14)&\sim&1-2\bar{\beta}_9,\nn
(15)&\sim&
1+\alpha_{16}^{p,1/m},\nn
(16)&\sim& -1 + \alpha_{25}^{p}+\alpha_{26}^{p}, \nn 
(17)&\sim& -2 +  \alpha_{25}^{p}+\alpha_{26}^{p} \, .
\eeqa
Again, our standard choice of the unitary phases leads to certain simplifications. In particular,  
it eliminates the contributions from
diagram (16).
The explicit result for the $1/m$-corrections to the
one-pion-exchange pseudoscalar NN current operator has the form 
\beqa
\label{CurrentPs1piRel}
P_{{\rm 2N:} \, 1\pi , \, 1/m}^{a \, (Q^0)}&=&i \frac{g_A M_\pi^2}{16
  m_q F_\pi^2 \, m}\bigg\{i [ \fet \tau_1 \times
\fet \tau_2 ]^a
\bigg[-\frac{\vec{B}_1\cdot\vec{k}}{(q_1^2+M_\pi^2)^2(k^2+M_\pi^2)} + \frac{1}{q_1^2+M_\pi^2}\bigg(\frac{\vec{B}_2\cdot\vec{k}}{k^2(k^2+M_\pi^2)^2} 
+ \frac{\vec{B}_3\cdot\vec{k}}{k^2(k^2+M_\pi^2)}\bigg)  \bigg]\nn
&+& \tau_1^a\bigg[ -\frac{ \vec{B}_5\cdot\vec{k}}{(q_1^2+M_\pi^2)^2(k^2+M_\pi^2)}+ \frac{1}{q_1^2+M_\pi^2}\bigg(\frac{\vec{B}_6\cdot\vec{k}}{k^2(k^2+M_\pi^2)^2} 
+ \frac{\vec{B}_7\cdot\vec{k}}{k^2(k^2+M_\pi^2)}\bigg)\bigg] \bigg\} +
1 \; \leftrightarrow \; 2\,,
\eeqa
where the vector quantities $\vec{B}_i$ 
are defined in
Eq.~(\ref{OPEOneOvermNuclFFs}).
Similarly to the previously considered contributions, one notices that the pseudoscalar current $P_{{\rm 2N:} \,
  1\pi}^{a \, (Q^0)}$ is directly related to the pion-pole terms in the corresponding axial current operator $\vec
A_{{\rm 2N:} \, 1\pi , \, 1/m}^{a \, (Q)}$ in Eq.~(\ref{Current1piRel}). 

Last but not least, there are also contributions coming from time-derivatives of unitary
transformations in diagrams $(3)$ and $(4)$ of 
Fig.~\ref{fig:tree} 
which are proportional to $k_0$, see Eq.~(\ref{temp22}). For the standard choice
of the unitary phases, these are given by 
\beq
P_{{\rm 2N:\,}1\pi, {\rm UT^\prime} }^{a \; ({Q^0})} = i
\frac{M_\pi^2}{m_q k^2} \, \vec k \cdot\vec {A}_{{\rm 2N:\,}1\pi, {\rm UT^\prime} }^{a \; ({Q})} \,,
\label{NNCurrent1piUTPrimePs}
\eeq
where $\vec {A}_{{\rm 2N:\,}1\pi, {\rm UT^\prime} }^{a \; ({Q})}$ is given
in Eq.~(\ref{NNCurrent1piUTPrime}).

\subsubsection{Two-pion-exchange contributions}
\label{sec:2N2piPS}
We now turn to the two-pion exchange contributions emerging from 
diagrams $(4-6),(10-12),(17),(18),(20)$ of
 Fig.~\ref{fig:tpe}. 
We found that graphs $(4-6)$ depend on the unitary phases via
\beqa
(2),(3) &\sim& 1+2\alpha_{1}^{ax}, \nn 
(5)&\sim& -3 + 2\alpha_{25}^{p}+2\alpha_{26}^{p},\nn
(4),(6)&\sim& (-2+\alpha_{4}^{p}-\alpha_5^{p}-\alpha_6^{p}+\alpha_{10}^{p}+\alpha_{13}^{p})\frac{1}{(k^2+M_\pi^2)^2}\big[ \ldots \big]+(-1+\alpha_{12}^{p}+\alpha_{15}^{p})\frac{1}{k^2+M_\pi^2}.
\eeqa
Notice that the contributions involving second-order pion-pole terms 
resulting from  diagrams (4), (6) vanish for our standard choice
of the unitary phases. The resulting pseudoscalar current can be
expressed in terms of pion-pole contributions to the 
axial current operator $\vec A_{{\rm 2N:} \, 2\pi}^{a\, (Q)}$ given in
Eq.~(\ref{Current2pi}) via 
\beq
\label{Current2piPs}
P_{{\rm 2N:} \, 2\pi}^{a\, (Q^0)} = i
\frac{M_\pi^2}{m_q k^2} \, \vec k \cdot \vec A_{{\rm 2N:} \, 2\pi}^{a\,
  (Q)} \bigg|_{g_{13} = g_{14} = g_{15} = g_{16} = g_{17} = 0} \,,
\eeq
with the scalar functions $g_1(q_1), \ldots , g_{12}(q_1)$  being
defined in Eq.~(\ref{giOfq1TPEStrF}).

\subsubsection{Short-range contributions to the pseudoscalar current at order $Q^0$}

We first consider the static contributions and begin with tree-level
diagrams, which could emerge from the derivative-less terms in the
effective Lagrangian ${\cal L}_{NN}$ involving a single insertion of
the pseudoscalar source. It is, however, not possible to build such
structures in the Lagrangian which fulfill all the required
symmetries. Thus, there are no static tree-level contributions to the
pseudoscalar current at this order. 

Next, one-loop diagrams $(3-5)$, $(8)$, $(9)$, $(17)$, $(18)$, $(25-30)$ 
of Fig.~\ref{fig:contact} involving a single insertion of the leading
contact interactions from $\mathcal{L}_{NN}^{(0)}$ are found to yield
nonvanishing contributions to the pseudoscalar current. 
The ones emerging from graphs 
$(3-5)$, $(25)$ and $(27-30)$ depend on the unitary phases via
\beqa
(3)&\sim& (-2 + \alpha_4^{p} -
\alpha_5^{p} - \alpha_6^{p}+\alpha_{10}^{p}+\alpha_{13}^{p})\frac{1}{(k^2+M_\pi^2)^2}\big[ \ldots \big] +
(-1+\alpha_{12}^{p}+\alpha_{15}^{p})\frac{1}{k^2+M_\pi^2}\big[ \ldots \big],\nn
(4)&\sim& (-4 +  \alpha_4^{p} - \alpha_5^{p} - \alpha_6^{p} + \alpha_{10}^{p}
+\alpha_{13}^{p}+\alpha_{21}^{p}+\alpha_{22}^{p} + \alpha_{23}^{p} - \alpha_{24}^{p})
\frac{1}{(k^2+M_\pi^2)^2}\big[ \ldots \big] \nn
&+& (-1 +
\alpha_{12}^{p}+\alpha_{15}^{p})\frac{1}{k^2+M_\pi^2}\big[ \ldots \big],\nn
(5),(24),(28-30)&\sim& -2 + \alpha_{21}^{p}+\alpha_{22}^{p} + \alpha_{23}^{p}-\alpha_{24}^{p},
\nn 
(21)&\sim& 1+\alpha_1^{ax},\nn 
(23)&\sim& \alpha_1^{ax}, \nn
(25)&\sim& (-2 + \alpha_4^{p} - \alpha_5^{p} - \alpha_6^{p} +
\alpha_{10}^{p} + \alpha_{13}^{p})\frac{1}{(k^2+M_\pi^2)^2}\big[ \ldots \big] + (-2 +
\alpha_{12}^{p}+\alpha_{15}^{p})\frac{1}{k^2 + M_\pi^2}\big[ \ldots \big], \nn
(27)&\sim& (-4 + \alpha_4^{p} - \alpha_5^{p} - \alpha_6^{p} + \alpha_{10}^{p}
+ \alpha_{13}^{p}+\alpha_{21}^{ap}+\alpha_{22}^{p} + \alpha_{23}^{p}-\alpha_{24}^{p})
\frac{1}{(k^2+M_\pi^2)^2}\big[ \ldots \big] \nn
&+&
(\alpha_{12}^{p}+\alpha_{15}^{p})\frac{1}{k^2+M_\pi^2}\big[ \ldots \big].
\eeqa
For our standard choice, the total contribution of these diagrams is found
to vanish. Given that the static one-pion-contact terms in the 3NF at N$^3$LO 
vanish  after antisymmetrization, our result is consistent with the
matching condition. Thus we conclude that there are no static contributions
to the short-range pseudoscalar current operator at order $Q^0$.

In addition to the static terms, one also encounters relativistic $1/m$-corrections involving a
single insertion of the LO contact interactions from
$\mathcal{L}_{NN}^{(0)}$ as shown in
Fig.~\ref{fig:relativistic_contact}. The unitary phase
dependence of the corresponding contributions is given by
\beqa
(1)&\sim& 1+\alpha_{16}^{p, 1/m},\nn 
(2),(4),(5)&\sim&
\alpha_{21}^{p}+\alpha_{22}^{p}+\alpha_{23}^{p} -\alpha_{24}^{p} - 2, \nn 
(3)&\sim&
\alpha_{17}^{p}+\alpha_{18}^{p}+\alpha_{19}^{p}-\alpha_{20}^{p}
+\alpha_{21}^{p}+\alpha_{22}^{p}+\alpha_{23}^{p}-\alpha_{24}^{p} - 4,\nn
(6)&\sim& \alpha_{17}^{p}+\alpha_{18}^{p}+\alpha_{19}^{p}-\alpha_{20}^{p} - 2.
\eeqa 
For our standard choice, the pseudoscalar current is expressed in
terms of the axial current $\vec A_{{\rm 2N: \, cont,}\, 1/m}^{a \,
  (Q)}$ defined in Eq.~(\ref{CurrentContRel}) via 
\beq 
\label{CurrentContRelPs}
P_{{\rm 2N: \, cont,}\, 1/m}^{a \, (Q^0)} = i \frac{M_\pi^2}{m_q k^2}
\, \vec k \cdot \vec A_{{\rm 2N: \, cont,}\, 1/m}^{a \,  (Q)} \,.
\eeq

Finally there are also contributions from diagram $(5)$ of Fig.~\ref{fig:tree}
which are proportional to $k^0$ and for this reason are demoted to
order $Q^0$. For the standard choice of unitary phases, we find the contribution
\beqa
\label{NNCurrentPsContUTPrime}
P_{{\rm 2N:\,cont,\,UT^\prime}}^{a \; ({Q^0})} &=& i \frac{M_\pi^2}{m_q k^2} \, \vec k \cdot \vec A_{{\rm 2N:\,cont,\,UT^\prime}}^{a \; ({Q})} \,,
\eeqa
where $\vec A_{{\rm 2N:\,cont,\,UT^\prime}}^{a \; ({Q})} $ is
specified in Eq.~(\ref{NNCurrentContUTPrime}). 

\subsection{Three-nucleon pseudoscalar operators}
\def\theequation{\arabic{section}.\arabic{equation}}
\label{sec:3N:Ps}

We now discuss 3N contributions to the pseudoscalar current
operator which start to contribute at the same order $Q^0$ and emerge
from tree-level diagrams constructed solely from the lowest-order
vertices. 3N diagrams which
yield non-vanishing contributions and
which do not involve contact interactions are $(8-18)$ and $(22-26)$ of Fig.~\ref{fig:tpe_3N}. 
Out of these diagrams, 
graphs $(8-18)$ yield expressions which depend on the unitary phases as follows:
\beqa
(12),(13)&\sim& -1+\alpha_{25}^{p}+\alpha_{26}^{p}, \nn 
(17),(18)&\sim& -2+\alpha_{25}^{p}+\alpha_{26}^{p}, \nn
(8-11)&\sim&
(-2+\alpha_4^{p}-\alpha_5^{p}-\alpha_6^{p}+\alpha_{10}^{p}+\alpha_{13}^{p})\frac{1}{(k^2+M_\pi^2)^2}
\big[ \ldots \big] + (-2+\alpha_{12}^{p}+\alpha_{15}^{p})\frac{1}{k^2+M_\pi^2}\big[ \ldots \big],\nn
(14-16)&\sim&
(-2+\alpha_4^{p}-\alpha_5^{p}-\alpha_6^{p}+\alpha_{10}^{p}+\alpha_{13}^{p})\frac{1}{(k^2+M_\pi^2)^2}
\big[ \ldots \big] + (\alpha_{12}^{p}+\alpha_{15}^{p})\frac{1}{k^2+M_\pi^2}\big[ \ldots \big] \,.
\eeqa
For our standard choice of the unitary phases, the contributions from diagrams
$(12-16)$ as well as all expressions involving second-order pion-pole
terms turn out  to vanish.
In order to facilitate a comparison with the four-nucleon force at
N$^3$LO, we write the resulting expression for the 3N pseudoscalar current $P_{{\rm 3N:}\,
  \pi}^{a \, (Q^0)}$ in the form 
\beq
\label{Current3NPiPs}
P_{{\rm 3N:}\,
  \pi}^{a \, (Q^0)} =- i\frac{2\, F_\pi^2 M_\pi^2}{g_A m_q} \sum_{i=5}^{8}   \frac{\vec C_i^a\cdot\vec{k}}{k^2}
+
5\,{\rm permutations} \; 
= \; i \frac{M_\pi^2}{m_q k^2} \, \vec k \cdot \vec A_{{\rm 3N:}\,
  \pi}^{a \, (Q)} \bigg|_{\rm pion-pole \; terms}\,,
\eeq
where $\vec C_i^a$ are defined in Eq.~(\ref{Ci3NDefinition}).

Finally, diagrams $(5-19)$ of Fig.~\ref{fig:contact_3N} involving 
one or
more insertions of the the leading 2N contact interactions also 
contribute to the 3N pseudoscalar current. 
All these graphs yield expressions which depend upon the unitary phases:
\beqa
(7),(9),(10),(13-19)&\sim& -2 + \alpha_{21}^{p} + \alpha_{22}^{p} + \alpha_{23}^{p}-\alpha_{24}^{p},
\nn
(5),(6)&\sim&
(-2+\alpha_4^{p}-\alpha_5^{p}-\alpha_6^{p}+\alpha_{10}^{p}+\alpha_{13}^{p})\frac{1}{(k^2+M_\pi^2)^2}
\big[ \ldots \big] + (-2+\alpha_{12}^{p}+\alpha_{15}^{p})\frac{1}{k^2+M_\pi^2}\big[ \ldots \big], \nn
(8)&\sim&
(-2+\alpha_4^{p}-\alpha_5^{p}-\alpha_6^{p}+\alpha_{10}^{p}+\alpha_{13}^{p})\frac{1}{(k^2+M_\pi^2)^2}
\big[ \ldots \big] + (\alpha_{12}^{p} + \alpha_{15}^{p})\frac{1}{k^2+M_\pi^2}\big[ \ldots \big], \nn 
(12)&\sim& -2 + \alpha_{25}^{p} + \alpha_{26}^{p}, \nn
(11)&\sim&
(-4+\alpha_4^{p}-\alpha_5^{p}-\alpha_6^{p}+\alpha_{10}^{p}+\alpha_{13}^{p}
+ \alpha_{21}^{p} + \alpha_{22}^{p} + \alpha_{23}^{p} - \alpha_{24}^{p})\frac{1}{(k^2+M_\pi^2)^2}
\big[ \ldots \big] \nn
&+& (\alpha_{12}^{p} + \alpha_{15}^{p})\frac{1}{k^2+M_\pi^2}\big[ \ldots \big].
\eeqa
For our standard choice, the contributions of graphs
(7-11), (13-19)  and all expressions involving second-order pion-pole terms vanish
yielding the result:
\beq
\label{Current3NContPs}
P_{\rm 3N:\, cont}^{a \, (Q)} =- i\frac{2\, F_\pi^2 M_\pi^2}{g_A m_q}\sum_{i=2}^{3}   \frac{\vec D_i^a\cdot\vec{k}}{k^2}
+
5\,{\rm permutations} \; = \; 
i \frac{M_\pi^2}{m_q k^2} \, \vec k \cdot \vec A_{{\rm 3N: \, cont}}^{a \, (Q)} \bigg|_{\rm pion-pole \; terms}\,,
\eeq
with $\vec D_i^a$ defined in Eq.~(\ref{DiFF3NDefinition}).
\section{Summary of the derived pseudoscalar contributions}
\def\theequation{\arabic{section}.\arabic{equation}}
\label{sec:summaryPsCurrents}

We now summarize the derived contributions to
the nuclear pseudoscalar current operator
\beq
{\fet P} = {\fet P}_{\rm 1N}   + {\fet
    P}_{\rm 2N}   +  {\fet P}_{\rm 3N}   + \ldots \,.
\eeq  
The chiral power counting implies that
$n$-nucleon operators are, in general, suppressed by two powers of the expansion
parameter relative to $n-1$-nucleon operators so the one expects the
hierarchy\footnote{The leading contribution to the two-nucleon current
operator at order $Q^{-2}$ turns out to vanish.}
\beq
 {\fet P}_{\rm 1N}   \sim \mathcal{O}(Q^{-4}), \quad \quad 
 {\fet P}_{\rm 2N}   \sim \mathcal{O} (Q^{-2}), \quad  \quad
{\fet P}_{\rm 3N}   \sim \mathcal{O} (Q^{0})\,, \quad \quad \ldots \,.
\eeq
Thus, at fourth order in the chiral expansion relative to the dominant
one-body contribution at order, it is necessary and sufficient to include  single-,
two- and three-nucleon operators. 

\begin{table}[t]
\caption{Chiral expansion of the nuclear pseudoscalar operator up to N$^3$LO. 
\label{tab_sum_current_ps}}
\smallskip
\begin{ruledtabular}
\begin{tabular}{@{\extracolsep{\fill}}lrrr}
\noalign{\smallskip}
 order &  single-nucleon  &  two-nucleon  &
                                                            three-nucleon  
\smallskip
 \\
\hline \hline
&&& \\[-7pt]
LO ($Q^{-4}$) & $P^{a}_{{\rm 1N:  \, static}}, \;$
                Eq.~(\ref{LOSingleNCurrentPExpr})  & --- & --- \\ [5pt] \hline
&&& \\[-9pt]
NLO ($Q^{-2}$) & $P^{a}_{{\rm 1N:  \, static}}, \;$
                Eq.~(\ref{SingleNPQtoMinus2StaticExpr})  & --- & --- \\
[5pt] \hline
&&& \\[-9pt]
N$^2$LO ($Q^{-1}$) & --- & $P^{a}_{{\rm 2N:  \, 1\pi}}, \;$ Eq.~(\ref{PsCurrentTreeOPE})  & --- \\ [2pt]
& & \hskip -0.35 true cm $+ \; P^{a}_{{\rm 2N:  \, cont}}, \;$
    Eq.~(\ref{PsCurrentTreeCont})  &\\ [5pt] \hline
&&& \\[-9pt]
N$^3$LO ($Q^0$) & $P^{a}_{{\rm 1N:  \, static}}, \;$
                Eq.~(\ref{PseudoscalarTwoLoopConjecture}) & $P^{a}_{{\rm
                                                 2N:  \, 1\pi}}, \;$
                                                 Eq.~(\ref{CurrentPs1pi})
                                            &  $P^{a}_{{\rm 3N:
                                              \, \pi}}, \;$
                                              Eq.~(\ref{Current3NPiPs})
  \\ [2pt]
& \hskip -0.35 true cm $+ \; P^{ a}_{{\rm 1N:\,}1/m, {\rm
    UT^\prime}}, \;$  Eq.~(\ref{k0OvermNuclcorrP}) &
 \hskip -0.35 true cm $+ \;  P_{{\rm 2N:\,}1\pi, {\rm
                                                    UT^\prime} }^{a} ,
                                                    \;$
                                                    Eq.~(\ref{NNCurrent1piUTPrimePs})
                                            &
  \hskip -0.35 true cm $+ \;  P^{a}_{{\rm
    3N: \, cont}}, \;$ Eq.~(\ref{Current3NContPs})   \\ [2pt]
& \hskip -0.35 true cm $+ \; P^{ a }_{{\rm 1N: \, 1/m^2}} , \;$
  Eq.~(\ref{oneovermTo2corrP}) &
  \hskip -0.35 true cm $+ \;  P^{a}_{{\rm 2N:} \, 1\pi , \, 1/m },
                                                    \;$
                                Eq.~(\ref{CurrentPs1piRel}) & \\ [2pt]
&&  \hskip -0.35 true cm $+ \;  P^{a}_{{\rm 2N:} \, 2\pi },
                                                    \;$
                                Eq.~(\ref{Current2piPs}) & \\ [2pt]
&&  \hskip -0.35 true cm $+ \;    P_{{\rm 2N:\,cont,\,UT^\prime}}^{a} ,
                                                    \;$
                                Eq.~(\ref{NNCurrentPsContUTPrime}) & \\
  [2pt]
&&  \hskip -0.35 true cm $+ \;    P_{{\rm 2N:\,cont,\, 1/m}}^{a} ,
                                                    \;$
                                Eq.~(\ref{CurrentContRelPs}) &  \\ &&& \\[-11pt]
\end{tabular}
\end{ruledtabular}
\end{table}

In Table~\ref{tab_sum_current_ps}, we summarize all derived
contributions to the nuclear pseudoscalar current up to N$^3$LO based
on our counting scheme for the nucleon mass and the standard
choice of the unitary phases.  As it is already clear from the
previous section, the chiral expansion of the pseudoscalar current
closely resembles that of the axial-vector current operator, see Table \ref{tab_sum_current},
with the corresponding contributions appearing one order lower.   
Finally, notice that isospin breaking corrections to the exchange
pseudoscalar current operator start contributing at order $Q$ which is
beyond the accuracy of our analysis.

\section{Current conservation}
\def\theequation{\arabic{section}.\arabic{equation}}
\label{sec:results}

Current conservation leads to the relation (\ref{continuityeqmomspace}) 
between the axial current, charge, pseudoscalar density
$\hat{\vec{A}}^a(\vec{k},k_0)$, $\hat{A}^{0, a}(\vec{k},k_0)$, $\hat{P}^{a}(\vec{k},k_0)$ and 
the contributions to the nuclear Hamiltonian
\beq
\hat H = \hat H_{\rm 1N}^{(Q^0)}  + \hat V_{{\rm 2N:} \, 1\pi}^{(Q^0)} + \hat
V_{{\rm 2N: \, cont}}^{(Q^0)} + \mathcal{O} \big( Q^2 \big) \,,
\eeq
 where $\hat X$ means that the quantity $X$ is to be taken as an
 operator rather than matrix element with respect to the nucleon
 momenta, and $\hat H_{\rm 1N}^{(Q^0)}$ refers to the nonrelativistic kinetic
 energy operator. Notice further that in order to get a correct chiral order
 for any sequence of operators in the convention we are using, one should take into account the 
 suppression factor of $Q^3$ for every intermediate nucleonic
 state. For example, the chiral order of the operator $\hat H_{\rm 1N}^{(Q^0)}  \, \hat A^{0, a \, (Q^{-1})}_{{\rm 1N:}\,
  1/m}$ is $Q^2$, while that of  
$\hat V_{{\rm 2N:} \, 1\pi}^{(Q^0)} \, \hat V_{{\rm 2N:} \, 1\pi}^{(Q^0)} \,\hat{A}^{0, a \,
  (Q^{-1})}_{{\rm 1N:\,UT^\prime}}$ is $Q^5$. Alternatively, one can,
of course, also explicitly verify the chiral order of any sequence of
operators by adding together the inverse mass dimension $\kappa$ of
all vertices as explained in section \ref{sec:formalism}. 
Last but not least, we remind the
reader that within the adopted counting scheme for the nucleon mass with $m \sim
\Lambda_b^2/M_\pi$, the energy-transfer $k_0$ is counted as $k_0 \sim
Q^2/m = \mathcal{O} (Q^3)$. 

In the following, we will explicitly verify the
 validity of the continuity equation for all derived contributions to
 the charge and current operators. 

\begin{itemize}
\item \emph{Single-nucleon current operator}\\
Requiring the continuity equation (\ref{continuityeqmomspace}) to hold
true at all considered orders in the chiral expansion, we obtain the
relations 
\beqa
\vec k \cdot \hat{\vec A}^{a \, (Q^{-3})}_{{\rm 1N:\,
    static}} - m_q i\,\hat{P}_{{\rm 1N:\, static}}^{a\,(Q^{-4})} &=&  0\,,\label{comm1_singleN}\\
\vec k \cdot \hat{\vec A}^{a \, (Q^{-1})}_{{\rm 1N:\,
    static}} - m_q i\,\hat{P}_{{\rm 1N:\, static}}^{a\,(Q^{-2})}&=&  0\,,\label{comm2_singleN}\\
\vec k \cdot \hat{\vec A}^{a \, (Q)}_{{\rm 1N:\,}
    1/m^2} - m_q i\,\hat{P}_{{\rm 1N:}\, 1/m^2}^{a\,(Q^{0})}&=& \Big[ \hat H_{\rm 1N}^{(Q^0)}  , \; \hat A^{0, a \, (Q^{-1})}_{{\rm 1N:}\,
  1/m}  -\frac{\partial}{\partial k_0}  
\vec k \cdot \hat{\vec A}^{a \, (Q)}_{{\rm 1N:\,}1/m, 
    {\rm UT^\prime}} 
+ \Big[ \hat H_{\rm 1N}^{(Q^0)}  , \;
\frac{\partial}{\partial k_0} \hat A^{0, a \, (Q^{-1})}_{{\rm 1N:}\,
   {\rm
    UT^\prime}}\Big]\nn
&+& m_q i\frac{\partial}{\partial k_0} \hat{P}^{ a \, (Q^0)}_{{\rm
    1N:\,}1/m, {\rm UT^\prime}} \Big],\label{comm4_singleN}\\
\vec k \cdot \hat{\vec A}^{a \, (Q)}_{{\rm 1N:\,
    static}} - m_q i\,\hat{P}_{{\rm 1N:\, static}}^{a\,(Q^{0})} &=&  0\,.\label{comm5_singleN}
\eeqa
It is easy to verify that the derived contributions fulfill the first
three equations. Notice further that the last  equation implies the
Goldberger-Treiman-like relation between the LECs $f_1^A$ and $f_1^P$: 
\beq
f_1^P=-4 f_1^A\,.
\eeq
It is then straightforward to verify the validity of the last relation in
Eq.~(\ref{comm5_singleN}) using Eqs.~(\ref{SingleNtwoLoopExpr}) and (\ref{PseudoscalarTwoLoopConjecture}). 

\item \emph{Two-nucleon current operator}\\
At leading order, the continuity equation (\ref{continuityeqmomspace}) 
leads to the relations 
\beqa 
\label{comm1}
\vec k \cdot \hat{\vec A}^{a \, (Q^0)}_{{\rm 2N:} \,
  1\pi} - m_q i\,\hat{P}_{{\rm 2N:\,}1\pi}^{a\,(Q^{-1})}&=&  0 \,, \\[4pt]
\label{comm2}
\vec k \cdot \hat{\vec A}^{a \, (Q^0)}_{\rm 2N: \,
  cont} - m_q i\,\hat{P}_{{\rm 2N:\,cont}}^{a\,(Q^{-1})}&=&  0 \,,
\eeqa
which are trivially fulfilled, see Eqs.~(\ref{PsCurrentTreeOPE}) and
(\ref{PsCurrentTreeCont}). 

At order $Q^2$, the continuity equation induces a number or
relations between the various contributions. Identifying all
order-$Q^2$ terms of the one-pion range  in 
Eq.~(\ref{continuityeqmomspace}),  we
obtain the relation 
\beq
\vec k \cdot \hat{\vec A}^{a \, (Q)}_{{\rm 2N:} \,
  1\pi} - m_q i\,\hat{P}_{{\rm 2N:\,}1\pi}^{a\,(Q^{0})} = 
\mbox{shorter-range terms} \,.
\eeq
Here, we used the fact that there are no static contributions to the single-nucleon charge operator
for $k_0=0$ up to order $Q^0$. For this reason, there is no commutator of the one-pion-exchange
potential with the single-nucleon charge operator at the considered order. Further, there are
no static $k_0$-dependent contributions to the single-nucleon axial
and pseudoscalar currents at orders $Q$ and $Q^{0}$, respectively. 
For this reason, there are no terms of the one-pion range  on the left-hand side of
Eq.~(\ref{continuityeqmomspace}) involving a single commutator at
order $Q^2$. Finally, the
double-commutator term involving a $k_0$-derivative of the single nucleon
charge operator cannot give rise to operators of the one-pion range in
the static limit. Notice further that the shorter-range terms on the right-hand side of
Eq.~(\ref{comm3}) reflect the ambiguity in separating the one- and
two-pion exchange contributions. Using the derived expressions for 
$\hat{\vec A}^{a \, (Q)}_{{\rm 2N:} \,  1\pi}$ and $\hat{P}_{{\rm
    2N:\,}1\pi}^{a\,(Q^{0})}$ given in Eqs.~(\ref{Current1pi}) and (\ref{CurrentPs1pi}), respectively, we obtain 
\beq
\label{comm3}
\vec k \cdot \hat{\vec A}^{a \, (Q)}_{{\rm 2N:} \,
  1\pi} - m_q i\,\hat{P}_{{\rm
    2N:\,}1\pi}^{a\,(Q^{0})} =   \frac{g_A^3}{32 \pi F_\pi^4} \tau_2^a
\vec{q}_2\cdot\vec{\sigma}_2\Big( (1+g_A^2) M_\pi +(2 M_\pi^2+q_1^2)A(q_1)\Big) \; + \; 1 \leftrightarrow 2\,.
\eeq
As expected, the expression on the right-hand side of Eq.~(\ref{comm3}) has no pole at
$q_1^2=-M_\pi^2$ and can be cast into the form of a spectral integral
taken over the region $\mu \ge 2 M_\pi$. It is important to keep these
terms in
mind  when verifying the continuity equation for the two-pion
exchange contributions. More precisely, the sum of the static one- and two-pion exchange
contributions to the current operator, being unaffected by 
the above mentioned ambiguity in separating the terms according to
the range, is expected to fulfill the relation 
\beqa
\label{comm5}
\vec k \cdot \hat{\vec A}^{a \, (Q)}_{{\rm 2N:} \,
  1\pi} - m_q i\,\hat{P}_{{\rm
    2N:\,}1\pi}^{a\,(Q^{0})} + \vec k \cdot \hat{\vec A}^{a \, (Q)}_{{\rm 2N:} \,
  2\pi} - m_q i\,\hat{P}_{{\rm
    2N:\,}2\pi}^{a\,(Q^{0})} 
&=&  \Big[ \hat V_{{\rm 2N:} \, 1\pi}^{(Q^0)} , \;  \hat A^{0,
  a \, (Q^{-1})}_{{\rm 2N: \, 1\pi}} 
- \frac{\partial}{\partial k_0}
\vec{k}\cdot\hat{\vec A}^{a \, (Q)}_{{\rm 2N:}\,
  1\pi, {\rm UT^\prime}}  \\
&+& 
\frac{\partial}{\partial k_0}\Big[\hat V_{{\rm 2N:} \, 1\pi}^{(Q^0)} ,\;\hat{A}^{0, a \,
  (Q^{-1})}_{{\rm 1N:\,UT^\prime}}\Big]
+m_q
i\frac{\partial}{\partial k_0} \hat{P}^{ a \, (Q^0)}_{{\rm 2N:\,}1\pi,
  {\rm UT^\prime}}\Big] \,.
\nonumber
\eeqa
Using the derived expressions for the various charge and current
operators and applying dimensional regularization to evaluate the
integrals appearing on the right-hand side, we have explicitly
verified that this equation indeed holds true.  

Consider now static contributions to the continuity equation at
order $Q^2$, which involve contact interactions.
Since we have only vanishing terms in both the
short-range axial and pseudoscalar current operators at orders $Q$ and
$Q^2$, respectively, the right-hand side of
Eq.~(\ref{continuityeqmomspace}) vanishes trivially. On the other
hand, the left-hand side of this equation does contain non-vanishing
terms. Notice, however, that since there is no static
single-nucleon charge operator at order $Q^{-1}$ for
$k_0=0$, we do not have any commutator with it. Similarly, we have $\hat A^{0,
  a \, (Q^{-1})}_{{\rm 2N: \,cont}}=0$ so that there cannot
be any commutator with this operator. Furthermore, due to the absence
of static $k_0$-dependent contributions to the single-nucleon pseudoscalar
current up to order $Q^{-1}$, there is also no
commutator of this operator with the effective Hamiltonian. Collecting
the remaining static order-$Q^2$ short-range contributions on the left-hand side of 
Eq.~(\ref{continuityeqmomspace}), we end up with the relation  
\beqa
0 &=&  \Big[ \hat V_{{\rm 2N: \, cont}}^{(Q^0)} , \;  \hat A^{0,
  a \, (Q^{-1})}_{{\rm 2N:}\, 1\pi} 
- \frac{\partial}{\partial k_0}
\vec{k}\cdot\hat{\vec A}^{a \, (Q)}_{{\rm 2N:}\,
  1\pi, {\rm UT^\prime}} 
+\frac{\partial}{\partial k_0}\Big[\hat V_{{\rm 2N:} \, 1\pi}^{(Q^0)} ,\;\hat{A}^{0, a \,
  (Q^{-1})}_{{\rm 1N: \, UT^\prime}}\Big]
+m_q
i\frac{\partial}{\partial k_0} \hat{P}^{ a \, (Q^0)}_{{\rm 2N:\,}1\pi,
  {\rm UT^\prime}}\Big]\nn
&+&\Big[\hat V_{{\rm 2N:} \,
  1\pi}^{(Q^0)}, \;- \frac{\partial}{\partial k_0} 
\vec{k}\cdot\hat{\vec A}^{a \, (Q)}_{{\rm 2N: \,cont, \,UT^\prime}} 
+\frac{\partial}{\partial k_0}\Big[\hat V_{{\rm 2N:\, cont}}^{(Q^0)} ,\;\hat{A}^{0, a \,
  (Q^{-1})}_{{\rm 1N:\,UT^\prime}}\Big]
+m_q i\frac{\partial}{\partial k_0} \hat{P}^{ a \, (Q^0)}_{{\rm 2N:\,cont}, {\rm UT^\prime}}\Big]\,. \label{comm6}
\eeqa
Using the derived contributions to the current operators, we have
verified that this relation is indeed fulfilled.  In fact, we even find  
\beq
0 =  
- \frac{\partial}{\partial k_0}
\vec{k}\cdot\hat{\vec A}^{a \, (Q)}_{{\rm 2N: \,cont, \,UT^\prime}} 
+\frac{\partial}{\partial k_0}\Big[\hat V_{{\rm 2N:\, cont}}^{(Q^0)} ,\;\hat{A}^{0, a \,
  (Q^{-1})}_{{\rm 1N:\,UT^\prime}}\Big]
+m_q i\frac{\partial}{\partial k_0} \hat{P}^{ a \, (Q^0)}_{{\rm
    2N:\,cont}, {\rm UT^\prime}},
\label{vanishingModifiedChargeCont}
\eeq
so that Eq.~(\ref{comm6}) 
simplifies to
\beq
0 =  \Big[ \hat V_{{\rm 2N: \, cont}}^{(Q^0)} , \;  \hat A^{0,
  a \, (Q^{-1})}_{{\rm 2N:}\, 1\pi}
- \frac{\partial}{\partial k_0} 
\vec{k}\cdot\hat{\vec A}^{a \, (Q)}_{{\rm 2N:}\,
  1\pi, {\rm UT^\prime}} +\frac{\partial}{\partial k_0}\Big[\hat V_{{\rm 2N:} \, 1\pi}^{(Q^0)} ,\;\hat{A}^{0, a \,
  (Q^{-1})}_{{\rm 1N: \, UT^\prime}}\Big]
+m_q
i\frac{\partial}{\partial k_0} \hat{P}^{ a \, (Q^0)}_{{\rm 2N:\,}1\pi,
  {\rm UT^\prime}}\Big].
\eeq

Next, for the $1/m$-corrections to one-pion-exchange two-nucleon
axial-vector current, the continuity equation implies the relation 
\beqa
\label{comm7}
\vec k \cdot \hat{\vec A}^{a \, (Q)}_{{\rm 2N:} \,
  1\pi, \, 1/m} - m_q i\,\hat{P}_{{\rm
    2N:\,}1\pi, 1/m}^{a\,(Q^{0})}&=&  \Big[ \hat H_{\rm 1N}^{(Q^0)}  , \; \hat A^{0, a \, (Q^{-1})}_{{\rm 2N:}\,
  1\pi} 
-\frac{\partial}{\partial k_0}
\vec{k}\cdot\hat{\vec A}^{a \, (Q)}_{{\rm 2N:}\,
  1\pi, {\rm UT^\prime}} +  
\frac{\partial}{\partial k_0}
\Big[\hat V_{{\rm 2N:} \, 1\pi}^{(Q^0)} , \;\hat{A}^{0, a \,
  (Q^{-1})}_{{\rm 1N:\, UT^\prime}}\Big]\nn
&+& m_q i\frac{\partial}{\partial k_0}\hat P^{a \, (Q^{0})}_{{\rm 2N:}\,
  1\pi, {\rm UT^\prime}} \Big] 
\; + \;  \Big[ \hat V_{{\rm 2N:} \, 1\pi}^{(Q^0)} , \;  \hat A^{0,
  a \, (Q^{-1})}_{{\rm 1N:} \, 1/m}   
-\frac{\partial}{\partial k_0}  \vec{k}\cdot\hat{\vec A}^{
  a \, (Q)}_{{\rm 1N:} \, 1/m, {\rm UT^\prime}} \nn
&+&\frac{\partial}{\partial k_0}
\Big[\hat H_{\rm 1N}^{(Q^0)}, \;\hat{A}^{0, a \,
  (Q^{-1})}_{{\rm 1N: \, UT^\prime}}\Big]
+  m_q i\frac{\partial}{\partial k_0}\hat P^{
  a \, (Q^{0})}_{{\rm 1N:} \, 1/m, {\rm UT^\prime}} \Big]
\,,\quad\quad
\eeqa
which is indeed fulfilled for the derived operators. We have also
explicitly verified the validity of the continuity equation for 
the $1/m$-corrections to the 2N axial vector current involving contact
interactions:
\beqa
\vec k \cdot \hat{\vec A}^{a \, (Q)}_{{\rm 2N: \,
  cont}, \, 1/m}- m_q i\,\hat{P}_{{\rm
    2N:\,cont}, 1/m}^{a\,(Q^{0})} &=& \Big[ \hat V_{{\rm 2N:\,cont}}^{(Q^0)} , \;  \hat A^{0,
  a \, (Q^{-1})}_{{\rm 1N:} \, 1/m}   
-\frac{\partial}{\partial k_0}  \vec{k}\cdot\hat{\vec A}^{
  a \, (Q)}_{{\rm 1N:} \, 1/m, {\rm UT^\prime}} 
+\frac{\partial}{\partial k_0}
\Big[\hat H_{\rm 1N}^{(Q^0)}, \;\hat{A}^{0, a \,
  (Q^{-1})}_{{\rm 1N: \, UT^\prime}}\Big]\nn
&+&  m_q i\frac{\partial}{\partial k_0}\hat P^{
  a \, (Q^{0})}_{{\rm 1N:} \, 1/m, {\rm UT^\prime}}
\Big]\,.
\label{comm8}
\eeqa
One may also expect further contributions to the right-hand side of
Eq.~(\ref{comm8}) stemming from the commutator
with the kinetic energy. However, as already pointed out, those contributions
turn out to vanish, see Eq.~(\ref{vanishingModifiedChargeCont}).

\item \emph{Three-nucleon current operator}\\
Finally, we have verified that the following two relations are
fulfilled for the derived three-nucleon current operator: 
\beqa
\label{comm9}
\vec k \cdot \hat{\vec A}^{a \, (Q)}_{{\rm 3N:} \,
  \pi} 
- m_q i\,\hat{P}^{a \, (Q^0)}_{{\rm 3N:} \,
  \pi}&=&  \Big[ \hat V_{{\rm 2N:} \, 1\pi}^{(Q^0)} , \;  \hat A^{0,
  a \, (Q^{-1})}_{{\rm 2N:}\, 1\pi} 
- \frac{\partial}{\partial k_0}
\vec{k}\cdot\hat{\vec A}^{a \, (Q)}_{{\rm 2N:}\,
  1\pi, {\rm UT^\prime}} 
+\frac{\partial}{\partial k_0}\Big[\hat V_{{\rm 2N:} \, 1\pi}^{(Q^0)} ,\;\hat{A}^{0, a \,
  (Q^{-1})}_{{\rm 1N: \, UT^\prime}}\Big]\nn
&+& m_q i\frac{\partial}{\partial k_0}\hat P^{
  a \, (Q^{0})}_{{\rm 2N:}\, 1\pi,  {\rm UT^\prime}} \Big]\,, \\[4pt]
\label{comm10}
\vec k \cdot \hat{\vec A}^{a \, (Q)}_{\rm 3N: \,
  cont} - m_q i\,\hat{P}^{a \, (Q^0)}_{{\rm 3N:\,
    cont}}
&=&  \Big[ \hat V_{{\rm 2N: \, cont}}^{(Q^0)} , \;  \hat A^{0,
  a \, (Q^{-1})}_{{\rm 2N: \, 1\pi}}
- \frac{\partial}{\partial k_0}
\vec{k}\cdot\hat{\vec A}^{a \, (Q)}_{{\rm 2N:}\,
  1\pi, {\rm UT^\prime}}  
+\frac{\partial}{\partial k_0}\Big[\hat V_{{\rm 2N:} \, 1\pi}^{(Q^0)} ,\;\hat{A}^{0, a \,
  (Q^{-1})}_{{\rm 1N: \,UT^\prime}}\Big] \nn
&+& m_q i\frac{\partial}{\partial k_0}\hat P^{
  a \, (Q^{0})}_{{\rm 2N:}\, 1\pi,  {\rm UT^\prime}}
\Big]\,. 
\eeqa
Again, we made use of Eq.~(\ref{vanishingModifiedChargeCont}) to
simplify the right-hand side of Eq.~(\ref{comm10}). 
\end{itemize}
To summarize, we have explicitly verified the validity of the
continuity equation for \emph{all} derived contributions to the
current operators.

\section{Comparison with earlier work}
\def\theequation{\arabic{section}.\arabic{equation}}
\label{sec:Baroni}

We are now in the position to compare our results with the earlier
derivations and begin with the single-nucleon contributions. 
To the best of our knowledge, the most complete
expressions for the single-nucleon axial charge and current operators
up-to-and-including $1/m^2$-corrections 
are given in Ref.~\cite{Shen:2012xz} in terms of the axial and pseudoscalar form
factors of the nucleon. Our results for the single-nucleon charge and
current operators agree with the ones of that work up-to-and-including
$1/m$-terms. More precisely, our
expression for the charge operator in the first line of
Eq.~(\ref{OneNucleonFF}) agrees with the expressions in Eq.~(3.8) and
(3.14) of  Ref.~\cite{Shen:2012xz}. Also the static contributions to
the current operator given by the first two terms in the right-hand side
of the second equation in (\ref{OneNucleonFF}) coincide with the
corresponding terms in Eqs.~(3.9) and  (3.14) of  that work. 
On the other hand, for a general choice of the unitary phases $\bar
\beta_{8,9}$, we find non-vanishing contributions to the current
operator $\vec{A}^{ a \, (Q)}_{{\rm 1N:\,}1/m, {\rm
    UT^\prime}}$ proportional to $k_0/m$, see
Eq.~(\ref{k0OvermNuclcorr}). The absence of such
contributions in Ref.~\cite{Shen:2012xz} is consistent with 
the choice of $\bar \beta_8 = \bar \beta_9 =
-1/2$. On the other hand, our results for $1/m^2$-contributions $\vec{A}^{ a \,
  (Q)}_{{\rm 1N: \, 1/m^2}}  $ in Eq.~(\ref{oneovermTo2corr}) appear
to differ from the ones in Eq.~(3.9) of Ref.~\cite{Shen:2012xz} even for the
choice $\bar \beta_8 = \bar \beta_9 = -1/2$.

We now turn to the exchange axial current operators. 
As already pointed out, their first derivation within the framework of 
chiral EFT has been carried out by Park et
al.~\cite{Park:1993jf,Park:2002yp}. Given the incompleteness of this
calculation, which did not include some of the reducible-like
diagrams and ignored pion-pole contributions, we refrain from a
detailed comparison with that work. A more complete recent derivation
of the exchange axial currents has been carried out by Baroni et
al.~\cite{Baroni:2015uza,Baroni:2016xll} in the framework of TOPT.
As already pointed out above, the dominant contributions to the
two-nucleon charge and current operators ${A}^{0,  a \,
  (Q^{-1})}_{{\rm 2N: \, 1\pi}}$ and $\vec{A}^{a \,
  (Q^{0})}_{{\rm 2N: \, 1\pi}} + \vec{A}^{a \,
  (Q^{0})}_{{\rm 2N: \, cont}}$ are well known, and our results for
these terms agree with the ones derived using TOPT in
Ref.~\cite{Baroni:2015uza}. The first corrections to the dominant
two-nucleon terms emerge at order $Q$, see Tables~\ref{tab_sum_current} and \ref{tab_sum_charge}. 
While the authors of \cite{Baroni:2015uza} count the nucleon mass in a
different way as $m \sim \Lambda_b$, which implies that the relativistic corrections
are promoted to lower orders in the chiral expansion as compared to
our approach based on the assignment $m \sim \Lambda_b^2/M_\pi$, they 
have not considered the $1/m$-contributions to the exchange operators and focused
entirely on the static terms at order $Q$.\footnote{Notice that in
  order to define the static two-pion exchange current operator
  via matching to the $S$-matrix, one needs to specify the 
  $1/m$-corrections to the one-pion exchange current. Given that the
  authors of \cite{Baroni:2015uza} have neglected the non-static corrections to the energy
denominators when calculating time-ordered diagrams, it is not clear
to us that their result for the static two-pion exchange
contributions is complete.} 
They also do  not
give the contributions proportional to the energy
transfer. For the remaining static contributions to the two-nucleon
current operator  $\vec{A}^{a \,
  (Q)}_{{\rm 2N: \, 1\pi}} + \vec{A}^{a \,
  (Q)}_{{\rm 2N: \, 2\pi}}$, our expressions differ from the ones
given in  Refs.~\cite{Baroni:2015uza,Baroni:2016xll}. Given that the
explicit expressions for these operators are rather involved, we restrict ourselves
to the case of zero momentum transfer, $\vec{k}=0$, and compare the
expressions for the sum of the one- and two-pion exchange operators at
order $Q$.\footnote{The separation into one- and two-pion exchange terms
is ambiguous.} Notice that due to to a different counting of the
nucleon mass, which results in the NLO contributions to
$\fet{A}^{\mu}_{\rm 1N}$ appearing already at order $Q^{-2}$ rather
than at order $Q^{-1}$ as in our counting scheme, the authors of
\cite{Baroni:2015uza,Baroni:2016xll} regard the order-$Q$
contributions as being N$^4$LO. The expressions for the corresponding one- and
more-pion exchange (MPE) operators  
$\vec{j}_{a}^{{\rm \; N4LO}}({\rm MPE},  \vec{q}_1)$ and
$\vec{j}_{a}^{{\rm \; N4LO}}({\rm MPE},  \vec{q}_1)$ are given in
Eqs.~(16) and (17) of Ref.~\cite{Baroni:2016xll}. Only the 
MPE part of their current operator depends on the loop functions 
\beqa
\vec{j}_{a}^{{\rm \; N4LO}}({\rm MPE}, \vec{q}_1)&=&\frac{g_A^3}{32\pi
  F_\pi^4}\tau_2^a\bigg[W_1(q_1)\vec{\sigma}_1 + W_2(q_1)\vec{q}_1
\,\vec{\sigma}_1\cdot\vec{q}_1
+Z_1(q_1)\left(2\,\vec{q}_1 \,\vec{\sigma}_2\cdot\vec{q}_1
  \frac{1}{q_1^2+M_\pi^2}-\vec{\sigma}_2\right)\bigg]
\nn
&+&\frac{g_A^5}{32\pi F_\pi^4}\tau_1^a
W_3(q_1)(\vec{\sigma}_2\times\vec{q}_1)\times\vec{q}_1-\frac{g_A^3}{32\pi
F_\pi^4}[\fet\tau_1\times\fet\tau_2]^a
Z_3(q_1)\vec{\sigma}_1\times\vec{q}_1
\vec{\sigma}_2\cdot\vec{q}_1\frac{1}{q_1^2+M_\pi^2}\nn
&+&
1\leftrightarrow 2\,,
\eeqa
where the loop functions in our notation are given by
\beqa
W_1(q_1)&=&\frac{1}{2}A(q_1)\Big[4\left(1-2g_A^2\right)M_\pi^2+\left(1-5g_A^2
\right)q_1^2\Big]+\frac{1}{2}M_\pi\bigg[g_A^2\left(\frac{4M_\pi^2}{4M_\pi^2+q_1^2}
-9\right)+1\bigg],\nn
W_2(q_1)&=&\frac{M_\pi\left(4\left(2g_A^2+1\right)M_\pi^2+\left(3g_A^2+1
\right)q_1^2\right)}{2q_1^2\left(4M_\pi^2+q_1^2
\right)}-\frac{A(q_1)\left(4\left(2g_A^2+1\right)M_\pi^2+\left(g_A^2-1
\right)q_1^2\right)}{2q_1^2},\nn
W_3(q_1)&=&-\frac{4A(q_1)}{3}-\frac{1}{6M_\pi},\nn
Z_1(q_1)&=&2 A (q_1) \left(2 M_\pi^2 + q_1^2 \right) + 2 M_\pi ,\nn
Z_3(q_1)&=&\frac{1}{2}A(q_1)\left(4M_\pi^2+q_1^2\right)+\frac{M_\pi}{2}.
\eeqa
In strong contrast with our results, we note that due to the appearance of
the function $W_3(q_1)$, the current operator of
Baroni et al.~{\em does not exist in the chiral limit}.
 Even relaxing
all matching and renormalizability constraints on the unitary phases and
requiring only the factorization constraints of Eq.~(\ref{standard1}), we
are unable to find a choice of unitary phases which would bring our results
in agreement with the ones of Baroni et al..\footnote{The constraints of
Eq.~(\ref{standard1}) have to be imposed since the results Baroni et
al.~are given in a factorizable form.} In the restrictive
kinematics with $\vec{k}=0$, the dependence of only one unitary phase,
namely $\alpha_1^{ax}$, survives for the
axial-vector currents. Subtracting our result from the one of Baroni et
al., we obtain for the difference 
\beqa
\vec{j}_{a}^{{\rm \; N4LO}}({\rm MPE}, \vec{q}_1) - \vec {
  A}_{{\rm 2N:} \, 2\pi}^{a\, (Q)}-\vec {A}_{{\rm 2N:} \,
  1\pi}^{a\, (Q)}&=& A(q_1)\bigg(\frac{\alpha_1^{\rm
      ax}g_A^5}{256\pi 
F_\pi^4q_1^2}\bigg[ 4q_1^2[\fet\tau_1\times\fet\tau_2]^a(-\vec{q}_1\vec{q}_1\cdot\vec{\sigma}_1
\times\vec{\sigma}_2-\vec{q}_1\times\vec{\sigma}_2\vec{q}_1\cdot\vec{\sigma}_1)\nn
&+&
\tau_1^a\Big(-2\vec{q}_1\left(8M_\pi^2-q_1^2\right)\vec{q}_1\cdot\vec{
\sigma}_2 +\left(8M_\pi^2+q_1^2\right)\left(2q_1^2\vec{\sigma}_2+\vec{q}_1\vec{q}_1\cdot\vec{
\sigma}_1\right) \nn
&-&\vec{\sigma}_1\left(24M_\pi^2q_1^2+11 q_1^4\right) \Big)\bigg]
-\frac{g_A^5\left(\vec{\sigma}_2
  \tau_1^a  q_1^4+2
\vec{q}_1\left(6M_\pi^2+q_1^2\right)\vec{q}_1\cdot\vec{\sigma}_2\tau_1^a\right)}{96\pi
F_\pi^4q_1^2}\bigg)\nn
&+&{\rm rational\, function \,of} \,\vec{q}_1 \; + \; 1\leftrightarrow 2\;. \label{diffBaroniAndOurs}
\eeqa 
As one can see from the above equation,  we have to set
$\alpha_1^{ax}=0$ in order to eliminate the term proportional to
$[\fet\tau_1\times\fet\tau_2]^a$, which in turn is consistent with the
renormalizability constraints. However, after setting
$\alpha_1^{ax}=0$, we still obtain a difference:
\beqa
\vec{j}_{a}^{{\rm \; N4LO}}({\rm MPE}, \vec{q}_1) - \vec {
  A}_{{\rm 2N:} \, 2\pi}^{a\, (Q)}-\vec {A}_{{\rm 2N:} \,
  1\pi}^{a\, (Q)}\bigg|_{\alpha_1^{ax}=0}&=& -\frac{g_A^5 A(q_1)\left(\vec{\sigma}_2
  \tau_1^a  q_1^4+2
\vec{q}_1\left(6M_\pi^2+q_1^2\right)\vec{q}_1\cdot\vec{\sigma}_2\tau_1^a\right)}{96\pi
F_\pi^4q_1^2} \nn
&+& {\rm rational\, function \,of} \,\vec{q}_1 \; + \; 1\leftrightarrow 2\;. 
\eeqa
Thus, no choice of unitary phases makes our results
agree with the ones of Baroni et al.. We
conclude that our current operator and that of Baroni et al.~are unitary
non-equivalent (within the set of unitary transformations employed in
our analysis).\footnote{We can, however, not exclude the possibility
  of existence of a different set of
  unitary transformations, which would relate the two expressions for
  the current operator.}  Since the loop function $A(q_1)$ affects the long-range behavior of
the current, we get a disagreement even for the long-range terms. 

Concerning the axial charge operator, our results for the two-pion
exchange and short-range contributions $A^{0, a}_{{\rm 2N:}\, 2\pi}$
and $A^{0, a }_{\rm 2N:\, cont}$ at order $Q$ agree with the ones of
Ref.~\cite{Baroni:2015uza}. 
For the one-pion exchange contributions, we find
the same expressions for the chiral logarithms, i.e. those terms in Eq.~(\ref{Charge1piQTo1})
which involve the loop function $L(q)$. This is  not
surprising given that they  
originate solely from the irreducible topologies (26) and (28) in
Fig.~\ref{fig:ope}  and thus can be
obtained by evaluating the corresponding Feynman diagrams. 
On the other hand, the contributions to the scalar functions
$h_{6,7,8}$ in Eq.~(\ref{Charge1piQTo1}), which are given by rational functions of momenta and the
pion mass, differ from the corresponding terms found by Baroni and collaborators. 

Finally, what concerns the three-nucleon axial current operator, whose
leading terms  emerge at order $Q$, Baroni et
al.~only consider
in Ref.~\cite{Baroni:2016xll} the contributions of diagrams (21) and
(25) in Fig.~\ref{fig:tpe_3N}.
To the best of our knowledge, our results in Eqs.~(\ref{Current3NPi}) and (\ref{Current3NCont})
represent the first complete derivation of the dominant contributions
to the three-nucleon axial current operator 
$\vec{A}^{a }_{\rm 3N:\, \pi} + \vec{A}^{a }_{\rm 3N:\, cont}$.

\section{Summary and conclusions}
\def\theequation{\arabic{section}.\arabic{equation}}
\label{sec:summary}

In this paper, we have analyzed in detail the nuclear axial-vector
charge and current operators as well as the pseudoscalar currents in
the framework of heavy-baryon chiral effective field theory. The main
results of our study can be summarized below. 
\begin{itemize}
\item
First, we have worked out the general form of the continuity equation
for the nuclear iso-triplet vector and axial-vector current operators
based on the effective chiral Lagrangian involving (first-order)
time derivatives of external sources. The resulting continuity
equations (\ref{continuityeqmomspacevector}) and (\ref{continuityeqmomspace}) differ from their commonly
assumed form by terms involving energy-transfer-dependent 
contributions to the charge and current operators. 
\item
We have worked out Poincar\'{e} invariance constraints on the
axial-vector charge and current operators which manifest themselves in
the on-shell relation (\ref{FourVectorCondMomentumSpaceOnShell}) between the effective Hamiltonian, boost and
current operators. We have extended a formal proof of Ref.~\cite{Kruger:1997aw}
that the generators of the  Poincar\'{e} group acting in the Fock
space of nucleons and mesons are simultaneously block diagonalized by
the Okubo unitary transformation to the case of general 
interactions between the particles. This makes the proof valid for the
effective operators derived in the framework of chiral EFT. We have also proposed an
efficient way of calculating the effective boost operator which acts on
the nucleonic part of the Fock space. Using this approach, we
were able to explicitly verify  Poincar\'{e} invariance constraints
for the derived current operators. 
\item
We have performed a \emph{complete} derivation of the nuclear
axial-vector charge and current operators to order $Q$ in the chiral expansion, i.e.~to fourth order relative to
the dominant one-body contribution, using the method
of unitary transformation. The resulting currents are, per
construction, consistent with the nuclear forces worked out in
Refs.~\cite{Epelbaum:2002vt,Epelbaum:2006eu,Epelbaum:2007us, 
Bernard:2007sp,Bernard:2011zr,Krebs:2012yv,Krebs:2013kha,Epelbaum:2014sea} within the same approach. To render the loop integrals
finite by the counterterms in the effective Lagrangian, we had to exploit the unitary
ambiguity of the current operator in a systematic way. To this aim, we
have considered a large class of unitary transformations on the nucleonic subspace
of the Fock space, which are compatible with the chiral order we are
working at and reduce to the identity operation when the external axial
sources are switched off. The renormalized expressions for
the current operators are found to feature a substantial degree of
unitary ambiguity. We have argued that it is advantageous to choose the 
unitary phases, undetermined by the renormalizability constraint, in such a
way that the pion-pole contributions to the axial current operator
match the corresponding irreducible pion production amplitudes
in the nuclear potentials. This particular choice of the unitary phases, which we refer to
as standard, does not only appear natural, but is
expected to simplify the regularization of the current operators in
the way consistent with regularization of the nuclear forces. 
After matching to the nuclear forces, we end up with unambiguous
expressions for the axial charge and current operators which are
summarized in Tables \ref{tab_sum_current} and \ref{tab_sum_charge}.  To the best of our knowledge,
the energy-transfer-dependent terms and relativistic corrections to
the exchange operators have never been studied before in the framework
of chiral EFT. Furthermore, for three-body operators, only
irreducible-like topologies have been considered in the past, which,
in fact, only constitute a small subset of the relevant diagrams.  
\item
Using the same approach, we have independently derived the iso-triplet 
pseudoscalar current operator to fourth order relative to the dominant
one-body contribution. After renormalization and matching to the
nuclear forces, the expressions for the pseudoscalar currents do not
show any unitary ambiguity. Our final results for the pseudoscalar
current operator are summarized in Table \ref{tab_sum_current_ps}. To the best of our
knowledge, the pseudoscalar nuclear current operators have never been
studied before in the framework of chiral EFT. 
\item
We have explicitly verified that the continuity equation (\ref{continuityeqmomspace}) is
valid for all derived contributions to the charge and current
operators. 
\item
We have compared our results for energy-transfer-independent static
contributions to the two-body axial charge and current operators with
the recent calculation by Baroni et al.~\cite{Baroni:2015uza,Baroni:2016xll}. While our expressions
for the static two-pion exchange and short-range contributions to the 2N
charge operator at order $Q$ agree with the ones of Ref.~\cite{Baroni:2015uza},
there are differences in terms of the one-pion range. For the axial
current operator, our static expressions differ strongly from the ones found
by Baroni et al. We have illustrated this by considering a particular
case of the threshold kinematics with $\vec k = 0$. Notice that we have not
succeeded to reproduce the expressions of Ref.~\cite{Baroni:2016xll}
even by relaxing our
constraints on the unitary phases. We further emphasize that in
contrast to our results, the current operator derived by Baroni et
al.~does not exist in the chiral limit.  
\end{itemize}

The results of our work provide a solid basis for theoretical
investigations of weak processes in few- and many-nucleon
systems. The derived expressions for the currents are consistent with
the chiral 2N potentials of
Refs.~\cite{Epelbaum:2004fk,Epelbaum:2014efa,Epelbaum:2014sza}\footnote{To maintain
  consistency at the level of relativistic corrections, one will have
  to refit the N$^3$LO and N$^4$LO potentials of Refs.~\cite{Epelbaum:2014efa,Epelbaum:2014sza} using
  the relativistic version of the Schr\"odinger equation (\ref{SE}) similar to
  what has been done in Ref.~\cite{Epelbaum:2004fk}. This work is in progress.} and 3N forces
given in Refs.~\cite{Bernard:2007sp,Bernard:2011zr,Krebs:2012yv,Krebs:2013kha}. It would be interesting to test the novel
current operators by calculating the triton $\beta$-decay and muon
capture on $^3$He, for which precise experimental data are
available. Work along these lines is in progress, see also
Ref.~\cite{Baroni:2016xll,Skibinski:2016dve} for related recent studies. Notice further 
that the tritium half-life would come out as a
parameter-free prediction up to N$^3$LO once the LEC $D$, which  
governs the short-range behavior of both the axial current and
three-nucleon force at N$^2$LO, is determined in the strong sector. 
It would also be interesting to analyze muon capture on the deuteron
which is currently being measured by the MuSun experiment at PSI
\cite{Andreev:2007wg}. We also expect our findings  to be
relevant for a better understanding of the quenching of $g_A$ in
nuclei, which is important for analyzing the ongoing and future
experiments on neutrinoless double beta decay \cite{Menendez:2011qq}.

\section*{Acknowledgments}

We would like to thank Vadim Baru, Arseniy Filin, Jacek Golak, Fred
Myhrer and Roman Skibi{\'n}ski for sharing their insights into the
discussed topics and useful comments on the manuscript. 
This work was supported by DFG (SFB/TR 110, ``Symmetries and the Emergence of Structure in
QCD'') 
and BMBF (contract No.~05P2015 - NUSTAR R\&D). Further, this
research was supported in part by the National Science Foundation
under Grant No.~NSF PHY11-25915. 
The work of UGM was supported in part by The Chinese Academy of Sciences 
(CAS) President's International Fellowship Initiative (PIFI) grant no. 2015VMA076.

\appendix

\section{Block diagonalization of the generators of the Poincar\'{e} group}
\def\theequation{\Alph{section}.\arabic{equation}}
\setcounter{equation}{0}
\label{BlockDiagonalizationProof}

In this Appendix we will sketch the proof that Poincar\'{e} algebra
gets simultaneously block diagonalized by the Okubo unitary
transformation. The main steps are already described
in~\cite{Kruger:1997aw}. However, in~\cite{Kruger:1997aw} only a
special case of Yukawa-like interactions has been considered. Here, we
extend the proof to the case of arbitrary interactions. We will not repeat
the proof of translational and rotational invariance of the operator
$A$, which parametrizes the Okubo UT. 
These statements are independent of the form of the interactions and, for this
reason, the proof can be taken from~\cite{Kruger:1997aw}. 

We begin with decomposing the boost and Hamiltonian operators into the
free and interacting parts
\beq
K^j=K_0^j + K_I^j, \quad H_s=H_0 + H_I.
\eeq
From Poincar\'{e} algebra,  one immediately obtains the relations 
\beq
\big[K_0^j,H_0\big]=i\,P^j, \quad\big[K_I^j,H_0] +
\big[K_0^j,H_I\big]=0,\quad \big[K_I^j, H_I\big]=0, \quad \big[K_0^j,
F(H_0)\big] = i\,P^j \frac{\partial }{\partial H_0}F(H_0),
\label{PoincareAlgebraFollowUpRelations}
\eeq
where $F$ is some (analytic) function of $H_0$ only
(see~\cite{Kruger:1997aw} for details). The difficult part
is to show the simultaneous block diagonalization of the boost and
Hamilton operators. 

Starting from the Okubo block-diagonalization condition
\beq
\lambda\big(H_I + \big[H_0, A\big] + \big[H_I, A\big] - A\, H_I\,
A\big)\eta=0, 
\label{HamiltonianBlockDiagonalizationRelationGeneral}
\eeq
we have to show that the block-diagonalization condition  is also
valid for the boost operator:
\beq
\lambda\big(K_I^j + \big[K_0^j, A\big] + \big[K_I^j, A\big] - A\,
K_I^j\, A\big)\eta=0.
\label{BoostBlockDiagonalizationRelationGeneral}
\eeq 
Our proof makes use of a perturbative expansion. For this reason, we rescale all
couplings $c_j$ in a given Hamiltonian by
\beq
c_j\to g\,c_j, 
\eeq
and write  the operator $A$ in terms of the expansion in powers of the universal coupling constant $g$ via
\beq
A=\sum_{n=1}^\infty g^n\,A_n.
\eeq
At leading order in $g$, we get
\beq
A_1\eta|\beta\rangle = \frac{1}{E_\beta - H_0}\lambda H_I \eta|\beta\rangle
\quad{\rm and}\quad
\langle \alpha |\lambda A_1 = \langle \alpha | \lambda H_I \eta \frac{1}{H_0 - E_\alpha},
\eeq
where we assume that the states $|\alpha\rangle$ and $|\beta\rangle$
are eigenstates of the free Hamiltonian,
\beq
H_0 |\alpha\rangle = E_\alpha |\alpha\rangle \quad{\rm and}\quad H_0 |\beta\rangle =
E_\beta |\beta\rangle.
\eeq
In the following, we will often make  use of the relation 
\beq
\langle\alpha|\lambda \Big(K_0^j \frac{1}{E_\beta - H_0} X - X
\frac{1}{H_0 - E_\alpha} K_0^j\Big)\eta|\beta\rangle =
\frac{1}{E_\beta - E_\alpha}\langle\alpha|\lambda\big[K_0^j, X\big]\eta|\beta\rangle,
\label{CommutatorWithBoostAndX}
\eeq
where $X$ is some translationally invariant operator. We used here the
last relation of Eq.~(\ref{PoincareAlgebraFollowUpRelations}) along with
the translational invariance of the operators $H_0$ and $X$.
With these relations we are ready to consider the commutator of the free boost
with $A$:
\beq
\langle\alpha|\lambda\big[K_0^j, A_1\big]\eta|\beta\rangle = 
\frac{1}{E_\beta - E_\alpha}\langle\alpha|\lambda\big[K_0^j,
H_I\big]\eta|\beta\rangle=
\frac{1}{E_\beta - E_\alpha}\langle\alpha|\lambda\big[H_0,K_I^j
\big]\eta|\beta\rangle = -\langle\alpha|\lambda K_I^j
\eta|\beta\rangle,\label{K0A1Commutator}
\eeq
which is the desired relation of
Eq.~(\ref{BoostBlockDiagonalizationRelationGeneral}) to order
$g$. So far, there was no difference to ~\cite{Kruger:1997aw}. The
difference starts to show up from order $g^2$. To verify Eq.~(\ref{BoostBlockDiagonalizationRelationGeneral}) at order
$g^2$, we start with
Eq.~(\ref{HamiltonianBlockDiagonalizationRelationGeneral}). At 
order $g^2$, we get
\beq
\lambda\big( \big[H_0, A_2\big] + \big[H_I, A_1\big]\big)\eta=0,\label{H0A2Commutator}
\eeq
which is equivalent to
\beq
A_2\eta|\beta\rangle = \frac{1}{E_\beta - H_0}\lambda\big[H_I,
A_1\big]\eta|\beta\rangle \quad{\rm and}\quad\langle\alpha|\lambda
A_2=\langle\alpha|\lambda
\big[H_I, A_1\big]\eta\frac{1}{H_0 - E_\alpha}.
\eeq
Starting from these relations we consider the commutator 
\beqa
\langle\alpha|\lambda\big[K_0^j,
A_2\big]\eta|\beta\rangle&=&\frac{1}{E_\beta -
  E_\alpha}\langle\alpha|\lambda\big[K_0^j,\big[H_I,A_1\big]\big]\eta|\beta\rangle\nn
&=&
\frac{1}{E_\beta -
  E_\alpha}\Big(\langle\alpha|\lambda\big[H_I,\big[K_0^j,A_1\big]\big]\eta|\beta\rangle+
\langle\alpha|\lambda\big[A_1,\big[H_I,K_0^j\big]\big]\eta|\beta\rangle\Big).
\eeqa
Using Eq.~(\ref{K0A1Commutator}) and the second relation in
Eq.~(\ref{PoincareAlgebraFollowUpRelations}), we get
\beqa
\langle\alpha|\lambda\big[K_0^j,
A_2\big]\eta|\beta\rangle&=&\frac{1}{E_\beta -
  E_\alpha}\Big(\langle\alpha|\lambda\big[\lambda K_I^j\eta,H_I\big]\big]\eta|\beta\rangle+
\langle\alpha|\lambda\big[A_1,\big[K_I^j,H_0\big]\big]\eta|\beta\rangle\Big)\nn
&=&
\frac{1}{E_\beta -
  E_\alpha}\Big(\langle\alpha|\lambda\big[\lambda K_I^j\eta,H_I\big]\big]\eta|\beta\rangle+
\langle\alpha|\lambda\big[K_I^j,\big[A_1,H_0\big]\big]\eta|\beta\rangle
+
\langle\alpha|\lambda\big[H_0,\big[K_I^j,A_1\big]\big]\eta|\beta\rangle\Big).\quad\quad
\eeqa
Using Eq.~(\ref{HamiltonianBlockDiagonalizationRelationGeneral}),
restricted to order $g$, we get
\beqa
\langle\alpha|\lambda\big[K_0^j,
A_2\big]\eta|\beta\rangle&=&
\frac{1}{E_\beta -
  E_\alpha}\Big(\langle\alpha|\lambda\big[ \lambda K_I^j\eta, H_I\big]\big]\eta|\beta\rangle+
\langle\alpha|\lambda\big[K_I^j,\lambda H_I\eta\big]\eta|\beta\rangle
+
\langle\alpha|\lambda\big[H_0,\big[K_I^j,A_1\big]\big]\eta|\beta\rangle\Big)\nn
&=&
\frac{1}{E_\beta -
  E_\alpha}\Big(\langle\alpha|\lambda\big[ K_I^j, H_I\big]\big]\eta|\beta\rangle
+
\langle\alpha|\lambda\big[H_0,\big[K_I^j,A_1\big]\big]\eta|\beta\rangle\Big).
\eeqa
Using now the third relation in Eq.~(\ref{PoincareAlgebraFollowUpRelations}) we finally get
\beqa
\langle\alpha|\lambda\big[K_0^j,
A_2\big]\eta|\beta\rangle&=&
-\langle\alpha|\lambda \big[K_I^j,A_1\big]\eta|\beta\rangle.\label{K0A2Commutator}
\eeqa
Note that in Ref.~\cite{Kruger:1997aw}, the authors assumed that $\eta H_I \eta=0$, which leads to a simplified version of
Eq.~(\ref{H0A2Commutator}) and Eq.~(\ref{K0A2Commutator}) (see Eqs.~(40) and (43)
of~\cite{Kruger:1997aw}). It is also important to note that for the derivation of Eq.~(\ref{K0A2Commutator}),
the relation $\big[K_I^j,H_I\big]=0$ is essential.

For orders higher than $g^3$,
Eq.~(\ref{HamiltonianBlockDiagonalizationRelationGeneral}) becomes 
\beq
\lambda\Big(\big[H_0,A_{n+1}\big] + \big[H_I,A_n\big] -
\sum_{\nu=1}^{n-1}A_\nu H_I A_{n-\nu}\Big)\eta = 0, \quad n\geq 2.\label{HBlockDiagonalizationPerturbativ}
\eeq
Starting from this relation we have to show the validity of the
corresponding equation for the boost operator:
\beq
\lambda\Big(\big[K_0^j,A_{n+1}\big] + \big[K_I^j,A_n\big] -
\sum_{\nu=1}^{n-1}A_\nu K_I^j A_{n-\nu}\Big)\eta = 0, \quad n\geq 2. \label{KBlockDiagonalizationPerturbativ}
\eeq
We will show this by induction. To start the proof by induction, we need to
consider order-$g^3$ terms which corresponds to the case of $n=2$. We rewrite
Eq.~(\ref{HBlockDiagonalizationPerturbativ}) for $n=2$ to
\beq
A_3\eta|\beta\rangle = \frac{1}{E_\beta - H_0} \lambda X_3\eta|\beta\rangle \quad{\rm and}\quad\langle\alpha|\lambda
A_3=\langle\alpha|\lambda
X_3\eta\frac{1}{H_0 - E_\alpha},
\eeq
with 
\beq
X_3 =\big[H_I, A_2\big] - A_1 H_I A_1.
\eeq
For the commutator of the free boost with $A_3$ we get
\beqa
\langle\alpha|\lambda\big[K_0^j,
A_3\big]\eta|\beta\rangle&=&
\frac{1}{E_\beta -
  E_\alpha}\langle\alpha|\lambda\big[ K_0^j, X_3\big]\eta|\beta\rangle,
\eeqa
where we used Eq.~(\ref{CommutatorWithBoostAndX}). For further
simplification, we make use of the general commutation-relation
identities valid for
arbitrary operators $A,B,C,D$
\beqa
\big[A,\big[B,C\big]\big]&=&\big[B,\big[A,C\big]\big]+\big[C,\big[B,A\big]\big],\label{ABCCommutatorRelation}\\
\big[A,B C D\big]&=&\big[A,B\big]C D + B\big[A,C\big]D + B C\big[A, D\big],\label{ABCDCommutatorRelation}
\eeqa
to get
\beq
\lambda\big[K_0^j,
X_3\big]\eta=\lambda\Big(\big[H_I,\big[K_0^j, A_2\big]\big] +
\big[A_2,\big[H_I,K_0^j\big]\big] -\big[K_0^j,A_1\big] H_I A_1 - A_1
\big[K_0^j,H_I\big] A_1 - A_1 H_I\big[K_0^j,A_1]\Big)\eta.
\eeq
Using Eq.~(\ref{K0A2Commutator}), Eq.~(\ref{K0A1Commutator})
and the second relation in Eq.~(\ref{PoincareAlgebraFollowUpRelations}),
we get
\beq
\lambda\big[K_0^j,
X_3\big]\eta=\lambda\Big(-\big[H_I,\lambda\big[K_I^j, A_1\big]\eta\big] +
\big[A_2,\big[K_I^j,H_0\big]\big] +K_I^j\eta H_I A_1 +A_1
\big[K_I^j,H_0\big] A_1 + A_1 H_I\lambda K_I^j\Big)\eta.\label{K0X3CommRel}
\eeq
Using Eq.~(\ref{ABCCommutatorRelation}) and Eq.~(\ref{H0A2Commutator}), we get
\beq
\big[A_2,\big[K_I^j,H_0\big] \big]= \big[K_I^j,\big[A_2,H_0\big]\big]
+ \big[H_0,\big[K_I^j,A_2\big]\big] =  \big[K_I^j,\lambda\big[H_I,A_1\big]\eta\big]
+ \big[H_0,\big[K_I^j,A_2\big]\big].\label{A2KIH0CommRel}
\eeq
Using Eq.~(\ref{ABCDCommutatorRelation}) and the order-$g$ restriction
of 
Eq.~(\ref{HamiltonianBlockDiagonalizationRelationGeneral}), we obtain
\beqa
A_1
\big[K_I^j,H_0\big] A_1 &=& \big[A_1 K_I^j A_1, H_0\big] + A_1
K_I^j\big[H_0,A_1\big] + \big[H_0,A_1\big]K_I^j A_1\nn
&=&
\big[A_1 K_I^j A_1, H_0\big] - A_1
K_I^j\lambda H_I\eta - \lambda H_I\eta K_I^j A_1.\label{A1KIH0CommRelA1}
\eeqa
Putting Eq.~(\ref{A2KIH0CommRel}) and Eq.~(\ref{A1KIH0CommRelA1}) in
Eq.~(\ref{K0X3CommRel}) and rearranging the summands we get
\beqa
\lambda\big[K_0^j,
X_3\big]\eta&=&\lambda\Big(\big[H_0,\big[K_I^j,A_2\big]\big] +
\big[A_1 K_I^j A_1, H_0\big] + \big[A_1,\big[H_I, K_I^j\big]\big]\Big),
\eeqa
where the last term vanishes due to $\big[H_I,K_I^j\big]=0$. With this
we finally get
\beqa
\langle\alpha|\lambda\big[K_0^j,
A_3\big]\eta|\beta\rangle&=&-\langle\alpha|\lambda\big[K_I^j,
A_2\big]\eta|\beta\rangle + \langle\alpha|\lambda A_1 K_I^j
A_1\eta|\beta\rangle.
\eeqa
This completes the proof for starting the induction. 

We now make an induction
assumption, that Eq.~(\ref{KBlockDiagonalizationPerturbativ}) is valid
for some arbitrary $n\ge 3$. In the induction step, we have to proof the same
relation for $n+1$. As before, we start with the application of $A_{n+2}$
to the initial and final states 
\beq
A_{n+2}\eta|\beta\rangle=
\frac{1}{E_\beta - H_0}\lambda
X_{n+2}\eta|\beta\rangle\quad{\rm and}
\quad
\langle\alpha|\lambda A_{n+2} =  \langle\alpha|\lambda
X_{n+2}\eta\frac{1}{H_0-E_\alpha},
\eeq
with
\beq
X_{n+2}=\big[H_I, A_{n+1}\big] - \sum_{\nu=1}^n A_\nu H_I A_{n+1-\nu}.
\eeq
For the matrix element of the free boost commutator with $A_{n+2}$, we
get, as previously, 
\beqa
\langle\alpha|\lambda\big[K_0^j,
A_{n+2}\big]\eta|\beta\rangle&=&\frac{1}{E_\beta - E_\alpha}
\langle\alpha|\lambda\big[K_0^j,
X_{n+2}\big]\eta|\beta\rangle.\label{K0AnPlus2CommutatorRelation}
\eeqa
We now use Eq.~(\ref{ABCCommutatorRelation}) 
to obtain
\beqa
\lambda\big[K_0^j,
X_{n+2}\big]\eta&=&\lambda\bigg(\big[H_I,\big[K_0^j,A_{n+1}\big]\big] +
\big[A_{n+1},\big[H_I,K_0^j\big]\big]-\sum_{\nu=1}^n\Big(\big[K_0^j,A_\nu\big]
H_I A_{n+1-\nu}+A_\nu\big[K_0^j,H_I\big]A_{n+1-\nu}\nn
 &+& A_\nu H_I\big[K_0^j,A_{n+1-\nu}\big]\Big)\bigg)\eta.\label{K0XnPlus2CommutatorStart}
\eeqa
Using the second relation in Eq.~(\ref{PoincareAlgebraFollowUpRelations}),
Eq.~(\ref{ABCCommutatorRelation}) and Eq.~(\ref{ABCDCommutatorRelation}), we get
\beqa
\big[A_{n+1},\big[H_I,K_0^j\big]\big] &=&
\big[K_I^j,\big[A_{n+1},H_0\big]\big] +
\big[H_0,\big[K_I^j,A_{n+1}\big]\big],\\
A_\nu\big[K_0^j,H_I\big]A_{n+1-\nu} &=&\big[H_0,A_\nu K_I^j
A_{n+1-\nu}\big] + \big[A_\nu, H_0\big] K_I^j A_{n+1-\nu} + A_\nu
K_I^j \big[A_{n+1-\nu}, H_0\big].
\eeqa
Putting these relations back into Eq.~(\ref{K0XnPlus2CommutatorStart}),
we obtain
\beq
\lambda\big[K_0^j,
X_{n+2}\big]\eta=\lambda\bigg(\big[H_0,\big[K_I^j,A_{n+1}\big]\big]-\sum_{\nu=1}^n \big[H_0,A_\nu K_I^j
A_{n+1-\nu}\big] + R_1  + R_2 \bigg)\eta,
\eeq
with
\beqa
R_1&=&\big[H_I,\big[K_0^j,A_{n+1}\big]\big] +
\big[K_I^j,\big[A_{n+1},H_0\big]\big], \nn
R_2&=&-\sum_{\nu=1}^n\Big(\big[K_0^j,A_\nu\big]
H_I A_{n+1-\nu}
+  \big[A_\nu, H_0\big] K_I^j A_{n+1-\nu} + A_{n+1-\nu}
K_I^j \big[A_{\nu}, H_0\big]
 + A_{n+1-\nu} H_I\big[K_0^j,A_{\nu}\big]\Big).\quad\quad
\eeqa
In the following, we need to show that 
\beq
\lambda (R_1 + R_2)\eta = 0.
\eeq 
The sum in $R_2$ has been rearranged in order to consider separately the cases
$\nu=1,2$ and the rest of the sum, where we can apply the induction
assumption. After application of the induction assumption, we obtain  
a double-sum contribution to $R_2$,  which, however, vanishes:
\beqa
&&\sum_{\nu=3}^n\sum_{\mu=1}^{\nu-2}\Big(A_\mu K_I^j
A_{\nu-1-\mu} H_I A_{n+1-\nu}-A_\mu H_I A_{\nu-1-\mu} K_I^j
A_{n+1-\nu}
-A_{n+1-\nu} K_I^j A_\mu H_I A_{\nu-1-\mu} \nn
&&+ A_{n+1-\nu} H_I 
A_\mu K_I^j
A_{\nu-1-\mu}\Big) = 0.
\eeqa
Due to the vanishing of the double sum, we can take the $\nu=2$ case into the
rest sum, such that we need to consider only the $\nu=1$ and $\nu\ge 2$
cases separately. With this we get 
\beqa
R_2&=&\lambda K_I^j\eta H_I A_n -\lambda H_I\eta K_I^j A_n - A_n K_I^j
\lambda H_I \eta + A_n H_I\lambda K_I^j \eta\nn 
&+& \sum_{\nu=1}^{n-1}\Big(\lambda
\big[K_I^j,A_\nu\big]\eta H_I A_{n-\nu} -\lambda \big[H_I, A_\nu\big]\eta
K_I^j A_{n-\nu}
-A_{n-\nu} K_I^j\lambda \big[H_I,A_\nu\big]\eta+A_{n-\nu} H_I\lambda\big[K_I^j, A_\nu\big]\eta\Big).\label{R2CommutatorFollowup}
\eeqa
After applying the induction
assumption to $R_1$, we get
\beqa
R_1
&=&-\big[H_I,\lambda\big[K_I^j,A_n\big]\eta\big] +
\big[K_I^j,\lambda\big[H_I,A_{n}\big]\eta\big] +
\sum_{\nu=1}^{n-1}\Big(\big[H_I, A_\nu K_I^j A_{n-\nu}\big] -\big[K_I^j, A_\nu H_I A_{n-\nu}\big]\Big). \label{R1CommutatorFollowup}
\eeqa
Adding Eqs.~(\ref{R2CommutatorFollowup}) and
(\ref{R1CommutatorFollowup}) together, we obtain
\beqa
\lambda (R_1+R_2)\eta
&=&\lambda\big[A_n,\big[H_I, K_I^j\big]\big] \eta+ \sum_{\nu=1}^{n-1}
A_\nu \big[H_I, K_I^j\big] A_{n-\nu}=0.
\eeqa
For the last equation, we again used the third relation in 
Eq.~(\ref{PoincareAlgebraFollowUpRelations}). With this, we get for
Eq.~(\ref{K0AnPlus2CommutatorRelation})
\beq
\langle\alpha|\lambda\big[K_0^j,A_{n+2}\big]\eta|\beta\rangle =
-\langle \alpha|\lambda\big[K_I^j, A_{n+1}\big]\eta|\beta\rangle +
\sum_{\nu=1}^n\langle\alpha|\lambda A_\nu K_I^j A_{n+1-\nu} \eta|\beta\rangle.
\eeq
This completes the proof of the induction step.

It is instructive to discuss the consequence of the simultaneous block
diagonalization in the symmetric energy momentum tensor
(Belinfante) notation. In this notation the boost operator has a
simple form at
$x_0=0$ 
\beq
K^j=-\int d^3 x \,x^j \Theta^{0 0}(x_0=0,\vec{x}).
\eeq
The Okubo block-diagonalization condition for the Hamiltonian,
expressed in terms of symmetric energy momentum tensor, is given by
\beq
\int d^3 x \langle\alpha|\lambda\left( \Theta^{0
    0}(x)-\big[A,\Theta^{0 0}(x)\big]-A\, \Theta^{0 0}(x) A\right)\eta
|\beta\rangle = 0.\label{BelinfanteBlockDiagonalizationCondition}
\eeq
Here we used the relation  
\beq
H_s=\int d^3x \,\Theta^{0 0}(x).
\eeq
Using translational  invariance of the operator $A$, we can rewrite Eq.~(\ref{BelinfanteBlockDiagonalizationCondition}) to
\beq
\int d^3 x \langle\alpha|\exp(-i\,\vec{P}\cdot\vec{x})\lambda\left( \Theta^{0
    0}(0)-\big[A,\Theta^{0 0}(0)\big]-A\, \Theta^{0 0}(0) A\right)\eta
\exp(i\,\vec{P}\cdot\vec{x})|\beta\rangle = 0,
\eeq
which means that
\beq
(2\pi)^3\delta(\vec{P}_\alpha-\vec{P}_\beta) \langle\alpha|\lambda\left( \Theta^{0
    0}(0)-\big[A,\Theta^{0 0}(0)\big]-A\, \Theta^{0 0}(0) A\right)\eta\label{MomentumSpaceDecouplingConstr}
|\beta\rangle = 0.
\eeq
We follow that for $\vec{P}_\alpha=\vec{P}_\beta$, we have
\beq
\langle\alpha|\lambda\left( \Theta^{0
    0}(0)-\big[A,\Theta^{0 0}(0)\big]-A\, \Theta^{0 0}(0) A\right)\eta\label{MomentumSpaceDecouplingWithoutDeltaFunction}
|\beta\rangle = 0.
\eeq
On the other hand, due to the simultaneous block diagonalization of the
boost, we get also
\beq
(2\pi)^3\left(\frac{\partial}{\partial P_\alpha^j}\delta(\vec{P}_\alpha-\vec{P}_\beta)\right) \langle\alpha|\lambda\left( \Theta^{0
    0}(0)-\big[A,\Theta^{0 0}(0)\big]-A\, \Theta^{0 0}(0) A\right)\eta
|\beta\rangle = 0.
\eeq
Integrating this equation over $\vec{P}_\alpha$, we get at
$\vec{P}_\alpha=\vec{P}_\beta$
\beq
\frac{\partial}{\partial P_\alpha^j} 
\langle\alpha|\lambda\left( \Theta^{0
    0}(0)-\big[A,\Theta^{0 0}(0)\big]-A\, \Theta^{0 0}(0)
  A\right)\eta
|\beta\rangle = 0.
\eeq
So, if we would denote
\beq
f(\vec{P}_\alpha)=\langle\alpha|\lambda\left( \Theta^{0
    0}(0)-\big[A,\Theta^{0 0}(0)\big]-A\, \Theta^{0 0}(0)
  A\right)\eta
|\beta\rangle
\eeq
the simultaneous block diagonalization means that the function
$f(\vec{P}_\alpha)$ is not only zero at $\vec{P}_\alpha=\vec{P}_\beta$
but has also an extremum at this momentum:
\beq
f(\vec{P}_\alpha=\vec{P}_\beta)=\frac{\partial}{\partial
  P_\alpha^j} f(\vec{P}_\alpha) \bigg|_{\vec{P}_\alpha=\vec{P}_\beta}=0,
\quad j=1,2,3.
\eeq

\section{Additional unitary transformations}
\label{sec:appen}
At leading order in the chiral expansion, there is one possibility for an additional unitary
transformation
\beq
\exp(S_{16}^{ax,{\rm LO}}-{\rm h.c.})=1+S_{16}^{ax,{\rm LO}}-{\rm
  h.c.} + \mathcal{O} \Big( \big( S_{16}^{ax,{\rm LO}} \big)^2 \Big),
\eeq
with
\beq
S_{16}^{ax,{\rm LO}}=\alpha_{16}^{ax,{\rm LO}} \eta  A_{0,1}^{(-1)}  
\lambda^1  \frac{1}{E_\pi^{2}}  H_{2,1}^{(1)} \eta .
\eeq
Here $H_{a,b}^{(\kappa)}$ denotes an interaction from the Hamiltonian with
$a$ nucleon and $b$ pion fields with the index $\kappa$ specified in
Eq.~(\ref{DefKappa}).  The kinetic energy operator in
this notation is
denoted by $H_{2, 0}^{(2)}$. Further, $A_{a,b}^{(\kappa)}$
denotes an interaction from the Hamiltonian with one axial vector
current, $a$ nucleon and $b$ pion fields.

At N$^2$LO, the following unitary transformations are possible:
\beq
\exp \bigg(\sum_{i=1}^{30}S_i^{ax} - {\rm h.c.} \bigg)=1+\sum_{i=1}^{30}S_i^{ax} -
{\rm h.c.}
+ \mathcal{O} \Big( \big( S_{i}^{ax} \big)^2 \Big) ,
\eeq
where
\beqa
S_{1}^{ax}&=&\alpha_{1}^{ax} \eta  A_{2,0}^{(0)}  \eta  H_{2,1}^{(1)}  
\lambda^1  \frac{1}{E_\pi^{3}}  H_{2,1}^{(1)}  \eta,\nn
S_{2}^{ax}&=&\alpha_{2}^{ax} \eta  H_{2,1}^{(1)}  \lambda^1  
\frac{1}{E_\pi^{2}}  A_{2,0}^{(0)}  \lambda^1  \frac{1}{E_\pi}  
H_{2,1}^{(1)}  \eta,\nn
S_{3}^{ax}&=&\alpha_{3}^{ax} \eta  H_{2,1}^{(1)}  \lambda^1  
\frac{1}{E_\pi^{2}}  A_{2,1}^{(1)}  \eta,\nn
S_{4}^{ax}&=&\alpha_{4}^{ax} \eta  H_{2,1}^{(1)}  \lambda^1  
\frac{1}{E_\pi^{2}}  A_{0,1}^{(-1)}  \lambda^2  \frac{1}{E_\pi}  
H_{2,1}^{(1)}  \lambda^1  \frac{1}{E_\pi}  H_{2,1}^{(1)}  \eta,\nn
S_{5}^{ax}&=&\alpha_{5}^{ax} \eta  H_{2,1}^{(1)}  \lambda^1  
\frac{1}{E_\pi}  A_{0,1}^{(-1)}  \lambda^2  \frac{1}{E_\pi^{2}}  
H_{2,1}^{(1)}  \lambda^1  \frac{1}{E_\pi}  H_{2,1}^{(1)}  \eta,\nn
S_{6}^{ax}&=&\alpha_{6}^{ax} \eta  H_{2,1}^{(1)}  \lambda^1  
\frac{1}{E_\pi}  A_{0,1}^{(-1)}  \lambda^2  \frac{1}{E_\pi}  
H_{2,1}^{(1)}  \lambda^1  \frac{1}{E_\pi^{2}}  H_{2,1}^{(1)}  \eta,\nn
S_{7}^{ax}&=&\alpha_{7}^{ax} \eta  A_{0,1}^{(-1)}  \lambda^1  
\frac{1}{E_\pi}  H_{2,1}^{(1)}  \lambda^2  \frac{1}{E_\pi}  
H_{2,1}^{(1)}  \lambda^1  \frac{1}{E_\pi^{2}}  H_{2,1}^{(1)}  \eta,\nn
S_{8}^{ax}&=&\alpha_{8}^{ax} \eta  A_{0,1}^{(-1)}  \lambda^1  
\frac{1}{E_\pi}  H_{2,1}^{(1)}  \lambda^2  \frac{1}{E_\pi^{2}}  
H_{2,1}^{(1)}  \lambda^1  \frac{1}{E_\pi}  H_{2,1}^{(1)}  \eta,\nn
S_{9}^{ax}&=&\alpha_{9}^{ax} \eta  A_{0,1}^{(-1)}  \lambda^1  
\frac{1}{E_\pi^{2}}  H_{2,1}^{(1)}  \lambda^2  \frac{1}{E_\pi}  
H_{2,1}^{(1)}  \lambda^1  \frac{1}{E_\pi}  H_{2,1}^{(1)}  \eta,\nn
S_{10}^{ax}&=&\alpha_{10}^{ax} \eta  H_{2,1}^{(1)}  \lambda^1  
\frac{1}{E_\pi^{3}}  A_{0,1}^{(-1)}  \eta  H_{2,1}^{(1)}  \lambda^1  
\frac{1}{E_\pi}  H_{2,1}^{(1)}  \eta,\nn
S_{11}^{ax}&=&\alpha_{11}^{ax} \eta  H_{2,1}^{(1)}  \lambda^1  
\frac{1}{E_\pi^{2}}  A_{0,1}^{(-1)}  \eta  H_{2,1}^{(1)}  \lambda^1  
\frac{1}{E_\pi^{2}}  H_{2,1}^{(1)}  \eta,\nn
S_{12}^{ax}&=&\alpha_{12}^{ax} \eta  H_{2,1}^{(1)}  \lambda^1  
\frac{1}{E_\pi}  A_{0,1}^{(-1)}  \eta  H_{2,1}^{(1)}  \lambda^1  
\frac{1}{E_\pi^{3}}  H_{2,1}^{(1)}  \eta,\nn
S_{13}^{ax}&=&\alpha_{13}^{ax} \eta  A_{0,1}^{(-1)}  \lambda^1  
\frac{1}{E_\pi^{3}}  H_{2,1}^{(1)}  \eta  H_{2,1}^{(1)}  \lambda^1  
\frac{1}{E_\pi}  H_{2,1}^{(1)}  \eta,\nn
S_{14}^{ax}&=&\alpha_{14}^{ax} \eta  A_{0,1}^{(-1)}  \lambda^1  
\frac{1}{E_\pi^{2}}  H_{2,1}^{(1)}  \eta  H_{2,1}^{(1)}  \lambda^1  
\frac{1}{E_\pi^{2}}  H_{2,1}^{(1)}  \eta,\nn
S_{15}^{ax}&=&\alpha_{15}^{ax} \eta  A_{0,1}^{(-1)}  \lambda^1  
\frac{1}{E_\pi}  H_{2,1}^{(1)}  \eta  H_{2,1}^{(1)}  \lambda^1  
\frac{1}{E_\pi^{3}}  H_{2,1}^{(1)}  \eta,\nn
S_{16}^{ax}&=&\alpha_{16}^{ax} \eta  A_{0,1}^{(-1)}  \lambda^1  
\frac{1}{E_\pi^{2}}  H_{2,1}^{(3)}  \eta,\nn
S_{17}^{ax}&=&\alpha_{17}^{ax} \eta  A_{0,1}^{(-1)}  \lambda^1  
\frac{1}{E_\pi^{2}}  H_{2,0}^{(2)}  \lambda^1  \frac{1}{E_\pi}  
H_{2,1}^{(1)}  \eta,\nn
S_{18}^{ax}&=&\alpha_{18}^{ax} \eta  A_{0,1}^{(-1)}  \lambda^1  
\frac{1}{E_\pi}  H_{2,0}^{(2)}  \lambda^1  \frac{1}{E_\pi^{2}}  
H_{2,1}^{(1)}  \eta,\nn
S_{19}^{ax}&=&\alpha_{19}^{ax} \eta  H_{2,0}^{(2)}  \eta  A_{0,1}^{(-1)}  
\lambda^1  \frac{1}{E_\pi^{3}}  H_{2,1}^{(1)}  \eta,\nn
S_{20}^{ax}&=&\alpha_{20}^{ax} \eta  A_{0,1}^{(-1)}  \lambda^1  
\frac{1}{E_\pi^{3}}  H_{2,1}^{(1)}  \eta  H_{2,0}^{(2)}  \eta,\nn
S_{21}^{ax}&=&\alpha_{21}^{ax} \eta  A_{0,1}^{(-1)}  \lambda^1  
\frac{1}{E_\pi^{2}}  H_{4,0}^{(2)}  \lambda^1  \frac{1}{E_\pi}  
H_{2,1}^{(1)}  \eta,\nn
S_{22}^{ax}&=&\alpha_{22}^{ax} \eta  A_{0,1}^{(-1)}  \lambda^1  
\frac{1}{E_\pi}  H_{4,0}^{(2)}  \lambda^1  \frac{1}{E_\pi^{2}}  
H_{2,1}^{(1)}  \eta,\nn
S_{23}^{ax}&=&\alpha_{23}^{ax} \eta  H_{4,0}^{(2)}  \eta  A_{0,1}^{(-1)}  
\lambda^1  \frac{1}{E_\pi^{3}}  H_{2,1}^{(1)}  \eta,\nn
S_{24}^{ax}&=&\alpha_{24}^{ax} \eta  A_{0,1}^{(-1)}  \lambda^1  
\frac{1}{E_\pi^{3}}  H_{2,1}^{(1)}  \eta  H_{4,0}^{(2)}  \eta,\nn
S_{25}^{ax}&=&\alpha_{25}^{ax} \eta  A_{0,1}^{(-1)}  \lambda^1  
\frac{1}{E_\pi^{2}}  H_{2,2}^{(2)}  \lambda^1  \frac{1}{E_\pi}  
H_{2,1}^{(1)}  \eta,\nn
S_{26}^{ax}&=&\alpha_{26}^{ax} \eta  A_{0,1}^{(-1)}  \lambda^1  
\frac{1}{E_\pi}  H_{2,2}^{(2)}  \lambda^1  \frac{1}{E_\pi^{2}}  
H_{2,1}^{(1)}  \eta,\nn
S_{27}^{ax}&=&\alpha_{27}^{ax} \eta  A_{0,1}^{(-1)}  \lambda^1  
\frac{1}{E_\pi^{2}}  H_{2,1}^{(1)}  \lambda^2  \frac{1}{E_\pi}  
H_{2,2}^{(2)}  \eta,\nn
S_{28}^{ax}&=&\alpha_{28}^{ax} \eta  A_{0,1}^{(-1)}  \lambda^1  
\frac{1}{E_\pi}  H_{2,1}^{(1)}  \lambda^2  \frac{1}{E_\pi^{2}}  
H_{2,2}^{(2)}  \eta,\nn
S_{29}^{ax}&=&\alpha_{29}^{ax} \eta  H_{2,1}^{(1)}  \lambda^1  
\frac{1}{E_\pi}  A_{0,1}^{(-1)}  \lambda^2  \frac{1}{E_\pi^{2}}  
H_{2,2}^{(2)}  \eta,\nn
S_{30}^{ax}&=&\alpha_{30}^{ax} \eta  H_{2,1}^{(1)}  \lambda^1  
\frac{1}{E_\pi^{2}}  A_{0,1}^{(-1)}  \lambda^2  \frac{1}{E_\pi}  
H_{2,2}^{(2)}  \eta. \label{add:unitary:transf}
\eeqa
The operator $S_{16}^{ax}$ can have three possible contributions: the
first one is due to the interaction with pion-nucleon vertex from ${\cal
  L}_{\pi N}^{(3)}$ in the static limit, the second one is due to the
interaction with the $1/m$ correction to the leading order
pion-nucleon vertex and the third one is due to the interaction with a
tadpole. For these three possibilities, we have different unitary phases.
For this reason, we introduce instead of a single parameter $\alpha_{16}^{ax}$
three different phases, which we denote by
\beq
\alpha_{16}^{ax,{\rm Static}}, \quad \alpha_{16}^{ax,1/m}, \quad
\alpha_{16}^{ax,{\rm Tadpole}}.
\eeq 

We also considered all possible unitary transformations of the form
\beq
\exp\bigg(i \Big( \sum_i \tilde \alpha_i \tilde S_i^{ax} + {\rm h.c.} \Big) \bigg),
\eeq
where the operators $\tilde S_i^{ax}$ are again given by a sequence of
vertices from the Hamiltonian and energy denominators. 
These transformations, however, lead  to non-factorizable effective
interactions. For this reason, we set all the phases of these
transformations to zero. At the considered order, one exception is
given by the unitary transformation considered in Appendix
\ref{timederivOfaxialSourceAppendix}, whose generator does not involve
pion fields.

Last but not least, one may consider unitary transformations, whose
generators involve a time derivative acting on the external axial
source. At fourth order in the chiral expansion, such UTs would
generate static contributions to the single-nucleon axial charge and
current operators $\propto k_0^2$. We found, however, that such UTs
lead to non-factorizable operators unless all corresponding phases are set to zero.

\section{Vertices with time-derivatives of the axial-vector source}
\label{timederivOfaxialSourceAppendix}
In this appendix we provide some details concerning our treatment of 
the $\bar d_{22}$-vertex which involves a  time derivative of the axial-vector source
\beq
-d_{22} N^\dagger S_\mu
\Big(\partial^2\fet{a}^{\mu}-\partial^\mu\partial_\nu\fet{a}^\nu\Big)\cdot
\fet{\tau} N=d_{22} N^\dagger S_\mu
(\partial^\mu\dot{\fet{a}}^0)\cdot{\fet \tau} N+ \ldots\,.
\eeq
In the formulation presented so far, all contributions depending on 
the energy transfer were generated solely from time derivatives of the
additional unitary transformations involving external axial-vector sources. 
Indeed, we are free to perform partial
time-integration in the action and eliminate this kind of terms in
favor of time-derivatives of the nucleon fields. In the next step, we
can eliminate time derivatives of the nucleon fields in the effective
Lagrangian by applying the equation of motion. After these two steps, 
there cannot be any $k_0$-dependent contributions to the
nuclear axial charge operators  stemming from the $\bar d_{22}$ vertex. 
However, we are still free to apply an additional unitary transformation of the form
\beq
\exp\Big(i\beta_1^{ax} \bar{d}_{22} N^\dagger S_\mu (\partial^\mu\fet{a}^{0})\cdot
\fet{\tau} N\Big)
\eeq
on the effective Hamiltonian. Here, $\beta_1^{ax}$ is an arbitrary
dimensionless parameter.  Due to its explicit time dependence through
the appearance of the external source $\fet{a}^{0}$, this
unitary transformation induces a $k_0$-dependent contribution to the
single-nucleon charge operator of the form
\beq
A^{0,  a \, (Q)}_{{\rm 1N}: {\rm \,static, \,UT}_1^\prime}=\beta_1^{ax} \bar{d}_{22} k_0  
\frac{\tau_i^a}{2}
\vec{k}\cdot \vec{\sigma}_i.
\eeq
In addition, when applied to the free Hamiltonian, it generates a
relativistic correction to the single-nucleon charge operator 
\beq
\label{relativisticbeta1}
A^{0,  a \, (Q)}_{{\rm 1N}: 1/m, \,{\rm UT}_2}=\beta_1^{ax}
\bar{d}_{22}  \frac{\tau_i^a}{2}
\vec{k}\cdot \vec{\sigma}_i \frac{p_i^2-p_i^{\prime\, 2}}{2m}.
\eeq 
Further, when acting on the one-pion
exchange and the leading contact interaction potential, it also
induces the contributions to the $2N$ charge operator of the kind 
\beqa
\label{beta1ope}
A^{0,  a \, (Q)}_{{\rm 2N:\,static,\, UT}_3}&=&\beta_1^{ax}
\frac{\bar{d}_{22}}{2}  \Big[\tau_1^a
\vec{k}\cdot \vec{\sigma}_1 , V_{{\rm 2N:}\, 1\pi}^{(Q^0)}\Big] +
1\leftrightarrow 2,\\
\label{beta1cont}
A^{0,  a \, (Q)}_{{\rm 2N:\,static,\, UT}_4}&=&\beta_1^{ax}
\frac{\bar{d}_{22}}{2}  \Big[\tau_1^a
\vec{k}\cdot \vec{\sigma}_1 , V_{{\rm 2N:\, cont}}^{(Q^0)}\Big] +
1\leftrightarrow 2\,.
\eeqa

On the other hand, it is instructive to trace back the contributions of the $\bar d_{22}$-vertex  to the nuclear
axial charge operator.  To eliminate the dependence on the time
derivative in the term 
\beq
\label{timederNd22}
-\bar{d}_{22} N^\dagger S_\mu (\partial^\mu \fet{a}^0)\cdot{\fet \tau} \dot{N} + h.c.,
\eeq
we make a redefinition of the nucleon field via
\beq
N\to N - i\, \bar{d}_{22} S_\mu (\partial^\mu \fet{a}^0)\cdot{\fet \tau} N\,.
\eeq
This eliminates the term in Eq.~(\ref{timederNd22}) from the
effective Lagrangian at the cost of introducing the new vertices 
\beqa
&& i \frac{\bar{d}_{22}}{F_\pi} g_A N^\dagger S_\mu (\partial^\mu
\fet{a}^0)\cdot{\fet \tau} S_\nu(\partial^\nu {\fet\pi})\cdot{\fet \tau} N
-i \,\bar{d}_{22} N^\dagger S_\mu (\partial^\mu \fet{a}^0)\cdot{\fet \tau} \frac{\nabla^2}{2m}N\nn
&&+ i\,C_S \bar{d}_{22} N^\dagger S_\mu (\partial^\mu \fet{a}^0)\cdot{\fet \tau} N
N^\dagger N-4 i\, C_T \bar{d}_{22} N^\dagger S_\mu (\partial^\mu
\fet{a}^0)\cdot{\fet \tau} S_\nu N N^\dagger S^\nu N + h.c.\,,
\label{tempo5}
\eeqa
which generate contributions to the 1N  and
2N charge operators. Adding the resulting 1N  terms to the one
specified  in Eq.~(\ref{relativisticbeta1}),
we finally get the corresponding relativistic correction:
\beq
(1+\beta_1^{ax})
\bar{d}_{22}  \frac{\tau_i^a}{2}
\vec{k}\cdot \vec{\sigma}_i \frac{p_i^2-p_i^{\prime\, 2}}{2m}.
\eeq
Further, adding the 2N contributions to the static one-pion
exchange and contact axial charge operator generated by the vertices in
Eq.~(\ref{tempo5}) to the terms in Eqs.~(\ref{beta1cont})
and (\ref{beta1ope}) we obtain
\beq
(1+\beta_1^{ax})
\frac{\bar{d}_{22}}{2}  \Big[\tau_1^a
\vec{k}\cdot \vec{\sigma}_1 , V_{{\rm 2N:}\, 1\pi}^{(Q^0)}+V_{{\rm 2N:\, cont}}^{(Q^0)}\Big] +
1\leftrightarrow 2\,.
\eeq 
We see that choosing the unitary phase $\beta_1^{ax}$ according to $\beta_1^{ax}=-1$ results in the
absence of contributions $\propto \bar d_{22}$ to the nuclear axial
charge operator apart from the 1N term 
\beq
- \bar{d}_{22} k_0  
\frac{\tau_i^a}{2}
\vec{k}\cdot \vec{\sigma}_i.
\eeq
Thus, for this particular choice, the expression for the single-nucleon charge
operator agrees with the on-shell result.  This is the choice of the
unitary phase $\beta_1^{ax}$ we adopt in our derivation, see Eq.~(\ref{SingleNChargeStaticUTPrimeQTo1}). It is
conceivable that the choice $\beta_1^{ax}=-1$ is compatible with 
renormalizability of the 1N current operator at the two-loop level
which, however, goes beyond the scope of our work.

\end{document}